\documentclass[12pt,oneside,english]{amsart}
\usepackage[T1]{fontenc}
\usepackage[latin9]{inputenc}
\usepackage{geometry}
\usepackage{amstext}
\usepackage{amsthm}
\usepackage{amssymb}
\usepackage{stmaryrd}
\usepackage{graphicx}
\usepackage{mathtools}
\usepackage{wasysym}
\usepackage{mathrsfs}
\usepackage{bbm}
\usepackage{esint}
\usepackage{color}
\usepackage{verbatim}
\usepackage{subfigure}
\usepackage{url}
\usepackage[percent]{overpic}

\makeatletter

\newcommand{\sertit}{Slit-strip Ising boundary conformal field theory}
\newcommand{\shorttit}{Slit-strip Ising BCFT}
\newcommand{\partnum}{2}
\newcommand{\parttit}{Scaling limits of fusion coefficients}

\numberwithin{equation}{section}
\numberwithin{figure}{section}
\theoremstyle{plain}
\newtheorem{thm}{\protect\theoremname}
\newtheorem*{thm*}{\protect\theoremname}
\newtheorem{prop}[thm]{\protect\propositionname}
\newtheorem*{prop*}{\protect\propositionname}
\newtheorem{lem}[thm]{\protect\lemmaname}

\newtheorem{cor}[thm]{\protect\corollaryname}
\newtheorem{rmk}[thm]{\protect\remarkname}

\numberwithin{thm}{section}

\definecolor{kallecol}{rgb}{.99,.1,.5}
\definecolor{davidcol}{rgb}{.5,.1,.99}
\definecolor{sketchcol}{rgb}{.4,.4,.8}
\definecolor{outlinecol}{rgb}{.8,.4,.3}

\newcommand{\term}[1]{{\bf #1}} % to be used for new terms when defined

\newcommand{\lftsym}{\mathrm{L}}
\newcommand{\rgtsym}{\mathrm{R}}
\newcommand{\topsym}{\mathrm{T}}
\newcommand{\botsym}{\mathrm{B}}
\newcommand{\slitsym}{\mathrm{S}}
\newcommand{\wildsym}{\star}%{\sharp}
\newcommand{\wildsymbis}{\star'}%{\sharp'}
\newcommand{\kT}{k_{\topsym}}
\newcommand{\kL}{k_{\lftsym}}
\newcommand{\kR}{k_{\rgtsym}}

\newcommand{\ii}{\mathbbm{i}}
\newcommand{\re}{\Re\mathfrak{e}}
\newcommand{\im}{\Im\mathfrak{m}}
\newcommand{\sgn}{\mathrm{sgn}}
\newcommand{\eps}{\varepsilon}
\newcommand{\bdry}{\partial}

\newcommand{\tccw}{\tau}

\newcommand{\cconj}[1]{\overline{#1}}
\newcommand{\ccshort}[1]{{#1}^*}
\newcommand{\cclong}[1]{{\left( #1 \right)}^*}
\newcommand{\innprod}[2]{\langle #1 , #2 \rangle }%{\lambda}
\newcommand{\Pfaff}{\mathrm{Pf}}

\newcommand{\finsubset}{\Subset}

\newcommand{\half}{\frac{1}{2}}
\newcommand{\mhalf}{\frac{-1}{2}}

\newcommand{\cstone}{\mathsf{1}}

\newcommand{\SymmGrp}{\mathfrak{S}}
\newcommand{\perm}{\varsigma}

\newcommand{\const}{\mathrm{const.}}

\newcommand{\aaa}{-\frac{1}{2}}
\newcommand{\bbb}{\frac{1}{2}}

\newcommand{\OO}{\mathcal{O}}

\newcommand{\Est}{\mathrm{E}}
\newcommand{\Wst}{\mathrm{W}}

\newcommand{\NE}{\mathrm{NE}}
\newcommand{\NW}{\mathrm{NW}}
\newcommand{\SW}{\mathrm{SW}}
\newcommand{\SE}{\mathrm{SE}}

\newcommand{\eighthroot}{e^{\ii \frac{\pi}{4}}}%{[e^{\ii \frac{\pi}{4}}]}%{\lambda}
\newcommand{\eighthrootbar}{e^{-\ii \frac{\pi}{4}}}%{[e^{-\ii \frac{\pi}{4}}]}%{\lambda}
\newcommand{\eighthrootthree}{e^{\ii \frac{3 \pi}{4}}}
\newcommand{\id}{\mathsf{id}}

\newcommand{\dmn}{\mathrm{dim}}
\newcommand{\spn}{\mathrm{span}}
\newcommand{\End}{\mathrm{End}}

\newcommand{\anticomm}[2]{[ #1 , #2 ]_+}

\newcommand{\ldual}{\langle}
\newcommand{\rdual}{\rangle}

\newcommand{\C}{\mathbb{C}} % complex plane
\newcommand{\R}{\mathbb{R}} % reals
\newcommand{\Z}{\mathbb{Z}} % integers
\newcommand{\Znn}{\Z_{\geq 0}} % non-negative integers
\newcommand{\Znp}{\Z_{\leq 0}} % non-positive integers
\newcommand{\Zpos}{\Z_{> 0}} % positive integers
\newcommand{\N}{\mathbb{N}} % natural numbers
\newcommand{\UHP}{\mathbb{H}} % upper half-plane
\newcommand{\bC}{\C} % alternative macro for complex plane
\newcommand{\bR}{\R} % alternative macro for reals
\newcommand{\bZ}{\Z} % alternative macro for integers
\newcommand{\bN}{\N} % alternative macro for natural numbers
\newcommand{\bH}{\UHP} % alternative macro for upper half-plane

\newcommand{\dcint}[1]{\int^{\sharp}_{{#1}}} % discrete contour integral symbol
\newcommand{\dcoint}[1]{\sqint^{\sharp}_{{#1}}} 
\newcommand{\dd}[1]{\ud^\sharp {#1}}
\newcommand{\ud}{\mathrm{d}} % upright d for ordinary differential
\newcommand{\contour}{\gamma}%{\boldsymbol{\gamma}}
\newcommand{\contourT}{\contour^{\topsym}}
\newcommand{\contourL}{\contour^{\lftsym}}
\newcommand{\contourR}{\contour^{\rgtsym}}
\newcommand{\contourX}{\contour^{\wildsym}}
\newcommand{\plaquettecontour}{\contour_{\Box}}
\newcommand{\vertedgecontour}{\contour_{\mid}}
\newcommand{\horcontour}{\eta}%{\boldsymbol{\eta}}

\newcommand{\chorcontour}[1]{\horcontour({#1})}

\newcommand{\width}{\ell}%{\delta}
\newcommand{\widthL}{\ell_\lftsym}%{\ell^\lftsym}%{\delta}
\newcommand{\widthR}{\ell_\rgtsym}%{\ell^\rgtsym}%{\delta}
\newcommand{\height}{h}%{\delta}
\newcommand{\heightT}{h_\topsym}%{\delta}
\newcommand{\heightB}{h_\botsym}%{\delta}

\newcommand{\cstrip}{{\mathbb{S}}}
\newcommand{\gengraph}{G}%{{\mathbb{G}}}
\newcommand{\dstrip}{{\cstrip^{(\width)}}}

\newcommand{\dstriptrTB}[2]{{\cstrip^{(\width;#1,#2)}}}
\newcommand{\slit}{\mathrm{slit}}
\newcommand{\cslitstrip}{\cstrip_{\slit}}
\newcommand{\dslitstrip}{{\cstrip_{\slit}^{(\width)}}}

\newcommand{\dslitstriptrTB}[2]{{\cstrip_{\slit}^{(\width;#1,#2)}}}
\newcommand{\dslitstripT}{{\cstrip_{\slit}^{\topsym;(\width)}}}
\newcommand{\dslitstripL}{{\cstrip_{\slit}^{\lftsym;(\width)}}}
\newcommand{\dslitstripR}{{\cstrip_{\slit}^{\rgtsym;(\width)}}}
\newcommand{\dslitstripX}{{\cstrip_{\slit}^{\wildsym;(\width)}}}
\newcommand{\cslitstripT}{{\cstrip_{\slit}^{\topsym}}}
\newcommand{\cslitstripL}{{\cstrip_{\slit}^{\lftsym}}}
\newcommand{\cslitstripR}{{\cstrip_{\slit}^{\rgtsym}}}

\newcommand{\dE}{\mathrm{E}}
\newcommand{\dV}{\mathrm{V}}
\newcommand{\edgeof}[2]{{\{ #1 , #2 \}}}

\newcommand{\Cliff}{\mathrm{Cliff}}%{\mathfrak{Cliff}}
\newcommand{\CliffGen}{\mathrm{CliffGen}}%{\mathfrak{CliffGen}}
\newcommand{\Creation}{\mathrm{Cre}}%{\mathfrak{Cre}}
\newcommand{\Annihilation}{\mathrm{Ann}}%{\mathfrak{Ann}}
\newcommand{\spinoper}{\mathsf{S}}
\newcommand{\spinoperS}{\hat{\mathsf{S}}}
\newcommand{\ferhol}{\psi}%{\boldsymbol{\psi}}
\newcommand{\ferbar}{\psi^*}%{\boldsymbol{\psi}^*}
\newcommand{\ferholS}{\hat{\psi}}%{\hat{\boldsymbol{\psi}}}
\newcommand{\ferbarS}{\hat{\psi}^*}%{\hat{\boldsymbol{\psi}}{}^*}
\newcommand{\poshalfint}{\mathcal{K}}
\newcommand{\dposhalfint}{\mathcal{K}^{(\width)}}
\newcommand{\dposhalfintW}[1]{\mathcal{K}^{(#1)}}
\newcommand{\dposhalfintL}{\dposhalfintW{\widthL}}
\newcommand{\dposhalfintR}{\dposhalfintW{\widthR}}

\newcommand{\dfermode}[1]{\mathsf{f}^{(\width)}_{#1}}
\newcommand{\dfermodeL}[1]{\mathsf{f}^{(\width)}_{\lftsym;#1}}
\newcommand{\dfermodeR}[1]{\mathsf{f}^{(\width)}_{\rgtsym;#1}}
\newcommand{\dpoleopT}[1]{\mathsf{p}^{(\width)}_{\topsym;#1}}
\newcommand{\dpoleopL}[1]{\mathsf{p}^{(\width)}_{\lftsym;#1}}
\newcommand{\dpoleopR}[1]{\mathsf{p}^{(\width)}_{\rgtsym;#1}}

\newcommand{\cffunhol}{\mathfrak{m}}
\newcommand{\cffunbar}{\mathfrak{m}^*}
\newcommand{\ccffun}{e}%{\mathfrak{e}}
\newcommand{\ccffunL}{\ccffun^{\lftsym}}
\newcommand{\ccffunR}{\ccffun^{\rgtsym}}
\newcommand{\ccffunT}{\ccffun^{\topsym}}
\newcommand{\ccffunX}{\ccffun^{{\wildsym}}}
\newcommand{\ccffunXbis}{\ccffun^{{\wildsymbis}}}

\newcommand{\ccfFun}{E}%{\mathfrak{E}}

\newcommand{\crit}{\mathrm{c}}
\newcommand{\invtemp}{{{\beta}}}
\newcommand{\invtempcrit}{\invtemp^{\bZ^2}_{\crit}}
\newcommand{\Hamiltonian}{\mathcal{H}}
\newcommand{\PartF}{\mathcal{Z}}
\newcommand{\PartFren}{\mathfrak{z}}
\newcommand{\spinconf}{\sigma}%{\underline{\sigma}}

\newcommand{\EX}{\mathsf{E}}%{\mathbb{E}}
\newcommand{\PR}{\mathsf{P}}%{\mathbb{P}}
\newcommand{\PRspace}{\Omega}%{\boldsymbol{\Omega}}

\newcommand{\bcssMPP}{{\scriptscriptstyle -++}}

\newcommand{\PRspPlus}[1]{\PRspace^{\plussym}_{#1}}
\newcommand{\PRspMinus}[1]{\PRspace^{\minussym}_{#1}}
\newcommand{\PRspPM}[2]{\PRspace^{\plussym / \minussym}_{{#1};{#2}}}

\newcommand{\PRspLocmono}[1]{\PRspace^{\locmonosym}_{#1}}

\newcommand{\set}[1]{\left\{ #1 \right\}}

\newcommand{\lft}{a}
\newcommand{\rgt}{b}
\newcommand{\mdpt}{0}
\newcommand{\lftd}{\lft'}
\newcommand{\rgtd}{\rgt'}
\newcommand{\crosssecsym}{\mathcal{I}}
\newcommand{\crosssec}{\crosssecsym}
\newcommand{\crosssecL}{\crosssecsym_{\lftsym}}
\newcommand{\crosssecR}{\crosssecsym_{\rgtsym}}

\newcommand{\crosssecdual}{\crosssec^*}
\newcommand{\crosssecLdual}{\crosssec_{\lftsym}^*}
\newcommand{\crosssecRdual}{\crosssec_{\rgtsym}^*}
\newcommand{\dinterval}[2]{\llbracket #1 , #2 \rrbracket}
\newcommand{\dintervaldual}[2]{\llbracket #1 , #2 \rrbracket^*}
\newcommand{\ccrosssec}{I}
\newcommand{\ccrosssecL}{\ccrosssec_{\lftsym}}
\newcommand{\ccrosssecR}{\ccrosssec_{\rgtsym}}

\newcommand{\plussym}{\boldsymbol{+}}
\newcommand{\minussym}{\boldsymbol{-}}
\newcommand{\locmonosym}{\mathrm{mono}}%{\boldsymbol{mono}}

\newcommand{\statesp}{V}%{\mathcal{S}}
\newcommand{\statespdbl}{\widetilde{\statesp}}
\newcommand{\statespXX}[1]{\widetilde{\statesp}_{{\scriptscriptstyle {#1}}}}
\newcommand{\statespXXx}[1]{\widetilde{\statesp}_{{{#1}}}}
\newcommand{\statespPP}{\widetilde{\statesp}_{{\scriptscriptstyle ++}}}
\newcommand{\statespMP}{\widetilde{\statesp}_{{\scriptscriptstyle -+}}}
\newcommand{\statespPM}{\widetilde{\statesp}_{{\scriptscriptstyle +-}}}
\newcommand{\statespMM}{\widetilde{\statesp}_{{\scriptscriptstyle --}}}

\newcommand{\statespS}[1]{\widetilde{\statesp}_{{\scriptscriptstyle {#1}}}}

\newcommand{\spinrow}{\rho}%{\boldsymbol{\rho}}
\newcommand{\spinrowfolded}{\spinrow'}
\newcommand{\spinrowalt}{\tau}%{\boldsymbol{\tau}}

\newcommand{\fold}[1]{\mathtt{fold}_{<#1}}
\newcommand{\basisvec}[1]{{u}_{#1}}%{\boldsymbol{u}_{#1}}
\newcommand{\stateIn}{{v}_{\mathrm{in}}}%{\boldsymbol{v}_{\mathrm{in}}}
\newcommand{\stateOut}{{v}_{\mathrm{out}}}%{\boldsymbol{v}_{\mathrm{out}}}
\newcommand{\Tmat}{\mathsf{T}}
\newcommand{\TmatVer}{\Tmat_{\mathrm{ver}}}
\newcommand{\TmatVerLR}{\Tmat^{\slit}_{\mathrm{ver}}}

\newcommand{\TmatHorSqrt}{\Tmat_{\mathrm{hor}}^{1/2}}

\newcommand{\TmatBasic}{\mathsf{T}^{[\, \width\,]}}
\newcommand{\TmatLR}{\mathsf{T}^{[\widthL|\widthR]}}%{\mathsf{T}^{\slit}}
\newcommand{\Teigval}[1]{{\mu}_{#1}^{(\width)}}%{\boldsymbol{\mu}_{#1}^{(\width)}}
\newcommand{\TeigvalLR}[2]{{\mu}_{#1,#2}^{(\widthL,\widthR)}}%{\boldsymbol{\mu}_{#1,#2}^{(\widthL,\widthR)}}
\newcommand{\genvec}{v}%{\boldsymbol{v}}

\newcommand{\rvvec}[1]{\genvec^{(#1)}}

\newcommand{\vac}{\genvec_{\emptyset}}

\newcommand{\PFvecEven}{\genvec^{\mathrm{PF}}_{{ \scriptscriptstyle ++}}}

\newcommand{\PFvecXXx}[1]{\genvec^{\mathrm{PF}}_{{ {#1}}}}
\newcommand{\PFvecS}[1]{\genvec^{\mathrm{PF}}_{{ \scriptscriptstyle {#1}}}}
\newcommand{\PFvecEvenSlit}{%
    \genvec^{\mathrm{PF;\slit}}_{{ \scriptscriptstyle +++ }}}

\newcommand{\PFevOdd}{%
    {\mu}^{\mathrm{PF}}_{{ \scriptscriptstyle -+}}}
\newcommand{\PFevEven}{%
    {\mu}^{\mathrm{PF}}_{{ \scriptscriptstyle ++}}}
\newcommand{\PFevXX}[1]{%
    {\mu}^{\mathrm{PF}}_{{ \scriptscriptstyle {#1}}}}
\newcommand{\PFevXXx}[1]{%
    {\mu}^{\mathrm{PF}}_{{ {#1}}}}
\newcommand{\PFevS}[1]{%
    {\mu}^{\mathrm{PF}}_{{ \scriptscriptstyle {#1}}}}
\newcommand{\PFevEvenSlit}{%
    {\mu}^{\mathrm{PF;\slit}}_{{\scriptscriptstyle +++}}}

\newcommand{\geneigval}{\mu}%{\boldsymbol{\mu}}
\newcommand{\vaceigval}{\geneigval^{(\width)}_{\emptyset}}
\newcommand{\state}[1]{\genvec_{#1}}

\newcommand{\vacLR}{\genvec^{\slit}_{\emptyset;\emptyset}}
\newcommand{\vaceigvalLR}%
	{\geneigval^{(\widthL,\widthR)}_{\emptyset,\emptyset}}
\newcommand{\stateLR}[2]{\genvec^{\slit}_{#1;#2}}
\newcommand{\parts}{{\alpha}}%{\boldsymbol{\alpha}}
\newcommand{\partsL}{\parts_{\lftsym}}
\newcommand{\partsR}{\parts_{\rgtsym}}
\newcommand{\pSign}[2]{{\epsilon}_{#1}(#2)}%{\boldsymbol{\epsilon}_{#1}(#2)}
\newcommand{\pLen}[1]{{\# #1}}%{\boldsymbol{s}(#1)}

\newcommand{\fusionLimit}[3]{{\Phi}_{#1 ; \, #2, #3}}
\newcommand{\fusionLimitTuple}[3]{{\Psi}_{#1 ; \, #2, #3}}
\newcommand{\fusionIsing}[3]{{\Phi}_{#1 ; \, #2, #3}^{(\width)}}
\newcommand{\fusionIsingW}[4]{{\Phi}_{#2 ; \, #3, #4}^{(#1)}}
\newcommand{\recIsingIC}{(${\mathrm{REC0}}^{(\width)}$)}
\newcommand{\recIsingT}{(${\mathrm{REC}}^{(\width)}_\topsym$)}
\newcommand{\recIsingL}{(${\mathrm{REC}}^{(\width)}_\lftsym$)}
\newcommand{\recIsingR}{(${\mathrm{REC}}^{(\width)}_\rgtsym$)}
\newcommand{\recLimitIC}{($\mathrm{REC0}$)}
\newcommand{\recLimitT}{($\mathrm{REC}_\topsym$)}
\newcommand{\recLimitL}{($\mathrm{REC}_\lftsym$)}
\newcommand{\recLimitR}{($\mathrm{REC}_\rgtsym$)}

\newcommand{\functionsp}{\mathscr{F}}

\newcommand{\cfunctionsp}{\mathscr{L}^2}
\newcommand{\cfunctionspL}{\cfunctionsp_{\lftsym}}
\newcommand{\cfunctionspR}{\cfunctionsp_{\rgtsym}}
\newcommand{\cfspTzero}{\cfunctionsp_{\topsym;\mathrm{zero}}}
\newcommand{\cfspTpole}{\cfunctionsp_{\topsym;\mathrm{pole}}}
\newcommand{\cfspLzero}{\cfunctionsp_{\lftsym;\mathrm{zero}}}
\newcommand{\cfspLpole}{\cfunctionsp_{\lftsym;\mathrm{pole}}}
\newcommand{\cfspRzero}{\cfunctionsp_{\rgtsym;\mathrm{zero}}}
\newcommand{\cfspRpole}{\cfunctionsp_{\rgtsym;\mathrm{pole}}}
\newcommand{\cprTzero}{\Pi_{\topsym;\mathrm{zero}}}
\newcommand{\cprTpole}{\Pi_{\topsym;\mathrm{pole}}}

\newcommand{\cprLpole}{\Pi_{\lftsym;\mathrm{pole}}}

\newcommand{\cprRpole}{\Pi_{\rgtsym;\mathrm{pole}}}
\newcommand{\dprTzero}{\Pi_{\topsym;\mathrm{zero}}^{(\width)}}
\newcommand{\dprTpole}{\Pi_{\topsym;\mathrm{pole}}^{(\width)}}
\newcommand{\dprLzero}{\Pi_{\lftsym;\mathrm{zero}}^{(\width)}}
\newcommand{\dprLpole}{\Pi_{\lftsym;\mathrm{pole}}^{(\width)}}
\newcommand{\dprRzero}{\Pi_{\rgtsym;\mathrm{zero}}^{(\width)}}
\newcommand{\dprRpole}{\Pi_{\rgtsym;\mathrm{pole}}^{(\width)}}
\newcommand{\deigvalsym}{\Lambda}

\newcommand{\eigval}[1]{\deigvalsym_{#1}^{(\width)}}
\newcommand{\eigvalW}[2]{\deigvalsym_{#2}^{(#1)}}
\newcommand{\eigvalL}[1]{\deigvalsym_{#1}^{(\width_\lftsym)}}
\newcommand{\eigvalR}[1]{\deigvalsym_{#1}^{(\width_\rgtsym)}}
\newcommand{\eigf}[1]{\mathfrak{f}_{#1}}
\newcommand{\eigfL}[1]{\mathfrak{f}_{\lftsym;#1}}
\newcommand{\eigfR}[1]{\mathfrak{f}_{\rgtsym;#1}}
\newcommand{\eigfT}[1]{\mathfrak{f}_{\topsym;#1}}
\newcommand{\eigfX}[1]{\mathfrak{f}_{{\wildsym};#1}}
\newcommand{\eigfXbis}[1]{\mathfrak{f}_{{\wildsymbis};#1}}
\newcommand{\eigF}[1]{\mathfrak{F}_{#1}}
\newcommand{\eigFL}[1]{\mathfrak{F}_{\lftsym;#1}}
\newcommand{\eigFR}[1]{\mathfrak{F}_{\rgtsym;#1}}
\newcommand{\poleL}[1]{\mathfrak{p}_{\lftsym;#1}}
\newcommand{\poleR}[1]{\mathfrak{p}_{\rgtsym;#1}}
\newcommand{\poleT}[1]{\mathfrak{p}_{\topsym;#1}}
\newcommand{\poleX}[1]{\mathfrak{p}_{{\wildsym};#1}}
\newcommand{\poleXbis}[1]{\mathfrak{p}_{{\wildsymbis};#1}}
\newcommand{\PoleL}[1]{\mathfrak{P}_{\lftsym;#1}}
\newcommand{\PoleR}[1]{\mathfrak{P}_{\rgtsym;#1}}
\newcommand{\PoleT}[1]{\mathfrak{P}_{\topsym;#1}}
\newcommand{\cpoleL}[1]{{p}^{\lftsym}_{#1}}%{\mathfrak{q}^{\lftsym}_{#1}}
\newcommand{\cpoleR}[1]{{p}^{\rgtsym}_{#1}}%{\mathfrak{q}^{\rgtsym}_{#1}}
\newcommand{\cpoleT}[1]{{p}^{\topsym}_{#1}}%{\mathfrak{q}^{\topsym}_{#1}}
\newcommand{\cpoleX}[1]{{p}^{{\wildsym}}_{#1}}%{\mathfrak{q}^{{\wildsym}}_{#1}}
\newcommand{\cpoleXbis}[1]{{p}^{{\wildsymbis}}_{#1}}%{\mathfrak{q}^{{\wildsymbis}}_{#1}}
\newcommand{\cPoleL}[1]{{P}^{\lftsym}_{#1}}%{\mathfrak{Q}^{\lftsym}_{#1}}
\newcommand{\cPoleR}[1]{{P}^{\rgtsym}_{#1}}%{\mathfrak{Q}^{\rgtsym}_{#1}}
\newcommand{\cPoleT}[1]{{P}^{\topsym}_{#1}}%{\mathfrak{Q}^{\topsym}_{#1}}

\newcommand{\dfunctionsp}{\functionsp^{(\width)}}
\newcommand{\dfunctionspW}[1]{\functionsp^{(#1)}}
\newcommand{\dfunctionspL}{\functionsp^{(\width)}_{\lftsym}}
\newcommand{\dfunctionspR}{\functionsp^{(\width)}_{\rgtsym}}
\newcommand{\dfspTzero}{\functionsp_{\topsym;\mathrm{zero}}^{(\width)}}
\newcommand{\dfspTpole}{\functionsp_{\topsym;\mathrm{pole}}^{(\width)}}
\newcommand{\dfspLzero}{\functionsp_{\lftsym;\mathrm{zero}}^{(\width)}}
\newcommand{\dfspLpole}{\functionsp_{\lftsym;\mathrm{pole}}^{(\width)}}
\newcommand{\dfspRzero}{\functionsp_{\rgtsym;\mathrm{zero}}^{(\width)}}
\newcommand{\dfspRpole}{\functionsp_{\rgtsym;\mathrm{pole}}^{(\width)}}

\newcommand{\refl}{\mathsf{R}}

\newcommand{\voavar}{\boldsymbol{\xi}}

\newcommand{\confmap}{\varphi}
\newcommand{\mapSS}{\confmap}
\newcommand{\intkerSSsym}{\mathbf{k}}
\newcommand{\intkerSS}[1]{\intkerSSsym_{#1}}

\newcommand{\gencfFun}{\mathfrak{K}}

\newcommand{\matkerSSsym}{\mathbf{K}}
\newcommand{\matkerSS}[1]{\matkerSSsym_{#1}}

\newcommand{\tple}[1]{\underline{#1}}

\usepackage{babel}

  \providecommand{\corollaryname}{Corollary}
  \providecommand{\definitionname}{Definition}
  \providecommand{\lemmaname}{Lemma}
  \providecommand{\propositionname}{Proposition}
  \providecommand{\remarkname}{Remark}
\providecommand{\theoremname}{Theorem}

\usepackage{babel}

  \providecommand{\corollaryname}{Corollary}
  \providecommand{\definitionname}{Definition}
  \providecommand{\lemmaname}{Lemma}
  \providecommand{\propositionname}{Proposition}
  \providecommand{\remarkname}{Remark}
\providecommand{\theoremname}{Theorem}

\usepackage{babel}
\providecommand{\corollaryname}{Corollary}
  \providecommand{\definitionname}{Definition}
  \providecommand{\lemmaname}{Lemma}
  \providecommand{\propositionname}{Proposition}
  \providecommand{\remarkname}{Remark}
\providecommand{\theoremname}{Theorem}

\makeatother

\usepackage{babel}
\providecommand{\corollaryname}{Corollary}
\providecommand{\definitionname}{Definition}
\providecommand{\lemmaname}{Lemma}
\providecommand{\propositionname}{Proposition}
\providecommand{\remarkname}{Remark}
\providecommand{\theoremname}{Theorem}

\begin{document}

\title[\shorttit{} \partnum{}]%
{{\large\scshape\bfseries \sertit{} \partnum{}: \\
\parttit{}}}

\author[Ameen \& Kyt\"ol\"a \& Park
]
{Taha Ameen, Kalle Kyt\"ol\"a, and S.C. Park
}

\address{Department of Electrical and Computer Engineering, University of Illinois at Urbana-Champaign, Urbana, IL}

\email{tahaa3@illinois.edu}

\address{Department of Mathematics and Systems Analysis, Aalto University,
P.O. Box 11100, FI-00076 Aalto, Finland}

\email{kalle.kytola@aalto.fi}

\address{School of Mathematics, Korea Institute of Advanced Study, 85 Hoegi-ro, 
Dongdaemun-gu, Seoul 02455, Republic of Korea}

\email{scpark@kias.re.kr}

\begin{abstract}
This is the second in a series of three articles about
recovering the full algebraic structure of 
a boundary conformal field theory (CFT) from
the scaling limit of the critical Ising model in slit-strip 
geometry. Here we study the fusion coefficients of
the Ising model in the lattice slit-strip, with 
locally monochromatic boundary conditions. The fusion 
coefficients are certain renormalized limits of
boundary correlation functions at the three extremities of the
truncated lattice slit-strips, in a basis of random variables
whose correlation functions have an essentially exponential 
dependence on the truncation heights.
The key technique is to associate operator valued discrete
$1$-forms to certain discrete holomorphic functions.
This provides a direct analogy with currents in boundary conformal
field theory. For two specific applications of this technique,
we use distinguished discrete holomorphic functions 
from the first article of the series.
First, we rederive the known diagonalization of the
Ising transfer matrix in a form that parallels boundary conformal
field theory. Second, we characterize the Ising model fusion coefficients 
by a recursion written purely in terms of inner products of
the distinguished discrete holomorphic functions.
The convergence result for the discrete holomorphic functions
proven in the first part can then be used to derive the convergence 
of the fusion coefficients in the scaling limit.
In the third article of the series, it will be shown that
up to a transformation that accounts for our chosen slit-strip geometry,
the scaling limits of the fusion coefficients become the
structure constants of the vertex operator algebra of
a fermionic conformal field theory. 
\end{abstract}

\maketitle

\tableofcontents

\newpage

\section{Introduction}%
\label{sec: intro}

% *******************************************************************
% **   \section{Introduction}                                      **
% *******************************************************************

\subsection{Critical Ising model on the slit-strip}

This article is the second in a three part series about
the boundary conformal field theory describing the
critical Ising model with locally monochromatic boundary conditions
in its scaling limit. It is this second part which concerns 
the Ising model itself~--- the first part 
was primarily about necessary discrete holomorphic functions
and their limits, and the last part will be primarily
about the algebraic formulation of the conformal field theory.

\begin{figure}[tb]
\centering
\subfigure[A sample from Ising model on the square grid strip~$\dstrip$.] 
{
    \includegraphics[width=.35\textwidth]{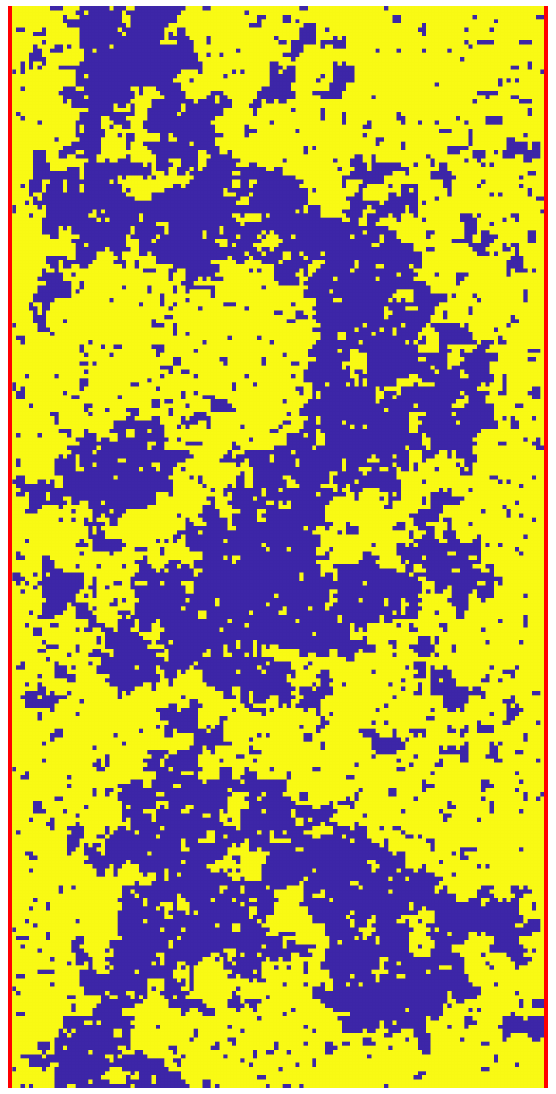}
	\label{sfig: Ising in strip}
}
\hspace{2.5cm}
\subfigure[A sample from Ising model on the square grid 
slit strip~$\dslitstrip$.] 
{
    \includegraphics[width=.35\textwidth]{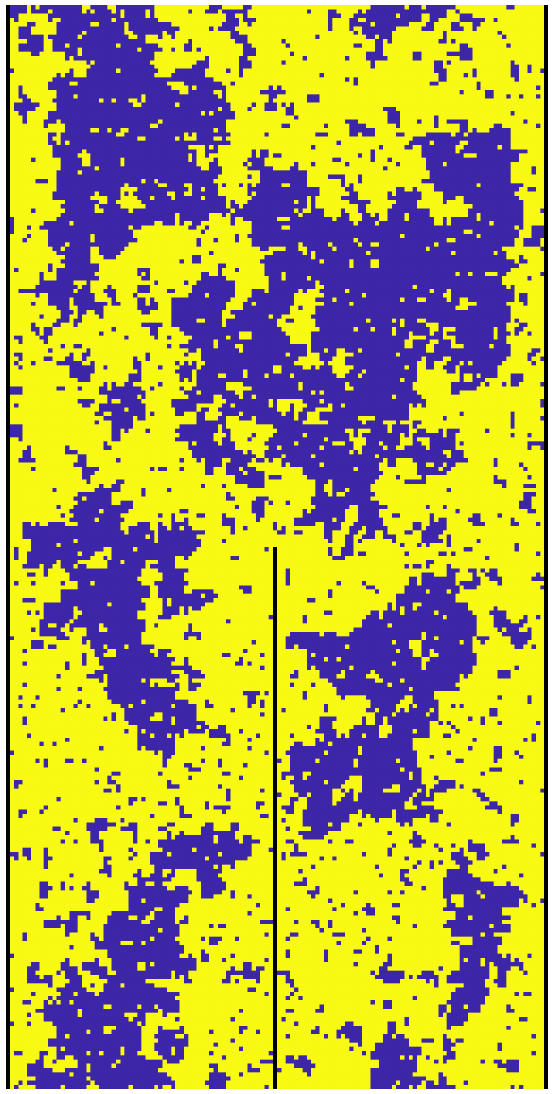}
	\label{sfig: Ising in slit-strip}
}
\caption{Ising model samples in the two geometries.
}
\label{fig: Ising samples}
\end{figure}

The Ising model will be defined precisely in 
Sections~\ref{sub: Ising on finite graphs}~%
--~\ref{sub: Ising model in infinite strip and slit-strip}, but
we describe it informally right away to be able to emphasize
aspects that are consequential for the overall picture.
For more comprehensive treatments of the Ising model,
see~\cite{MW-two_dimensional_Ising_model,
Palmer-planar_Ising_correlations}, and for some of the notable
recent progress in rigorous conformal invariance results for it, 
see~\cite{Smirnov-conformal_invariance_in_RCM_1,
Hongler-thesis, Izyurov-thesis, Dubedat-exact_bosonization,
CS-universality_in_Ising, HS-energy_density, HK-Ising_interfaces_and_free_bc,
HKV-CFT_at_the_lattice_level, CDHKS-convergence_of_ising_interfaces,
CHI-conformal_invariance_of_spin_correlations, 
CGN-planar_Ising_magnetization_field, CGN-planar_Ising_magnetization_field_2,
BDH-crossing_probabilities_with_free_bc, KS-conformal_invariance_of_RCM_2,
KS-configurations_of_FK_Ising_interfaces,
Izyurov-Ising_interfaces_in_multiply_connected_domains,
PW-crossing_probabilities_of_multiple_Ising_interfaces,
BH-scaling_limit_of_Ising_interfaces, GHP-Ising_local_spin_correlations,
KS-boundary_touching_loops, CHI-primary_field_correlations}
or the review~\cite{Chelkak-state_of_the_art_and_perspectives}.

The Ising model on a graph~$\gengraph = (\dV , \dE)$
with vertex set~$\dV$ and edge set~$\dE$
is a random assignment of $\pm 1$ spins to the vertices of the
graph: the sample space consists of possible spin configurations
$\spinconf = (\sigma_{z})_{z \in \dV}$, where $\sigma_z \in \set{-1, +1}$
denotes the spin at vertex~$z\in \dV$.
Given an inverse temperature parameter~$\invtemp \geq 0$
(the temperature is~$\propto 1/\invtemp$), 
the Ising model probability 
measure~$\PR_{\invtemp ; \gengraph}$ 
is informally such that
\begin{align*}
\PR_{\invtemp ; \gengraph} \big[ \set{\spinconf} \big]
\; \propto \;
    \exp \Big( \invtemp \sum_{\set{z,w} \in \dE} \sigma_z \sigma_w \Big) ,
\end{align*}
so that configurations with more alignment among neighboring spins
have higher probabilities, and the strength of this local alignment 
tendency increases with~$\invtemp$ (which is to say that it
decreases with the temperature). This preference for local alignment
is also why the model was originally introduced as a model of
ferromagnetism~\cite{Lenz-beitrag, Ising-beitrag}.
With physics developments in renormalization group and
universality~\cite{Wilson-renormalization_group_and_critical_phenomena_1_and_2},
this model was understood to correctly describe the critical
behavior and phase transition in uniaxial ferromagnets (not all
ferromagnets), as well as a variety of a priori different phenomena such as
the liquid-vapor transition. For finite graphs the formula 
above directly serves as the definition of the 
Ising model probability measure, but for infinite graphs
the proper definition requires a limit from finite graphs instead.

Specifically, we consider the Ising
model on graphs~$\gengraph$, 
which are square grid approximations of the infinite
strip and the infinite slit-strip domains
illustrated in Figures~\ref{fig: square grid strip and slit-strip}.
On the infinite square grid,
a phase transition occurs at the critical 
point~$\invtempcrit = \frac{1}{2} \log (\sqrt{2} + 1)$
\cite{Peierls-on_Ising_model_of_ferromagnetism,
KW-statistics_of_the_2D_ferromagnet}:
at high temperatures~$\invtemp<\invtempcrit$ the
spins decorrelate exponentially with distance (paramagnetic phase),
whereas in low temperatures~$\invtemp>\invtempcrit$ 
a uniformly positive correlation
among spins remains at arbitrarily large distances 
(ferromagnetic phase). 
Conformal field theory should describe the behavior
at the critical point~$\invtemp=\invtempcrit$, so we focus
on this case throughout.

Boundary conditions are imposed by declaring a subset
$\bdry \dV \subset \dV$ of vertices as the boundary vertices, 
and by conditioning on the values of the spins of these.
We mainly impose boundary conditions on the following separate
segments of the boundaries of the strip and slit-strip.
The left and the right boundaries of the strip are considered the
separate boundary segments of the strip geometry. In the 
slit-strip geometry, the left, the right, and the slit
are considered the separate boundary segments.
A plus boundary condition on a segment of the boundary
amounts to conditioning the spins on that segment to~$+1$,
minus boundary condition similarly to~$-1$, and by locally 
monochromatic boundary conditions we mean merely conditioning
on the spins being constant on each boundary component, so that
any combinations of plus and minus boundary values on the separate
segments of the boundary can occur.
Figures~\ref{fig: Ising samples} illustrate typical samples
of the Ising model on strip and slit-strip graphs,
at the critical inverse temperature~$\invtemp = \invtempcrit$,
and with plus boundary conditions on all of the above 
boundary segments.

Since the Ising model consists of a random spin configuration
$\spinconf = (\sigma_z)_{z \in \dV}$, random
variables in the model are functions
\begin{align*} 
X = f(\spinconf)
\end{align*}
of the configuration. 
Given real-valued random variables 
$X_1 = f_1(\spinconf)$, \ldots, $X_n = f_n(\spinconf)$,
the expected value of their product
\begin{align*}
\EX \big[ X_1 \cdots X_n \big]
\end{align*}
with respect to the probability 
measure~$\PR = \PR_{\invtemp ; \gengraph}$,
(or appropriately conditioned version of it 
if boundary conditions are imposed) is called
a correlation function. We will consider in particular
correlation functions of the following type,
which capture the idea of three point boundary 
correlation functions associated to the three infinite extremities
of the slit-strip geometry.
The Ising model is considered on square grid 
approximations of the
slit-strip of fixed width~$\width$,
truncated from both above and below at finite 
heights~$\pm\height$.
We take random variables 
\begin{align*}
T = f_{\topsym}(\spinconf) , \qquad 
R = f_{\rgtsym}(\spinconf) , \qquad
L = f_{\lftsym}(\spinconf) ,
\end{align*}
where the function~$f_{\topsym}$ 
depends only on the spins on the top row of the truncated 
slit-strip graph, 
$f_{\rgtsym}$~
depends only on the spins on the right half of the bottom
row of the truncated slit-strip graph, and
$f_{\lftsym}$~
depends only on the spins on the left half of the bottom
row of the truncated slit-strip graph.
The correlation functions
\begin{align*}
\EX^{(\width;\height)} \big[ T R L \big]
\end{align*}
of such triples of random variables at the three boundary
segments are the physical quantities of interest to us.
Among all possible random variables of the above type,
we choose a particular basis for which these correlation
functions (or more precisely the numerators, when the correlation
functions are expressed so that the partition function appears as
their denominator) have a purely exponential
dependence on the truncation height~$\height$.
For such basis random variables 
$T_{\parts} = f^{(\parts)}_{\topsym}(\spinconf)$,
$R_{\partsR} = f^{(\partsR)}_{\rgtsym}(\spinconf)$,
$L_{\partsL} = f^{(\partsL)}_{\lftsym}(\spinconf)$,
the fusion 
coefficients~$\fusionIsing{\parts}{\partsR}{\partsL}$
of the Ising model are defined so that they capture the
renormalized limits of the boundary correlation functions 
in the sense that 
\begin{align*}
\lim_{\height \to \infty} 
    e^{a\height} \; 
    \EX^{(\width;\height)} \Big[ T_{\parts} L_{\partsL} R_{\partsR} \Big]
\; = \; \frac{\fusionIsing{\parts}{\partsR}{\partsL}}{Z} ,
\end{align*}
where $a = a(\parts, \partsR, \partsL)$ is the appropriate rate to
renormalize the exponential dependence, and the factor~$Z$ itself includes
a certain simple fusion coefficient
(see Section~\ref{sec: fusion coefficients} for details).
If one instead considers for example plus boundary conditions
or any mixed monochromatic boundary conditions (i.e., prescribed spin values
on each of the boundary components), some details need to be adjusted:
the rate $a = a(\parts, \partsR, \partsL)$ of exponential dependence
and the overall normalizing constant~$Z$ depend on the boundary
conditions, and naturally only the subspace of functions that are supported
on configurations allowed by the boundary conditions remains relevant.
However the limits obtained for such boundary correlation functions are still
proportional to the same fusion coefficients. For this reason, we view the
fusion coefficients as giving the the proper description of boundary
correlation functions at the three infinite extremities of the slit-strip.

Our general goal in this series of articles is to show that the
full algebraic structure of a certain conformal field theory
can be recovered from these fusion coefficients
$\fusionIsing{\parts}{\partsR}{\partsL}$ in the scaling
limit~$\width \to \infty$. 
In the remaining part of this introduction we try to provide context
for why such a result should be expected in the first place,
what have been some of the mathematical difficulties in
formulating and proving such a result, 
which details of the probabilistic questions are believed
to be consequential for the correct statement,
and to what extent similar results should hold
more generally for other models of statistical physics.

\subsection{Statistical physics and quantum field theory}

In the paradigm of constructive quantum field theory, one seeks 
to obtain rigorous constructions of specific quantum field 
theories via probabilistic and analytical 
techniques~\cite{ID-statistical_field_theory, FFS-triviality_in_QFT}.
If a probability measure on fields in a Euclidean space
can be constructed subject to certain axioms, an analytic
continuation of one spatial dimension from real to imaginary provides
the physical time dimension, and yields
an actual quantum field theory  in Minkowski space-time
\cite{OS-axioms_for_Euclidean_Greens_functions_1,
OS-axioms_for_Euclidean_Greens_functions_2}. 
The construction of
suitable probability measures 
on fields is itself a major undertaking in mathematical physics.
A typical approach is to start from lattice
discretizations which serve to regularize the field theories and
make them a priori well-defined, and 
to then try to show the existence
and desired properties of their scaling limits. This makes the 
constructive field theory approach at its core essentially
equivalent to scaling limit questions in mathematical 
statistical mechanics.

The underlying idea that quantum field theory
and statistical physics can be done in the same formalism
still justifiably retains an element of surprise,
although it has long had the status of quite uncontroversial 
folklore~\cite{McCoy-connection_between_statistical_mechanics_and_QFT}.
The idea was certainly present in the work of Wilson
on the renormalization 
group~\cite{Wilson-renormalization_group_and_critical_phenomena_1_and_2}.
The archetype of lattice model of statistical physics,
the square-lattice Ising model, was in fact formulated
in an essentially quantum field theoretic formalism
as early as in 1949 by Kaufman~\cite{Kaufman-crystal_statistics_2}.

In two dimensions, massless quantum field theories can be argued
by general grounds to enjoy conformal invariance
properties, and consequently be very stringently algebraically 
constrained~\cite{BPZ-infinite_conformal_symmetry_in_QFT}.
In view of the general connection between 
statistical physics and quantum field theories,
conformal field theory (CFT) is thus argued to apply to
a wide variety of planar statistical physics models at their critical
points of continuous phase 
transitions~\cite{BPZ-infinite_conformal_symmetry_of_critical_fluctuations}.
The critical planar Ising model is the prime example of
this picture. 
Conformal field theory
predicts values of critical exponents, functional
forms of scaling limits of multi-point correlation functions,
as well as a vast number of
other intricate features of the Ising model at and 
near criticality~\cite{DMS-CFT, Mussardo-statistical_field_theory}.
Such predictions are also not just 
excellent approximations: due to their algebraic underpinnings,
they are supposed to be exactly correct.
The entirety of CFT predictions forms a sound and appealing 
overall picture, whose specifics agree exquisitely with
numerics, simulations, and alternative methods of theoretical 
physics. Moreover, some of these predictions for the
critical Ising model have even been verified
rigorously with the sophisticated mathematical
methods that have been developed in the century of research~---
including with the transfer matrix formalism,
dimer representations, Kac-Ward matrices,
and recently by discrete complex analysis methods.
Indeed, such is the success of the CFT picture for the 
Ising model~--- and perhaps the familiarity of this
most prominent example case of the general picture~---
that it has become necessary to carefully specify whether
by the Ising model one means the probabilistic 
lattice model
or a certain conformal field theory!

Yet, the vast majority of the conformal field theory ideas
have remained very elusive to a rigorous mathematical approach,
even in the case of the Ising model~---
despite the spectacular progress in rigorous conformal invariance 
results in the past decade.
% ~\cite{%
% Smirnov-conformal_invariance_in_RCM_1,
% Hongler-thesis, Izyurov-thesis, Dubedat-exact_bosonization,
% CS-universality_in_Ising, HS-energy_density, HK-Ising_interfaces_and_free_bc,
% HKV-CFT_at_the_lattice_level, CDHKS-convergence_of_ising_interfaces,
% CHI-conformal_invariance_of_spin_correlations, 
% CGN-planar_Ising_magnetization_field, CGN-planar_Ising_magnetization_field_2,
% BDH-crossing_probabilities_with_free_bc, KS-conformal_invariance_of_RCM_2,
% KS-configurations_of_FK_Ising_interfaces,
% Izyurov-Ising_interfaces_in_multiply_connected_domains,
% PW-crossing_probabilities_of_multiple_Ising_interfaces,
% BH-scaling_limit_of_Ising_interfaces, GHP-Ising_local_spin_correlations,
% KS-boundary_touching_loops, CHI-primary_field_correlations}.
It is still difficult to find even a precisely phrased conjecture
in the literature of the totally commonplace assertion that the 
scaling limit of the critical Ising model is a conformal field 
theory~--- let alone a proof.\footnote{The very recent
work~\cite{Hongler_et_al-CFT_book} starts to address the question
of the proper mathematical formulation of this general statement
seriously.}
Our goal in this series is to give a precise formulation
and proof of such a statement, which at least recovers the full
algebraic structure of a boundary conformal field theory 
in the scaling limit of the Ising model.
The conclusion of our main result will be stated in the last
part of the series, and it is only there that we really 
need the detailed definitions about conformal field theory.
We nevertheless should address at least what type of a mathematical
object we mean when talking about conformal field theory.

\subsection{On definitions of conformal field theory}

Axiomatic approaches to conformal field theory
incorporate the strong constraints that arise from
conformal invariance into the general properties of quantum 
field theories.
A number of different mathematical ways of doing so have been
put forward~--- we focus on two, which have arguably been the most 
successful and the most influential.

The notion of the chiral symmetry algebra of a CFT was in essence
formulated already in the pioneering physics literature, and
the mathematical definition of a vertex operator algebra (VOA)
\cite{FLM-VOAs_and_Monster, Kac-vertex_algebras_for_beginners,
LL-introduction_to_VOAs}
building from the work of Borcherds and Frenkel--Lepowsky--Meurman
fully captures its precise meaning.
An algebraic approach to the definition of conformal field theory 
can then be formulated in terms of representation theory of 
VOAs~\cite{Huang-CFT_and_VOA}.
Despite being somewhat involved, VOAs are nevertheless sufficiently 
concrete that many relevant
examples of them can be constructed,
and while one can not yet conclude all the
desired properties of interest to physics, there is
steady progress towards using VOAs as a starting point for the
more quantum field theoretically formulated CFTs as well.

By contrast, 
the functorial axiomatization of CFTs by 
G.~Segal~\cite{Segal-definition_of_CFT, Segal-definition_of_CFT-new}
places focus on the geometry of the space-time 
(or the space in Euclidean formulation),
and on the time evolution semigroup of operators (or the semigroup generated by 
the transfer matrix in Euclidean lattice formulation)
as well as operators that generalize these.
This axiomatization is thus closer to the language of
quantum field theory generally, and far reaching
conclusions can be derived starting from it.
Verifying these functorial axioms in specific cases is unfortunately
difficult, and few examples of CFTs in this sense are known to exist.

The very definition of conformal field theory therefore
still poses mathematical challenges. The choice of the appropriate
definition involves matters of mathematical taste, as well as 
trade-offs between where the difficulties should lie:
in constructing examples of CFTs, or in deriving desired 
conclusions about them. 
Our choice in this series is to draw geometric inspiration from Segal's 
functorial approach, but to accept VOAs as the
definition of (the algebraic structure of) conformal field theories.

Whatever is taken as the mathematical axiomatization of CFTs,
a fundamental question is to connect the constructive quantum
field theory approach to the definition of CFTs, which incorporates
all the structure that makes CFTs so remarkably powerful.
In other words, one should show that starting from a given probabilistic
lattice model, by passing to the scaling limit in which the lattice
spacing is let tend to zero, one recovers objects satisfying the
axioms of a (specific) conformal field theory.

\subsection{The correct CFT for the scaling limit of the Ising model}

In the above we have still
disregarded all subtle issues stemming from 
the fact that conformal field theory is
supposed to apply in a few 
different general situations,
which require adapting the definitions appropriately.
The algebraic approach of VOAs and the functorial approach by Segal
thus represent just the two main frameworks of definitions.

A specific issue relevant to the present work, and generally
to any application of CFT to statistical physics, is the difference
between a \emph{boundary CFT} and a \emph{bulk CFT}.
From the statistical physics point of view the difference is
whether we consider the models in domains which have physical
boundaries or not. Algebraically the difference is the same
as that of a \emph{chiral CFT} and of a \emph{full CFT}:
the latter has separate holomorphic and antiholomorphic chiral 
algebras while the former only has one.
Both the holomorphic and antiholomorphic fields of a CFT
should be equally meaningful\footnote{Some readers will
undoubtedly view this as too generous, because actual physical
fields are typically of neither chirality strictly.
Only in a formal sense does the physical field behave as if it was
a product of a holomorphic and an antiholomorphic part. 
An acceptable less forgiving reading of this sentence
is therefore that the holomorphic and antiholomorphic fields are equally 
\emph{meaningless} both with and without boundaries\ldots
The rest of the conclusion remains unaffected.}
in domains with and without physical boundaries,
but in the presence of boundaries the associated holomorphic and
antiholomorphic chiral algebras become coupled ultimately 
due to the 
boundary conditions in the statistical physics model.
Our results will feature a boundary CFT.

A general definition of CFTs should of course admit
many specific instances;
in particular to any critical 
statistical physics model there should correspond 
a CFT specifically describing its scaling limit.
The folklore about the Ising
model turns out to be somewhat curious regarding this point.
In any reasonable sense 
(for example as VOAs)
there are in fact two 
different CFTs routinely claimed to do the job.
These two CFTs of the Ising model are
the unitary $(4,3)$-minimal model
(with the rational Virasoro VOA as its
chiral algebra) and the massless free fermion
(with a certain simple super-VOA as its chiral algebra).
Both are well-known to be ``the'' scaling limit of the same model,
while unmistakably not the same CFTs!
Serious consideration of 
whether the scaling limit of
the Ising model is in fact
an interacting bosonic quantum field theory
or a free fermionic one could get 
philosophical~\cite{McCoy-connection_between_statistical_mechanics_and_QFT}.
Ultimately it necessarily comes down to what questions do we seek
to answer, i.e., what specific quantities are considered
in the scaling limit.

In this series of articles, we will recover specifically the super-VOA of
the free fermionic boundary CFT in the scaling limit from
the Ising model fusion coefficients. So in our setup,
what is it that dictates that the correct CFT is the free fermion 
rather than the unitary $(4,3)$-minimal model?
In a similar vein, what dictates that we obtain the VOA only,
and not also some (twisted) modules for it, 
as generally expected to be the case?

In fusion coefficients
we allow for the most general observables depending on
spins on the parts of the boundary representing the 
three infinite extremities
of the slit-strip. The rest of the domain boundary is
taken to have locally monochromatic
boundary conditions.
With other choices of boundary conditions or other
quantities considered, the details of the statement
and conclusion should
be modified.
We want to highlight exactly that: in any precise 
formulation of the scaling limit of any statistical 
physics model as a CFT, such choices
(boundaries or not, allowed boundary conditions, 
allowed observables) will inevitably
be involved, and the correct formulation of the statement 
itself has to depend on these!\footnote{To put it bluntly,
the often encountered statement that the scaling
limit of the Ising model is a particular CFT 
(pick your favorite)
is meaningless without further elaboration!
}

\subsection{The role of the slit-strip geometry}

As stated in the very beginning, we specifically consider 
the Ising model on graphs~$\gengraph$, 
which are square lattice\footnote{As 
emphasized by the renormalization group picture,
the core features of a scaling limit should be
universal, and independent of details
such as the choice of lattice or even the 
the microscopic interactions in the model itself.
A CFT scaling limit describes such a universality class. 
Our insistence on the importance of certain other details
(for example the allowed observables) is merely
the statement that those details are not among the irrelevant ones
for the determination of the universality class.}
approximations
of the infinite slit and the infinite slit-strip domains
illustrated in Figure~\ref{fig: square grid strip and slit-strip}.
Two reasons in particular stand out for why we use such a 
slit-strip geometry.
First, in a Segal-type approach to boundary CFT,
the slit-strip plays the role analogous to the pair-of-pants
surface in bulk CFT: it is the fundamental building block of
more general geometries.
Second, we expect that for critical models of 
statistical physics in general, the vertex operator algebra
(as well as Segal type vertex operators) is
the scaling limit of the slit-strip transfer matrix
operators in a manner entirely parallel to 
our main result.\footnote{The exact VOA should be changed to the VOA
appropriate for the CFT in question, and (depending on boundary
conditions) other modules for the VOA and intertwining operators
between them should in general appear also, besides the VOA itself.}
Indeed, for certain loop models, the
transfer matrix formalism in lattice discretizations
of the slit-strip has already been successfully used 
to probe 
fusion products of modules
of the conjecturally associated very intricate 
CFTs~\cite{GJRSV-inspiration}.
With the idea expected to be so generally valid,
working out the case of the Ising model in full
detail should provide a valuable prototype.

\subsection{Related work and original contribution}
We will analyze the fusion coefficients using 
the very classical method of
transfer matrices, the idea of which 
in fact goes back to Ising's original work 
on the one-dimensional Ising model~\cite{Ising-beitrag},
and which for the two-dimensional Ising model was used 
by Onsager in his foundational work~\cite{Onsager-crystal_statistics}.
With the locally monochromatic boundary conditions that
we specifically use, the transfer matrix
has been analyzed in~\cite{AM-transfer_matrix_for_a_pure_phase}.
We rely particularly on the fermionic nature of the 
transfer matrix formalism for the Ising model
in that our main calculations
are done with certain Clifford algebra valued discrete
one-forms.
The fermionic nature of the transfer matrix was
first observed by Kaufmann~\cite{Kaufman-crystal_statistics_2}, 
and was also the subject of the influential work of
Schultz \& Mattis \& Lieb~\cite{SML-Ising_model_as_a_problem_of_fermions}.
The transfer matrix formalism in its most common form is used
for calculations in rectangles, strips, cylinders, and tori,
but it quite readily adapts to our calculations in the 
slit-strip as well.
Underlying our calculations is the recently observed close 
connection~\cite{HKZ-discrete_holomorphicity_and_operator_formalism}
between the transfer matrix and the 
analytic continuation of certain discrete holomorphic
functions solving a Riemann boundary value problem, i.e.,
the key techniques pioneered by 
Smirnov~\cite{Smirnov-towards_conformal_invariance}
which have enabled a breakthrough in rigorous progress and spectacular mathematical results
on conformal invariance of the scaling limit of the
Ising model during the past decade. Earlier uses of discrete
complex analysis in the Ising model include
\cite{KC-determination_of_an_operator_algebra,
Perk-quadratic_identities_for_Ising_model_correlations,
Mercat-discrete_Riemann_surfaces},
and~\cite{Smirnov-discrete_complex_analysis_and_probability,
CCK-revisiting, Chelkak-state_of_the_art_and_perspectives}
can serve as reviews of various aspects of 
the more recent uses of discrete holomorphicity method.

Apart from a few auxiliary calculations contained in the first
article~\cite{part-1} of this series and 
in~\cite{HKZ-discrete_holomorphicity_and_operator_formalism},
we provide a self-contained diagonalization of the critical
Ising model transfer matrix in the strip with locally 
monochromatic boundary conditions. This result is certainly 
not new, it has been 
obtained in~\cite{AM-transfer_matrix_for_a_pure_phase}
and is reviewed in, e.g., \cite{Palmer-planar_Ising_correlations}~--- 
and it is in essence merely a variation on 
the theme of~\cite{Kaufman-crystal_statistics_2,
SML-Ising_model_as_a_problem_of_fermions}.
We write the calculation in slightly different terms, and
the power of this approach becomes evident in domains of
more general geometries than just the strip~---
in particular in the slit-strip.
Our derivation is based on Clifford algebra
valued discrete one-forms, which have two crucial properties:
closedness and slidability along vertical boundaries.
We show that such forms can be constructed using
s-holomorphic functions with Riemann boundary values as 
coefficients of two natural fermion field components.
With such forms, contour deformation arguments
entirely parallel to boundary conformal field theory can
be performed. Thus our method is first of all
directly reminiscent of conformal field theory. 
Moreover, with it the diagonalization of the
transfer matrix immediately reduces to a simple
question of discrete complex analysis:
namely of finding the s-holomorphic solutions to Riemann boundary
value problem in the strip which are eigenfunctions of vertical 
translations.

The real advantage of our method of closed and vertically
slidable Clifford algebra valued discrete one-forms, however,
only becomes evident in the slit-strip.
At first sight, the fusion 
coefficients appear to involve such one-forms, which are 
only locally defined. But with the discrete complex analysis
results in the first part of the series~\cite{part-1}, 
we can trade unwanted singular parts of such one-forms to 
globally defined one-forms, which can then be contour deformed
to those extremities of the slit-strip in which they have
no singularities. This is at the core of our characterization
of the fusion coefficients in terms of a recursion.
The coefficients in the recursion involve inner products among
distinguished discrete holomorphic functions, and
from the results~\cite{part-1} about the convergence
of these functions in the scaling limit, we deduce the 
convergence of the fusion coefficients themselves in the scaling 
limit.
The scaling limit statement for the fusion
coefficients may have been foreseen by experts for many decades,
but its derivation seems to genuinely require the novel techniques
of discrete complex 
analysis~\cite{Smirnov-towards_conformal_invariance,
CS-discrete_complex_analysis_on_isoradial_graphs}.

\subsection{Organization of this work}

Part one~\cite{part-1} of this series was concerned with spaces 
of holomorphic functions with Riemann boundary values in the
strip and the slit-strip, as well as lattice analogues
of these. In this article, the necessary results from the first
part are recalled as they are needed:
Sections~\ref{sec: discrete complex analysis},
\ref{sec: distinguished functions}, and 
\ref{sec: limit distinguished functions} recall the definitions
of discrete complex analysis, specific discrete holomorphic
functions, and scaling limits of such functions, respectively.

The original results of this article are contained in 
Sections~\ref{sec: operator valued forms},
\ref{sec: transfer matrices}, 
and~\ref{sec: continuum fusion coefs}.

In Section~\ref{sec: operator valued forms}
we first review the Clifford algebra acting on the
transfer matrix state space, and select a basis of Clifford
generators consisting of discrete holomorphic and discrete
anti-holomorphic fermions. 
Quoting the result of the conjugation of these fermions
by the transfer matrix 
from~\cite{HKZ-discrete_holomorphicity_and_operator_formalism},
we present a construction of closed and vertically slidable 
Clifford algebra valued $1$-forms from s-holomorphic 
solutions to the Riemann boundary value problem.
The integration and contour deformation of such operator valued
$1$-forms is a direct discrete counterpart of the way that
current modes of the chiral symmetry algebra in boundary
conformal field theory are treated. This section can therefore be regarded
as an exact lattice realization of a key algebraic technique
of boundary conformal field theory. It is crucial that the technique
works not only in the lattice strip, but also in other domains,
including the lattice slit-strip, so it can ultimately be
applied to the fusion coefficients.

In Section~\ref{sec: transfer matrices} we define in detail our
setups for the Ising model in the strip and the slit-strip. In this
section we include a review how the transfer matrix allows for the 
calculation of partition functions and correlation functions in the 
strip and the slit-strip. 
This part is straightforward and mostly well-known, so we do not 
provide proofs here. We instead make sure that the definitions
and statements are self-contained and sufficient for
our applications, and we arrange the sequence of 
statements so that even a reader without prior familiarity with 
the transfer matrix formalism should be able to
fill in the missing proofs with relative ease.
In the remaining part of the section,
the transfer matrix method is used in conjunction with the
method of Clifford-algebra valued discrete $1$-forms
of Section~\ref{sec: operator valued forms}.
As a first illustration of the method, we use discrete holomorphic
vertical translation eigenfunctions as coefficient functions in the
forms, and obtain a self-contained diagonalization of the Ising
transfer matrix in the strip. As the main result of the section,
we use discrete holomorphic functions adapted to the slit-strip
geometry to obtain a recursion that determines the Ising model
fusion coefficients.

In Section~\ref{sec: continuum fusion coefs} we introduce continuum
fusion coefficients as certain explicit integrals with Pfaffian-form 
kernels and coefficient functions which are quarter-integer Fourier
modes, i.e., vertical translation eigenfunctions in the continuum.
We derive a recursion for these continuum fusion coefficients, 
entirely parallel to the recursion for the Ising model fusion 
coefficients. Numerical constants in the two recursions only
differ because the continuum version involves inner
products among distinguished continuum holomorphic functions,
while the original Ising model version involves inner
products among distinguished discrete holomorphic functions.
The convergence in the scaling limit of the inner products
among the functions is then sufficient to conclude that
the Ising model fusion coefficients converge in the scaling
limit to these continuum fusion coefficients.

In the final part~\cite{part-3} of our series it will
be shown that the continuum fusion coefficients are
essentially (i.e., up to
a transformation of necessitated by the slit-strip geometry)
equal to the structure constants of the vertex operator 
algebra of a fermionic conformal field theory. 
Combined with the result of this second part, 
we can conclude that the VOA structure constants~---
and thus indeed the full algebraic structure of the associated
boundary conformal field theory~--- can be recovered 
from the Ising model fusion coefficients in the scaling limit,
and vice versa.

{\bf Acknowledgments:}
S.P. is supported by a KIAS Individual Grant (MG077201) at Korea Institute 
for Advanced Study.
T.A. was affiliated with the Department of Mathematics and Statistics at the 
American University of Sharjah, and summer intern at Aalto University, and 
recognizes their financial support.

\bigskip

\section{Lattice domains and notions of discrete complex analysis}%
\label{sec: discrete complex analysis}

% *******************************************************************
% **   \section{S-hol., RBVs, and distinguished functions}         **
% *******************************************************************

In this section we define the infinite strip and slit-strip lattice
domains, and the two notions of discrete complex analysis that we rely
on: s-holomorphicity and Riemann boundary 
values~\cite{Smirnov-towards_conformal_invariance,
CS-discrete_complex_analysis_on_isoradial_graphs}.
Our specific conventions
are as in~\cite{part-1}.

\subsection{The infinite lattice strip and slit-strip graphs}

All the graphs we consider are essentially subgraphs of the
square grid~$\bZ^2$ (with its graph structure of nearest neighbor edges).
We consider the square grid as embedded in the complex plane,
$\bZ^2 \subset \bC$. Without comment, we identify vertices of any of
our graphs with the corresponding complex numbers, edges with the
complex numbers that are the midpoints of their two endpoint 
vertices\footnote{The slit-strip is a multi-graph with doubled edges
between certain vertices, so it is a slight
abuse of notation to sometimes identify two edges with the same complex
number, but we trust the correct choice is always clear from the context.},
and plaquettes with the complex numbers
at the midpoint of the corresponding faces of the square grid.

Fix $\lft , \rgt \in \bZ$ such that $\lft < \mdpt < \rgt$, and
denote~$\width = \rgt - \lft \in \bN$.
The parameters~$\lft$ and~$\rgt$ will serve as the horizontal
positions of the left and the right boundaries of the vertical
lattice strip and slit-strip, respectively; $\mdpt$~will be the horizontal
position of the slit when appropriate; $\width$~will be the total
width of the strip or slit-strip. Although our lattice
definitions depend on $\lft , \rgt$, we usually only
indicate the discretization by a superscript~$\width$,
for simplicity.

\begin{figure}[tb]
\centering
\subfigure[The square grid strip~$\dstrip$.] 
{
  \includegraphics[width=.35\textwidth]{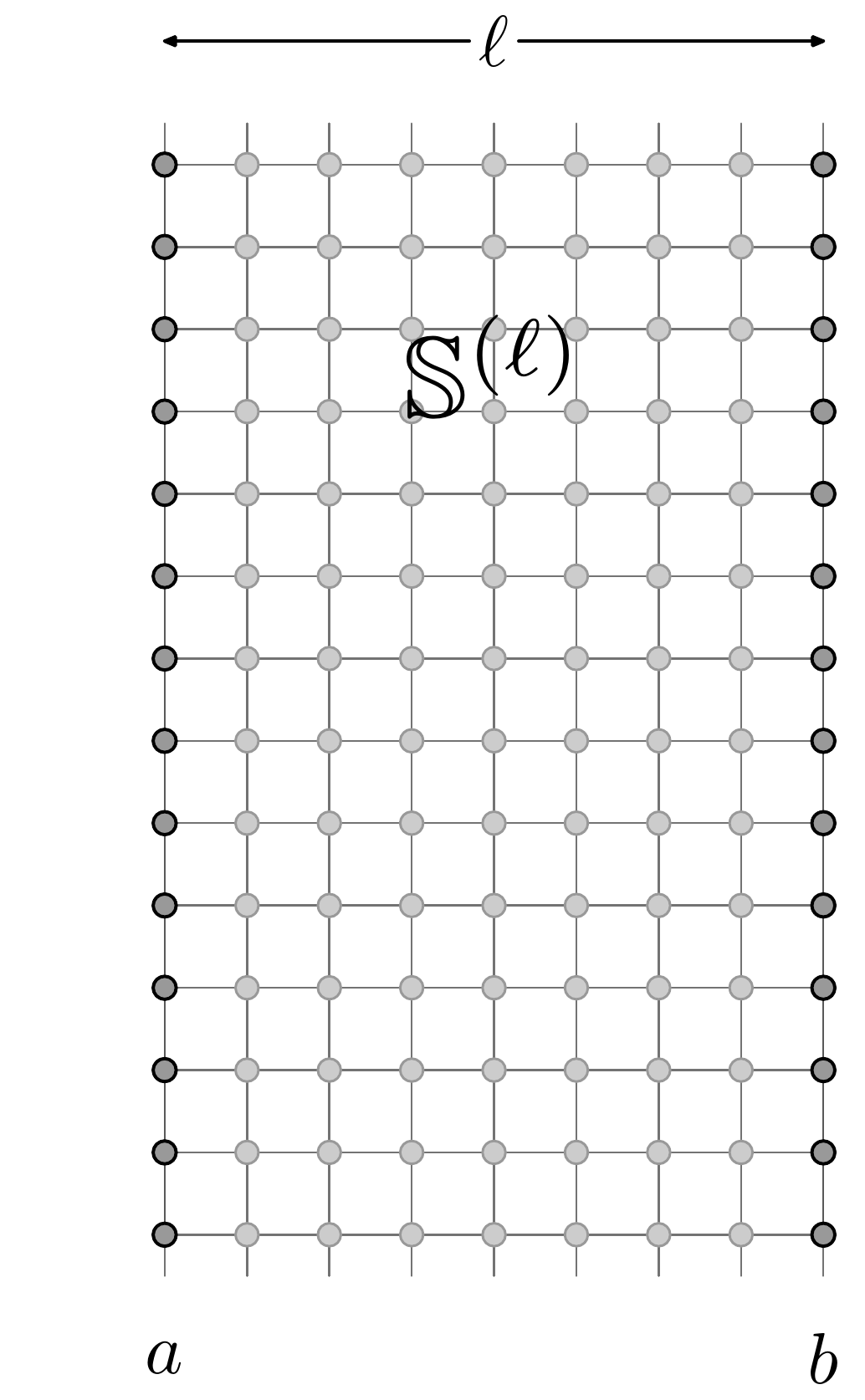}
  \label{sfig: lattice strip}
}
\hspace{2.5cm}
\subfigure[The square grid slit-strip~$\dslitstrip$.] 
{
  \includegraphics[width=.35\textwidth]{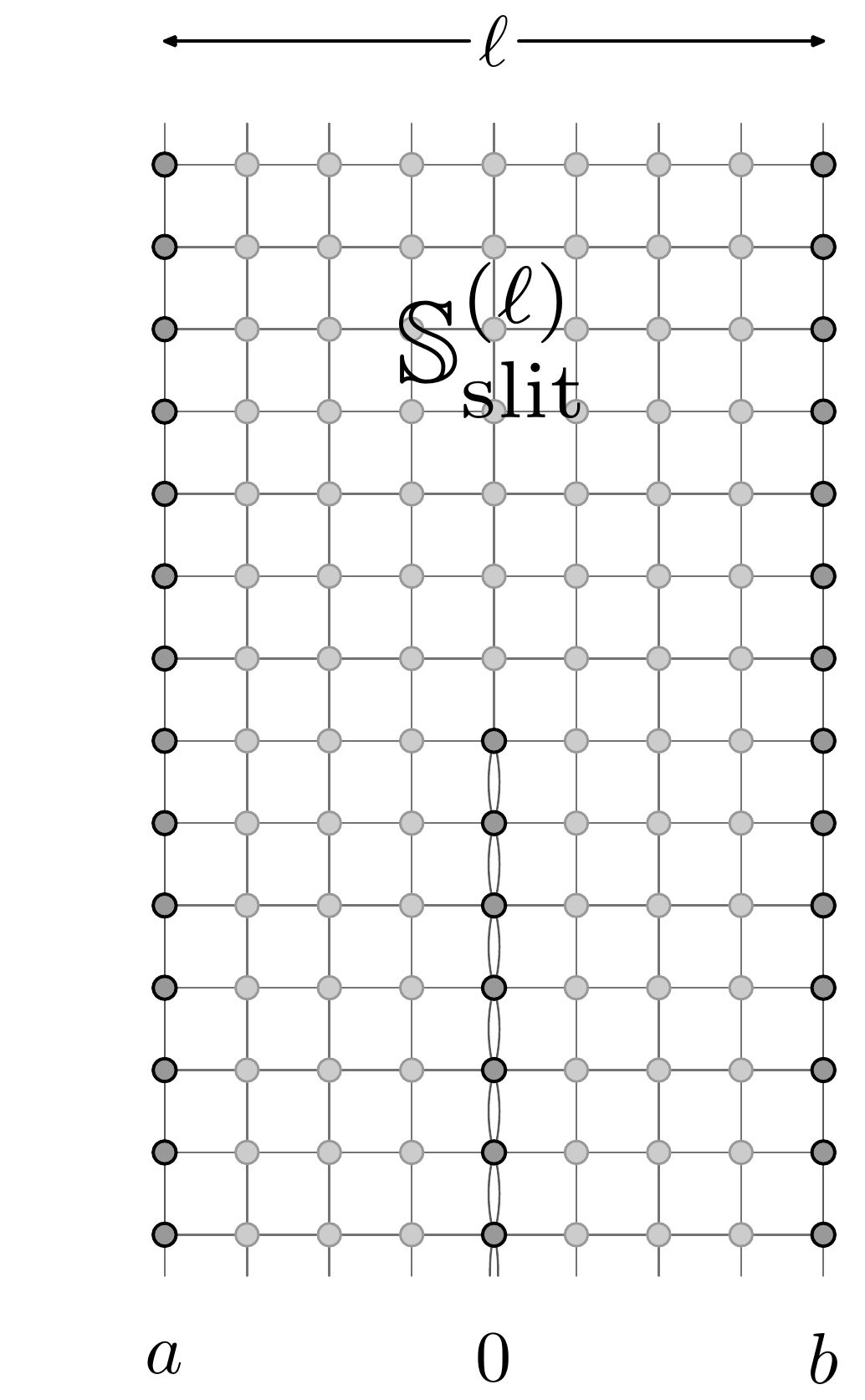}
  \label{sfig: lattice slit-strip}
}
\caption{The vertically infinite lattice strip and slit-strip.
}
\label{fig: square grid strip and slit-strip}
\end{figure}

The vertically infinite lattice strip is a graph with vertex set
\begin{align*}
\dstrip
\; := \; \dinterval{\lft}{\rgt} \times \bZ ,
\end{align*}
where $\dinterval{\lft}{\rgt} = \set{\lft, \lft+1, \ldots, \rgt-1, \rgt}$
is the integer interval from~$\lft$ to~$\rgt$. This is considered as
an induced subgraph of the square grid~$\bZ^2$, i.e., with nearest
neighbor edges
\begin{align*}
\dE(\dstrip) \; = \; 
    \set{ \{u,v\} \; \Big| \; u,v \in \dstrip , \; \|v-u\| = 1 } .
\end{align*}
Figure~\ref{fig: square grid strip and slit-strip}(a) illustrates
the graph~$\dstrip$.

The vertically infinite lattice slit-strip is a multi-graph with the
same vertex set
\begin{align*}
\dslitstrip
\; := \; \dinterval{\lft}{\rgt} \times \bZ .
\end{align*}
It is also taken to have only nearest neighbor edges, but the edge
between $-\ii \, y$ and $-\ii \, (y+1)$ for $y \in \Znn$ is doubled,
with one of the edges interpreted to be on the left and the other
on the right of the negative imaginary axis. The set of edges of this
multi-graph is denoted by~$\dE(\dslitstrip)$.
Figure~\ref{fig: square grid strip and slit-strip}(b) illustrates
the graph~$\dslitstrip$, with the ``slit'' along the negative imaginary
axis.

\subsection{Functions on the lattice domains}

The discrete functions of interest to us will be defined in
the lattice strip~$\dstrip$ and the lattice 
slit-strip~$\dslitstrip$. More precisely, we consider
complex-valued functions
\begin{align*}
F \colon \dE \to \bC
\end{align*}
defined on the set~$\dE$ of edges of the corresponding graph,
$\dE = \dE(\dstrip)$ or $\dE = \dE(\dslitstrip)$.
Occasionally we also consider functions only locally defined
on one of the following three pieces of the lattice
slit-strip,
\begin{align}
\label{eq: subdomains of discrete slit-strip}
\dslitstripT = \; & \dinterval{\lft}{\rgt} \times \Znn , &
\dslitstripL = \; & \dinterval{\lft}{\mdpt} \times \Znp , &
\dslitstripR = \; & \dinterval{\mdpt}{\rgt} \times  \Znp ,
\end{align}
interpreted as subgraphs of the lattice slit-strip so that their
nearest neighbor edge sets are
\begin{align*}
\dE(\dslitstripT), \; \dE(\dslitstripR), \; \dE(\dslitstripL) \; 
\subset \; \dE(\dslitstrip)
\end{align*}
(the only overlap is on horizontal edges at height~$0$).

Key notions to us are suitable holomorphicity properties
and boundary conditions of such functions.
Specifically, we will use 
discrete holomorphicity in a specific $\bR$-linear sense of
s-holomorphicity
(and in Section~\ref{sec: operator valued forms} we will encounter
two complexifications of this real-linear notion).
We will also use Riemann boundary conditions which specify the
arguments of the functions
(up to multiples of~$\pi$) on the boundaries.

\subsubsection*{S-holomorphicity}
Let~$\dE$ be the set of edges of $\dstrip$ or $\dslitstrip$,
or of any of the subgraphs~\eqref{eq: subdomains of discrete slit-strip}.
A function
\begin{align*}
F \colon \dE \to \bC
\end{align*}
is said to be \term{s-holomorphic}, if for any two edges $z_1 , z_2 \in \dE$
adjacent to the same vertex~$v$ and face~$p$, we have
\begin{align}\label{eq: s-holomorphicity}
F(z_1) + \frac{\ii \, |v-p|}{v-p} \, \overline{F(z_1)}
= F(z_2) + \frac{\ii \, |v-p|}{v-p} \, \overline{F(z_2)} .
\end{align}

\subsubsection*{Riemann boundary values}
The Riemann boundary values we impose require the values 
of the functions on boundary points~$z$ of the domain to be
real multiples of~$\ii \, \tccw(z)^{-1/2}$, where
$\tccw(z)$ is the tangent to the boundary of the domain, with
positive (i.e., counterclockwise) orientation.
Since all of the boundaries of interest to us are vertical
(with tangents~${\tccw(z) = \pm \ii}$), we can phrase this
boundary conditions explicitly as follows.

A discrete function in the strip
\begin{align*}
F \colon \dE(\dstrip) \to \bC
\end{align*}
is said to have \term{Riemann boundary values}
if on the left and the right boundaries its values satisfy
\begin{align}\label{eq: dRBV in strip}
F \big( \lft + \ii y' \big) \in e^{-\ii \pi / 4} \, \bR
\qquad \text{ and } \qquad
F \big( \rgt + \ii y' \big) \in e^{+\ii \pi / 4} \, \bR .
\end{align}
In the slit-strip, boundary conditions are also applied along the
slit:
if~$\dE$ is the set of edges of the slit-strip~$\dslitstrip$
or one of the subdomains~\eqref{eq: subdomains of discrete slit-strip},
then a function
\begin{align*}
F \colon \dE \to \bC
\end{align*}
is said to have \term{Riemann boundary values} if it
satisfies~\eqref{eq: dRBV in strip} on the edges of
the left and the right boundaries, and in addition
\begin{align}\label{eq: dRBV on the slit part}
F \big( \mdpt^- + \ii y' \big) \in e^{+\ii \pi / 4} \, \bR 
\qquad \text{ and } \qquad
F \big( \mdpt^+ + \ii y' \big) \in e^{-\ii \pi / 4} \, \bR
\end{align}
where~$\mdpt^\pm + \ii y'$ stand for the edges along the slit
on the two sides.

\subsubsection*{Restrictions to a cross-section}
We focus particularly on the restriction of the functions
to the horizontal cross-section at height~$0$, and we form
real vector spaces of such functions, as in~\cite{part-1}.

In the discrete setting, we therefore study functions
defined on the set
\begin{align}
\crosssecdual = \dintervaldual{\lft}{\rgt} 
    = \set{ \lft + \half , \lft + \frac{3}{2} , \ldots
        , \rgt - \frac{3}{2} , \rgt - \half } 
\end{align}
of horizontal edges on the cross-section. We use the 
real vector space
\begin{align}\label{eq: discrete function space}
\dfunctionsp := \bC^{\crosssecdual} 
\end{align}
of complex-valued functions on~$\crosssecdual$,
which is of dimension~$\dmn_\bR(\dfunctionsp) = 2 \width$.
We equip it with the
real Hilbert space structure such that inner product and norm are
\begin{align}\label{eq: discrete inner product}
\innprod{f}{g} := \; &
    \re \left( \sum_{x' \in \crosssecdual} f(x') \, \overline{g(x')} \right) 
    , &
\|f\| := \; &
    \Big( \sum_{x' \in \crosssecdual} |f(x')|^2 \Big)^{1/2} .
\end{align}

We furthermore define a unitary
transformation~$\refl \colon \dfunctionsp \to \dfunctionsp$ by
\begin{align}\label{eq: discrete reflection operator}
(\refl f)(x') \; = \; -\ii \, \overline{f(x')} 
\qquad \text{ for } x' \in \crosssecdual .
\end{align}
This transformation~$\refl$ has the interpretation of a reflection across
the horizontal cross-section; see~\cite[Remark~3.2]{part-1}.

% : if $F \colon \dE(\dstrip) \to \bC$ is s-holomorphic and
% has Riemann boundary values in the discrete strip, then so is the
% ``reflected function''~$z \mapsto -\ii \, \overline{F(\overline{z})}$.
% % The operator~$\refl$ maps the restriction to~$\crosssecdual$ 
% % %the cross section 
% % of the former to the latter.
% % The operator~$\refl$ relates the restrictions of these to
% % the cross section~$\crosssecdual$.
% % %the cross section of the former to the latter.
% The restrictions of these function to~$\crosssecdual$
% are related by~$\refl$.
% %the cross section of the former to the latter.

\bigskip

\section{Clifford algebra valued discrete 1-forms}
\label{sec: operator valued forms}

% *******************************************************************
% **   \section{Clifford algebra valued 1-forms}                   **
% *******************************************************************

In this section we develop discretizations of boundary
conformal field theory contour integration and contour deformation
manipulations \cite{Cardy-surface_critical_behavior,
Cardy-effect_of_boundary_conditions,
Cardy-Verlinde_formula} (see also~\cite{DMS-CFT,Cardy-BCFT_encyclopedia})
relevant for the Ising model on the lattice
strip and slit-strip.\footnote{More general lattice domains could be
considered with only additional complications to the notation.
We nevertheless focus only on the strip and the slit-strip,
where convergence results of distinguished functions
can be used to reconstruct the 
vertex operator algebra stucture in the scaling limit.}

We start this section by briefly reviewing the Clifford algebra action
on the transfer matrix state space, and recall the notion of
discrete holomorphic and antiholomorphic fermions~$\ferhol, \ferbar$
from~\cite{HKZ-discrete_holomorphicity_and_operator_formalism}.
These form a pair of Clifford generator valued functions on
the edges of the lattice strip or lattice slit-strip,
which satisfy complexifications of the 
s-holomorphicity and Riemann boundary values.

The main result in this section is that
if one introduces discrete $1$-forms with values
in the space of Clifford generators, using coefficient
functions for~$\ferhol$ and~$\ferbar$, which form a pair that
is imaginary complexified s-holomorphic (ICSH) and has imaginary 
complexified Riemann boundary values (ICRBV), then the $1$-forms
are closed and have vanishing integrals along vertical boundaries.
This enables contour deformation manipulations exactly analogous
to boundary conformal field theory.

The simplest application of this observation is a reformulation of 
the calculation to diagonalize the Ising transfer matrix
in a strip, which we will present in 
Section~\ref{sub: diagonalization of strip transfer matrix}.
This diagonalization becomes straightforward
by choosing coefficient functions
in the $1$-forms that are eigenfunctions of vertical translations,
as given in Section~\ref{sec: distinguished functions}.
 
A more interesting and novel prospect, however,
is choosing coefficient functions
in the $1$-forms to be 
globally defined functions in the slit-strip, which have
prescribed singularities, see 
Section~\ref{sec: distinguished functions}.
Employing such coefficient functions, the results of this section
will be used to derive a recursive
characterization of the fusion coefficients of the Ising model
in Section~\ref{sec: fusion coefficients}.

\subsection{Clifford algebra action on a state space}

In this section we consider operators and operator valued forms
on the state space of the Ising model transfer matrix formalism.
We use only rudimentary Clifford algebra theory;
e.g., \cite{Palmer-planar_Ising_correlations}
is an appropriate reference.
In this first subsection, we define the state space and
introduce a Clifford algebra action on it.

\subsubsection*{The state space}

The state space for the transfer matrix is associated to the horizontal 
cross-section~$\crosssec = \dinterval{\lft}{\rgt}$
of the lattice strip (or slit-strip).
It has basis vectors~$\basisvec{\spinrow}$
indexed by configurations~$\spinrow \in \set{\pm 1}^{\crosssec}$
of $\pm 1$-spins in a cross-section row~$\crosssec$.
The (full) state space~$\statespdbl$ is defined as the complex vector
space with 
basis~$( \basisvec{\spinrow} )_{\spinrow \in \set{\pm 1}^{\crosssec}}$,
i.e.,
\begin{align}\label{eq: full state space}
\statespdbl 
= \, \bC^{\set{\pm 1}^{\crosssec}} \,
= \; & \spn_\bC \set{ \basisvec{\spinrow} \; \Big| \; 
		\spinrow \in \set{\pm 1}^{\crosssec} } .
\end{align}
At various stages, we employ different subspaces of the
state space~$\statespdbl$. For the purposes of the present section,
the most important is the subspace~$\statesp \subset \statespdbl$
\begin{align}\label{eq: irreducible state space}
\statesp = \spn_\bC \set{ \basisvec{\spinrow} \; \Big| \;
    \spinrow \in \set{\pm 1}^{\crosssec} , \; \spinrow_{\rgt} = +1}
\end{align}
spanned by the basis vectors, whose rightmost spin is~$+1$.
We call~$\statesp$ the irreducible state-space, because it will be
an irreducible representation of the Clifford algebra action below.

We make the state space (and its subspaces)
a complex inner product space in such a way that the basis
vectors~$\basisvec{\spinrow}$, $\spinrow \in \set{\pm 1}^{\crosssec}$
are orthonormal. The conjugate transpose with respect to this basis is denoted 
by superscript~$\dagger$.
In order to clearly distinguish the complex inner 
product on the state space from the real inner products used in the function 
spaces in Section~\ref{sec: distinguished functions}
and~\cite{part-1}, 
we will write the inner product of 
vectors~$\genvec_1, \genvec_2 \in \statespdbl$ in matrix notation as
\begin{align*}
\genvec_1^\dagger \, \genvec_2
    .
\end{align*}
In this notation, orthonormality of the basis in particular amounts to
$\basisvec{\spinrowalt}^\dagger \, \basisvec{\spinrow} 
= \delta_{\spinrowalt,\spinrow}$
(Kronecker delta).

\subsubsection*{Clifford generators}
The 
state space carries a representation of a Clifford algebra. 
We directly choose a convenient basis of generators of the Clifford
algebra for the present purposes by introducing discrete holomorphic and 
discrete antiholomorphic fermion operators on the state space.

For any $x' \in \crosssecdual = \dintervaldual{\lft}{\rgt}$, we define the 
involution
\begin{align*}
\fold{x'} \colon \; & \set{\pm 1}^{\crosssec} \to \set{\pm 1}^{\crosssec}
\end{align*}
which flips the values of the spins to the left of~$x'$:
\begin{align*}
\fold{x'}(\spinrow) = \; & \spinrowfolded ,
\qquad \text{ where } &
\spinrowfolded_x = \; & \begin{cases}
	- \spinrow_x & \text{ if $x<x'$} \\
	\phantom{-} \spinrow_x & \text{ if $x>x'$}.
    \end{cases}
\end{align*}
With the help of this, we define 
linear maps on the state space
\begin{align*}
\ferhol_{x'} , \ferbar_{x'} \colon \statespdbl \to \statespdbl
\end{align*}
by setting their values on the basis vectors $\basisvec{\spinrow}$, $\spinrow 
\in \set{\pm 1}^{\crosssec}$, to be
\begin{align*}
\ferhol_{x'} \basisvec{\spinrow} 
= \frac{-\spinrow_{x'-\half} + \ii \, \spinrow_{x'+\half}}{\sqrt{2}} \, 
        \basisvec{\fold{x'}(\spinrow)}
\qquad \text{ and } \qquad
\ferbar_{x'} \basisvec{\spinrow} 
= \frac{-\ii \, \spinrow_{x'-\half} + \spinrow_{x'+\half}}{\sqrt{2}} \, 
        \basisvec{\fold{x'}(\spinrow)} .
\end{align*}
When $\fold{x'}(\spinrow) = \spinrowfolded$, we have
$\spinrow_{\rgt} = \spinrowfolded_{\rgt}$, so
these are also well-defined linear
maps of the irreducible state space~$\statesp$,
\begin{align*}
\ferhol_{x'} , \ferbar_{x'} \colon \statesp \to \statesp .
\end{align*}
The linear span of these operators,
\begin{align*}
\CliffGen = \spn_\bC \Big( \set{\ferhol_{x'} \; \big| \; x' \in \crosssecdual}
        \cup \set{\ferbar_{x'} \; \big| \; x' \in \crosssecdual} \Big)
\end{align*}
is called the space of Clifford generators.
We can interpret either
$\CliffGen \subset \End(\statespdbl)$ or 
$\CliffGen \subset \End(\statesp)$~---
all relevant properties remain identical in both cases.

\subsubsection*{Properties of the Clifford generators}
On the state space we use the inner product with respect to which the basis 
$(\basisvec{\spinrow})_{\spinrow \in \set{\pm 1}^{\crosssec}}$ is orthonormal. 
The Hilbert space adjoints of the above operators are easily calculated: 
$\ferbar_{x'}$ is self-adjoint and $\ferhol_{x'}$ anti-self-adjoint.
\begin{lem}\label{lem: adjoints of fermions}
For any $x' \in \crosssecdual$, we have
\begin{align*}
(\ferhol_{x'})^\dagger = - \ferhol_{x'} 
\qquad \text{ and } \qquad 
(\ferbar_{x'})^\dagger = \ferbar_{x'} .
\end{align*}
\end{lem}

The \term{anticommutator} of two linear 
operators~$A,B \colon V \to V$ on a vector 
space~$V$ is
\begin{align*}
\anticomm{A}{B} = A B + B A .
\end{align*}
The anticommutators among the operators above are also straightforwardly 
calculated.
\begin{lem}\label{lem: anticommutators of height zero fermions}
For any $x'_1 , x'_2 \in \crosssecdual$, we have
\begin{align*}
\anticomm{\ferhol_{x'_1}}{\ferhol_{x'_2}}
    = - 2 \, \delta_{x'_1 , x'_2} \, \id , \qquad
\anticomm{\ferbar_{x'_1}}{\ferbar_{x'_2}}
    = + 2 \, \delta_{x'_1 , x'_2} \, \id , \qquad
\anticomm{\ferhol_{x'_1}}{\ferbar_{x'_2}}
    = 0 .
\end{align*}
\end{lem}
This explicitly shows that 
the anticommutator defines a nondegenerate bilinear form
$(\cdot, \cdot)$ on~$\CliffGen$
via
\[ \anticomm{\phi_1}{\phi_2} = (\phi_1,\phi_2) \; \id
\qquad \text{for $\phi_1 , \phi_2 \in \CliffGen$.}
\]

It follows that the subalgebra 
$\Cliff \subset \End(\statesp)$ generated by 
the elements of~$\CliffGen$ is a Clifford algebra of dimension 
${\dmn(\Cliff) = 2^{2 \width}}$.

A pair of subspaces
$ \Creation , \Annihilation \subset \CliffGen $
is said to be a polarization if
\[ \CliffGen = \Creation \oplus \Annihilation \]
and
\[ \anticomm{\Creation}{\Creation} = 0 , \qquad
\anticomm{\Annihilation}{\Annihilation} = 0 .
\]
Given any polarization, the unique irreducible representation of 
the Clifford algebra~$\Cliff$ can be identified with
the exterior algebra of the 
subspace~$\Creation \subset \CliffGen $,
\[ \bigwedge \Creation 
    = \bigoplus_{d=0}^{\width} \wedge^d \, \Creation , \]
and the one-dimensional subspace
$\wedge^0 \, \Creation \subset \bigwedge \Creation$
within this exterior algebra
consists of vectors annihilated simultaneously by all
of~$\Annihilation$. A non-zero vector with this property is called a
vacuum with respect to the polarization.
The unique irreducible representation in particular has
dimension~$2^{\width}$, and consequently the 
state space~$\statesp$ must be
isomorphic to this irreducible representation.

\begin{rmk}
Note that although we have the direct sum decomposition
\[ \CliffGen = \spn_\bC \set{\ferhol_{x'}\; \big| \; x' \in \crosssecdual} 
    \, \oplus \, \spn_\bC \set{\ferbar_{x'}\; \big| \; x' \in \crosssecdual} ,
\]
this pair of subspaces is \underline{not} a polarization~--- the 
reason for our choice of basis was altogether different;
see Section~\ref{sub: fermions in the strip},
Proposition~\ref{prop: complexified SH and RBV for fermions},
in particular.
\end{rmk}

\subsection{The strip transfer matrix and fermions in the lattice strip}
For the purposes of defining a physically relevant Clifford algebra
generator valued functions on the lattice strip, we need
the Ising transfer matrix.
We fix a parameter
\begin{align*}
\invtemp \; = \; \invtempcrit \, = \; \frac{1}{2} \log (\sqrt{2} + 1) ,
\end{align*}
the critical inverse temperature of the Ising model on the square 
grid.\footnote{The transfer matrices make sense and can be used
at any inverse temperature~$\beta \ge 0$, but the discrete complex
analysis properties are specific to the critical point.}

The transfer matrix is a $\bC$-linear map
\begin{align*} 
\Tmat \colon \statespdbl \to \statespdbl
\end{align*}
constructed out of two constituent matrices, a diagonal matrix
$\TmatHorSqrt \colon \statespdbl \to \statespdbl$
(accounting for the Ising model interactions on horizontal
edges, with weight half), and a symmetric matrix 
$\TmatVer \colon \statespdbl \to \statespdbl$
(accounting for the Ising model interactions on vertical edges as well as
locally monochromatic boundary conditions on the left and right boundaries).
The matrix~$\TmatHorSqrt$ is defined by the matrix elements
\begin{align}\label{eq: horsqrt transfer matrix entries}
\basisvec{\spinrowalt}^\dagger \TmatHorSqrt \, \basisvec{\spinrow}
\; = \; \delta_{\spinrowalt,\spinrow} \;
        \exp \Big( \frac{\invtemp}{2} 
                \sum_{x=\lft}^{\rgt-1} \spinrow_{x} \spinrow_{x+1} \Big) ,
\end{align}
and the matrix~$\TmatVer$ is defined by the matrix elements
\begin{align}\label{eq: ver transfer matrix entries}
\basisvec{\spinrowalt}^\dagger \TmatVer \, \basisvec{\spinrow}
\; = \; \exp \Big( \invtemp
                \sum_{x=\lft}^{\rgt} \spinrow_{x} \spinrowalt_{x} \Big)
        \; \delta_{\spinrowalt_{\lft},\spinrow_{\lft}}
        \, \delta_{\spinrowalt_{\rgt},\spinrow_{\rgt}} .
\end{align}
The transfer matrix $\Tmat$ is then defined as
\begin{align}\label{eq: Ising transfer matrix}
\Tmat \; = \; \TmatHorSqrt \, \TmatVer \, \TmatHorSqrt \, .
\end{align}

Conjugation by the transfer matrix preserves the space of Clifford 
generators.\footnote{For conjugation, we already use the simple
fact that the transfer matrix~$\Tmat$ is invertible. We explicitly
state this later along with other well-known properties, in
Theorem~\ref{thm: transfer matrix properties strip}(i).}
This is made explicit in the following.
\begin{prop}\label{prop: induced rotation}
Denote $\lambda = \frac{1+\ii}{\sqrt{2}}$.
For $x' \in \crosssecdual$, we have
\begin{align*}
\Tmat^{-1} \ferhol_{x'} \Tmat
    = \; & \begin{cases}
        (1+\frac{1}{\sqrt{2}}) \ferhol_{\lftd} 
        + (\lambda^3+\frac{\lambda^{-3}}{\sqrt{2}}) \ferbar_{\lftd} 
        + \frac{\lambda^3}{\sqrt{2}} \ferhol_{\lftd+1}
        + \frac{1}{\sqrt{2}} \ferbar_{\lftd+1}
            & \text{ if } x' = \lftd \\
        2 \ferhol_{x'} 
        - \sqrt{2} \ferbar_{x'} 
        + \frac{\lambda^3}{\sqrt{2}} \ferhol_{x'+1}
        + \frac{1}{\sqrt{2}} \ferbar_{x'+1}
        + \frac{\lambda^{-3}}{\sqrt{2}} \ferhol_{x'-1}
        + \frac{1}{\sqrt{2}} \ferbar_{x'-1}
            & \text{ if } x' \neq \lftd, \rgtd \\
        (1+\frac{1}{\sqrt{2}}) \ferhol_{\rgtd} 
        + (\lambda^{-3}+\frac{\lambda^{3}}{\sqrt{2}}) \ferbar_{\rgtd} 
        + \frac{\lambda^{-3}}{\sqrt{2}} \ferhol_{\rgtd-1}
        + \frac{1}{\sqrt{2}} \ferbar_{\rgtd-1}
            & \text{ if } x' = \rgtd
      \end{cases} \\
\Tmat^{-1} \ferbar_{x'} \Tmat
    = \; & \begin{cases}
        (1+\frac{1}{\sqrt{2}}) \ferbar_{\lftd} 
        + (\lambda^{-3}+\frac{\lambda^{3}}{\sqrt{2}}) \ferhol_{\lftd} 
        + \frac{\lambda^{-3}}{\sqrt{2}} \ferbar_{\lftd+1}
        + \frac{1}{\sqrt{2}} \ferhol_{\lftd+1}
            & \text{ if } x' = \lftd \\
        2 \ferbar_{x'} 
        - \sqrt{2} \ferhol_{x'} 
        + \frac{\lambda^{-3}}{\sqrt{2}} \ferbar_{x'+1}
        + \frac{1}{\sqrt{2}} \ferhol_{x'+1}
        + \frac{\lambda^{3}}{\sqrt{2}} \ferbar_{x'-1}
        + \frac{1}{\sqrt{2}} \ferhol_{x'-1}
            & \text{ if } x' \neq \lftd, \rgtd \\
        (1+\frac{1}{\sqrt{2}}) \ferbar_{\rgtd} 
        + (\lambda^{3}+\frac{\lambda^{-3}}{\sqrt{2}}) \ferhol_{\rgtd} 
        + \frac{\lambda^{3}}{\sqrt{2}} \ferbar_{\rgtd-1}
        + \frac{1}{\sqrt{2}} \ferhol_{\rgtd-1}
            & \text{ if } x' = \rgtd .
      \end{cases}
\end{align*}
\end{prop}
\begin{proof}
The proof is an explicit calculation. The details of the case of general 
inverse temperature~$\invtemp$ can be found 
in~\cite{HKZ-discrete_holomorphicity_and_operator_formalism}, and upon 
specializing to the critical 
case~$\invtemp = \invtempcrit$, they yield the above formulas.
\end{proof}

\subsubsection*{Holomorphic and antiholomorphic fermions in the strip}
\label{sub: fermions in the strip}

We now define two Clifford algebra generator valued functions on the midpoints 
of the edges of the discrete strip~$\dstrip$ of
Figure~\ref{sfig: lattice strip}.
The horizontal edges at height zero are naturally 
identified with points~$x' \in \crosssecdual$ of the dual cross-section, and 
horizontal edges at height~$y \in \bZ$ are identified with points~$x' + \ii y$ 
with~$x' \in \crosssecdual$. 
On such horizontal edges we define
\begin{align}\label{eq: def fermions on horizontal edges}
\ferhol(x'+\ii y) = \Tmat^{-y} \, \ferhol_{x'} \, \Tmat^{y}
\qquad \text{ and } \qquad 
\ferbar(x'+\ii y) = \Tmat^{-y} \, \ferbar_{x'} \, \Tmat^{y} .
\end{align}
At zero height, we therefore simply have
$\ferhol(x') = \ferhol_{x'}$ and $\ferbar(x') = \ferbar_{x'}$,
and extending to other heights~$y$ is done as usual
in the transfer matrix formalism;
compare, e.g., with~\eqref{eq: height dependent spin operator in the strip}.
It follows from Proposition~\ref{prop: induced rotation} 
that we have $\ferhol(x'+\ii y), \ferbar(x'+\ii y) \in \CliffGen$
in general.

Vertical edges of the strip~$\dstrip$, likewise, can be identified with
their midpoints, which are of the form
$x + \ii y'$, with $x \in \crosssec$
and $y' \in \bZ + \frac{1}{2}$. The vertical edges 
with~$x = \lft$ constitute the left boundary of the strip~$\dstrip$,
and those with~$x = \rgt$ constitute the right boundary.
The following proposition summarizes the
key principle behind our 
extension of the definitions of~$\ferhol$ and
$\ferbar$ from horizontal edges to vertical edges.

\begin{prop}
\label{prop: complexified SH and RBV for fermions}
There exists unique extensions
\begin{align*}
\ferhol , \ferbar \colon \dE(\dstrip) \to \CliffGen
\end{align*}
of~\eqref{eq: def fermions on horizontal edges} to
vertical edges with the following properties:

The pair~$(\ferhol,\ferbar)$ is 
\term{complexified s-holomorphic (CSH)} in the sense that
for any edges~$z_1, z_2 \in \dE(\dstrip)$
adjacent to a vertex~$v$ and a face~$p$ we have
\begin{align}
\label{eq: CSH of fermions}
\ferhol(z_1) + \frac{\ii \, |v-p|}{v-p} \, \ferbar(z_1)
= \ferhol(z_2) + \frac{\ii \, |v-p|}{v-p} \, \ferbar(z_2) ,
\end{align}
and it has \term{complexified Riemann boundary values (CRBV)}
in the sense that
\begin{align}
\label{eq: CRBV of fermions}
\ferhol(\lftsym) + \ii \, \ferbar(\lftsym) = \; & 0 &
\ferhol(\rgtsym) - \ii \, \ferbar(\rgtsym) = \; & 0 .
\end{align}
for $\rgtsym = \rgt + \ii y'$,
$\lftsym = \lft + \ii y'$ with~$y' \in \bZ + \half$.
\end{prop}

We will prove this statement by giving a number of formulas
for the fermions on vertical edges in terms of those on horizontal
edges, and showing that the various formulas agree.
There will be differences 
between the treatment of the boundary edges and the rest,
stemming ultimately from the cases in
Proposition~\ref{prop: induced rotation}.

Before delving into the details, let us emphasize the following.
Having these discrete complex analysis 
properties, \eqref{eq: CSH of fermions} and~\eqref{eq: CRBV of fermions},
valid simultaneously with the 
propagation~\eqref{eq: def fermions on horizontal edges}, is
a nontrivial property of the transfer matrix~$\Tmat$
of the critical Ising model and of the chosen basis
$\set{\ferhol_{x'} \, \big| \, x \in \dintervaldual{\lft}{\rgt}} \cup
\set{\ferbar_{x'} \, \big| \, x \in \dintervaldual{\lft}{\rgt}}$
of the Clifford generators.
The relationship of the transfer matrix formalism 
to discrete complex analysis 
was observed in~\cite{HKZ-discrete_holomorphicity_and_operator_formalism}.

Let us then look at the calculations needed
for the extension of the fermions to vertical edges.

\begin{figure}[tb]
\centering
\subfigure[The horizontal edges adjacent to a given vertical edge.]
{
    \includegraphics[width=.35\textwidth]{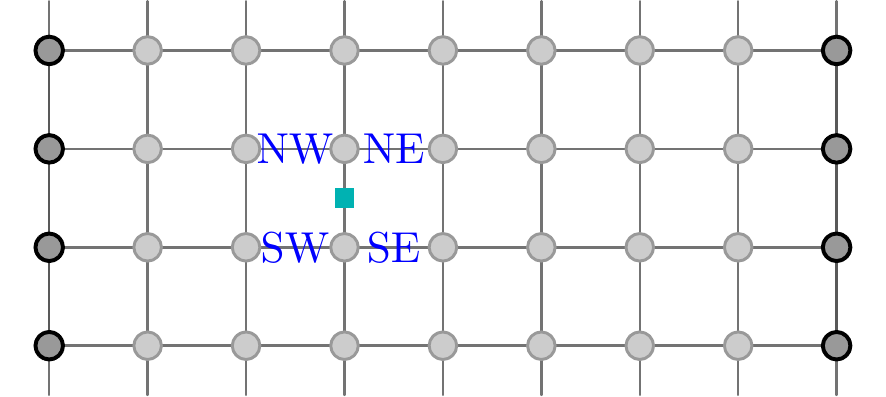}
	\label{sfig: neighbors to a vertical edge}
}
\hspace{2.5cm}
\subfigure[The edges surrounding a plaquette.]
{
    \includegraphics[width=.35\textwidth]{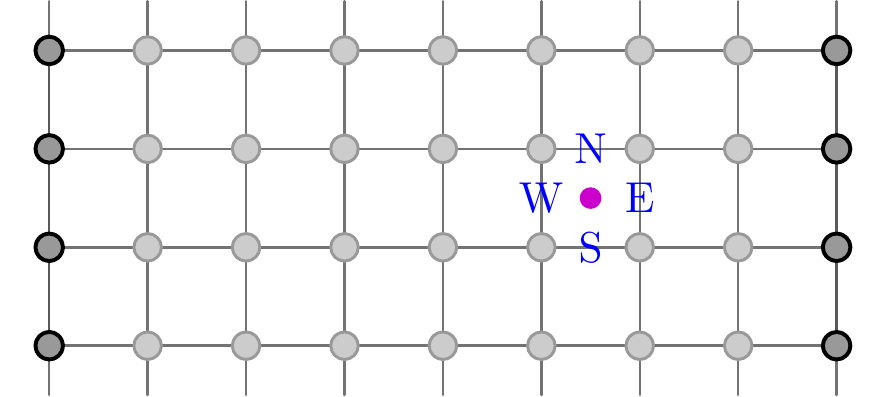}
	\label{sfig: neighbors to a plaquette}
}
\caption{Illustrations of the notation for neighboring edges.
}
\label{fig: neighbors by directions}
\end{figure}

Let $x \in \crosssec$
and $y' \in \bZ + \frac{1}{2}$.
Consider the vertical edge $z = x + \ii y'$, and denote
\begin{align*}
\NW = \; & z -\frac{1}{2}+\frac{\ii}{2} &
\NE = \; & z +\frac{1}{2}+\frac{\ii}{2} \\
\SW = \; & z -\frac{1}{2}-\frac{\ii}{2} &
\SE = \; & z +\frac{1}{2}-\frac{\ii}{2} 
\end{align*}
as in 
Figure~\ref{sec: operator valued forms}.\ref{sfig: neighbors to a vertical edge}.
Note that $\NW$ and $\SW$ are horizontal edges of~$\dstrip$ 
whenever~$x \neq \lft$, and $\NE$ and $\SE$
are horizontal edges of~$\dstrip$ whenever~$x \neq \rgt$.

Denote $\lambda = \frac{1+\ii}{\sqrt{2}}$.
If~$x \neq \rgt$, let
\begin{align}
\label{eq: ferhol vertical edge right side expression}
\ferhol_{(\Est)}(z) := & \;
  \frac{\lambda}{\sqrt{2}} \ferhol (\NE)
 +\frac{1}{\sqrt{2}} \ferbar (\NE)
 -\frac{\lambda^3}{\sqrt{2}} \ferhol (SE)
 -\frac{1}{\sqrt{2}} \ferbar (\SE) , \\
\label{eq: ferbar vertical edge right side expression}
\ferbar_{(\Est)}(z) := & \;
  \frac{\lambda^{-1}}{\sqrt{2}} \ferbar (\NE)
 +\frac{1}{\sqrt{2}} \ferhol (\NE)
 -\frac{\lambda^{-3}}{\sqrt{2}} \ferbar (\SE)
 -\frac{1}{\sqrt{2}} \ferhol (\SE) ,
\end{align}
and if~$x \neq \lft$, let
\begin{align}
\label{eq: ferhol vertical edge left side expression}
\ferhol_{(\Wst)}(z) := & \;
\frac{\lambda^{-1}}{\sqrt{2}} \ferhol (\NW)
+\frac{1}{\sqrt{2}} \ferbar (\NW)
-\frac{\lambda^{-3}}{\sqrt{2}} \ferhol (\SW)
-\frac{1}{\sqrt{2}} \ferbar (\SW) \\
\label{eq: ferbar vertical edge left side expression}
\ferbar_{(\Wst)}(z) := & \;
  \frac{\lambda}{\sqrt{2}} \ferbar (\NW)
 +\frac{1}{\sqrt{2}} \ferhol (\NW)
 -\frac{\lambda^{3}}{\sqrt{2}} \ferbar (\SW)
 -\frac{1}{\sqrt{2}} \ferhol (\SW) .
\end{align}

If $x=\lft$, let also
\begin{align}
\label{eq: ferhol left boundary vertical edge expression}
\ferhol_{(\lftsym)}(z) := & \;
   \frac{1-\lambda}{2\sqrt{2}} \Big(
    \lambda \, \ferhol (\NE)
    + \ferbar (\NE) \Big)
 + \frac{1-\lambda^{-1}}{2\sqrt{2}} \Big(
    \lambda^{-1} \, \ferhol (\SE)
    - \ferbar (\SE) \Big) \\
\label{eq: ferbar left boundary vertical edge expression}
\ferbar_{(\lftsym)}(z) := & \;
   \frac{1-\lambda^{-1}}{2\sqrt{2}} \Big(
    \lambda^{-1} \, \ferbar (\NE)
    + \ferhol (\NE) \Big)
 + \frac{1-\lambda}{2\sqrt{2}} \Big(
    \lambda \, \ferbar (\SE)
    - \ferhol (\SE) \Big) ,
\end{align}
and if $x=\rgt$, let also
\begin{align}
\label{eq: ferhol right boundary vertical edge expression}
\ferhol_{(\rgtsym)}(z) := & \;
   \frac{1-\lambda^{-1}}{2\sqrt{2}} \Big(
    \lambda^{-1} \, \ferhol (\NW)
    + \ferbar (\NW) \Big)
 + \frac{1-\lambda}{2\sqrt{2}} \Big(
    \lambda \, \ferhol (\SW)
    - \ferbar (\SW) \Big) \\
\label{eq: ferbar right boundary vertical edge expression}
\ferbar_{(\rgtsym)}(z) := & \;
   \frac{1-\lambda}{2\sqrt{2}} \Big(
    \lambda \, \ferbar (\NW)
    + \ferhol (\NW) \Big)
 + \frac{1-\lambda^{-1}}{2\sqrt{2}} \Big(
    \lambda^{-1} \, \ferbar (\SW)
    - \ferhol (\SW) \Big) .
\end{align}
\begin{lem}
\label{lem: equality of various expressions for vertical edge fermions}
If $x \neq \lft, \rgt$, then at the vertical edge $z = x + \ii y'$
the following equalities hold:
\begin{align*}
\ferhol_{(\Est)}(z) = \; & \ferhol_{(\Wst)}(z) , &
\ferbar_{(\Est)}(z) = \; & \ferbar_{(\Wst)}(z) .
\end{align*}
If $x=\lft$, then at the edge $z = x + \ii y'$ of the left boundary
the following equalities hold:
\begin{align*}
\ferhol_{(\Est)}(z) = \; & \ferhol_{(\lftsym)}(z) , &
\ferbar_{(\Est)}(z) = \; & \ferbar_{(\lftsym)}(z) .
\end{align*}
If $x=\rgt$, then at the edge $z = x + \ii y'$ of the right boundary
the following equalities hold:
\begin{align*}
\ferhol_{(\Wst)}(z) = \; & \ferhol_{(\rgtsym)}(z) , &
\ferbar_{(\Wst)}(z) = \; & \ferbar_{(\rgtsym)}(z) .
\end{align*}
\end{lem}
\begin{proof}
The last two terms in each of the expressions
\eqref{eq: ferhol vertical edge right side expression}~--~
\eqref{eq: ferbar right boundary vertical edge expression}
involve horizontal edge 
fermions in the row at height~$y'-\half$, while the first two terms 
correspondingly involve horizontal edge fermions in the row above,
at height~$y'+\half$.
Using the definition~\eqref{eq: def fermions on horizontal edges}
and the explicit expression of conjugation of Clifford generators
by~$\Tmat$ in Proposition~\ref{prop: induced rotation}, we may write
the fermions at height~$y'+\half$ in terms of those at height~$y'-\half$.

For~$x \neq \lft, \rgt$, we explicitly find that 
\eqref{eq: ferhol vertical edge right side expression}
and~\eqref{eq: ferhol vertical edge left side expression} are both equal to
\begin{align*}
  \frac{\lambda^{-1}}{\sqrt2} \, \ferhol (\SW)
+ \frac{\lambda}{\sqrt2} \, \ferhol (\SE)
+ \frac{\ii}{\sqrt2} \, \ferbar (\SW)
- \frac{\ii}{\sqrt2} \, \ferbar (\SE) ,
\end{align*}
and that \eqref{eq: ferhol vertical edge right side expression}
and~\eqref{eq: ferhol vertical edge left side expression} are both equal to
\begin{align*}
  \frac{\lambda}{\sqrt2} \, \ferbar (\SW)
+ \frac{\lambda^{-1}}{\sqrt2} \, \ferbar (\SE)
- \frac{\ii}{\sqrt2} \, \ferhol (\SW)
+ \frac{\ii}{\sqrt2} \, \ferhol (\SE) ,
\end{align*}
proving the first two asserted equalities.

Similarly, for~$x = \lft$, we find that 
\eqref{eq: ferhol vertical edge right side expression}
and \eqref{eq: ferhol left boundary vertical edge expression} 
are both equal to
\begin{align*}
(\ii+\lambda^{-1}) \ferhol (\SE)
+ (-1+\lambda^{-1}) \ferbar (\SE) ,
\end{align*}
while
\eqref{eq: ferbar vertical edge right side expression}
and \eqref{eq: ferbar left boundary vertical edge expression} 
are both equal to
\begin{align*}
(-\ii+\lambda) \ferbar (\SE)
+ (-1+\lambda) \ferhol (\SE) .
\end{align*}

Finally, for~$x = \rgt$, we find that 
\eqref{eq: ferhol vertical edge left side expression}
and \eqref{eq: ferhol right boundary vertical edge expression} 
are both equal to
\begin{align*}
(-\ii+\lambda) \ferhol (\SW)
+ (-1+\lambda) \ferbar (\SW) ,
\end{align*}
while
\eqref{eq: ferbar vertical edge left side expression}
and \eqref{eq: ferbar right boundary vertical edge expression} 
are both equal to
\begin{align*}
(\ii+\lambda^{-1}) \ferbar (\SW)
+ (-1+\lambda^{-1}) \ferhol (\SW) .
\end{align*}
These direct calculations prove the assertion.
\end{proof}

\begin{proof}[Proof of 
Proposition~\ref{prop: complexified SH and RBV for fermions}]
The uniqueness of the extension of $\ferhol, \ferbar$ to vertical
edges is clear: Equations~\eqref{eq: CSH of fermions}
and~\eqref{eq: CRBV of fermions} can be used to solve for the 
values~$\ferhol(z)$ and $\ferbar(z)$
on a vertical edge~$z$ in terms of the fermions on any two
adjacent horizontal edges. 
Equations~\eqref{eq: ferhol vertical edge right side expression}
~--~\eqref{eq: ferbar right boundary vertical edge expression}
are (some of) the expressions thus obtained.

Conversely, since the various expressions agree by
Lemma~\ref{lem: equality of various expressions for vertical edge fermions},
we can 
use~\eqref{eq: ferhol vertical edge right side expression}~--~
\eqref{eq: ferbar right boundary vertical edge expression}
to extend the definition of the holomorphic and antiholomorphic fermions
$\ferhol, \ferbar$ from horizontal 
edges~\eqref{eq: def fermions on horizontal edges}
to the vertical edges at heights ${y' \in \bZ + \half}$, by
setting
\begin{align*}
\ferhol(x+\ii y') 
  = \; & \text{\eqref{eq: ferhol vertical edge right side expression}} 
  =      \text{\eqref{eq: ferhol left boundary vertical edge expression}} , &
\ferbar(x+\ii y') 
  = \; & \text{\eqref{eq: ferbar vertical edge right side expression}} 
  =      \text{\eqref{eq: ferbar left boundary vertical edge expression}} , &
& \text{if $x = \lft$} \\
\ferhol(x+\ii y') 
  = \; & \text{\eqref{eq: ferhol vertical edge right side expression}}
  =      \text{\eqref{eq: ferhol vertical edge left side expression}} , &
\ferbar(x+\ii y') 
  = \; & \text{\eqref{eq: ferbar vertical edge right side expression}}
  =      \text{\eqref{eq: ferbar vertical edge left side expression}} , &
& \text{if $x \in \dinterval{\lft+1}{\rgt-1}$} \\
\ferhol(x+\ii y') 
  = \; & \text{\eqref{eq: ferhol vertical edge left side expression}}
  =      \text{\eqref{eq: ferhol right boundary vertical edge expression}} , &
\ferbar(x+\ii y') 
  = \; & \text{\eqref{eq: ferbar vertical edge left side expression}}
  =      \text{\eqref{eq: ferbar right boundary vertical edge expression}} , &
& \text{if $x = \rgt$}.
\end{align*}
It is straightforward to verify that the equalities of the various
expressions above are equivalent to the discrete complex analysis
properties~\eqref{eq: CSH of fermions}
and~\eqref{eq: CRBV of fermions}.
\end{proof}

\subsection{Clifford algebra valued $1$-forms in the strip}
\label{sub: one-forms in the strip}

We next define Clifford generator valued discrete 
$1$-forms and their discrete line integrals.
The crucial observation is that such forms are closed when their coefficient 
functions satisfy a property closely related to s-holomorphicity, and they have 
vanishing integrals along vertical boundaries when the coefficient functions 
satisfy a property closely related to Riemann boundary values.

\subsubsection*{Clifford algebra generator valued $1$-forms}
By a Clifford generator valued discrete $1$-form on
the lattice strip~$\dstrip$
we mean a formal expression
\begin{align*}
\cffunhol(z) \, \ferhol(z) \, \dd{z} 
+ \cffunbar(z) \, \ferbar(z) \, \dd{\bar{z}} ,
\end{align*}
where $\cffunhol, \cffunbar \colon \dE(\dstrip) \to \bC$ are two
complex-valued functions on the edges of the strip.
A discrete contour
on~$\dstrip$ is an ordered finite sequence 
$\contour = (w_0,w_1,\ldots,w_m)$ of vertices
$w_0,w_1,\ldots,w_m \in \dV(\dstrip)$
such that $z_j = \edgeof{w_{j-1}}{w_j} \in \dE(\dstrip)$ for all 
$j=1,\ldots,m$.
We define the discrete contour integral along~$\contour$ of a Clifford 
generator valued $1$-form as
\begin{align*}
& \dcint{\contour} \Big( \cffunhol(z) \, \ferhol(z) \, \dd{z} 
    + \cffunbar(z) \, \ferbar(z) \, \dd{\bar{z}} \Big) \\
:= \; & \sum_{j=1}^m \Big( 
  \cffunhol(z_j) \, \ferhol(z_j) \, \big( w_j - w_{j-1} \big)
  + \cffunbar(z_j) \, \ferbar(z_j) \, \big(\cconj{w}_j - \cconj{w}_{j-1} \big)   
  \Big)
\; \in \; \CliffGen .
\end{align*}
The integral and differential notations~$\dcint{}$ and~$\dd{}$ 
are meant to emphasize that integration is done on the square grid.
Cases where integrals are taken along discrete contours that
start and end at the same vertex are
furthermore highlighted using the notation~$\dcoint{}$.

\subsubsection*{Closed $1$-forms}
\begin{figure}[tb]
    \centering
    \includegraphics[width=.35\textwidth]{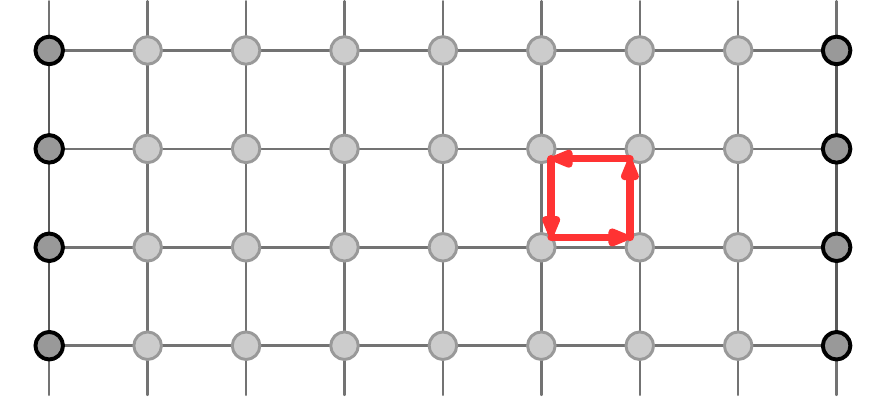}
	\label{sfig: bulk plaquette integral}
    \caption{A counterclockwise oriented plaquette contour.}
    \label{fig: integrals around plaquettes}
\end{figure}
If $w_1,w_2,w_3,w_4 \in \dV(\dstrip)$ are the vertices of a square face
of~$\dstrip$ in counterclockwise order, then
the discrete contour $\plaquettecontour = (w_0,w_1,w_2,w_3,w_4)$ with $w_0=w_4$
is said to be a counterclockwise oriented plaquette of~$\dstrip$;
see Figure~\ref{fig: integrals around plaquettes}.
We say that a Clifford generator valued discrete $1$-form
$\cffunhol(z) \, \ferhol(z) \, \dd{z}
+ \cffunbar(z) \, \ferbar(z) \, \dd{\bar{z}}$
is \term{closed} if for all 
counterclockwise oriented plaquettes~$\plaquettecontour$ 
of~$\dstrip$ we have
\begin{align}\label{eq: vanishing integral around plaquette}
& \dcoint{\plaquettecontour} \Big( \cffunhol(z) \, \ferhol(z) \, \dd{z} 
    + \cffunbar(z) \, \ferbar(z) \, \dd{\bar{z}} \Big) = 0 .
\end{align}

It turns out that a simple discrete complex analysis property of
the coefficient functions ensures the closedness of the one-form.
We say that a pair $(\cffunhol,\cffunbar)$ of 
functions~$\dE(\dstrip) \to \bC$ is
\term{imaginary complexified s-holomorphic}
\term{(ICSH)}, if whenever $z_1 , z_2 \in \dE$ are edges adjacent
to a vertex~$v$ and a face~$p$, we have
\begin{align}\label{eq: ICSH}
\cffunhol(z_1) - \frac{\ii \, |v-p|}{v-p} \, \cffunbar(z_1)
= \cffunhol(z_2) - \frac{\ii \, |v-p|}{v-p} \, \cffunbar(z_2) .
\end{align}

The terminology is explained by the following.
\begin{lem}\label{lem: SH and ICSH}
If $F \colon \dE(\dstrip) \to \bC$ is s-holomorphic and 
we set 
\begin{align*}
\cffunhol = \ii \, F 
\qquad \text{ and } \qquad
\cffunbar = - \ii \, \cconj{F} , 
\end{align*}
then $(\cffunhol,\cffunbar)$ is ICSH.
\end{lem}
\begin{proof}
This follows directly from the definition~\eqref{eq: s-holomorphicity}
of s-holomorphicity.
\end{proof}

The relevance of ICSH pairs stems from the following result, whose
proof essentially exploits the same algebraic relations as 
the quintessential trick with s-holomorphic functions:
the ``(well-definedness of the)
imaginary part of the integral of the square''~\cite{CS-universality_in_Ising}.
\begin{prop}\label{prop: ICSH implies closed}
If $(\cffunhol,\cffunbar)$ is ICSH, then the Clifford generator valued 
discrete $1$-form
$\cffunhol(z) \, \ferhol(z) \, \dd{z}
+ \cffunbar(z) \, \ferbar(z) \, \dd{\bar{z}}$
is closed.
\end{prop}
\begin{proof}
Consider a plaquette centered at face~$p$,
with vertices~$v_1, v_2, v_3, v_4 = v_0$
in counterclockwise orientation.
For $j=1,\ldots,4$, let $z_j = \frac{v_{j-1} + v_j}{2}$ denote the
edge from~$v_{j-1}$ to~$v_j$, and use again the cyclic
interpretation~$z_0 = z_4$.
The integral around the plaquette is
\begin{align*}
& \dcoint{\plaquettecontour} \Big( \cffunhol(z) \, \ferhol(z) \, \dd{z} 
    + \cffunbar(z) \, \ferbar(z) \, \dd{\bar{z}} \Big) \\
= \; & \sum_{j=1}^{4} \Big( \cffunhol(z_j) \ferhol(z_j) \, (v_j - v_{j-1})
    + \cffunbar(z_j) \ferbar(z_j) \, \overline{(v_j - v_{j-1})} \Big) \\
= \; & \sum_{j=0}^{3} \Big( 
    (v_j-p) \, \big( \cffunhol(z_j) \ferhol(z_j) 
            - \cffunhol(z_{j+1}) \ferhol(z_{j+1}) \big) \\
& \qquad + \overline{(v_j-p)} \, \big( \cffunbar(z_j) \ferbar(z_j) 
            - \cffunbar(z_{j+1}) \ferbar(z_{j+1}) \big) \Big) .
\end{align*}

Regarding the first of the two terms in the sum,
note that we can write
\begin{align*}
& \cffunhol(z_j) \ferhol(z_j) - \cffunhol(z_{j+1}) \ferhol(z_{j+1}) \\
= \; & \frac{1}{2} \Big( 
    \underbrace{\big( \cffunhol(z_{j}) - \cffunhol(z_{j+1}) \big) }_{
          \frac{\ii |v_j-p|}{v_j-p}
          ( \cffunbar(z_{j}) - \cffunbar(z_{j+1}) ) }
        \big( \ferhol(z_{j}) + \ferhol(z_{j+1}) \big)
  + \big( \cffunhol(z_{j}) + \cffunhol(z_{j+1}) \big) 
        \underbrace{\big( \ferhol(z_{j}) - \ferhol(z_{j+1}) \big) }_{
        \frac{- \ii |v_j-p|}{v_j-p}
          ( \ferbar(z_{j}) - \ferbar(z_{j+1}) ) }
    \Big) ,
\end{align*}
where the last expression indicates how the assumed ICSH 
equation~\eqref{eq: ICSH}
for the coefficient functions and the 
CSH equation~\eqref{eq: CSH of fermions}
for the fermions are used next. Namely, with this, we simplify 
one contribution to the integral around the plaquette,
\begin{align*}
& \sum_{j=0}^{3} 
    (v_j-p) \, \big( \cffunhol(z_j) \ferhol(z_j) 
            - \cffunhol(z_{j+1}) \ferhol(z_{j+1}) \big) \\
= \; & \sum_{j=0}^{3} \frac{\ii \, |v_j - p|}{2}
    \Big( 
    \big( \cffunbar(z_{j}) - \cffunbar(z_{j+1}) \big)
        \big( \ferhol(z_{j}) + \ferhol(z_{j+1}) \big) \\
  & \qquad\qquad\qquad - \big( \cffunhol(z_{j}) + \cffunhol(z_{j+1}) \big) 
        \big( \ferbar(z_{j}) - \ferbar(z_{j+1}) \big)
    \Big) \\
= \; &  \frac{\ii}{2\sqrt{2}} \sum_{j=0}^{3}
    \Big( \cffunbar(z_{j}) \ferhol(z_{j})
        + \cffunbar(z_{j}) \ferhol(z_{j+1})
        - \cffunbar(z_{j+1}) \ferhol(z_{j})
        - \cffunbar(z_{j+1}) \ferhol(z_{j+1}) \\
  & \qquad\qquad - \cffunhol(z_{j}) \ferbar(z_{j})
    + \cffunhol(z_{j}) \ferbar(z_{j+1})
    - \cffunhol(z_{j+1}) \ferbar(z_{j})
    + \cffunhol(z_{j+1}) \ferbar(z_{j+1}) \Big) \\
= \; &  \frac{\ii}{2\sqrt{2}} \sum_{j=0}^{3}
    \Big( \cffunbar(z_{j}) \ferhol(z_{j+1})
        - \cffunbar(z_{j+1}) \ferhol(z_{j})
        + \cffunhol(z_{j}) \ferbar(z_{j+1})
        - \cffunhol(z_{j+1}) \ferbar(z_{j}) \Big) ,
\end{align*}
where in the last step we observed telescopic cancellations.

By similar calculations the other contribution to the integral
can be written as
\begin{align*}
& \sum_{j=0}^{3} 
    \overline{(v_j-p)} \, \big( \cffunbar(z_j) \ferbar(z_j) 
            - \cffunbar(z_{j+1}) \ferbar(z_{j+1}) \big) \\
= \; &  \frac{\ii}{2\sqrt{2}} \sum_{j=0}^{3}
    \Big( - \cffunbar(z_{j}) \ferhol(z_{j+1})
        + \cffunbar(z_{j+1}) \ferhol(z_{j})
        - \cffunhol(z_{j}) \ferbar(z_{j+1})
        + \cffunhol(z_{j+1}) \ferbar(z_{j}) \Big) .
\end{align*}
We then see that the two contributions cancel, yielding
the desired conclusion
\begin{align*}
& \dcoint{\plaquettecontour} \Big( \cffunhol(z) \, \ferhol(z) \, \dd{z} 
    + \cffunbar(z) \, \ferbar(z) \, \dd{\bar{z}} \Big) = 0 .
\end{align*}
\end{proof}

\subsubsection*{Vertically slidable $1$-forms}
\begin{figure}[tb]
\centering
    \includegraphics[width=.35\textwidth]{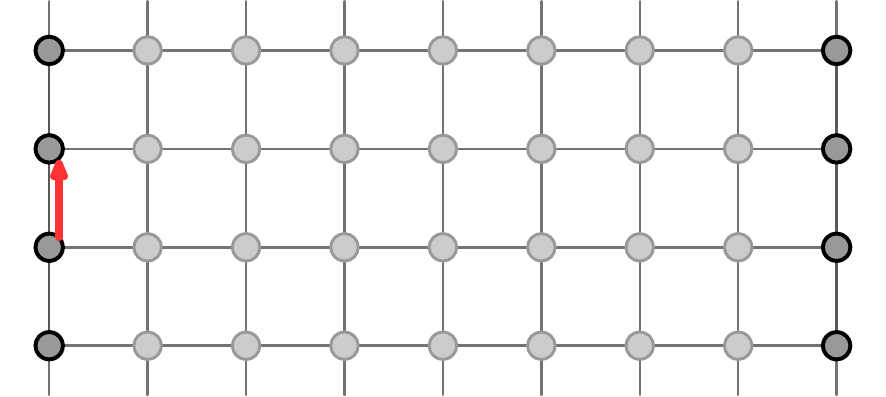}
    \caption{A vertical boundary edge.}
    \label{fig: integrals along vertical edges}
\end{figure}
A one-step discrete contour $\vertedgecontour = (w_0,w_1)$ 
is said to be an oriented vertical boundary edge of~$\dstrip$,
if $w_0,w_1 \in \dV(\dstrip)$ are such that
$\re(w_0) = \re(w_1) \in \set{\lft, \rgt}$ and $|\im(w_0) - \im(w_1)|=1$;
see Figure~\ref{fig: integrals along vertical edges}.
We say that a Clifford generator valued discrete $1$-form
$\cffunhol(z) \, \ferhol(z) \, \dd{z}
+ \cffunbar(z) \, \ferbar(z) \, \dd{\bar{z}}$
is \term{vertically slidable} if for all oriented vertical boundary 
edges~$\vertedgecontour$ 
of~$\dstrip$ we have
\begin{align}\label{eq: vanishing integral on vertical edge}
& \dcint{\vertedgecontour} \Big( \cffunhol(z) \, \ferhol(z) \, \dd{z} 
    + \cffunbar(z) \, \ferbar(z) \, \dd{\bar{z}} \Big) = 0 .
\end{align}

If $(\cffunhol,\cffunbar)$ is a pair of functions~$\dE(\dstrip) \to \bC$, such 
that for all vertical edges $\lftsym = \lft + \ii y'$ of the left 
boundary and all vertical edges $\rgtsym = \rgt + \ii y'$ of the right
boundary we have
\begin{align}\label{eq: ICRBV}
\cffunhol(\lftsym) - \ii \, \cffunbar(\lftsym) = \; & 0 &
\cffunhol(\rgtsym) + \ii \, \cffunbar(\rgtsym) = \; & 0 ,
\end{align}
then we say that the pair $(\cffunhol,\cffunbar)$ has
\term{imaginary complexified Riemann boundary values}
\term{(ICRBV)}. 

The terminology is explained by the following.
\begin{lem}\label{lem: RBV and ICRBV}
If $F \colon \dE(\dstrip) \to \bC$ has Riemann boundary values and 
we set 
\begin{align*}
\cffunhol = \ii \, F 
\qquad \text{ and } \qquad
\cffunbar = - \ii \, \cconj{F} , 
\end{align*}
then $(\cffunhol,\cffunbar)$ has ICRBV.
\end{lem}
\begin{proof}
This follows directly from the definition~\eqref{eq: dRBV in strip}
of Riemann boundary values.
\end{proof}

The relevance of pairs having ICRBV stems from the following.
\begin{prop}\label{prop: ICRBV implies vertically slidable}
If $(\cffunhol,\cffunbar)$ has ICRBV, then the Clifford generator valued 
discrete $1$-form
$\cffunhol(z) \, \ferhol(z) \, \dd{z}
+ \cffunbar(z) \, \ferbar(z) \, \dd{\bar{z}}$
is vertically slidable.
\end{prop}
\begin{proof}
Consider, e.g., an edge $\lftsym = \lft + \ii y'$
on the left boundary.
For the upwards oriented vertical edge
$\vertedgecontour = (\lftsym - \frac{\ii}{2}, \lftsym + \frac{\ii}{2})$,
we get directly from the ICRBV
equation $\cffunhol(\lftsym) - \ii \, \cffunbar(\lftsym) = 0$
and the property $\ferhol(\lftsym) + \ii \, \ferbar(\lftsym) = 0$ of the 
fermions that
\begin{align*}
\dcint{\vertedgecontour} \Big( \cffunhol(z) \, \ferhol(z) \, \dd{z} 
    + \cffunbar(z) \, \ferbar(z) \, \dd{\bar{z}} \Big) 
= \; & \ii \, \cffunhol(\lftsym) \, \ferhol(\lftsym) 
     - \ii \, \cffunbar(\lftsym) \, \ferbar(\lftsym) \\
= \; & \ii \, \cffunhol(\lftsym) \, \ferhol(\lftsym) 
     - \ii \, \cffunhol(\lftsym) \, \ferhol(\lftsym)
= 0 .
\end{align*}
The case of edges on the right boundary is similar.
\end{proof}

\subsubsection*{Integrals of one-forms across the strip}

\begin{figure}[tb]
    \centering
    \includegraphics[width=.35\textwidth]{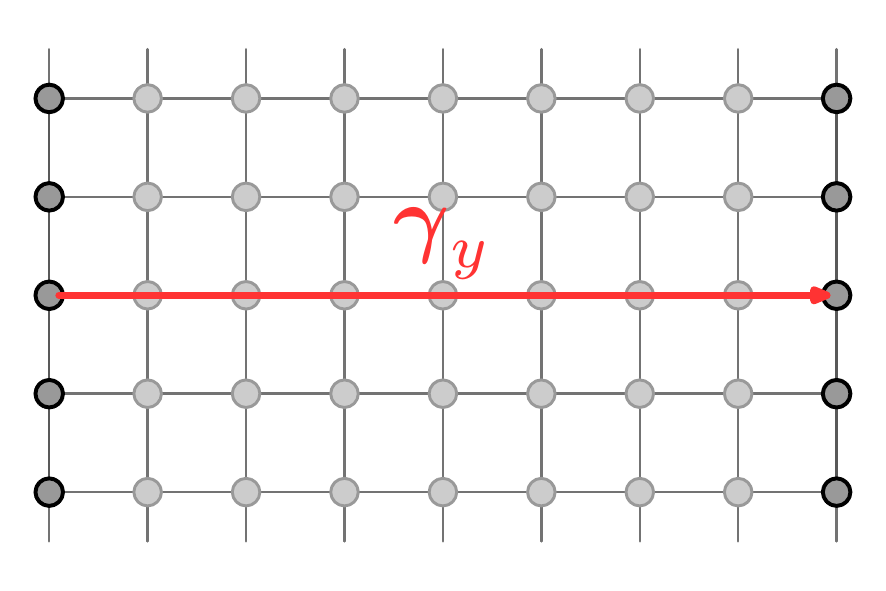}
    \caption{Creation and annihilation operators are defined as integrals of certain $1$-forms across the strip.
    }
    \label{fig: integrals across strip}
\end{figure}

Denote by~$\contour_y$ 
the discrete contour crossing the strip from left to right 
at height~$y \in \bZ$,
\begin{align*}
\contour_y := \big( \lft + \ii y , \; (\lft + 1) + \ii y , \;
        \ldots , \; (\rgt - 1) + \ii y , \; \rgt + \ii y \big)
\end{align*}
as in Figure~\ref{fig: integrals across strip}.
When $(\cffunhol,\cffunbar)$ is a pair of 
functions $\dE(\dstrip) \to \bC$, we consider in particular the integral across 
the strip at zero height
\begin{align}\label{eq: general strip integrated one form}
\dcint{\contour_0} \Big( \cffunhol(z) \, \ferhol(z) \, \dd{z} 
		+ \cffunbar(z) \, \ferbar(z) \, \dd{\bar{z}} \Big)
\; \in \; \CliffGen .
\end{align}

The Hilbert space adjoint of such an operator is an operator of the same form, 
given explicitly below.
\begin{prop}\label{prop: adjoint of integrated one-form}
The Hilbert space adjoint of 
\begin{align*}
\phi = \; \dcint{\contour_0} \Big( \cffunhol(z) \, \ferhol(z) \, \dd{z} 
		+ \cffunbar(z) \, \ferbar(z) \, \dd{\bar{z}} \Big) 
\end{align*}
is
\begin{align*}
\phi^\dagger 
= \; \dcint{\contour_0} \Big( -\overline{\cffunhol(z)} \, \ferhol(z) \, \dd{z} 
		+ \overline{\cffunbar(z)} \, \ferbar(z) \, \dd{\bar{z}} \Big) 
\end{align*}
\end{prop}
\begin{proof}
On the contour~$\contour_0$ at height zero, i.e., for $z \in \crosssecdual$, 
we have the anti-self-adjointness
$\big( \ferhol(z) \big)^\dagger = - \ferhol(z)$ of the holomorphic fermions
and the self-adjointness
$\big( \ferbar(z) \big)^\dagger = \ferbar(z)$ of the antiholomorphic fermions
according to Lemma~\ref{lem: adjoints of fermions}. 
The assertion follows directly from these and conjugate linearity of the 
Hilbert space adjoint and the fact that the steps of the contour~$\contour_0$
are horizontal (so~$\dd{z}, \dd{\bar{z}}$ are real).
\end{proof}
The anticommutators of such operators are scalar multiples of
the identity, where the scalar is obtained as a discrete 
integral of products of the coefficient functions.
\begin{prop}\label{prop: anticommutators of integrated one-forms}
Let
\begin{align*}
\phi_1 = \; \dcint{\contour_0} \Big( \cffunhol_1(z) \, \ferhol(z) \, \dd{z} 
		+ \cffunbar_1(z) \, \ferbar(z) \, \dd{\bar{z}} \Big) \\
\phi_2 = \; \dcint{\contour_0} \Big( \cffunhol_2(z) \, \ferhol(z) \, \dd{z} 
		+ \cffunbar_2(z) \, \ferbar(z) \, \dd{\bar{z}} \Big) .
\end{align*}
The anticommutator of these operators is
\begin{align*}
\anticomm{\phi_1}{\phi_2}
= \bigg( \dcint{\contour_0} \big( 
	  - 2 \, \cffunhol_1(z) \, \cffunhol_2(z) \, \dd{z}
	  + 2 \, \cffunbar_1(z) \, \cffunbar_2(z) \, \dd{\bar{z}} \big) \bigg)
	\; \id .
\end{align*}
\end{prop}
\begin{proof}
Lemma~\ref{lem: anticommutators of height zero fermions} gives the 
anticommutators among the holomorphic 
fermions~$\ferhol(z)$ and the antiholomorphic fermions~$\ferbar(z)$,
\begin{align*}
\anticomm{\ferhol(z_1)}{\ferhol(z_2)}
    = - 2 \, \delta_{z_1 , z_2} \, \id , \qquad
\anticomm{\ferbar(z_1)}{\ferbar(z_2)}
    = + 2 \, \delta_{z_1 , z_2} \, \id , \qquad
\anticomm{\ferhol(z_1)}{\ferbar(z_2)}
    = 0 .
\end{align*}
The assertion follows directly from these and the bilinearity of the 
anticommutators.
\end{proof}

The above integrals were taken at zero height, $y=0$. However,
under the assumptions on 
the coefficient functions~$\cffunhol , \cffunbar$ discussed in 
Section~\ref{sub: one-forms in the strip}, the integral 
across the strip at any other height~$y \in \bZ$ also yields the same result.
\begin{prop}\label{prop: height independence of integrals across strip}
Suppose that the pair~$(\cffunhol , \cffunbar)$ of coefficient functions in the 
strip~$\dstrip$ is ICSH and has ICRBV. Then for any~$y \in \bZ$ we have
\begin{align*}
\dcint{\contour_y} \Big( \cffunhol(z) \, \ferhol(z) \, \dd{z} 
		+ \cffunbar(z) \, \ferbar(z) \, \dd{\bar{z}} \Big) 
= \dcint{\contour_0} \Big( \cffunhol(z) \, \ferhol(z) \, \dd{z} 
		+ \cffunbar(z) \, \ferbar(z) \, \dd{\bar{z}} \Big) .
\end{align*}
\end{prop}
\begin{proof}
Using discrete contour deformation,
the difference of the two integrals in the asserted equality
can be written as a sum of integrals 
around plaquettes between heights~$0$ and~$y$ and of integrals along
vertical boundary edges between these heights.
According to 
Proposition~\ref{prop: ICSH implies closed} the former vanish under the 
assumption of ICSH, and according to 
Proposition~\ref{prop: ICRBV implies vertically slidable} the latter vanish 
under the assumption of ICRBV.
\end{proof}

\begin{rmk}\label{rmk: the convenient association of operators}
A convenient way to use the integrals across the strip,
and phrase the properties anticommutator and Hilbert space adjoint
properties is the following. Given an s-holomorphic function
$F \colon \dE(\dstrip) \to \bC$ with Riemann boundary values,
note that the pair
% $\cffunhol = \ii \, F$,
% $\cffunbar = -\ii \, \overline{F}$ is ICSH and has ICRBV,
% but by complex linearity of these conditions,
% the same holds for the pair
${\cffunhol = \frac{1}{2} \eighthrootthree \, F}$,
${\cffunbar = \frac{1}{2} \eighthrootbar \, \overline{F}}$
is ICSH and has ICRBV~--- by virtue of Lemmas~\ref{lem: SH and ICSH}
and~\ref{lem: RBV and ICRBV} and the complex linearity of these
conditions. The integral~\eqref{eq: general strip integrated one form}
across the strip with these coefficient functions only depends on the 
restriction~$f = F |_{\crosssecdual} \in \dfunctionsp$
of~$F$ to the cross-section~$\crosssecdual$, so let us denote it by
\begin{align}\label{eq: convenient strip integrated one form}
\phi(f) \; := \; 
\frac{\eighthroot}{2}
    \dcint{\contour_0} \Big( \ii \, f(z) \, \ferhol(z) \, \dd{z} 
		- \ii \, \overline{f(z)} \, \ferbar(z) \, \dd{\bar{z}} \Big) .
% \; \in \; \CliffGen .
\end{align}
The Hilbert space adjoint calculated in 
Proposition~\ref{prop: adjoint of integrated one-form} is then
easily seen to yield the function similarly associated with
the function~$\refl f = -\ii \, \overline{f}$ 
reflected using~\eqref{eq: discrete reflection operator}, i.e.,
\begin{align}\label{eq: convenient integrated one-form adjoint}
\phi(f)^\dagger \; = \; \phi(\refl f) .
% \; \in \; \CliffGen .
\end{align}
% Recalling the reflection operator~$\refl f = -\ii \, \overline{f}$ 
% in~\eqref{eq: discrete reflection operator}
Given two such functions
% $F_1, F_2 \colon \dE(\dstrip) \to \bC$
with respective
restrictions~$f , g \in \dfunctionsp $,
% restrictions~${f_1 = F_1 |_{\crosssecdual}}$ and
% ${f_2 = F_2 |_{\crosssecdual}}$, 
the anticommutator calculated in 
Proposition~\ref{prop: anticommutators of integrated one-forms} is
correspondingly simplified to the form
\begin{align}\label{eq: convenient integrated one-form anticommutator}
\anticomm{\phi (f)}{\phi (g)}
\; = \; \innprod{\refl f}{g} \; \id 
% \; \in \; \CliffGen .
\end{align}
which involves the inner product~\eqref{eq: discrete inner product}
and the reflection~\eqref{eq: discrete reflection operator}.
\end{rmk}

\subsection{The formalism in the slit-strip}
\label{sub: slit-strip Clifford}

We next turn to the slit-strip~$\dslitstrip$ of
Figure~\ref{sfig: lattice slit-strip},
and find analogous results. 
The statements and their proofs are mostly almost identical to the preceding 
sections, so we mainly content ourselves with indicating what needs to change. 

First of all, in the slit-strip the Ising model interactions are
different above and below the cross-section at height zero, and
we need separate transfer matrices for the strip-like top half,
and the bottom half with the slit.

\begin{figure}[tb]
\subfigure[Doubled vertical edges: left and right sides of the slit.] 
{
    \includegraphics[width=.35\textwidth]{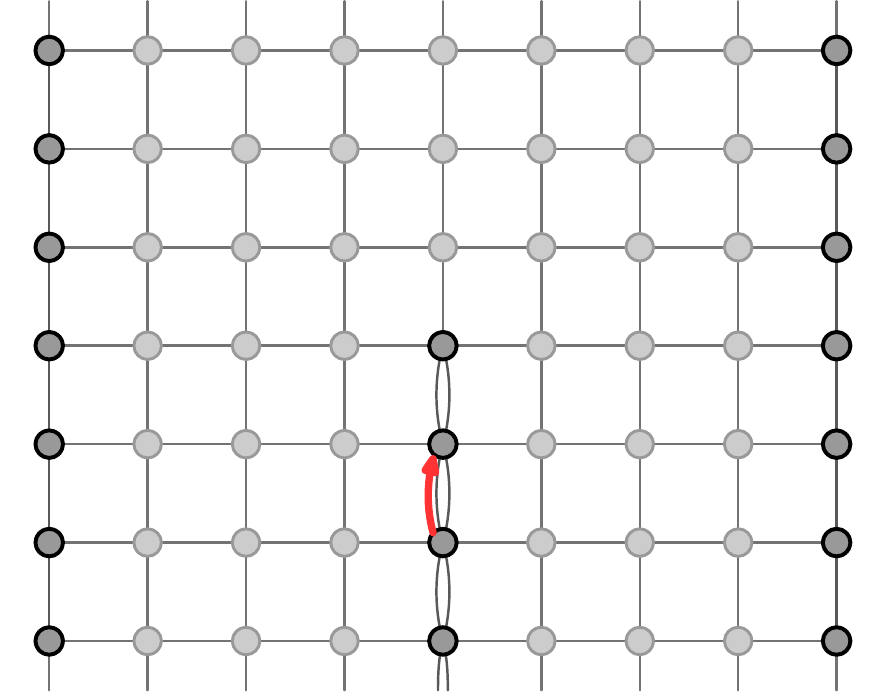}
	\label{sfig: doubled vertical edges in slit-strip}
}
\hspace{2.5cm}
\subfigure[Integration contour across the left substrip.] 
{
    \includegraphics[width=.35\textwidth]{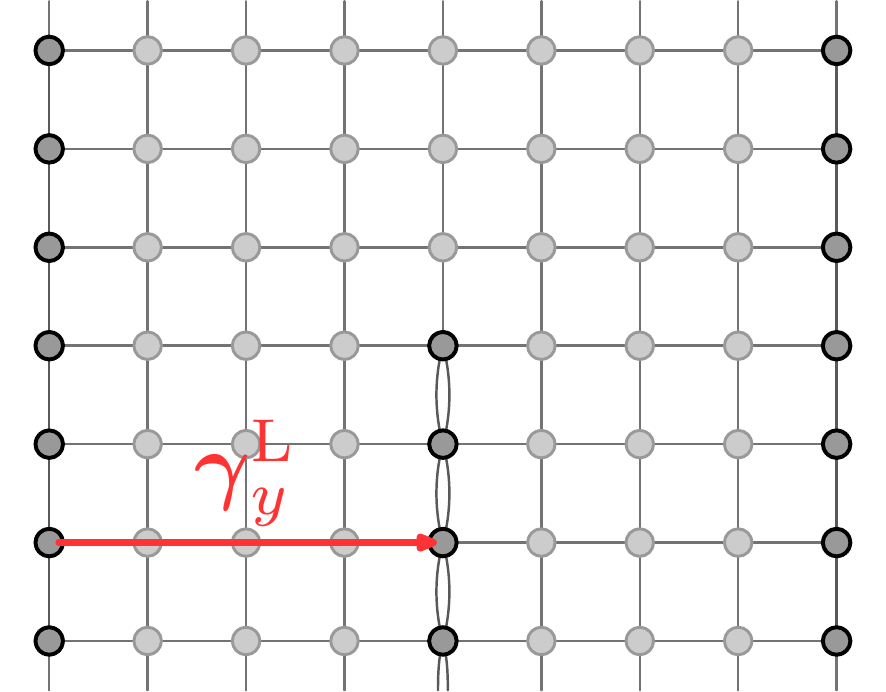}
	\label{sfig: integral across the left substrip}
}
\caption{
Illustration of some of the changes needed
for the slit-strip~$\dslitstrip$.
}
\label{fig: slit-strip illustrations}
\end{figure}

Minor differences arise since
the slit-strip~$\dslitstrip$ is now considered as a multi-graph: it 
has double edges between 
adjacent vertices on the slit~--- one edge interpreted as the left side of the 
slit and the other as the right side, as in 
Figure~\ref{sec: operator valued forms}.\ref{sfig: doubled vertical edges in slit-strip}. Plaquettes adjacent to 
the slit must
be defined so that they are using the correct one of these edges, and 
contours must not traverse the slit. To keep the presentation palatable, 
we trust such conventions to be evident and we furthermore abuse the notation 
by ``identifying'' also the edges on the slit with their embedded positions in 
the complex plane~$\bC$ (although two different edges are thus ``identified'' 
with the same point).

In the context of the slit-strip, we moreover consider one-forms 
which are only required to be locally defined. This could of course 
have been done for the strip already, but it really starts 
to play an essential role when 
we introduce the creation and annihilation operators 
associated to the three extremities 
of the slit-strip. 
Specifically, we will typically define the coefficient functions of 
the  one-forms on one of the three subgraphs 
in~\eqref{eq: subdomains of discrete slit-strip}:
the top half~$\dslitstripT$, the left leg~$\dslitstripL$, or the
right leg~$\dslitstripR$.
If we insisted on coefficient functions defined 
globally on the whole slit-strip~$\dslitstrip$, 
then functions relevant to the diagonalization of the transfer matrix
would fail to possess scaling limits. 
What we do instead can be interpreted as using naturally chosen local 
coordinates around each of the extremities,
while certain key calculations then require changing bases to 
globally defined functions. This is ultimately
how geometric notions enter our algebraic calculations.

\subsubsection*{The slit-strip transfer matrix}

In the lattice slit-strip~$\dslitstrip$, the local interactions of
the Ising model above the vertical position~$y=0$ are just like in the
strip. We correspondigly use the same transfer
matrix~\eqref{eq: Ising transfer matrix} for this part.
To clearly distinguish it from the one used in the slit part,
we now denote it by~$\TmatBasic := \Tmat$.

Below the vertical position~$y=0$, we have to take into account
locally monochromatic boundary conditions on the slit, and we
therefore use a different transfer matrix~$\TmatLR$.
It is again a $\bC$-linear map
\begin{align*} 
\TmatLR \colon \statespdbl \to \statespdbl ,
\end{align*}
constructed out of two constituent matrices. The diagonal matrix
$\TmatHorSqrt \colon \statespdbl \to \statespdbl$
accounting for interactions on horizontal edges (with weight half)
remains exactly the same as before.
The symmetric matrix 
$\TmatVerLR \colon \statespdbl \to \statespdbl$ still
accounts for interactions on vertical edges and
locally monochromatic boundary conditions on the left and right boundaries,
and now additionally for locally monochromatic boundary conditions on the slit.
It is defined by the matrix elements
\begin{align}\label{eq: ver slit transfer matrix entries}
\basisvec{\spinrowalt}^\dagger \TmatVerLR \, \basisvec{\spinrow}
\; = \; \exp \Big( \invtemp
                \sum_{x=\lft}^{\rgt} \spinrow_{x} \spinrowalt_{x} \Big)
        \; \delta_{\spinrowalt_{\lft},\spinrow_{\lft}}
        \, \delta_{\spinrowalt_{\mdpt},\spinrow_{\mdpt}}
        \, \delta_{\spinrowalt_{\rgt},\spinrow_{\rgt}} .
\end{align}
The transfer matrix~$\TmatLR$ for the slit part is then defined as
\begin{align}\label{eq: Ising transfer matrix slit part}
\TmatLR \; = \; \TmatHorSqrt \, \TmatVerLR \, \TmatHorSqrt \, .
\end{align}

\subsubsection*{Holomorphic and antiholomorphic fermions in the slit-strip}

Analogously to Section~\ref{sub: fermions in the strip},
we define two Clifford algebra generator valued functions on 
the edges of the discrete slit-strip~$\dslitstrip$.
We use a hat in the notation to distinguish them from the earlier
introduced operators in the discrete strip.

Again the horizontal edges at height zero are naturally 
identified with points~$x' \in \crosssecdual$ of the dual cross-section,
and the horizontal edges at height~$y \in \bZ$ are identified with 
points~$x' + \ii y$ with~$x' \in \crosssecdual$.
As with the spin operators in the 
slit-strip~\eqref{eq: height dependent spin operator in the slit-strip},
there is now a difference between the top half of the strip, $y \geq 0$, and 
the bottom half, $y \leq 0$ (zero height could be thought of as belonging to 
either the top or bottom).
Let therefore $x' \in \crosssecdual$, and $y \in \Znn$.
For horizontal edges in the top half we set
\begin{align}\label{eq: def fermions on horizontal edges of slit strip T}
\ferholS(x'+\ii y) = (\TmatBasic)^{-y} \, \ferhol_{x'} \, (\TmatBasic)^{y}
, \qquad
\ferbarS(x'+\ii y) = (\TmatBasic)^{-y} \, \ferbar_{x'} \, (\TmatBasic)^{y} ,
\end{align}
and for horizontal edges in the bottom half we set
\begin{align}\label{eq: def fermions on horizontal edges of slit strip B}
\ferholS(x'-\ii y) = (\TmatLR)^{y} \, \ferhol_{x'} \, (\TmatLR)^{-y}
, \qquad 
\ferbarS(x'-\ii y) = (\TmatLR)^{y} \, \ferbar_{x'} \, (\TmatLR)^{-y} .
\end{align}
From these values of the fermions on horizontal edges, the principle of 
extending to vertical edges is to require the following discrete 
complex analysis properties.
\begin{prop}
\label{prop: complexified SH and RBV for fermions in slit strip}
There exists unique extensions
\begin{align*}
\ferholS , \ferbarS \colon \dE(\dslitstrip) \to \CliffGen
\end{align*}
of~\eqref{eq: def fermions on horizontal edges of slit strip T}~--
~\eqref{eq: def fermions on horizontal edges of slit strip B} to
vertical edges with the following properties:

The pair~$(\ferholS,\ferbarS)$ is 
complexified s-holomorphic (CSH)
in the sense that
for any edges~$z_1, z_2 \in \dE(\dslitstrip)$
adjacent to a vertex~$v$ and a face~$p$ we have
\begin{align*}
\ferholS(z_1) + \frac{\ii \, |v-p|}{v-p} \, \ferbarS(z_1)
= \ferholS(z_2) + \frac{\ii \, |v-p|}{v-p} \, \ferbarS(z_2) ,
\end{align*}
and it has
complexified Riemann boundary values (CRBV)
in the sense that
\begin{align*}
\ferholS(\lftsym) + \ii \, \ferbarS(\lftsym) = \; & 0 &
\ferholS(\rgtsym) - \ii \, \ferbarS(\rgtsym) = \; & 0 .
\end{align*}
for any left boundary edge~$\lftsym$ 
(including edges on the slit which are 
the left boundary of the right substrip)
and any right boundary edge~$\rgtsym$ 
(including edges on the slit which are 
the right boundary of the left substrip).
\end{prop}
The proof is similar to that of 
Proposition~\ref{prop: complexified SH and RBV for fermions}:
formulas analogous
to~\eqref{eq: ferhol vertical edge right side expression}~--~
\eqref{eq: ferbar right boundary vertical edge expression}
(with just $\ferholS,\ferbarS$ in place of~$\ferhol,\ferbar$)
are used to extend the definition of the holomorphic and
antiholomorphic fermions
$\ferholS, \ferbarS$ from horizontal edges to vertical edges.

\subsubsection*{Clifford algebra valued $1$-forms in the slit-strip}

As in Section~\ref{sub: one-forms in the strip},
we define Clifford generator valued discrete 
$1$-forms and their discrete line integrals.

By a Clifford generator valued discrete $1$-form in
the slit-strip~$\dslitstrip$,
we again mean a formal expression
\begin{align*}
\cffunhol(z) \, \ferholS(z) \, \dd{z} 
+ \cffunbar(z) \, \ferbarS(z) \, \dd{\bar{z}} ,
\end{align*}
where $\cffunhol, \cffunbar \colon \dE \to \bC$ are two complex 
valued functions on a set of edges~$\dE$.
For globally defined $1$-forms we take $\dE = \dE(\dslitstrip)$,
and for locally defined $1$-forms we take
$\dE = \dE(\dslitstripT)$, $\dE = \dE(\dslitstripR)$, or
$\dE = \dE(\dslitstripL)$.

For a discrete contour~$\contour$ in the slit-strip~$\dslitstrip$ 
(or in one of the subgraphs~$\dslitstripT,\dslitstripL,\dslitstripR$), 
the integral
\[ \dcint{\contour} \Big( \cffunhol(z) \, \ferholS(z) \, \dd{z} 
    + \cffunbar(z) \, \ferbarS(z) \, \dd{\bar{z}} \Big)
\; \in \; \CliffGen \]
is defined by exactly the same formula as 
in Section~\ref{sub: one-forms in the strip}, but now instead using the 
slit-strip fermions~$\ferholS(z)$ and $\ferbarS(z)$ defined above.

The two crucial notions of such Clifford generator valued $1$-forms
are defined basically as before. 
The $1$-form is said to be \term{closed}
if for all counterclockwise oriented 
plaquettes~$\plaquettecontour$ of~$\dslitstrip$
(or of~$\dslitstripT,\dslitstripL,\dslitstripR$), 
its integral vanishes as in~\eqref{eq: vanishing integral around plaquette}. 
The $1$-form is said to be \term{vertically slidable} if 
for all oriented vertical 
boundary edges~$\vertedgecontour$ of~$\dslitstrip$
(or of~$\dslitstripT,\dslitstripL,\dslitstripR$)
its integral vanishes as 
in~\eqref{eq: vanishing integral on vertical edge}.

Also the two notions of coefficient functions for the $1$-forms
are essentially as before. 
A pair~$(\cffunhol,\cffunbar)$ of complex-valued
functions on the edges of $\dslitstrip$, $\dslitstripT$, $\dslitstripL$, or 
$\dslitstripR$, 
is said to be \term{ICSH} (in the slit-strip or a subgraph)
if Equations~\eqref{eq: ICSH} hold for each plaquette (of the 
corresponding graph).
The pair $(\cffunhol,\cffunbar)$ is said to have \term{ICRBV}
(in the slit-strip or a subgraph)
if Equations~\eqref{eq: ICRBV} hold for all left boundary vertical
edges~$\lftsym$ and all right boundary vertical edges~$\rgtsym$ (of the
corresponding graph).

The reasons for the terminology are just as in
Lemmas~\ref{lem: SH and ICSH} and~\ref{lem: RBV and ICRBV}.
\begin{lem}\label{lem: SH and ICSH in slit-strip}
If $F \colon \dE \to \bC$
is s-holomorphic and 
we set 
$\cffunhol = \ii \, F$ and $\cffunbar = - \ii \, \cconj{F}$, 
then $(\cffunhol,\cffunbar)$ is ICSH.
\end{lem}
\begin{lem}\label{lem: RBV and ICRBV in slit-strip}
If $F \colon \dE \to \bC$ 
has Riemann boundary values and 
we set $\cffunhol = \ii \, F$ and $\cffunbar = - \ii \, \cconj{F}$, 
then $(\cffunhol,\cffunbar)$ has ICRBV.
\end{lem}

Also the implications of ICSH and ICRBV are as in
Propositions~\ref{prop: ICSH implies closed} 
and~\ref{prop: ICRBV implies vertically slidable}.
\begin{prop}\label{prop: ICSH implies closed in slit-strip}
If $(\cffunhol,\cffunbar)$ is ICSH, then the Clifford generator valued 
discrete $1$-form
$\cffunhol(z) \, \ferholS(z) \, \dd{z}
+ \cffunbar(z) \, \ferbarS(z) \, \dd{\bar{z}}$
is closed.
\end{prop}
\begin{prop}\label{prop: ICRBV implies vertically slidable in slit-strip}
If $(\cffunhol,\cffunbar)$ has ICRBV, then the Clifford generator valued 
discrete $1$-form
$\cffunhol(z) \, \ferholS(z) \, \dd{z}
+ \cffunbar(z) \, \ferbarS(z) \, \dd{\bar{z}}$
is vertically slidable.
\end{prop}

\subsubsection*{Integrals of one-forms across the slit-strip}

At non-negative heights~$y \in \Znn$ we still consider the same 
contours~$\contour_y$ across the slit-strip from left to right as before.
The contour~$\contour_y$ lies in the subgraph~$\dslitstripT$,
and we also denote it by~$\contourT_y$ when convenient.

At non-positive heights~$y \in \Znp$, we consider 
contours~$\contourL_{y}$ and~$\contourR_{y}$ across the 
two halves of the slit-strip, from left to middle and from middle to right,
\begin{align*}
\contourL_{y} := \; & \big( \lft + \ii y , \; (\lft + 1) + \ii y , \;
        \ldots , \; (\mdpt - 1) + \ii y , \; \mdpt + \ii y \big) \\
\contourR_{y} := \; & \big( \mdpt + \ii y , \; (\mdpt + 1) + \ii y , 
\;
        \ldots , \; (\rgt - 1) + \ii y , \; \rgt + \ii y \big) 
\end{align*}
as in, e.g., Figure~\ref{sec: operator valued forms}.\ref{sfig: integral across the left substrip}.
The contour~$\contour^{\lftsym}_y$ lies in the subgraph~$\dslitstripL$
and the contour~$\contourR_y$ in the subgraph~$\dslitstripR$

Again, when $(\cffunhol,\cffunbar)$ is a pair of complex-valued functions 
on the edges, we consider in particular the integrals 
at zero height
\begin{align}\label{eq: general slit strip integrated one form}
\dcint{\contour^{\wildsym}_0} \Big( \cffunhol(z) \, \ferholS(z) \, \dd{z} 
		+ \cffunbar(z) \, \ferbarS(z) \, \dd{\bar{z}} \Big)
\; \in \; \CliffGen ,
\end{align}
where~$\contour^{\wildsym}_0$ is any of $\contourT_0$, $\contourL_{0}$,
$\contourR_{0}$.

The Hilbert space adjoints work out exactly as in 
Proposition~\ref{prop: adjoint of integrated one-form} and the 
anticommutators as in 
Proposition~\ref{prop: anticommutators of integrated one-forms};
the proofs go through verbatim.

Under the assumptions of ICSH and ICRBV for coefficient functions, the 
independence of the integrals on the chosen height~$y$ can again be derived by 
the closedness and vertical slidability. The precise statements are as 
follows.
\begin{prop}\label{prop: height independence of integrals across slit-strip}
Let $\wildsym$ stand for either~$\topsym$, $\lftsym$ or~$\rgtsym$.
Suppose that the pair~$(\cffunhol , \cffunbar)$ of complex-valued coefficient 
functions on the edges of the subgraph~$\dslitstripX$ 
is ICSH and has ICRBV. Then we have
\begin{align*}
\dcint{\contourX_y} \Big( \cffunhol(z) \, \ferholS(z) \, \dd{z} 
		+ \cffunbar(z) \, \ferbarS(z) \, \dd{\bar{z}} \Big) 
= \dcint{\contourX_0} \Big( \cffunhol(z) \, \ferholS(z) \, \dd{z} 
		+ \cffunbar(z) \, \ferbarS(z) \, \dd{\bar{z}} \Big)
\end{align*}
for any~$y \in \Znn$ if $\wildsym = \topsym$ and 
for any~$y \in \Znp$ if $\wildsym = \lftsym , \rgtsym$.
\end{prop}
The proof is as in
Proposition~\ref{prop: height independence of integrals across strip}.

\bigskip

\section{Distinguished functions on the lattice}
\label{sec: distinguished functions}

% *******************************************************************
% **   \section{S-hol., RBVs, and distinguished functions}         **
% *******************************************************************

In this section we discuss certain distinguished s-holomorphic
functions in the lattice strip and slit-strip,
constructed in~\cite{part-1}. Clifford algebra valued 
discrete $1$-forms of Section~\ref{sec: operator valued forms}
with these distinguished functions as the coefficient functions
will be our main tools for the analysis
of the Ising model in Section~\ref{sec: transfer matrices}.

\subsubsection*{Indexing sets}
Certain indexing sets of half-integers 
will be used throughout the rest of this article, as well as
the subsequent one~\cite{part-3}.

Let
\begin{align}
\label{eq: positive half integers} 
\poshalfint 
:= \big[ 0 , \infty \big) \cap \Big( \bZ + \half \Big)
= \set{\frac{1}{2} , \frac{3}{2} , \frac{5}{2} , \ldots }
\end{align}
be the set of all positive half-integers, and
\begin{align}
\pm\poshalfint 
:= \poshalfint \cup \big( -\poshalfint \big) = \bZ + \half
\end{align}
the set of all half-integers.

In the discrete setting, the width parameter~$\width$ 
serves as a truncation, so we denote
\begin{align}
\label{eq: discrete positive half integers} 
\dposhalfint 
:= \; & \big[ 0 , \width \big) \cap \Big( \bZ + \half \Big)
= \set{\frac{1}{2} , \frac{3}{2} , \ldots,
    \width - \frac{1}{2}} 
\end{align}
and
\begin{align}
\pm\dposhalfint 
:= \; & \dposhalfint \cup \big( -\dposhalfint \big) .
\end{align}
We also sometimes use the substrip widths~${\widthR = \rgt}$
and ${\widthL = - \lft}$ in place of the width~$\width$.
Moreover, the set~$\set{\parts \subset \dposhalfint}$
of all subsets of~$\dposhalfint$ will be our
indexing set for the transfer matrix eigenvectors
(see Section~\ref{sub: diagonalization of strip transfer matrix}),
and the sets~$\{\partsL \subset \dposhalfintL \}$ 
and~$\{\partsR \subset \dposhalfintR \}$ will play an analogous role
for the transfer matrix of the slit part
(see Section~\ref{sec: operators in the slit strip}).

\subsection{Distinguished functions in the strip}
The following distinguished functions in the
lattice strip were introduced in~\cite{part-1}.

For any $k \in \dposhalfint$, 
according to \cite[Lemma~3.6]{part-1},
there exists a unique
\begin{align*}
\omega^{(\width)}_k 
\in \Big( (k-\half) \pi / \width , \; k \pi / \width \Big)
\qquad \text{ satisfying } \qquad
\frac{\cos \big( (\width + \half) \, \omega^{(\width)}_k  \big)}
{\cos \big( (\width - \half) \, \omega^{(\width)}_k \big)} 
= \; & 3 - 2 \sqrt{2} .
\end{align*}
In terms of it, we define
\begin{align}\label{eq: eigenvalue of vertical translation}
\eigval{\pm k} := 
    \Big( 2 - \cos (\omega_k^{(\width)} ) 
    + \sqrt{ \big(3-\cos ( \omega_k^{(\width)} ) \big)
        \big(1-\cos (\omega_k^{(\width)}) \big)} \; \Big)^{\pm 1} .
\end{align}
Note that $\eigval{k} > 1$ and $\eigval{-k} = 1/\eigval{k} < 1$
for $k \in \dposhalfint$.

\begin{prop}[{\cite[Proposition~3.7]{part-1}}]
\label{prop: discrete vertical translation eigenfunctions}
For $k \in \dposhalfint$,
there exists unique functions
\begin{align*}
\eigF{k} , \eigF{-k} \colon \dE(\dstrip) \to \bC
\end{align*}
which are s-holomorphic in the lattice strip~$\dstrip$,
have Riemann boundary values on the left and right boundaries,
have the vertical translation eigenfunction property
\begin{align*}
\eigF{\pm k} (z + \ii h) = (\eigval{\pm k})^h \; \eigF{\pm k}(z)
\qquad \text{ for all } z \in \dE(\dstrip) \text{ and } h \in \bZ ,
\end{align*}
and satisfy the following normalization conditions:
the arguments on the left boundary are
$\eigF{\pm k}(\lft + \ii y') \in e^{-\ii \pi / 4} \, \bR_+$,
for $y' \in \bZ + \half$, and the
restrictions
\begin{align*}
\eigf{\pm k} := \eigF{\pm k} \big|_{\crosssecdual} \; \in \dfunctionsp
\end{align*}
to the cross-section~$\crosssecdual$ have unit norms,
$\| \eigf{\pm k} \| = 1$. 

The following reflection relations hold for these functions:
\begin{align}\label{eq: opposite index discrete functions}
\eigf{-k} (x') 
    = \; & -\ii \; \overline{\eigf{k} (x')} , &
\eigF{-k} (x + \ii y) 
    = \; & -\ii \; \overline{\eigF{k} (x - \ii y)} .
\end{align}

Moreover, the collection $(\eigf{k})_{k \in \pm \dposhalfint}$
of functions forms 
an orthonormal basis of the
real Hilbert
space~$\dfunctionsp = \bC^{\crosssecdual}$ of~\eqref{eq: discrete function space}.
\end{prop}

\subsubsection*{Decomposition to poles and zeroes at the top}

The functions~$\eigf{+k}$, for $k \in \dposhalfint$, are
exponentially growing in the upwards direction of the lattice 
strip~$\dstrip$, since~$\eigval{+k}>1$, while the functions~$\eigf{-k}$
are exponentially decaying. We interpret the growing ones as having a 
pole at the top extremity, and the decaying ones as having a zero.

We decompose the discrete function
space~$\dfunctionsp$ into the corresponding 
subspaces. More precisely, we define
\begin{align}\label{eq: discrete top poles and zeros}
\dfspTpole = \; & \spn \set{ \eigf{+k} \; \Big| \; k \in \dposhalfint} , &
\dfspTzero = \; & \spn \set{ \eigf{-k} \; \Big| \; k \in \dposhalfint} ,
\end{align}
and we denote by
\begin{align*}
\dprTpole \colon \; & \dfunctionsp \to \dfspTpole , &
\dprTzero \colon \; & \dfunctionsp \to \dfspTzero  ,
\end{align*}
the orthogonal projections onto these subspaces.

\subsection{Distinguished functions in the slit-strip}
\label{sub: distinguished functions in slit strip}

In the slit-strip there are three infinite extremities: 
the top, the right leg, and the left leg.
In the left and right substrips there are natural functions
defined analogously to the whole strip, which allow us to define
poles and zeroes, and obtain corresponding decompositions. 
The distinguished functions in the slit-strip will then be
globally defined
functions which have zeroes (i.e., regular behavior) in two
of the three infinite extremities, and have a pole of a given 
order in the third (i.e., a prescribed singular part).

\subsubsection*{Decomposition to poles and zeroes in the left and right legs}

We can apply 
Proposition~\ref{prop: discrete vertical translation eigenfunctions}
in the left and right substrips (replacing one of~$\lft, \rgt$ by~$\mdpt$)
to obtain s-holomorphic functions
\begin{align*}
\eigFL{\pm k} \colon \; & \dE(\dslitstripL) \to \bC \quad
\text{ for $k \in \dposhalfintL$ } , \quad &
\eigFR{\pm k} \colon \; & \dE(\dslitstripR) \to \bC \quad
\text{ for $k \in \dposhalfintR$ } , &
\end{align*}
with Riemann boundary values in the corresponding substrips,
and with vertical translation 
eigenvalues~$\eigvalW{\widthL}{\pm k}$
and~$\eigvalW{\widthR}{\pm k}$, respectively. 
Splitting the cross-section~$\crosssecdual = \dintervaldual{\lft}{\rgt}$
to two halves $\crosssecdual_\lftsym := \dintervaldual{\lft}{\mdpt}$
and $\crosssecdual_\rgtsym :=\dintervaldual{\mdpt}{\rgt}$, the discrete 
function space
splits naturally to orthogonally complemetary subspaces
\begin{align*}
\dfunctionspL 
:= \; & \bC^{\crosssecdual_\lftsym} \subset \dfunctionsp, &
\dfunctionspR 
:= \; & \bC^{\crosssecdual_\rgtsym} \subset \dfunctionsp ,
\end{align*}
consisting of functions supported on one of the two halves
of the cross-section.
Note that for functions 
in~$\dfunctionspL, \dfunctionspR$,
we use the norm and inner product defined 
exactly analogously to~\eqref{eq: discrete inner product},
and these coincide with the norm and inner product they inherit as
subspaces of~$\dfunctionsp$ when the functions on one half of the
cross-section are extended as zero to
the other half.

Applying Proposition~\ref{prop: discrete vertical translation eigenfunctions}
to the left and right substrips, one obtains
% substrips $\dinterval{\lft}{\mdpt} \times \bZ$ and 
% $\dinterval{\mdpt}{\rgt} \times \bZ$
% the following 
functions
%as restrictions to the two halves of the cross-section
\begin{align*}
\eigfL{\pm k}
    := \; & \eigFL{\pm k} \big|_{\crosssecdual_\lftsym} 
    \; \in \; \dfunctionspL , &
\eigfR{\pm k}
    := \; & \eigFR{\pm k} \big|_{\crosssecdual_\rgtsym} 
    \; \in \; \dfunctionspR ,
\end{align*}
which form orthonormal bases~$(\eigfL{k})_{k \in \pm \dposhalfintL}$
and~$(\eigfR{k})_{k \in \pm \dposhalfintR}$ of the
subspaces~$\dfunctionspL$ and~$\dfunctionspR$, respectively.
In terms of these, we define the subspaces\footnote{Note 
that poles and zeroes in the left and right legs are determined by 
the exponential growth or decay in the \emph{downwards} direction,
and thus the signs of the indices are the opposite compared 
to~\eqref{eq: discrete top poles and zeros}.}
\begin{align}\label{eq: discrete leg poles and zeros}
\dfspLpole = \; & \spn \set{ \eigfL{-k} \; \Big| \; k \in \dposhalfintL} , &
\dfspRpole = \; & \spn \set{ \eigfR{-k} \; \Big| \; k \in \dposhalfintR} , \\
\nonumber
\dfspLzero = \; & \spn \set{ \eigfL{+k} \; \Big| \; k \in \dposhalfintL} , &
\dfspRzero = \; & \spn \set{ \eigfR{+k} \; \Big| \; k \in \dposhalfintR} ,
\end{align}
and we denote by
\begin{align*}
\dprLpole \colon \; & \dfunctionsp \to \dfspLpole , &
\dprRpole \colon \; & \dfunctionsp \to \dfspRpole , \\
\dprLzero \colon \; & \dfunctionsp \to \dfspLzero , &
\dprRzero \colon \; & \dfunctionsp \to \dfspRzero ,
\end{align*}
the orthogonal projections onto these subspaces.

\subsubsection*{Distinguished discrete functions in the slit-strip}

For a function~$f \in \dfunctionsp$, we call the projections
$\dprTpole(f)$, $\dprTpole(f)$, $\dprTpole(f)$
its singular parts in the top, right, and left, respectively.
When a singular part vanishes, we say that the function admits
a regular extension in the corresponding extremity.
The distinguished functions are characterized by having
only one non-vanishing singular part as follows.
\begin{prop}[{\cite[Section~3.4]{part-1}}]
\label{prop: discrete pole functions}
For $\kT \in \dposhalfint , \kR \in \dposhalfintR , \kL \in \dposhalfintL$,
there exists unique functions
$\poleT{\kT} , \poleR{\kR} , \poleL{\kL} \in \dfunctionsp$ such that
\begin{align}
\nonumber
\dprTpole(\poleT{\kT}) = \; & \eigf{+\kT} , &
\dprLpole(\poleT{\kT}) = \; & 0 , &
\dprRpole(\poleT{\kT}) = \; & 0 , \\ 
\label{eq:cpole-asymptotic}
\dprTpole(\poleL{\kL}) = \; & 0 , &
\dprLpole(\poleL{\kL}) = \; & \eigfL{-\kL} , &
\dprRpole(\poleL{\kL}) = \; & 0 , \\ 
\nonumber
\dprTpole(\poleR{\kR}) = \; & 0 , &
\dprLpole(\poleR{\kR}) = \; & 0 , &
\dprRpole(\poleR{\kR}) = \; & \eigfR{-\kR} .
\end{align}
These functions are the restrictions to the cross-section~$\crosssecdual$
of unique s-holomorphic functions 
\begin{align*}
\PoleT{\kT} , \PoleL{\kL} , \PoleR{\kR} \colon \; & \dE(\dslitstrip) \to \bC 
\end{align*}
with Riemann boundary values. 
\end{prop}
Note that these discrete pole functions
$\PoleT{\kT} , \PoleL{\kL} , \PoleR{\kR}$
are defined globally on the whole lattice
slit-strip~$\dslitstrip$.

\bigskip

\section{Ising model and the transfer matrices}
\label{sec: transfer matrices}

% *******************************************************************
% **   \section{Ising model and transfer matrices}                 **
% *******************************************************************

In this section we define the Ising model in the lattice strip and lattice
slit-strip, and review the transfer matrix formalism to the calculation of
correlation functions. We then show how the method of Clifford algebra valued
$1$-forms can be used first of all to diagonalize the transfer matrix
in the strip~$\dstrip$ and, more interestingly, to calculate certain renormalized
boundary correlation functions
in the slit strip~$\dslitstrip$.

The Ising model on a graph is a random assignment of $\pm 1$ spins 
to the vertices of the graph.
Configurations of spins with more alignment among the spins of
neighboring vertices are given relatively higher probabilities,
and the strength of this tendency of local alignment is controlled by a
parameter~$\invtemp>0$ interpreted as the inverse temperature.
We always consider the Ising model with unit coupling constants and 
no external magnetic field. On a given graph, therefore,
$\invtemp$ is the only parameter in the model, and we moreover take 
it to be the critical value for the square lattice.

The definition of the Ising model probability measure
is straightforward when the 
graph is finite. Almost the only subtlety to pay attention to is
the choice of boundary conditions.
Our choice will be locally monochromatic boundary
conditions on vertical boundaries of the strip and slit-strip.
The commonly used plus and minus
boundary conditions can also be obtained straightforwardly
from our results.

On infinite graphs such as the lattice 
strip~$\dstrip$ and the lattice slit-strip~$\dslitstrip$,
the construction of
the Ising model probability measure requires
approximating the infinite graph 
with a sequence of increasingly
large finite subgraphs, and to consider the weak limit of the
associated probability measures. The existence of such infinite volume
limits is usually established using correlation inequalities 
(FKG-inequality in the case of plus boundary conditions, or Griffiths' 
inequality in the case of free boundary conditions), but in our 
setup also follows easily from the transfer matrix formalism 
(Section~\ref{sub: transfer matrix formalism}).

Figure~\ref{fig: Ising samples} 
illustrates samples of Ising spin configurations
in a lattice strip and a lattice slit-strip,
with locally monochromatic boundary conditions and
at the critical inverse temperature.
The two colors represent the two possible values~$\pm 1$ of the spins.

We take the point of view that the fundamental quantities
about the Ising model are its correlation functions, i.e., expected 
values of suitable random variables of the spin configuration.
For instance, the weak limit 
defining infinite volume Ising probability measure is itself formulated 
by means of such correlation functions. The key quantities featuring in 
our main result (and ultimately leading to vertex operator
algebra structure constants) will be certain renormalized limits of
boundary correlation functions, which we will call fusion coefficients.

\subsection{Ising model on finite graphs}
\label{sub: Ising on finite graphs}

We begin by defining the Ising model on a general finite graph,
and discussing boundary conditions.
After the general definitions, we specialize 
to the cases where the finite graph is taken to be a truncated lattice 
strip or a truncated lattice slit-strip, and locally
monochromatic boundary conditions on vertical
boundary components are used.

\subsubsection*{Ising model without boundary conditions}

Let $\gengraph = (\dV(\gengraph) , \dE(\gengraph))$ be a graph
with a finite set~$\dV(\gengraph)$ of vertices, and
a (finite) set~$\dE(\gengraph)$ of edges.
Let also the inverse temperature parameter~$\invtemp>0$ be fixed.

The sample space for the Ising model on~$\gengraph$ is the
set of $\pm 1$-valued configurations on the vertices,
\begin{align*}
\PRspace = \set{\pm 1}^{\dV(\gengraph)} .
\end{align*}
For a spin configuration 
$\spinconf \in \PRspace$,
\begin{align*}
\spinconf = (\sigma_z)_{z \in \dV(\gengraph)} ,
\qquad \text{ with } \qquad
\sigma_z \in \set{\pm 1} 
\text{ for each } z \in \dV(\gengraph) ,
\end{align*}
the energy (Hamiltonian) is defined as
\begin{align*}
\Hamiltonian_{\gengraph} (\spinconf) =
    - \sum_{\set{z,w} \in \dE(\gengraph)} \sigma_z \sigma_w .
\end{align*}
The Ising model probability measure $\PR_{\invtemp ; \gengraph}$
on~$\PRspace$ 
is then defined by setting the probabilities of spin configurations 
$\spinconf \in \set{\pm 1}^{\dV(\gengraph)}$
proportional to their Boltzmann weights, i.e.,
\begin{align}\label{eq: Ising proba general}
\PR_{\invtemp ; \gengraph} \big[ \set{\spinconf} \big]
    = \frac{1}{\PartF_\gengraph(\invtemp)} \, 
        e^{\minussym  \invtemp \, \Hamiltonian_{\gengraph} (\spinconf)} ,
\end{align}
where the partition function
\begin{align}\label{eq: Ising pf general}
\PartF_\gengraph(\invtemp)
    = \sum_{\spinconf \in \set{\pm 1}^{\dV(\gengraph)}}
        e^{\minussym  \invtemp \, \Hamiltonian_{\gengraph} (\spinconf)}
\end{align}
normalizes the total mass of~$\PR_{\invtemp ; \gengraph}$ to one.

\subsubsection*{Imposing boundary conditions}

A~priori, the Ising model
probability measure $\PR_{\invtemp ; \gengraph}$
is defined according to~\eqref{eq: Ising proba general}.
Imposing boundary conditions amounts to modifying the
measure by appropriate conditioning.

Let a subset of vertices $\bdry \gengraph \subset \dV(\gengraph)$
of a finite graph be declared as boundary.
The Ising model on~$\gengraph$ with plus boundary conditions
is then the conditional probability measure,
conditioned on the event
\begin{align*}
\PRspPlus{\bdry \gengraph} := \set{ \spinconf \in \PRspace 
        \; \Big| \; \spinconf |_{\bdry \gengraph} \equiv +1 }
\end{align*}
that all the spins on the boundary are~$+1$.
This conditional probability measure is denoted by
$\PR^{\plussym}_{\invtemp ; \gengraph} [\,\cdot\,]
= \PR_{\invtemp ; \gengraph} \big[ \,\cdot\, 
        \big| \, \PRspPlus{\bdry \gengraph} \big]$,
and it
is explicitly characterized by the probabilities
\begin{align*}
\PR^{\plussym}_{\invtemp ; \gengraph} \big[ \set{\spinconf} \big]
= \; & \frac{1}{\PartF^{\plussym}_\gengraph(\invtemp)} \, 
        e^{\minussym  \invtemp \, \Hamiltonian_{\gengraph} (\spinconf)} \qquad
\text{ for } \spinconf \in \PRspPlus{\bdry \gengraph} ,
\end{align*}
where the partition function for plus boundary conditions is the sum
\begin{align*}
\PartF^{\plussym}_\gengraph(\invtemp)
    = \sum_{\spinconf \in \PRspPlus{\bdry \gengraph}}
        e^{\minussym  \invtemp \, \Hamiltonian_{\gengraph} (\spinconf)}
\end{align*}
which is obtained from~\eqref{eq: Ising pf general}
by keeping only the terms that correspond to the required boundary conditions.

Minus boundary conditions can be imposed similarly by conditioning on
the event $\PRspMinus{\bdry \gengraph} \subset \PRspace$ that the
values of spins in the subset~$\bdry \gengraph \subset \dV(\gengraph)$
(declared as boundary) are all~$-1$.

It is also possible to impose mixed boundary conditions,
plus boundary conditions on one part of the boundary and
minus boundary conditions on another part simultaneously.
More precisely, this amounts to conditioning on the
event~$\PRspPM{\bdry^{\plussym} \gengraph}{\bdry^{\minussym} \gengraph}
\subset \PRspace$ that the
values of spins in the 
subset~$\bdry^{\plussym} \gengraph \subset \dV(\gengraph)$
are all~$+1$ and those in a
subset~$\bdry^{\minussym} \gengraph \subset \dV(\gengraph)$
are all~$-1$.

We mostly use the following locally monochromatic
boundary conditions. Given finitely many disjoint 
subsets $\bdry^1 \gengraph , \ldots , \bdry^n \gengraph 
\subset \dV(\gengraph)$, we condition on the event
\begin{align}
\PRspLocmono{\bdry^1 \gengraph , \ldots , \bdry^n \gengraph}
= \; & \set{  \spinconf \in \PRspace 
        \; \Big| \; \spinconf |_{\bdry^j \gengraph} \equiv \const
        \text{ for each } j = 1 , \ldots, n }
\end{align}
that the spins are constant
on each of these. The conditioned probability measure is
\begin{align*}
\PR^{\locmonosym}_{\invtemp ; \gengraph} [\,\cdot\,]
= \PR_{\invtemp ; \gengraph} \big[ \,\cdot\, 
        \big| \, \PRspLocmono{\bdry^1 \gengraph , \ldots , \bdry^n \gengraph} 
\big] ,
\end{align*}
and it is explicitly characterized by the probabilities
\begin{align*}
\PR^{\locmonosym}_{\invtemp ; \gengraph} \big[ \set{\spinconf} \big]
= \; & \frac{1}{\PartF^{\locmonosym}_\gengraph(\invtemp)} \, 
        e^{\minussym  \invtemp \, \Hamiltonian_{\gengraph} (\spinconf)} \qquad
\text{ for } \spinconf \in \PRspLocmono{\bdry^1 \gengraph , \ldots , \bdry^n 
\gengraph} ,
\end{align*}
where the partition function for plus boundary conditions is the sum
\begin{align*}
\PartF^{\locmonosym}_\gengraph(\invtemp)
    = \sum_{\spinconf \in \PRspLocmono{\bdry^1 \gengraph , \ldots , \bdry^n 
\gengraph}}
        e^{\minussym  \invtemp \, \Hamiltonian_{\gengraph} (\spinconf)} .
\end{align*}
Note that,
by virtue of the tower property of conditioning,
it is possible to recover any mixture of
plus and minus boundary conditions on the 
segments~$\bdry^1 \gengraph , \ldots , \bdry^n \gengraph$
by further conditioning the locally monochromatic probability
measure~$\PR^{\locmonosym}_{\invtemp ; \gengraph}$
on the event of having the desired specific constant values
on the segments.
This at least partly justifies focusing on the case of locally
monochromatic boundary conditions, as we will do.

\subsection{Ising model in the truncated strip and slit-strip}
\label{sub: Ising on strip and slit strip}

\begin{figure}[tb]
\centering
\subfigure[The truncated square grid
strip. 
] 
{
  \includegraphics[width=.4\textwidth]{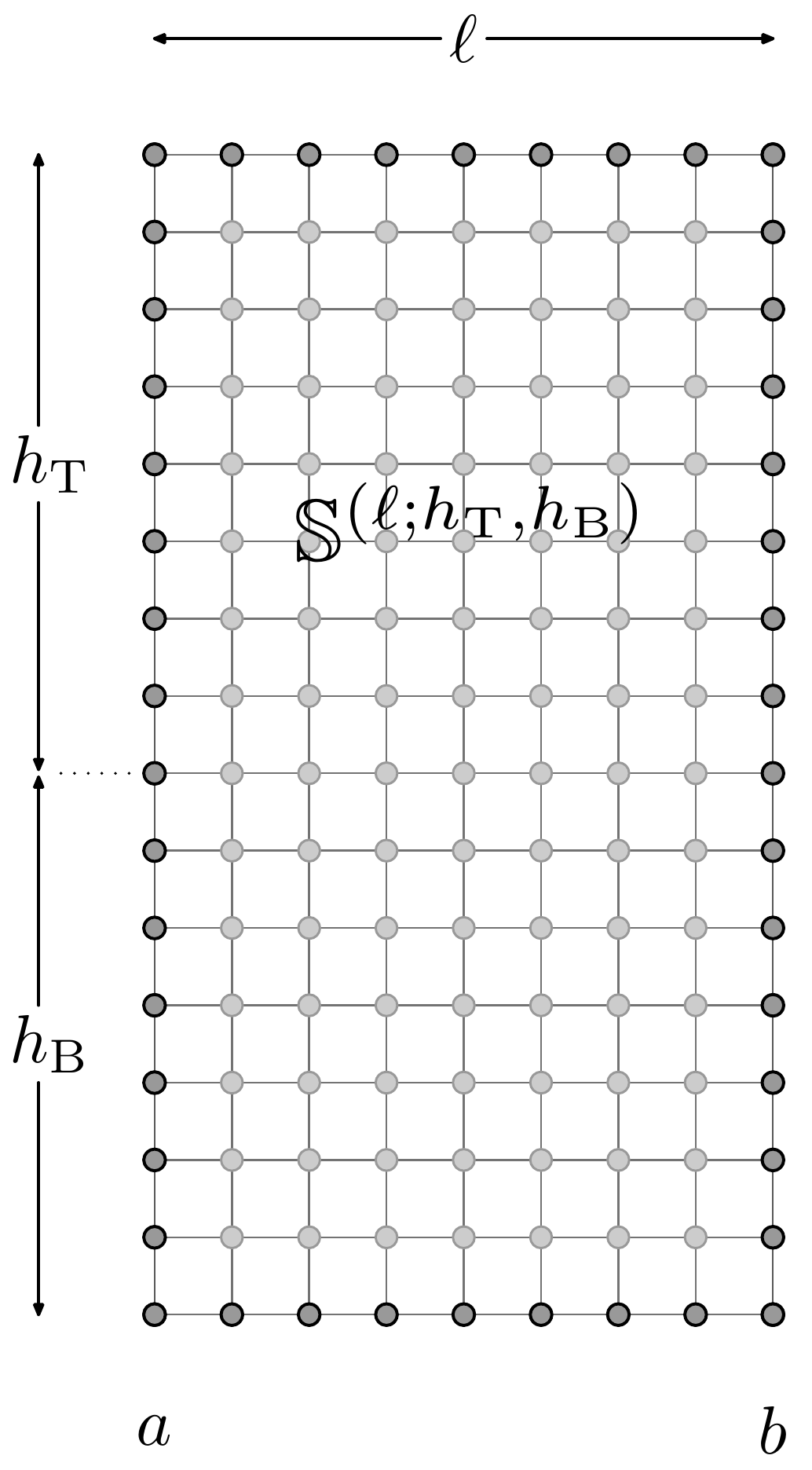}
  \label{sfig: lattice strip}
}
\hspace{1.5cm}
\subfigure[The truncated square grid 
slit-strip. 
] 
{
  \includegraphics[width=.4\textwidth]{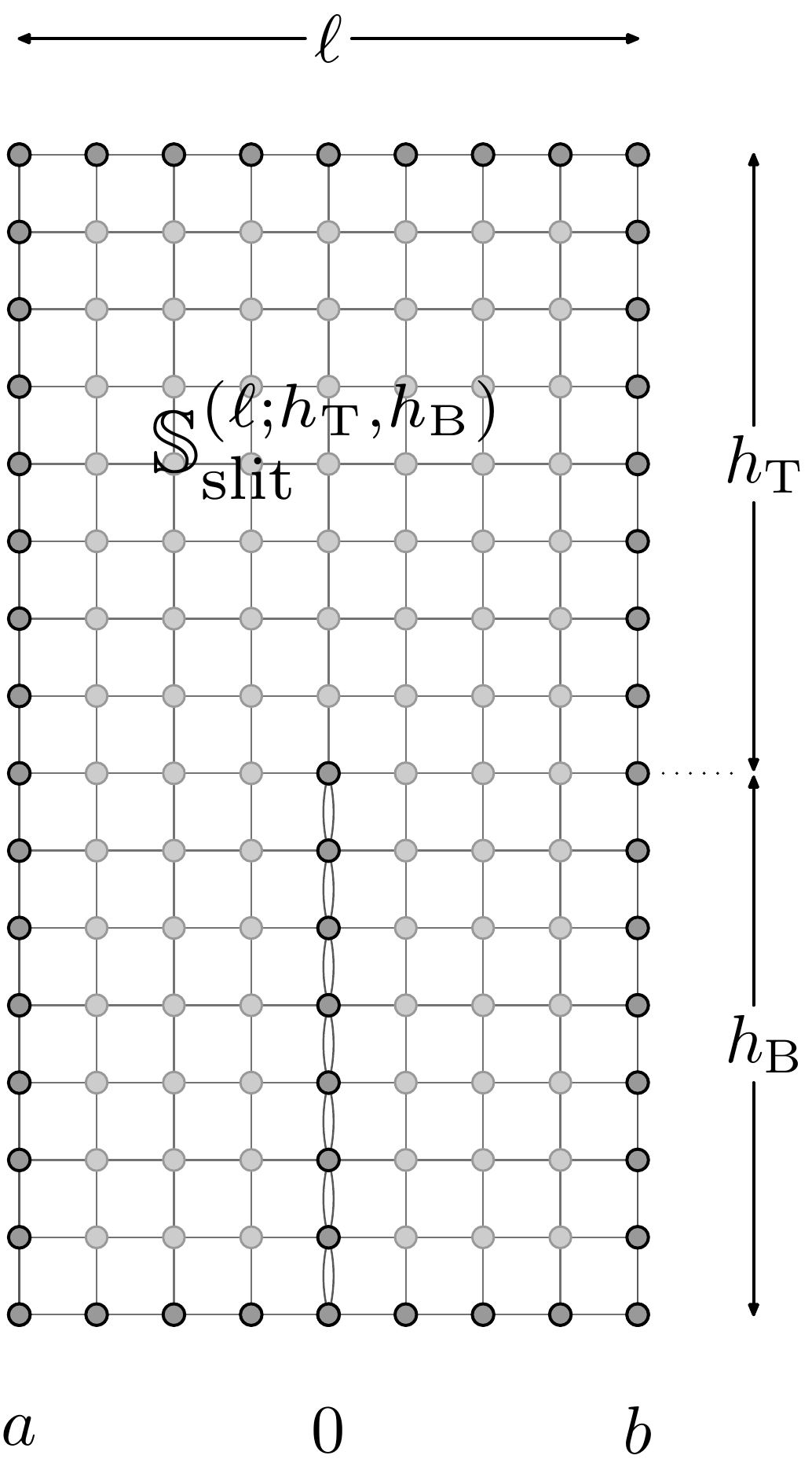}
  \label{sfig: lattice slit-strip}
}
\caption{The truncated discrete strip $\dstriptrTB{\heightT}{\heightB}$
and slit-strip~$\dslitstriptrTB{\heightT}{\heightB}$ graphs.}
\label{fig: truncated square grid strip and slit-strip}
\end{figure}

The finite graphs of interest to us are truncations of the 
lattice strip~$\dstrip$ and lattice slit-strip~$\dslitstrip$.
We now consider the Ising model with locally monochromatic
boundary conditions on these, and introduce in particular
notation for the main results of the article.

All of the graphs we consider in what follows
are subgraphs of the square lattice~$\bZ^2$.
We will therefore from here on without explicit mention fix the inverse 
temperature to the critical value for the square lattice,
\begin{align*}
\invtemp = \invtempcrit = \frac{1}{2} \log (\sqrt{2} + 1) ,
\end{align*}
and for simplicity we omit explicit references to
~$\invtemp$ from the notation.

\subsubsection*{The truncated strip}

We consider the positions $\lft , \rgt \in \bZ$
of the left and right boundaries fixed throughout so that
$\lft < \mdpt < \rgt$, and we denote 
by~$\width = \rgt - \lft \in \bN$ the width of the strip.
Let~$\heightT, \heightB \in \Zpos$ be given truncation heights
for the top and bottom parts.
The truncated strip, illustrated in 
Figure~\ref{sec: transfer matrices}.\ref{sfig: lattice strip},
is defined as
\begin{align*}
\dstriptrTB{\heightT}{\heightB}
:= \dinterval{\lft}{\rgt} \times \dinterval{-\heightB}{\heightT} ,
\end{align*}
where $\dinterval{\lft}{\rgt} = \set{\lft, \lft+1, \ldots, \rgt-1, \rgt}$
and $\dinterval{-\heightB}{\heightT} 
= \set{-\heightB, -\heightB+1 , \ldots, \heightT-1, \heightT}$
are integer intervals.
We view the truncated strip~$\dstriptrTB{\heightT}{\heightB}$
as a subgraph $\dstriptrTB{\heightT}{\heightB} \subset \dstrip$
of the lattice strip
(induced subgraph, also with nearest neighbor edges),
and note that as the truncation heights
increase, $\heightT, \heightB \to \infty$, the subgraphs exhaust
the whole lattice strip~$\dstrip$.
Boundary conditions will be imposed on the (disjoint)
subsets
\begin{align}\label{eq: left and right boundaries of the strip}
    \bdry_\lftsym \dstriptrTB{\heightT}{\heightB} = \; &
        \set{\lft} \times \dinterval{-\heightB}{\heightT} 
  & \bdry_\rgtsym \dstriptrTB{\heightT}{\heightB} = \; &
        \set{\rgt} \times \dinterval{-\heightB}{\heightT} \\
\nonumber
  & \text{(left boundary)} &
  & \text{(right boundary)}
\end{align}
constituting the vertical boundaries of the truncated
strip. In this setup, we denote the probability measure
with locally monochromatic boundary conditions on the left 
and right sides simply by
\begin{align*}
\PR^{(\width ; \heightT , \heightB)} ,
\end{align*}
and expected values with respect to it
by~$\EX^{(\width ; \heightT , \heightB)}$.

\subsubsection*{The truncated slit-strip}
Similarly, the truncated slit-strip
$\dslitstriptrTB{\heightT}{\heightB}$ of
Figure~\ref{sec: transfer matrices}.\ref{sfig: lattice slit-strip}
is taken to consist of the same 
vertices as~$\dstriptrTB{\heightT}{\heightB}$, but in this case 
we furthermore interpret the subset
\begin{align*}
\bdry_\slit \dstriptrTB{\heightT}{\heightB} = \; &
\set{\mdpt} \times \dinterval{-\heightB}{0}
\\
& \text{(slit)}
\end{align*}
as a part of the boundary where we will impose boundary conditions.
Again the truncated slit-strip~$\dstriptrTB{\heightT}{\heightB}$
is viewed as an induced subgraph 
$\dslitstriptrTB{\heightT}{\heightB} \subset \dslitstrip$
of the lattice slit-strip,
and as the truncation heights increase, $\heightT, \heightB \to \infty$,
these subgraphs exhaust the whole lattice slit-strip~$\dslitstrip$.

In this setup, we denote by
\begin{align*}
\PR_\slit^{(\width ; \heightT , \heightB)} 
\end{align*}
the probability measure
with locally monochromatic boundary conditions on the left 
and right sides as well as on the slit, and by
$\EX_\slit^{(\width ; \heightT , \heightB)}$ the expected values
with respect to it.
Arbitrary mixtures of plus and minus boundary conditions
on the left, right, and slit can again be straightforwardly
recovered by further conditioning.

We remark that for the Ising model with locally
monochromatic boundary conditions on the (truncated) slit-strip,
it makes no difference whether or not doubled edges along the slit
part are used. In a slight departure from the convention of the
previous sections, we therefore occasionally draw pictures without
the doubled edges.

\subsection{Ising model in the infinite strip and infinite slit-strip}
\label{sub: Ising model in infinite strip and slit-strip}

In the infinite lattice strip~$\dstrip$, the Ising model is
defined by an appropriate limit of the above models on
truncated strips~$\dstriptrTB{\heightT}{\heightB}$
as $\heightT, \heightB \to \infty$. The inclusion
$\dV(\dstriptrTB{\heightT}{\heightB}) \subset \dV(\dstrip)$,
allows us to interpret the sample space
$\set{\pm 1}^{\dV(\dstriptrTB{\heightT}{\heightB})}$
for the truncated strip as a subset of the sample space
$\PRspace = \set{\pm 1}^{\dV(\dstrip)}$ for the infinite strip, by
arbitrarily extending the spin configurations
(say as constant~$+1$ 
outside the truncated strip). 
The space $\PRspace = \set{\pm 1}^{\dV(\dstrip)}$ 
is a countable product of finite sets, naturally equipped with product topology.
We thus consider Borel probability measures
on this space,
and their weak convergence.

The basic infinite strip case is the 
Ising model probability measure on~$\dstrip$
with locally monochromatic boundary conditions,
defined as the following weak limit
\begin{align*}
\PR^{(\width)} 
\; = \; 
  \lim_{\heightT, \heightB \to \infty}  
\PR^{(\width;\heightT,\heightB)} 
.
\end{align*}
Expected values with respect to this measure are denoted 
by~$\EX^{(\width)}$. 
The existence of the weak limit above could be proven using
the Griffiths' correlation inequality,
whereas with plus boundary conditions the existence of the weak
limit could be proven using the FKG correlation inequality.
The existence of weak limits with all boundary 
conditions we consider is
also straightforwardly
obtained with the transfer matrix formalism, as in 
Corollary~\ref{cor: existence of infinite volume limits in the strip}
and Remark~\ref{rmk: pure boundary conditions}.\footnote{This includes the case of mixed boundary
conditions having, e.g., plus on the right and minus on the left.
Devising a proof based on correlation inequalities in this
case is not straightforward.}

Similarly in the infinite lattice slit-strip~$\dslitstrip$, the Ising model
with locally monochromatic boundary conditions is defined as 
\begin{align*}
\PR^{(\width)}_{\slit}
\; = \;
\lim_{\heightT, \heightB \to \infty} 
\PR^{(\width;\heightT,\heightB)}_{\slit}
\end{align*}
and the expected value with respect to it by~$\EX^{(\width)}_{\slit}$.
Similar comments apply to the existence of the weak limit, see 
Corollary~\ref{cor: existence of infinite volume limits in the slit-strip}
below, in particular.

Note that literal partition functions 
% play an important role in calculations, but as such, they 
are not meaningful for the Ising model on infinite 
domains such as~$\dstrip$ and~$\dslitstrip$~---
their defining sum~\eqref{eq: Ising pf general} would diverge and could not
be used to normalize probabilities as in~\eqref{eq: Ising proba general}.

Note also that 
while in finite volume, monochromatic and mixed boundary conditions
can be obtained by further conditioning from the locally monochromatic
boundary conditions, the same may fail in the infinite volume
limit if the corresponding events can 
have zero probability~---
and indeed for the genuinely mixed boundary conditions they do.
Therefore, while focusing on the case of locally monochromatic
boundary conditions, we occasionally 
make remarks about what to modify for plus, minus, and mixed
boundary conditions. We moreover emphasize that
our main objects of interest, the fusion coefficients,
actually do contain information about all of these boundary
conditions; this is ensured by the suitably chosen renormalization
in their very definition.

\subsection{Correlation functions}

We next briefly introduce the correlation functions of the Ising 
model that will be of primary interest to us.
The discussion is separated to bulk and boundary correlation functions,
and we moreover separately address the slit and the slit-strip cases.
Our main objects of interest, the fusion coefficients,
have an interpretation as boundary correlation functions
at the infinite extremities of the slit-strip.

\subsubsection*{Bulk correlation functions in the strip}

Consider first the Ising model with locally monochromatic
boundary conditions in 
the truncated strip~$\dstriptrTB{\heightT}{\heightB}$.
Given (finitely many) vertices
\begin{align*}
z_1 , \ldots , z_n \, \in \, \dV(\dstriptrTB{\heightT}{\heightB}) , 
\end{align*}
the associated spin correlation function is the quantity
\begin{align*}
\EX^{(\width;\heightT,\heightB)}
\Big[ \prod_{j=1}^n \spinconf_{z_j} \Big] \, .
\end{align*}
Apparently more generally, we could take an arbitrary 
(complex-valued) random variable on the probability space~$\PRspace$, i.e.,
an arbitrary function
$
g \colon \set{\pm 1}^{\dV(\dstriptrTB{\heightT}{\heightB})} \to \bC 
$,
and consider the expected value
$
\EX^{(\width;\heightT,\heightB)}
\big[ g(\spinconf) \big]
$.
However, any such function~$g$ is a linear combination of the functions 
of the form ${\spinconf \mapsto \prod_{j=1}^n \spinconf_{z_j}}$,
so focusing on the spin 
correlation functions entails no essential loss of generality
in the truncated strip.

In the infinite strip~$\dstrip$,  
we  still focus on similarly defined spin correlation functions, i.e., 
expected values of products of spins at finitely many vertices
\begin{align*} 
z_1 , \ldots , z_n \in \dV(\dstrip) . 
\end{align*}
Restricting to dependence on only finitely many spins here is justified
in particular by the fact that such cylinder functions are a 
measure determining and convergence determining class of functions
on the space~$\PRspace = \set{\pm 1}^{\dV(\dstrip)}$ equipped with its Borel 
sigma-algebra~\cite{Lanford, 
Billingsley-convergence_of_probability_measures, 
Georgii-Gibbs_measures}. In particular we have
\begin{align*}
\EX^{(\width)}
\Big[ \prod_{j=1}^n \spinconf_{z_j} \Big] \,
= \; & \lim_{\heightT, \heightB \to \infty} 
    \EX^{(\width;\heightT,\heightB)}
        \Big[ \prod_{j=1}^n \spinconf_{z_j} \Big] \, ,
\end{align*}
and moreover the weak limit~$\PR^{(\width)} =
\lim_{\heightT, \heightB \to \infty} 
    \PR^{(\width;\heightT,\heightB)}$
is characterized by such limits.
Exactly similar comments apply to other boundary conditions.

In the transfer matrix formalism, bulk correlation functions in the
strip involving a spin at $z = x + \ii y$
are expressed in terms of the following operators
on the state space~\eqref{eq: full state space}.
For $x \in \crosssec = \dinterval{\lft}{\rgt}$, let
\begin{align*}
\spinoper_{x} \colon \statespdbl \to \statespdbl
\end{align*}
be the diagonal matrix with matrix elements
\begin{align*}
\basisvec{\spinrowalt}^\dagger \, \spinoper_x \, \basisvec{\spinrow}
\; = \; \delta_{\spinrowalt , \spinrow} \, \spinrow_x \; \basisvec{\spinrow}.
\end{align*}
For $x \in \crosssec = \dinterval{\lft}{\rgt}$
and $y \in \bZ$, define the spin operator as
\begin{align}\label{eq: height dependent spin operator in the strip}
\spinoper(x + \ii y) = \Tmat^{-y} \; \spinoper_x \, \Tmat^{y} .
\end{align}
The precise statement of how to express bulk correlation functions in the
truncated strip~$\dstriptrTB{\heightT}{\heightB}$
in terms of the transfer matrix and these operators will be given in
Theorem~\ref{thm: transfer matrix properties strip}(6),
and a similar formula for bulk correlations in the infinite strip~$\dstrip$
will be given in Corollary~\ref{cor: transfer matrix and limit in strip}(2).

\subsubsection*{Bulk correlation functions in the slit-strip}
The definition of bulk correlation functions in the slit-strip
is exactly parallel.
In the transfer matrix formalism, bulk correlation functions 
involving the spin at $z = x + \ii y$ in the slit-strip
are expressed in terms of the following operators.
For $x \in \crosssec$ and $y \in \Znn$, we now define
\begin{align}\label{eq: height dependent spin operator in the slit-strip}
\spinoperS (x + \ii y)
    = (\TmatBasic)^{-y} \; \spinoper_x \, (\TmatBasic)^{y}  
\qquad \text{ and } \qquad
\spinoperS(x - \ii y)
    = (\TmatLR)^{+y} \; \spinoper_x \, (\TmatLR)^{-y} .
\end{align}
An expression for bulk correlation functions in the
truncated slit-strip~$\dslitstriptrTB{\heightT}{\heightB}$
in terms of the transfer matrices and these operators will be given in
Theorem~\ref{thm: transfer matrix properties slit-strip}(6),
and a similar formula for bulk correlations in the infinite 
slit-strip~$\dslitstrip$
will be given in Corollary~\ref{cor: transfer matrix and limit in slit strip}(2).

\subsubsection*{Boundary correlation functions in the strip}

Proper boundary correlation functions are
defined again in the finite, truncated strip and slit-strip~---
in the infinite strip and slit-strip one needs to form suitably
renormalized limits.
For definiteness, consider first the Ising model 
in the truncated strip, with locally monochromatic
boundary conditions on the left and right vertical boundary 
parts.
In addition to the left and right vertical bounderies,
the truncated strip has two horizontal boundary components
\begin{align*}
    \bdry_\topsym \dstriptrTB{\heightT}{\heightB} = \; &
        \dinterval{\lft}{\rgt} \times \set{\heightT}
  & \bdry_\botsym \dstriptrTB{\heightT}{\heightB} = \; &
        \dinterval{\lft}{\rgt} \times \set{-\heightB} \\
  & \text{(top boundary)} &
  & \text{(bottom boundary)},
\end{align*}
on which we have not imposed boundary conditions.
The configurations of all spins on the top and
bottom boundaries~$\bdry_\topsym \dstriptrTB{\heightT}{\heightB}$
and~$\bdry_\botsym \dstriptrTB{\heightT}{\heightB}$
are the random 
elements~$( \spinconf_{\, x+\ii \, \heightT} )_{x \in \crosssec}$
and~$( \spinconf_{\, x-\ii \, \heightB} )_{x \in \crosssec}$
of the row spin configuration space~$\set{\pm1}^{\crosssec}$,
where~$\crosssec = \dinterval{\lft}{\rgt}$.
Resorting to only a slight abuse of notation, we write these as
\begin{align*}
\spinconf \big|_{\bdry_\topsym \dstriptrTB{\heightT}{\heightB}}
   \; \in \, \set{\pm1}^{\crosssec}
\qquad \text{ and } \qquad
\spinconf \big|_{\bdry_\botsym \dstriptrTB{\heightT}{\heightB}}
   \; \in \, \set{\pm1}^{\crosssec} .
\end{align*}
Complex-valued random variables depending only on the boundary 
spins on the top 
, for example,
are therefore of the form
\begin{align*}
f \big( \spinconf |_{
			\bdry_\topsym \dstriptrTB{\heightT}{\heightB}} \big) ,
\qquad 
\text{ where $f \colon \set{\pm 1}^{\crosssec} \to \bC$ is a function.}
\end{align*}
By a boundary correlation function in the Ising model on the truncated 
strip~$\dstriptrTB{\heightT}{\heightB}$ with locally monochromatic
boundary conditions, we mean a quantity
\begin{align*}
\EX^{(\width;\heightT,\heightB)}
    \Big[ f_\topsym \big( \spinconf |_{
			\bdry_\topsym \dstriptrTB{\heightT}{\heightB}} \big)
	   \, f_\botsym \big( \spinconf |_{
			\bdry_\botsym \dstriptrTB{\heightT}{\heightB}} \big)
     \, \Big] ,
\end{align*}
where 
\begin{align*}
f_\topsym , f_\botsym \colon \set{\pm 1}^{\crosssec} \to \bC 
\end{align*}
are two given functions.

In the transfer matrix formalism, boundary correlation functions 
in the strip are expressed in terms of the following vectors
in the state space~\eqref{eq: full state space}.
For any $\spinrow \in \set{\pm 1}^{\crosssec}$, let us denote
the corresponding diagonal entry
of the matrix~$\TmatHorSqrt$ by
$c_{\spinrow} =
\exp \big( \frac{\invtemp}{2} 
    \sum_{x=\lft}^{\rgt-1} \spinrow_x \spinrow_{x+1} \big) > 0 $,
so that we have
$\TmatHorSqrt \; \basisvec{\spinrow} = c_{\spinrow} \, \basisvec{\spinrow}$.
For any function~$f \colon \set{\pm 1}^{\crosssec} \to \bC$, introduce the 
vector
\begin{align*}
\rvvec{f} :=
	\sum_{\spinrow \in \set{\pm 1}^{\crosssec}}
		c_{\spinrow} \, f(\spinrow) \; \basisvec{\spinrow}
\; \in \; \statespdbl .
\end{align*}
Denote by $\cstone$ the constant function~$1$
on~$\set{\pm 1}^{\crosssec}$. The corresponding vector
$\rvvec{\cstone} \in \statespdbl$ plays a special role,
in that it essentially encodes the free boundary conditions
(i.e., the absence of conditioning) on the top and bottom 
horizontal boundaries.\footnote{Other boundary conditions
(plus, minus, or mixed) on the top and
bottom can be achieved by selecting a different vector
as stated in the remark after 
Theorem~\ref{thm: transfer matrix properties strip}.
}
Precisely how to express the boundary correlation functions in
the truncated strip~$\dstriptrTB{\heightT}{\heightB}$ in terms of
the transfer matrix and such vectors will be stated in
Theorem~\ref{thm: transfer matrix properties strip}(7).
Renormalized limits boundary correlation functions
are addressed in Corollary~\ref{cor: transfer matrix and limit in strip}(3).

\subsubsection*{Boundary correlation functions in the slit-strip}
For the case of the truncated slit-strip, the top horizontal
boundary is as above, but the bottom horizontal
boundary is naturally split to two halves 
(which overlap at one vertex on the slit)
\begin{align*}
    \bdry_{\botsym;\lftsym} \dstriptrTB{\heightT}{\heightB} = \; &
        \dinterval{\lft}{\mdpt} \times \set{-\heightB}
  & \bdry_{\botsym;\rgtsym} \dstriptrTB{\heightT}{\heightB} = \; &
        \dinterval{\mdpt}{\rgt} \times \set{-\heightB} .\\
  & \text{(bottom left boundary)} &
  & \text{(bottom right boundary)} 
\end{align*}
Now boundary correlation functions of the Ising model
with locally monochromatic boundary conditions in the slit-strip
\begin{align*}
\EX^{(\width;\heightT,\heightB)}
    \Big[ f_\topsym \big( \spinconf |_{
			\bdry_\topsym \dstriptrTB{\heightT}{\heightB}} \big)
	   \, f_{\botsym;\lftsym} \big( \spinconf |_{
			\bdry_{\botsym;\lftsym} \dstriptrTB{\heightT}{\heightB}} \big)
	   \, f_{\botsym;\rgtsym} \big( \spinconf |_{
			\bdry_{\botsym;\rgtsym} \dstriptrTB{\heightT}{\heightB}} \big)
     \, \Big] 
\end{align*}
are defined
(with similar abuse of notation as above)
given three functions
\begin{align*}
f_\topsym \colon \set{\pm 1}^{\crosssec} \to \bC , \qquad
f_{\botsym;\lftsym} \colon \set{\pm 1}^{\crosssecL} \to \bC , \qquad
f_{\botsym;\rgtsym} \colon \set{\pm 1}^{\crosssecR} \to \bC , 
\end{align*}
where $\crosssec = \dinterval{\lft}{\rgt}$,
$\crosssecL = \dinterval{\lft}{\mdpt}$, and 
$\crosssecR = \dinterval{\mdpt}{\rgt}$.

For expressing boundary correlation functions in the slit-strip
in the transfer matrix formalism, we need in addition vectors
of the following form. Given two functions
$f_{\lftsym} \colon \set{\pm 1}^{\crosssecL} \to \bC$
$f_{\rgtsym} \colon \set{\pm 1}^{\crosssecR} \to \bC$,
we introduce the vector
\begin{align*}
\rvvec{f_\lftsym;f_\rgtsym} :=
	\sum_{\spinrow \in \set{\pm 1}^{\crosssec}} c_{\spinrow}
		\, f_{\lftsym}(\spinrow|_{\crosssecL})
		\, f_{\rgtsym}(\spinrow|_{\crosssecR})
		    \; \basisvec{\spinrow}
\; \in \; \statespdbl .
\end{align*}
Precisely how to express the boundary correlation functions in
the truncated slit-strip~$\dstriptrTB{\heightT}{\heightB}$ in terms of
the transfer matrix and such vectors will be stated in
Theorem~\ref{thm: transfer matrix properties slit-strip}(7).
Forming renormalized limits boundary correlation functions
in the slit-strip
is addressed in Corollary~\ref{cor: transfer matrix and limit in slit strip}(3).

\subsection{Transfer matrix formalism in the strip}
\label{sub: transfer matrix formalism}

The classical method of transfer matrices is very well suited for
the study of the Ising model in the lattice strip~$\dstrip$ and 
slit-strip~$\dslitstrip$, and it features crucially in our main result.
Transfer matrix methods for the
two-dimensional Ising model have a rich history, the pioneering
early contributions to which 
include~\cite{KW-statistics_of_the_2D_ferromagnet,
Onsager-crystal_statistics, 
Kaufman-crystal_statistics_2,
Yang-spontaneous_magnetization}.
Formulation of the transfer matrix formalism in terms of Clifford
algebra and fermions is due to Kaufman~\cite{Kaufman-crystal_statistics_2};
see also Schultz \& Mattis \& 
Lieb~\cite{SML-Ising_model_as_a_problem_of_fermions}. 
Transfer matrices for the Ising model with plus boundary conditions and the 
natural generalization of locally monochromatic boundary conditions
were studied in~\cite{AM-transfer_matrix_for_a_pure_phase}.
The textbook \cite{Palmer-planar_Ising_correlations} can serve as
a good reference about the transfer matrix method for the 
two-dimensional Ising model specifically, 
while~\cite{Baxter-exactly_solved_models} 
gives a broader view of 
transfer matrix methods for many different statistical mechanics models.

Mainly, our goal is to apply the Clifford algebra valued $1$-forms,
with coefficient functions related to those in 
Section~\ref{sec: distinguished functions},
to obtain formulas for the renormalized boundary correlation functions
in the slit-strip that we call the fusion coefficients.
As an illustration of our approach, which in our view has the
advantage of being conceptually close to boundary conformal field theory,
we also include a
self-contained proof of the diagonalization of the transfer matrix,
although the result itself is 
well-known~\cite{AM-transfer_matrix_for_a_pure_phase,
Palmer-planar_Ising_correlations}.

\subsubsection*{Subspaces of the state space for 
fixed boundary conditions in the strip}

The state space $\statespdbl
= \spn_\bC \big\{ \basisvec{\spinrow} \; \big| \; 
    \spinrow \in \set{\pm 1}^{\crosssec} \big\}$
defined in~\eqref{eq: full state space} decomposes into four subspaces
\begin{align*}
\statespdbl \, = \;\statespPP \, \oplus \, \statespMP \, 
    \oplus \, \statespPM \, \oplus \, \statespMM 
\end{align*}
defined by
\begin{align*}
\statespXXx{\epsilon_\lftsym \epsilon_\rgtsym}
= \; & \spn_\bC \set{ \basisvec{\spinrow} \; \Big| \; 
		\spinrow \in \set{\pm 1}^{\crosssec},
		\; \spinrow_{\lft} = \epsilon_{\lftsym},
		\; \spinrow_{\rgt} = \epsilon_{\rgtsym} } ,
\end{align*}
which are relevant for different monochromatic boundary conditions
on the left and right vertical boundaries (e.g., $\statespMP$
corresponds to minus boundary conditions on the left and 
plus boundary conditions on the right). Note that the
irreducible state space~\eqref{eq: irreducible state space}
considered in Section~\ref{sec: operator valued forms}
can be written as~$\statesp = \statespPP \oplus \statespMP$.

\subsubsection*{Main properties of the strip transfer matrix}

The following well-known theorem summarizes some key properties of the transfer 
matrix, and how it is used in calculations for the Ising model in 
the truncated strip.

\begin{thm}\label{thm: transfer matrix properties strip}
The transfer matrix~$\Tmat$ defined by~\eqref{eq: Ising transfer matrix}
has the following properties.
\begin{enumerate}
\item $\Tmat$ is an invertible symmetric matrix and
its entries are non-negative.
\item Each of the subspaces~$\statespXXx{\epsilon_\lftsym \epsilon_\rgtsym}
\subset \statespdbl$
is invariant for~$\Tmat$, and consequently also the irreducible state space 
$\statesp \subset \statespdbl$ is invariant for~$\Tmat$.
We may thus consider $\Tmat$
as an operator on any of these subspaces, by restriction.
\item The restriction of $\Tmat$ to any of the 
subspaces~$\statespXXx{\epsilon_\lftsym \epsilon_\rgtsym}$ satisfies the 
conditions of the Perron-Frobenius theorem. In particular,
it
has a unique normalized 
eigenvector $\PFvecXXx{\epsilon_\lftsym \epsilon_{\rgtsym}} \in
\statespXXx{\epsilon_\lftsym \epsilon_\rgtsym}$ with non-negative entries.
The corresponding 
eigenvalue~$\PFevXXx{\epsilon_\lftsym \epsilon_\rgtsym} > 0$
is the maximal eigenvalue of~$\Tmat$
on~$\statespXXx{\epsilon_\lftsym \epsilon_\rgtsym}$,
and it has multiplicity one.
\item The Perron-Frobenius eigenvalues satisfy
$\PFevEven \geq \PFevOdd$, and $\PFevXX{--} = \PFevEven$, and
${\PFevXX{+-} = \PFevOdd}$.
In particular, $\PFevEven$ is the maximal eigenvalue of~$\Tmat$  
on~$\statespdbl$ and~$\statesp$.
\item
The Ising model partition functions in the truncated 
strip~$\dstriptrTB{\heightT}{\heightB}$ with locally monochromatic
boundary conditions equals
\begin{align*}
\PartF_{\width;\heightT,\heightB} \, 
= \; \, & (\rvvec{\cstone})^\dagger 
            \, \Tmat^{\heightT+\heightB} \, \rvvec{\cstone} .
\end{align*}
\item
Let $z_1, \ldots, z_n \in \dV(\dstriptrTB{\heightT}{\heightB})$
be such that $\im(z_1) \leq \cdots \leq \im(z_n)$.
Then we have the following expressions for spin correlation functions of
the Ising model in the truncated strip 
with locally monochromatic boundary conditions
\begin{align*}
\EX^{(\width;\heightT,\heightB)}
\Big[ \prod_{j=1}^n \spinconf_{z_j} \Big] \, 
= \; \, & \frac{(\rvvec{\cstone})^\dagger \, \Tmat^{\heightT}
			  \; \spinoper(z_n) \, \cdots \, \spinoper(z_1)
              \; \Tmat^{\heightB} \, \rvvec{\cstone}}
          {(\rvvec{\cstone})^\dagger 
              \, \Tmat^{\heightT} 
              \, \Tmat^{\heightB} \, \rvvec{\cstone}} \; .
\end{align*}
\item Let
$f_\topsym , f_\botsym \colon \set{\pm 1}^{\crosssec} \to \bC$ be two 
functions.
Then we have the following expressions for boundary correlation functions
of the Ising model in the truncated strip with locally monochromatic
boundary conditions: 
\begin{align*}
\EX^{(\width;\heightT,\heightB)}
    \Big[ f_\topsym \big( \spinconf |_{
			\bdry_\topsym \dstriptrTB{\heightT}{\heightB}} \big)
	   \; f_\botsym \big( \spinconf |_{
			\bdry_\botsym \dstriptrTB{\heightT}{\heightB}} \big)
     \, \Big] 
= \; \, & \frac{(\rvvec{\cconj{f}_{\topsym}})^\dagger 
			  \, \Tmat^{\heightT+\heightB}
              \, \rvvec{f_{\botsym}}}
          {(\rvvec{\cstone})^\dagger 
			  \, \Tmat^{\heightT+\heightB}
              \, \rvvec{\cstone}} \; .
\end{align*}
\end{enumerate}
\end{thm}

The expressions in the last three items above make it
clear that the transfer matrix and its spectrum
give detailed information about the 
limit~$\heightB, \heightT \to \infty$ of infinite strip.\footnote{
In fact, an efficient proof of part~(4) of 
Theorem~\ref{thm: transfer matrix properties strip}
already already uses the property
\[ \PR^{(\width;\heightT,\heightB)}
  \Big[ \spinconf |_{\bdry_\lftsym \dstriptrTB{\heightT}{\heightB}}
    = \spinconf |_{\bdry_\rgtsym \dstriptrTB{\heightT}{\heightB}} \Big]
\; \geq \; \PR^{(\width;\heightT,\heightB)}
  \Big[ \spinconf |_{\bdry_\lftsym \dstriptrTB{\heightT}{\heightB}}
    \neq \spinconf |_{\bdry_\rgtsym \dstriptrTB{\heightT}{\heightB}} \Big] ,
\]
which is a consequence of the FKG inequality,
together with infinite height limits in the latter expressions.}
For the most succinct statements about the infinite height
limit, it is useful to note that in~(4) we in fact have
the strict inequality~$\PFevEven > \PFevOdd$ 
for any fixed width~$\width \in \bN$.
This strict inequality form of the statement will be obtained in 
Proposition~\ref{prop: eigenvector basis for transfer matrix} below.
With it, the following corollary about the 
infinite height limits is easily obtained.
\begin{cor}
\label{cor: transfer matrix and limit in strip}
There exists an~$\eps = \eps(\width)>0$
such that the infinite height asymptotics of the partition 
function and correlation functions of the Ising model in the
strip with locally monochromatic boundary conditions are
given by the following:\ \\
\begin{enumerate}
\item
As~$\heightB, \heightT \to \infty$ we have
\begin{align*}
\PartF_{\width;\heightT,\heightB}
= \; & (\PFevEven)^{\heightB + \heightT} 
        \Big( A + \OO(e^{-\eps \min(\heightB, \heightT)}) \Big) ,
\end{align*}
where 
$A = \, 2 \, \big((\PFvecEven)^\dagger \, \rvvec{\cstone} \big)^2 > 0$.
\item Let $z_1, \ldots, z_n \in \dV(\dstrip)$
be such that $\im(z_1) \leq \cdots \leq \im(z_n)$.
If $n$ is even,
then as~$\heightB, \heightT \to \infty$ we have
\begin{align*}
\EX^{(\width;\heightT,\heightB)}
		\Big[ \prod_{j=1}^n \spinconf_{z_j} \Big] \;
\; = \;\; & 
	\frac{(\PFvecEven)^\dagger \; \spinoper(z_n) \, \cdots \, 
			  \spinoper(z_1) \; \PFvecEven }
          {(\PFvecEven)^\dagger \; \PFvecEven} \;
          + \; \OO(e^{-\eps \min(\heightB, \heightT)}) .
\end{align*}
If $n$~is odd, the expression on the left  
vanishes by parity considerations.
\item If
$f_\topsym , f_\botsym \colon \set{\pm 1}^{\crosssec} \to \bC$
are such that $\rvvec{f_{\topsym}}, \rvvec{f_{\botsym}}$
are eigenvectors of~$\Tmat$ with 
eigenvalue~$\mu$ 
then as~$\heightB, \heightT \to \infty$ we have
\begin{align*}
& \EX^{(\width;\heightT,\heightB)}
    \Big[ f_\topsym \big( \spinconf |_{
			\bdry_\topsym \dstriptrTB{\heightT}{\heightB}} \big)
	   \; f_\botsym \big( \spinconf |_{
			\bdry_\botsym \dstriptrTB{\heightT}{\heightB}} \big)
     \, \Big] \\ 
= \; \, & 
    \Big( \frac{\mu}{\PFevEven} \Big)^{^{\heightT+\heightB}} \,
    \bigg( \frac{(\rvvec{\cconj{f}_{\topsym}})^\dagger \, \rvvec{f_{\botsym}}}
            {2 \, \big((\PFvecEven)^\dagger \, \rvvec{\cstone} \big)^2} 
        + \OO(e^{-\eps \min(\heightB, \heightT)}) \bigg) .
\end{align*}
If the eigenvectors have different eigenvalues, then
the expression on the left  
vanishes by orthogonality considerations.
\end{enumerate}
\end{cor}

Since the spin correlation functions cover a convergence determining 
collection for Borel probability measures on~$\set{\pm 1}^{\dV(\dstrip)}$,
property~(2) above in particular ensures the existence of the infinite 
volume limit probability measure.
\begin{cor}\label{cor: existence of infinite volume limits in the strip}
As $\heightT, \heightB \to \infty$,
the weak limit
$\PR^{(\width)}
\, = \, 
  \lim   
\PR^{(\width;\heightT,\heightB)} $
exists.
\end{cor}

\begin{rmk}\label{rmk: pure boundary conditions}
Statements entirely parallel to
Theorem~\ref{thm: transfer matrix properties strip}(5-7),
Corollary~\ref{cor: transfer matrix and limit in strip},
and Corollary~\ref{cor: existence of infinite volume limits in the strip}
hold with plus boundary conditions, minus boundary conditions,
and mixed boundary conditions (e.g., minus on the left and 
plus on the right boundary). They merely require replacing the 
vector~$\rvvec{\cstone}$ by its projection to the appropriate 
subspace~$\statespXX{\epsilon_\lftsym \epsilon_\rgtsym}$,
and using the corresponding Perron-Frobenius 
eigenvalue~$\PFevXX{\epsilon_\lftsym \epsilon_\rgtsym}$.
For these ``pure'' boundary conditions, we do not even need
the inequalities among the Perron-Frobenius eigenvalues in
the different subspaces, and the factors~$2$
and parity requirements
in Corollary~\ref{cor: transfer matrix and limit in strip}
are absent.
Also imposing boundary conditions on the top and bottom
boundaries is straightforward to handle by using different
vectors in place of~$\rvvec{\cstone}$.
\end{rmk}

\subsection{Diagonalization of the strip transfer matrix}
\label{sub: diagonalization of strip transfer matrix}

Having stated above the relevance of the transfer matrix and its
leading eigenvectors to the correlation functions of the Ising
model in the infinite strip, we now show how to diagonalize the
transfer matrix with the help of the Clifford algebra valued
$1$-forms of Section~\ref{sec: operator valued forms} and the
vertical translation eigenfunctions of 
Section~\ref{sec: distinguished functions}. The diagonalization
of this transfer matrix is a well-known 
result~\cite{AM-transfer_matrix_for_a_pure_phase,
Palmer-planar_Ising_correlations},
which we include mainly as a simple illustration of the use of the
boundary conformal field theory inspired method of
Section~\ref{sec: operator valued forms}.

\subsubsection*{Discrete fermion modes}

We now define specific operators of the general 
form~\eqref{eq: general strip integrated one form}, i.e., 
line integrals of the Clifford algebra valued $1$-forms across the strip.
For $k \in \pm\dposhalfint$, let
\[ \eigF{k} \colon \dE(\dstrip) \to \bC \]
denote the function on the edges of the discrete strip~$\dstrip$ as in
Proposition~\ref{prop: discrete vertical translation eigenfunctions}.
Recall that~$\eigF{k}$ is 
s-holomorphic, has Riemann boundary values, and is an eigenfunction of
vertical translations with eigenvalue~$\eigval{k}$.
Using this function, we define the corresponding 
pair~$(\cffunhol_k , \cffunbar_k)$ of coefficient functions by
\begin{align*}
\cffunhol_k (z) = \ii \, \eigF{k}(z) 
\qquad \text{ and } \qquad
\cffunbar_k(z) = -\ii \, \overline{\eigF{k}(z)} .
\end{align*}
The pair~$(\cffunhol_k , \cffunbar_k)$ is ICSH and has ICRBV by virtue of 
Lemmas~\ref{lem: SH and ICSH} and~\ref{lem: RBV and ICRBV}
and the s-holomorphicity and Riemann boundary values of~$\eigF{k}$.
We then define the $k$:th discrete fermion mode as
anticipated in Remark~\ref{rmk: the convenient association of operators},
\begin{align}\label{eq: discrete fermion mode}
\dfermode{k} 
= \frac{\eighthroot}{2} \,
    \dcint{\contour_0} \Big( \cffunhol_k(z) \, \ferhol(z) \, \dd{z} 
		+ \cffunbar_k(z) \, \ferbar(z) \, \dd{\bar{z}} \Big)
\; \in \; \CliffGen .
\end{align}

Let us first record the anticommutators and Hilbert space adjoints of 
the fermion modes.
\begin{prop}\label{prop: creation and annihilation algebra}
For $k \in \pm\dposhalfint$, we have
\begin{align*} 
(\dfermode{k})^\dagger = \dfermode{-k} , 
\end{align*}
and for $k_1 , k_2 \in \pm\dposhalfint$, we have
\begin{align*}
\anticomm{\dfermode{k_1}}{\dfermode{k_2}}
    = \delta_{k_1,-k_2} \, \id .
\end{align*}
\end{prop}
\begin{proof}
By~\eqref{eq: opposite index discrete functions}
we get the formula~$\refl \eigf{k} = \eigf{-k}$ for
the restrictions to the cross-section of the %vertical translation eigen
functions~$\eigF{k}$ and~$\eigF{-k}$
under the reflection~\eqref{eq: discrete reflection operator}.
The first asserted formula then follows from
Equation~\eqref{eq: convenient integrated one-form adjoint}
in Remark~\ref{rmk: the convenient association of operators}.
The second asserted formula follows from 
Equation~\eqref{eq: convenient integrated one-form anticommutator}
in the same remark, using furthermore that
$\innprod{\refl \eigf{k_1}}{\eigf{k_2}} =
\innprod{\eigf{-k_1}}{\eigf{k_2}} =\delta_{k_1, -k_2}$
by the orthonormality 
in Proposition~\ref{prop: discrete vertical translation eigenfunctions}.
\end{proof}

The vertical translation eigenfunction property gives the following.
\begin{lem}\label{lem: fermion mode eigenvalues}
For $k \in \pm\dposhalfint$, we have
\[ \Tmat \dfermode{k} \Tmat^{-1} = \eigval{k} \, \dfermode{k} , \]
where $\eigval{k}$ is as in~\eqref{eq: eigenvalue of vertical translation}.
\end{lem}
\begin{proof}
Directly from the definition~\eqref{eq: def fermions on horizontal edges}
of fermions on horizontal edges, we get that
\begin{align*} 
\Tmat \ferhol(z) \Tmat^{-1} = \ferhol(z-\ii)
\qquad \text{ and } \qquad
\Tmat \ferbar(z) \Tmat^{-1} = \ferbar(z-\ii) .
\end{align*}
With this observation and a change of variables by one lattice 
step vertically, the left hand side of the assertion becomes
an integral across the cross-section~$\contour_{-1}$ at height~$-1$
\begin{align*}
\Tmat \dfermode{k} \Tmat^{-1}
= \; & \frac{\eighthroot}{2} \,
    \dcint{\contour_0} \Big( \cffunhol_k(z) \, \ferhol(z-\ii) \, \dd{z} 
		+ \cffunbar_k(z) \, \ferbar(z-\ii) \, \dd{\bar{z}} \Big) \\
= \; & \frac{\eighthroot}{2} \,
    \dcint{\contour_{-1}} \Big( \cffunhol_k(z+\ii) \, \ferhol(z) \, \dd{z} 
		+ \cffunbar_k(z+\ii) \, \ferbar(z) \, \dd{\bar{z}} \Big) .
\end{align*}
Now the vertical translation eigenfunction property of~$\eigF{k}$
directly implies
\begin{align*} 
\cffunhol_k(z+\ii) = \eigval{k} \, \cffunhol_k(z)
\qquad \text{ and } \qquad
\cffunbar_k(z+\ii) = \eigval{k} \, \cffunbar_k(z) ,
\end{align*}
which then allows us to write the left hand side as
\begin{align*}
\Tmat \dfermode{k} \Tmat^{-1}
= \; & \eigval{k} \, \frac{\eighthroot}{2} \,
    \dcint{\contour_{-1}} \Big( \cffunhol_k(z) \, \ferhol(z) \, \dd{z} 
		+ \cffunbar_k(z) \, \ferbar(z) \, \dd{\bar{z}} \Big) .
\end{align*}
The assertion $\Tmat \dfermode{k} \Tmat^{-1} = \eigval{k} \, \dfermode{k}$
then follows by recalling from 
Proposition~\ref{prop: height independence of integrals across strip}
that the chosen integration height is irrelevant, since
the coefficient functions~$(\cffunhol_k, \cffunbar_k)$
are ICSH and have ICRBV.
\end{proof}

Due to the lemma above, the fermion modes shift eigenvalues of the Ising 
transfer matrix: if $\genvec$ is an 
eigenvector of~$\Tmat$ with 
eigenvalue~$\geneigval$, then $\dfermode{k} \genvec$ either vanishes or
is an eigenvector of~$\Tmat$ with eigenvalue~$\eigval{k} \geneigval$.
\begin{lem}\label{lem: shifting eigenvalues}
If $\genvec \in \statesp$ satisfies
$ \Tmat \genvec = \geneigval \, \genvec $,
then $\dfermode{k} \genvec \in \statesp$ satisfies
$\Tmat (\dfermode{k} \genvec) 
    = (\eigval{k} \geneigval) \, (\dfermode{k} \genvec) $.
\end{lem}
\begin{proof}
This follows from the calculation
\begin{align*}
\Tmat (\dfermode{k} \genvec )
= (\Tmat \dfermode{k} \Tmat^{-1}) (\Tmat \genvec)
= (\eigval{k} \, \dfermode{k}) (\geneigval \genvec)
= (\eigval{k} \geneigval) \, (\dfermode{k} \genvec) .
\end{align*}
\end{proof}

Since $\eigval{k} > 1$ for $k > 0$ and $\eigval{k} < 1$ for $k < 0$, the 
positive modes increase the magnitude of the transfer matrix eigenvalue and 
the negative modes decrease it. The positive modes
are called annihilation operators 
(the rationale for the terminology is
Proposition~\ref{prop: positive modes annihilate PF eigenvector} below)
and the negative modes are called 
creation operators.
Observe that indeed,
\begin{align}\label{eq: polarization cre and ann}
\CliffGen 
= \spn_\bC \set{\dfermode{k} \; \Big| \; k \in \dposhalfint} 
    \oplus \spn_\bC \set{\dfermode{-k} \; \Big| \; k \in \dposhalfint}
\end{align}
is a polarization of the Clifford generators:
for $k_1, k_2 \in \dposhalfint$ we have
$\anticomm{\dfermode{k_1}}{\dfermode{k_2}} = 0$
and $\anticomm{\dfermode{-k_1}}{\dfermode{-k_2}} = 0$ by 
Proposition~\ref{prop: creation and annihilation algebra}.\footnote{
This polarization property 
could alternatively be concluded directly from 
Lemma~\ref{lem: shifting eigenvalues}, since multiples of the
identity operator can neither increase nor decrease the magnitude of 
eigenvalues.}
% The rationale behind the terminology should become clear below.

\subsubsection*{Diagonalization of the transfer matrix}
In this section we combine the observations so far to provide a 
self-contained diagonalization of the transfer 
matrix~$\Tmat$ of the Ising model with locally monochromatic 
boundary conditions.

Denote by
\begin{align*}
\vac := \PFvecEven \in \statespPP \subset \statesp \subset \statespdbl 
\end{align*}
the Perron-Frobenius eigenvector of the Ising transfer 
matrix~$\Tmat$ in the plus-plus monochromatic sector~$\statespPP$,
as in Theorem~\ref{thm: transfer matrix properties strip}(3),
and by~$\vaceigval := \PFevEven$ the corresponding eigenvalue.

The discrete fermion modes~$\dfermode{k}$ with positive indices~$k>0$ are 
called annihilation operators for the following reason.
\begin{prop}\label{prop: positive modes annihilate PF eigenvector}
For any $k \in \dposhalfint$, we have
\[ \dfermode{k} \vac = 0 . \]
\end{prop}
\begin{proof}
According to Theorem~\ref{thm: transfer matrix properties strip}(5),
the Perron-Frobenius eigenvector~$\vac = \PFvecEven$ in~$\statespPP$ has the 
largest eigenvalue~$\vaceigval$
for~$\Tmat$, in all of~$\statesp$.
Now if $\dfermode{k} \vac$ were non-zero, then by 
Lemma~\ref{lem: shifting eigenvalues} it would be an eigenvector with a larger 
eigenvalue $\eigval{k} \vaceigval > \vaceigval$, which is a contradiction.
\end{proof}

Given a subset $\parts \subset \dposhalfint$, write it in the form
\begin{align}\label{eq: canonical ordering of a subset}
\parts = \set{k_1 , \ldots, k_m}
\qquad \text{ with } \qquad
0 < k_1 < \cdots < k_m .
\end{align}
We then set
\begin{align}\label{eq: transfer matrix eigenvectors}
\state{\parts} := \dfermode{-k_m} \cdots \dfermode{-k_1} \, \vac .
\end{align}
\begin{prop}\label{prop: eigenvector basis for transfer matrix}
The collection~$(\state{\parts})_{\parts \subset \dposhalfint}$
is an orthonormal basis of~$\statesp$ consisting of eigenvectors 
of~$\Tmat$.
Moreover, each 
$\state{\parts} \in \statesp$ is an eigenvector of~$\Tmat$
with eigenvalue
\begin{align}\label{eq: transfer matrix eigenvalues}
\Teigval{\parts} = \frac{\vaceigval}{\prod_{k \in \parts} \eigval{k}} ,
\end{align}
where $\eigval{k}$ are given by~\eqref{eq: eigenvalue of vertical translation}.
\end{prop}
\begin{proof}
Since~\eqref{eq: polarization cre and ann}
is a polarization and $\dfermode{k} \vac = 0$ for all $k \in \dposhalfint$,
the collection
$(\state{\parts})_{\parts \subset \dposhalfint}$ is a basis of a Fock
representation of the Clifford algebra, with vacuum vector~$\vac$. Since
$\state{\parts} \in V$ for each~$\parts$ and
the cardinality~$\# \set{\parts \subset \dposhalfint} = 2^{\width}$
of the indexing set coincides with the
dimension~$\dmn_\bC (\statesp) = 2^{\width}$, it follows that the 
collection is also a basis of~$\statesp$. Orthonormality follows by a routine 
calculation using the two properties in 
Proposition~\ref{prop: creation and annihilation algebra}
repeatedly.

Recalling that $\eigval{-k} = 1/\eigval{k}$,
a repeated application of Lemma~\ref{lem: shifting eigenvalues} gives
\[ \Tmat \state{\parts} = \Teigval{\parts} \state{\parts} , \]
so the basis vectors are indeed eigenvectors with the stated eigenvalues.
\end{proof}

A global spin flip (the linear map defined 
by~$\basisvec{\spinrow} \mapsto \basisvec{-\spinrow}$)
relates the eigenvectors of~$\Tmat$ in the irreducible 
state space~$\statesp = \statespPP \oplus \statespMP$ to the ones
in~$\statespMM \oplus \statespPM$, and one thus easily obtains an 
orthonormal basis of eigenvectors of~$\Tmat$ in~$\statespdbl$ as well.
By having found the eigenvectors~\eqref{eq: transfer matrix eigenvectors} 
and their eigenvalues~\eqref{eq: transfer matrix eigenvalues},
we have diagonalized the transfer matrix~$\Tmat$.\footnote{
Only the somewhat arbitrary overall multiplicative constant~$\vaceigval$ has 
not been explicitly determined here,
but a short calculation would also yield its value
$\vaceigval 
= \sqrt{\sqrt{2}-1} \, \big( 2 + \sqrt{2} \big)^{\width}
    \prod_{k \in \dposhalfint} (1 + 1/\eigval{k})^{-1}$.}

\subsection{Transfer matrix formalism in the slit-strip}

Recall from Section~\ref{sub: slit-strip Clifford}
that in the slit-strip, we use two separate transfer matrices:
$\TmatLR \colon \statespdbl \to \statespdbl$ given 
by~\eqref{eq: Ising transfer matrix slit part} for the bottom half with
the slit, and $\TmatBasic \colon \statespdbl \to \statespdbl$
still given by~\eqref{eq: Ising transfer matrix}
for the top half.

\subsubsection*{Subspaces of the state space for 
fixed boundary conditions in the strip}

When considering the bottom part of the slit-strip, it is natural
to decompose the state space~$\statespdbl$ to even further subspaces
\begin{align*}
\statespdbl \, = \; \statespS{+++} \, \oplus \, 
\statespS{-++} \, \oplus \, 
\statespS{+-+} \, \oplus \, 
\statespS{--+} \, \oplus \, 
\statespS{++-} \, \oplus \, 
\statespS{-+-} \, \oplus \, 
\statespS{+--} \, \oplus \, 
\statespS{---} 
\end{align*}
defined by
\begin{align*}
\statespXXx{\epsilon_\lftsym \epsilon_\slitsym \epsilon_\rgtsym}
= \; & \spn_\bC \set{ \basisvec{\spinrow} \; \Big| \; 
		\spinrow \in \set{\pm 1}^{\crosssec},
		\; \spinrow_{\lft} = \epsilon_{\lftsym},
		\; \spinrow_{\mdpt} = \epsilon_{\slitsym},
		\; \spinrow_{\rgt} = \epsilon_{\rgtsym} } ,
\end{align*}
which account for different monochromatic boundary conditions on the left and 
right vertical boundaries as well as on the slit.

\subsubsection*{Main properties of the slit-strip transfer matrices}

The following theorem, closely parallel to 
Theorem~\ref{thm: transfer matrix properties strip},
summarizes some key properties of the two transfer 
matrices above, and how they are used in calculations for the Ising 
model in the truncated slit-strip.

\begin{thm}\label{thm: transfer matrix properties slit-strip}
The transfer matrices~$\TmatLR$ and~$\TmatBasic = \Tmat$
have the following properties.
\begin{enumerate}
\item $\TmatLR$ is an invertible symmetric matrix and its entries are 
non-negative.
\item Each of the 
subspaces~$\statespS{\epsilon_\lftsym \epsilon_\slitsym \epsilon_\rgtsym} 
\subset \statespdbl$
is invariant for~$\TmatLR$, and consequently also the irreducible state space 
$\statesp \subset \statespdbl$ is invariant for~$\TmatLR$.
We may thus consider $\TmatLR$
as an operator on any of these subspaces, by restriction.
\item The restriction of $\TmatLR$ to any of the 
subspaces~$\statespS{\epsilon_\lftsym \epsilon_\slitsym \epsilon_\rgtsym}$ 
satisfies the conditions of the Perron-Frobenius theorem. In particular,
it
has a unique normalized 
eigenvector $\PFvecS{\epsilon_\lftsym \epsilon_\slitsym \epsilon_{\rgtsym}} \in
\statespS{\epsilon_\lftsym \epsilon_\slitsym \epsilon_\rgtsym}$ with 
non-negative entries.
The corresponding 
eigenvalue~$\PFevS{\epsilon_\lftsym \epsilon_\slitsym \epsilon_\rgtsym} > 0$
is the maximal eigenvalue of~$\TmatLR$
on~$\statespXXx{\epsilon_\lftsym  \epsilon_\slitsym \epsilon_\rgtsym}$,
and it has multiplicity one.
\item The maximal eigenvalue of~$\TmatLR$ 
on~$\statespdbl$ and~$\statesp$ is $\PFevS{+++}$.
\item The Ising model partition function in the truncated 
slit-strip~$\dslitstriptrTB{\heightT}{\heightB}$ with 
locally monochromatic boundary 
conditions can be expressed as follows:
\begin{align*}
\PartF^{(\width;\heightT,\heightB)}_{\slit} \, 
= \; \, & (\rvvec{\cstone})^\dagger 
            \, (\TmatBasic)^{\heightT} \, (\TmatLR)^{\heightB}
            \, \rvvec{\cstone} \; .
\end{align*}
\item
Let $z_1, \ldots, z_n \in \dV(\dstriptrTB{\heightT}{\heightB})$
be such that $\im(z_1) \leq \cdots \leq \im(z_n)$.
Then we have the following expressions for spin correlation
functions of the Ising model in the truncated slit-strip with
locally monochromatic boundary conditions:
\begin{align*}
\EX^{(\width;\heightT,\heightB)}_{\slit}
\Big[ \prod_{j=1}^n \spinconf_{z_j} \Big] \, 
= \; \, & \frac{(\rvvec{\cstone})^\dagger \, (\TmatBasic)^{\heightT}
			  \; \spinoperS(z_n) \, \cdots \, \spinoperS(z_1)
              \; (\TmatLR)^{\heightB} \, \rvvec{\cstone}}
          {(\rvvec{\cstone})^\dagger 
              \, (\TmatBasic)^{\heightT} 
              \, (\TmatLR)^{\heightB}
              \, \rvvec{\cstone}} \; ,\\
\end{align*}
\item Let
$f_{\topsym} \colon \set{\pm 1}^{\crosssec} \to \bC$,
$f_{\botsym; \lftsym} \colon \set{\pm 1}^{\crosssecL} \to \bC$,
$f_{\botsym; \rgtsym} \colon \set{\pm 1}^{\crosssecR} \to \bC$
be functions.
Then we have the following expressions for boundary correlation functions
of the Ising model in the truncated slit-strip, 
with locally monochromatic boundary conditions: 
\begin{align*}
& \EX^{(\width;\heightT,\heightB)}_{\slit}
    \Big[ f_\topsym \big( \spinconf |_{
			\bdry_\topsym \dstriptrTB{\heightT}{\heightB}} \big)
	   \; f_{\botsym;\lftsym} \big( \spinconf |_{
			\bdry_{\botsym;\lftsym} \dstriptrTB{\heightT}{\heightB}} \big)
	   \; f_{\botsym;\rgtsym} \big( \spinconf |_{
			\bdry_{\botsym;\rgtsym} \dstriptrTB{\heightT}{\heightB}} \big)
     \, \Big] \\
= \; \, & \frac{(\rvvec{\cconj{f}_{\topsym}})^\dagger 
              \, (\TmatBasic)^{\heightT} 
              \, (\TmatLR)^{\heightB}
              \, \rvvec{f_{\botsym; \lftsym};f_{\botsym; \rgtsym}}}
          {(\rvvec{\cstone})^\dagger 
              \, (\TmatBasic)^{\heightT} 
              \, (\TmatLR)^{\heightB}
              \, \rvvec{\cstone}} \; .
\end{align*}
\end{enumerate}
\end{thm}

The expressions in the last three items above again
allow for considering the
limit ${\heightB, \heightT \to \infty}$ of infinite 
slit-strip.
In order to simplify the statement, we
make use of the fact that~$\PFevXXx{+++} = \PFevXXx{---}$ are strictly
larger than the Perron-Frobenius eigenvalues in the other sectors,
which is seen as in
Proposition~\ref{prop: eigenvector basis for transfer matrix}.
\begin{cor}
\label{cor: transfer matrix and limit in slit strip}
There exists an $\eps = \eps(\width) > 0$ such that the 
infinite height asymptotics of the partition function
and correlation functions of the Ising model in the
slit-strip with locally monochromatic boundary conditions are
given by the following:\ \\
\begin{enumerate}
\item As~$\heightB, \heightT \to \infty$ we have
\begin{align*}
\PartF^{(\width;\heightT,\heightB)}_{\slit} \, 
= \; & (\PFevEven)^{\heightT} 
		\, (\PFevEvenSlit)^{\heightB} 
        \Big( A + \OO(e^{-\eps \min(\heightB, \heightT)}) \Big)
 ,
\end{align*}
where 
$A = 2 \, \big( (\rvvec{\cstone})^\dagger \, \PFvecEven \big) 
    \big( (\rvvec{\cstone})^\dagger \, \PFvecEvenSlit \big) > 0$.
\item Let $z_1, \ldots, z_n \in \dV(\dstrip)$
be such that $\im(z_1) \leq \cdots \leq \im(z_n)$.
If $n$~is even, then as~$\heightB, \heightT \to \infty$,
we have
\begin{align*}
\EX^{(\width;\heightT,\heightB)}_{\slit}
		\Big[ \prod_{j=1}^n \spinconf_{z_j} \Big]
\; = \;\; &
	\frac{(\PFvecEven)^\dagger \; \spinoperS(z_n) \, \cdots \, 
			  \spinoperS(z_1) \; \PFvecEvenSlit}
          {(\PFvecEven)^\dagger \; \PFvecEvenSlit}
    \; + \; \OO(e^{-\eps \min(\heightB, \heightT)}) . 
\end{align*}
If $n$~is odd, then the expression on the left
vanishes by parity considerations.
\item If~$f_\topsym \colon \set{\pm 1}^{\crosssec} \to \bC$
is such that $\rvvec{f_{\topsym}}$
is an eigenvector of~$\Tmat$ with eigenvalue~$\mu(f_{\topsym})$, and if 
$f_{\botsym; \lftsym} \colon \set{\pm 1}^{\crosssecL} \to \bC$,
$f_{\botsym; \rgtsym} \colon \set{\pm 1}^{\crosssecR} \to \bC$
are such that $\rvvec{f_{\botsym; \lftsym};f_{\botsym; \rgtsym}}$
is an eigenvector of~$\TmatLR$ with 
eigenvalue~$\mu(f_{\botsym; \lftsym};f_{\botsym; \rgtsym})$,
then as~$\heightB, \heightT \to \infty$ we have
\begin{align*}
& \EX^{(\width;\heightT,\heightB)}_{\slit}
    \Big[ f_\topsym \big( \spinconf |_{
			\bdry_\topsym \dstriptrTB{\heightT}{\heightB}} \big)
	   \; f_{\botsym;\lftsym} \big( \spinconf |_{
			\bdry_{\botsym;\lftsym} \dstriptrTB{\heightT}{\heightB}} \big)
	   \; f_{\botsym;\rgtsym} \big( \spinconf |_{
			\bdry_{\botsym;\rgtsym} \dstriptrTB{\heightT}{\heightB}} \big)
     \, \Big] \\
= \; \, & 
    \Big( \frac{\mu(f_{\topsym})}{\PFevEven} \Big)^{\heightT} \,
    \Big( \frac{\mu(f_{\botsym; \lftsym};f_{\botsym; \rgtsym})}
            {\PFevEvenSlit} \Big)^{\heightB} \,
    \bigg( \frac{(\rvvec{\cconj{f}_{\topsym}})^\dagger \, 
                \rvvec{f_{\botsym; \lftsym};f_{\botsym; \rgtsym}}}
            {2 \, \big( (\rvvec{\cstone})^\dagger \, \PFvecEven \big) 
            \big( (\rvvec{\cstone})^\dagger \, \PFvecEvenSlit \big)} 
        + \OO(e^{-\eps \min(\heightB, \heightT)}) \bigg) .
\end{align*}
\end{enumerate}
\end{cor}

Again since the spin correlation functions cover a convergence determining 
collection for Borel probability measures on~$\set{\pm 1}^{\dV(\dslitstrip)}$,
property~(2) above in particular ensures the existence of the infinite 
volume Ising probability measure and characterizes it.
\begin{cor}\label{cor: existence of infinite volume limits in the slit-strip}
As $\heightT, \heightB \to \infty$,
the weak limit
$\PR^{(\width)}_{\slit}
\, = \, 
  \lim 
\PR^{(\width;\heightT,\heightB)}_{\slit} $
exists.
\end{cor}

Plus, minus, and mixed boundary conditions on the vertical
boundary components, and boundary conditions on the top and
bottom horizontal boundaries can again be handled as
indicated in Remark~\ref{rmk: pure boundary conditions}.

% *******************************************************************
% **   \section{Determination of the fusion coefficients}          **
% *******************************************************************

\subsection{Fermionic operators for the slit-strip transfer matrix formalism}
\label{sec: operators in the slit strip}

We now introduce various specific choices of
integrals~\eqref{eq: general slit strip integrated one form}
of Clifford algebra valued $1$-forms across the slit-strip.
First of all, there will be three sets of 
discrete fermion modes, one for each extremity of the slit-strip,
that are exactly analogous to the creation and annihilation operators
in the strip, and they similarly yield a diagonalization
of the transfer matrix~$\TmatLR$ of the slit-part. The more novel
ones are operators better adapted to the slit-strip geometry, defined
using coefficient functions related to the globally defined
distinguished functions on the slit-strip given in 
Proposition~\ref{prop: discrete pole functions}. They will be crucial in our
characterization of the fusion coefficients in the next subsection.

\subsubsection*{Discrete fermion modes in the three extremities}

We first introduce fermion modes associated to each of the three extremities 
of the slit-strip~$\dslitstrip$. 

For the top extremity~$\dslitstripT$, we 
will in fact  use just the fermion 
modes~$\dfermode{k}$, $k \in \pm\dposhalfint$, 
introduced already in 
Section~\ref{sub: diagonalization of strip transfer matrix}.
The  functions~$\eigF{k}$ and the associated coefficient 
functions~$(\cffunhol_k,\cffunbar_k)$ are defined
on the edges of~$\dslitstripT$, and for $y \in \Znn$ the integration
along the contour~$\contour_y = \contourT_y$ from left to right across the 
slit-strip defines the appropriate fermion mode as 
in~\eqref{eq: discrete fermion mode}.

It therefore remains to introduce fermion modes associated with the left 
leg~$\dslitstripL$ and the right leg~$\dslitstripR$ of the slit-strip.
Let 
\begin{align*}
\eigFL{k} \colon \; & \dE(\dslitstripL) \to \bC &
\eigFR{k} \colon \; & \dE(\dslitstripR) \to \bC \\
		& \text{for $k \in \pm\dposhalfintL$} &
		& \text{for $k \in \pm\dposhalfintR$} 
\end{align*}
denote the functions as in
Section~\ref{sub: distinguished functions in slit strip}.
Recall that each of these is 
s-holomorphic, and has Riemann boundary values.
These are also effectively eigenfunctions of
vertical translations, in that the equations
\begin{align*}
\eigFL{k}(z + \ii) = \; & \eigvalL{k} \, \eigFL{k}(z) &
\eigFR{k}(z + \ii) = \; &\eigvalR{k} \, \eigFR{k}(z) 
\end{align*}
hold whenever both~$z$ and $z+\ii$ are in the domain of definition of the 
function.
In terms of these, we define coefficient functions by
\begin{align*}
\cffunhol_{\lftsym;k} (z) = \; & \ii \, \eigFL{k}(z) &
\cffunhol_{\rgtsym;k} (z) = \; & \ii \, \eigFR{k}(z) \\
\cffunbar_{\lftsym;k}(z) = \; & -\ii \, \overline{\eigFL{k}(z)} &
\cffunbar_{\rgtsym;k}(z) = \; & -\ii \, \overline{\eigFR{k}(z)} .
\end{align*}
The pairs  
$(\cffunhol_{\lftsym;k} , \cffunbar_{\lftsym;k})$ and
$(\cffunhol_{\rgtsym;k} , \cffunbar_{\rgtsym;k})$ are
ICSH and have ICRBV by virtue of 
Lemmas~\ref{lem: SH and ICSH in slit-strip} 
and~\ref{lem: RBV and ICRBV in slit-strip}.
We then define the discrete fermion modes in the 
left and right extremities as
\begin{align*}
\dfermodeL{k} 
= \; & \frac{\eighthroot}{2} \,
    \dcint{\contourL_0} \Big( \cffunhol_{\lftsym;k}(z) \, \ferholS(z) \, \dd{z} 
		+ \cffunbar_{\lftsym;k}(z) \, \ferbarS(z) \, \dd{\bar{z}} \Big)
    & \text{ for } k \in \pm\dposhalfintL  \\
\dfermodeR{k} 
= \; & \frac{\eighthroot}{2} \,
    \dcint{\contourR_0} \Big( \cffunhol_{\rgtsym;k}(z) \, \ferholS(z) \, \dd{z} 
		+ \cffunbar_{\rgtsym;k}(z) \, \ferbarS(z) \, \dd{\bar{z}} \Big)
    & \text{ for } k \in \pm\dposhalfintR .
\end{align*}

The anticommutators and Hilbert space adjoints of 
the fermion modes are as before.
\begin{prop}\label{prop: creation and annihilation algebras in extremities}
We have
\begin{align*} 
(\dfermodeL{k})^\dagger = \; & \dfermodeL{-k}  &
(\dfermodeR{k})^\dagger = \; & \dfermodeR{-k} 
\end{align*}
for 
$k \in \pm\dposhalfintL$ and $k \in \pm\dposhalfintR$, 
respectively.
Also, we have
\begin{align*}
\anticomm{\dfermodeL{k_1}}{\dfermodeL{k_2}}
    = \; & \delta_{k_1,-k_2} \, \id  &
\anticomm{\dfermodeR{k_1}}{\dfermodeR{k_2}}
    = \; & \delta_{k_1,-k_2} \, \id , &
\end{align*}
for  
$k_1,k_2 \in \pm\dposhalfintL$ and
$k_1,k_2 \in \pm\dposhalfintR$, respectively. Finally, we have
\begin{align*}
\anticomm{\dfermodeL{k}}{\dfermodeR{k'}} = \; & 0
\end{align*}
for  
$k \in \pm\dposhalfintL$ and $k' \in \pm\dposhalfintR$.
\end{prop}
\begin{proof}
The proofs of the first two assertions are similar to those of
Proposition~\ref{prop: creation and annihilation algebra}.
For the last assertion, note that
$\dfermodeL{k}$ is by construction a linear combination
of $\ferhol_{x'}$ and $\ferbar_{x'}$ 
for $x' \in \dintervaldual{\lft}{\mdpt}$,
whereas $\dfermodeR{k'}$ uses 
$x' \in \dintervaldual{\mdpt}{\rgt}$ instead.
These anticommute by 
Lemma~\ref{lem: anticommutators of height zero fermions}.
\end{proof}

The vertical translation eigenfunction property gives the following.
\begin{lem}
We have
\begin{align*}
\TmatLR \, \dfermodeL{k} \, (\TmatLR)^{-1} 
    = \; & \eigvalL{k} \, \dfermodeL{k} &
\TmatLR \, \dfermodeR{k} \, (\TmatLR)^{-1} 
    = \; & \eigvalR{k} \, \dfermodeR{k} 
\end{align*}
for 
$k \in \pm\dposhalfintL$ and $k \in \pm\dposhalfintR$, 
respectively.
\end{lem}
\begin{proof}
The proof is similar to Lemma~\ref{lem: fermion mode eigenvalues}.
\end{proof}

Due to this lemma, the fermion modes of the left and right leg shift 
eigenvalues of the transfer matrix~$\TmatLR$ for the slit part.
\begin{lem}\label{lem: shifting eigenvalues for slit transfer matrix}
If $\genvec \in \statesp$ satisfies
$ \TmatLR \genvec = \geneigval \, \genvec $,
then $\dfermodeL{k} \genvec \in \statesp$ and 
$\dfermodeR{k} \genvec \in \statesp$ satisfy
$\TmatLR (\dfermodeL{k} \genvec) 
    = (\eigvalL{k} \geneigval) \, (\dfermodeL{k} \genvec) $
and
$\TmatLR (\dfermodeR{k} \genvec) 
    = (\eigvalR{k} \geneigval) \, (\dfermodeR{k} \genvec) $,
respectively.
\end{lem}
\begin{proof}
The proof is similar to that of Lemma~\ref{lem: shifting eigenvalues}.
\end{proof}

Observe that
\begin{align}\label{eq: polarization cre and ann slit part}
\CliffGen = \; & \phantom{\oplus}\;\;
\spn_\bC \Big( \set{\dfermodeL{k} \; \Big| \; k \in \dposhalfintL}
	\cup \set{\dfermodeR{k} \; \Big| \; k \in \dposhalfintR} \Big) \\
\nonumber
& \oplus\; \spn_\bC \Big( \set{\dfermodeL{-k} \; \Big| \; k \in \dposhalfintL}
	\cup \set{\dfermodeR{-k} \; \Big| \; k \in \dposhalfintR} \Big)
\end{align}
is a polarization of the Clifford generators,
by Proposition~\ref{prop: creation and annihilation algebras in extremities}.
The modes $\dfermodeL{k}$ and $\dfermodeR{k}$ for positive~$k>0$ are again 
interpreted as annihilation operators and the modes for negative~$k<0$ as 
creation operators:
they now respectively raise and lower the magnitude of the eigenvalue of the 
transfer matrix~$\TmatLR$ of the slit part, according 
to Lemma~\ref{lem: shifting eigenvalues for slit transfer matrix}.

\subsubsection*{Diagonalization of the transfer matrix for the slit part}
With the fermion modes of the left and the right legs, we obtain the
following diagonalization of the transfer  matrix~$\TmatLR$ for the slit part.

Let
\begin{align*}
\vacLR = \PFvecEvenSlit \in \statespS{+++} \subset \statesp \subset \statespdbl 
\end{align*}
denote the Perron-Frobenius eigenvector of the transfer 
matrix~$\TmatLR$ in the plus-plus-plus monochromatic sector~$\statespS{+++}$,
as in Section~\ref{sec: transfer matrices}.

The modes with positive indices~$k>0$ serve
as annihilation operators.
\begin{prop}\label{prop: positive modes annihilate PF eigenvector of Tslit}
We have
\begin{align*} 
\dfermodeL{k} \vacLR = \; & 0 &
\dfermodeR{k} \vacLR = \; & 0 . 
\end{align*}
for 
$k \in \dposhalfintL$ and $k \in \dposhalfintR$, 
respectively.
\end{prop}
\begin{proof}
This follows directly from the maximality of the 
eigenvalue of the~$\vacLR = \PFvecEvenSlit$, 
Theorem~\ref{thm: transfer matrix properties slit-strip}(5).
\end{proof}

Given two subsets 
$\partsL \subset \dposhalfintL, \partsR \subset \dposhalfintR$, write them in 
the form
\begin{align*}
\partsL = \; & \set{k_1 , \ldots, k_m} &
\partsR = \; & \set{k'_1 , \ldots, k'_{m'}} \\
& 0 < k_1 < \cdots < k_m &
& 0 < k'_1 < \cdots < k'_{m'} .
\end{align*}
We then set 
\begin{align}\label{eq: slit transfer matrix eigenvectors}
\stateLR{\partsR}{\partsL} :=
\dfermodeR{-k'_{m'}} \cdots \dfermodeR{-k'_1} \,
\dfermodeL{-k_m} \cdots \dfermodeL{-k_1} \; \vacLR 
\end{align}
(note that we fix a specific ordering here, the choice of which
affects some signs later).
\begin{prop}\label{prop: eigenvector basis for slit transfer matrix}
The collection
\[ \big(\stateLR{\partsR}{\partsL} \big)_{\partsL \subset \dposhalfintL,
        \partsR \subset \dposhalfintR} \]
is an orthonormal basis of~$\statesp$ consisting of eigenvectors 
of~$\TmatLR$.
Moreover, each 
$\stateLR{\partsR}{\partsL} \in \statesp$ is an eigenvector of~$\TmatLR$
with eigenvalue
\begin{align}\label{eq: slit transfer matrix eigenvalues}
\TeigvalLR{\partsR}{\partsL}
= \frac{\vaceigvalLR}{\prod_{k \in \partsL} \eigvalL{k} \;
            \prod_{k' \in \partsR} \eigvalR{k'}} .
\end{align}
\end{prop}
\begin{proof}
In view of the polarization~\eqref{eq: polarization cre and ann slit part}, the 
proof is similar to that of 
Proposition~\ref{prop: eigenvector basis for transfer matrix}.
\end{proof}

\subsubsection*{Slit-strip adapted decompositions in the Clifford algebra}

For performing calculations with the creation and annihilation operators 
associated with each of the three extremities, we introduce yet another
set of operators for each extremity.
Informally speaking, these are constructed so that they have a specified
creation operator part, 
while the annihilation operator part
is chosen so as to make the underlying Clifford generator valued
one-form globally defined even in the scaling limit
(for the top extremity we in fact care about creation part for the
contragredient action, so we actually specify the annihilation part instead).
The primary advantage for calculations stems from 
the fact that global contour deformation is then possible,
unlike with 
the creation and annihilation operators themselves, which used
locally defined coefficient functions.

Let 
\begin{align*}
\PoleT{k} \colon \; & \dE(\dslitstrip) \to \bC &
\PoleL{k} \colon \; & \dE(\dslitstrip) \to \bC &
\PoleR{k} \colon \; & \dE(\dslitstrip) \to \bC \\
		& \text{for $k \in \dposhalfint$} &
		& \text{for $k \in \dposhalfintL$} &
		& \text{for $k \in \dposhalfintR$} 
\end{align*}
denote the functions as in
Proposition~\ref{prop: discrete pole functions}.
Recall that each of these is s-holomorphic, has Riemann boundary values
in the whole slit-strip~$\dslitstrip$.

From each of these s-holomorphic functions, we construct
globally defined coefficient function 
pairs as before, as in, e.g.,
Lemmas~\ref{lem: SH and ICSH in slit-strip}
and~\ref{lem: RBV and ICRBV in slit-strip}. For consistency with 
creation and annihilation operators, we moreover multiply 
by~${\eighthroot}/2$ before integration. The associated operators 
are defined as integrals across the full cross section 
contour~$\contour_0$ at zero height: we consider
\begin{align*}
\dpoleopT{k} := \; & \frac{\eighthroot}{2} \,
    \dcint{\contour_0} \Big( \ii \, \PoleT{k}(z) \, \ferholS(z) \, \dd{z} 
		- \ii \, \overline{\PoleT{k}(z)} \, \ferbarS(z) \, \dd{\bar{z}} \Big)
    & \text{ for } k \in \dposhalfint  , \\
\dpoleopL{k} := \; & \frac{\eighthroot}{2} \,
    \dcint{\contour_0} \Big( \ii \, \PoleL{k}(z) \, \ferholS(z) \, \dd{z} 
		- \ii \, \overline{\PoleL{k}(z)} \, \ferbarS(z) \, \dd{\bar{z}} \Big)
    & \text{ for } k \in \dposhalfintL  , \\
\dpoleopR{k} := \; & \frac{\eighthroot}{2} \,
    \dcint{\contour_0} \Big( \ii \, \PoleR{k}(z) \, \ferholS(z) \, \dd{z} 
		- \ii \, \overline{\PoleR{k}(z)} \, \ferbarS(z) \, \dd{\bar{z}} \Big)
    & \text{ for } k \in \dposhalfintR  .
\end{align*}
Since the integrations are a priori done on the cross-section at height zero,
these in fact only involve the restrictions
\begin{align*}
\PoleT{k} \big|_{\crosssecdual} = \; & \poleT{k} , &
\PoleL{k} \big|_{\crosssecdual} = \; & \poleL{k} , &
\PoleR{k} \big|_{\crosssecdual} = \; & \poleR{k} ,
\end{align*}
which are elements of the real Hilbert space
$\dfunctionsp = \bC^{\crosssecdual}$
of
complex-valued functions on the 
cross-section,~\eqref{eq: discrete function space}.

Similarly, e.g.,
$\dfermode{k}$ for a given~$k>0$ is defined by an 
integral~\eqref{eq: discrete fermion mode} across the 
cross-section at height zero, and only the restriction
\begin{align*}
\eigF{k} \big|_{\crosssecdual} = \; & \eigf{k}
\end{align*}
appears in the coefficient functions on this cross section.
The modes
$\dfermodeL{-k}$ and $\dfermodeR{-k}$ are defined by 
integrals across the left and right halves of
the cross-section at zero height,
and only the restrictions
\begin{align*}
\eigFL{-k} \big|_{\crosssecdual_\lftsym} = \; & \eigfL{-k} , &
\eigFR{-k} \big|_{\crosssecdual_\rgtsym} = \; & \eigfR{-k} ,
\end{align*}
to the two halves,
$\crosssecdual_\lftsym = \dintervaldual{\lft}{\mdpt}$ and 
$\crosssecdual_\rgtsym = \dintervaldual{\mdpt}{\rgt}$,
appear in the coefficient functions. Furthermore, we can extend the 
integration to range across the full cross-section provided that 
the coefficient functions are extended as zero on the complementary half. This 
allows us to still view the restrictions as elements of the function
space~$\dfunctionsp$.

Decompositions of the coefficient
functions in the function space~$\dfunctionsp$ therefore obviously yield 
corresponding decompositions of the operators. The following lemma phrases
the decompositions in such a way as to allow replacing any creation operators 
with linear combinations of annihilation operators. This, in turn, will yield 
recursions
for the fusion coefficients that we will study soon.
\begin{lem}\label{lem: decompositions of extremity creations}
For any $k \in \dposhalfint$, the operator~$\dfermode{k}$ can be decomposed as
\begin{align}\label{eq: decomposition of creation T}
\dfermode{k}
= \; & - \sum_{k' \in \dposhalfint} 
            \innprod{\eigf{-k'}}{\poleT{k}} \; \dfermode{-k'} \,
    + \sum_{k' \in \dposhalfintL} 
            \innprod{\eigfL{k'}}{\poleT{k}} \; \dfermodeL{k'} \,
    + \sum_{k' \in \dposhalfintR} 
            \innprod{\eigfR{k'}}{\poleT{k}} \; \dfermodeR{k'} \, .
\end{align}
For any $k \in \dposhalfintL$, the operator~$\dfermodeL{-k}$ can be decomposed 
as
\begin{align}\label{eq: decomposition of creation L}
\dfermodeL{-k}
= \; & \sum_{k' \in \dposhalfint} 
            \innprod{\eigf{-k'}}{\poleL{k}} \; \dfermode{-k'} \,
    - \sum_{k' \in \dposhalfintL} 
            \innprod{\eigfL{k'}}{\poleL{k}} \; \dfermodeL{k'} \,
    - \sum_{k' \in \dposhalfintR} 
            \innprod{\eigfR{k'}}{\poleL{k}} \; \dfermodeR{k'} \, .
\end{align}
For any $k \in \dposhalfintR$, the operator~$\dfermodeR{-k}$ can be decomposed 
as
\begin{align}\label{eq: decomposition of creation R}
\dfermodeR{-k}
= \; & \sum_{k' \in \dposhalfint} 
            \innprod{\eigf{-k'}}{\poleR{k}} \; \dfermode{-k'} \,
    - \sum_{k' \in \dposhalfintL} 
            \innprod{\eigfL{k'}}{\poleR{k}} \; \dfermodeL{k'} \,
    - \sum_{k' \in \dposhalfintR} 
            \innprod{\eigfR{k'}}{\poleR{k}} \; \dfermodeR{k'} \, .
\end{align}
\end{lem}
\begin{proof}
By Proposition~\ref{prop: discrete pole functions}, the 
functions~$\eigf{k}$ and~$\poleT{k}$
(or more precisely their extensions~$\eigF{k}$ and~$\PoleT{k}$)
have the same singular part in the top extremity. 
We can therefore expand their difference 
in terms of functions~$\eigf{-k'}$, $k'>0$, which are regular in the top 
extremity,
\begin{align*}
\eigf{k} - \poleT{k}
= \; & \sum_{k' \in \dposhalfint} c_{k;k'} \; \eigf{-k'} ,
\end{align*}
with certain real coefficients~$c_{k;k'} \in \bR$. Taking inner products of
both sides with~$\eigf{-k'}$, $k' \in \dposhalfint$, we see that the 
coefficients are given by
$c_{k;k'} = - \innprod{\eigf{-k'}}{\poleT{k}} $.

In view of the defining formula~\eqref{eq: discrete fermion mode}
and the above expansion, we find
\begin{align*}
\dfermode{k}
= \; & \dpoleopT{k}
	- \sum_{k' \in \dposhalfint} 
	            \innprod{\eigf{-k'}}{\poleT{k}} \; \dfermode{-k'} .
\end{align*}

On the other hand, the function~$\poleT{k}$ (or its extension~$\PoleT{k}$)
has no singularities in the left and the right bottom extremities
of the slit-strip. We can therefore
expand the restrictions of~$\poleT{k}$ to the left and right halves of the 
cross section in terms of the functions~$\eigfL{k'}$ and~$\eigfR{k'}$, $k'>0$, 
which are regular in the respective extremities,
\begin{align*}
\poleT{k}\big|_{\crosssecLdual}
= \; & \sum_{k' \in \dposhalfintL} 
	  \innprod{\eigfL{k'}}{\poleT{k}}  \; \eigfL{k'} , &
\poleT{k}\big|_{\crosssecRdual}
= \; & \sum_{k' \in \dposhalfintR} 
	  \innprod{\eigfR{k'}}{\poleT{k}}  \; \eigfR{k'} .
\end{align*}
If we now split the integration in~$\dpoleopT{k}$
across the cross section~$\contour_0$
in two parts, $\contour_0^\lftsym$~across the left substrip and 
$\contour_0^\rgtsym$~across the right substrip, and use the above 
decompositions, we get
\begin{align*}
\dpoleopT{k}
= \; & \sum_{k' \in \dposhalfintL} 
            \innprod{\eigfL{k'}}{\poleT{k}} \; \dfermodeL{k'}
    + \sum_{k' \in \dposhalfintR} 
            \innprod{\eigfR{k'}}{\poleT{k}} \; \dfermodeR{k'} .
\end{align*}
Combining the above, we obtain the first asserted formula.
The other two are similar.
\end{proof}

\subsection{Fusion coefficients}
\label{sec: fusion coefficients}

In Theorem~\ref{thm: transfer matrix properties slit-strip}(7),
boundary correlation functions in the truncated slit-strip were written in 
terms of quantities
\begin{align}\label{eq: truncated slit-strip boundary correlation}
\stateOut^\dagger \, (\TmatBasic)^{\heightT} (\TmatLR)^{\heightB} \, \stateIn ,
\end{align}
where $\stateIn, \stateOut \in \statespdbl$ were suitably chosen vectors 
encoding either the boundary conditions or the correlation functions in 
question.

In the infinite volume limit
$\heightT, \heightB \to \infty$, the 
quantities~\eqref{eq: truncated slit-strip boundary correlation} are
dominated by the eigenvectors of $\TmatBasic$ and $\TmatLR$ of largest 
eigenvalues 
onto which $\stateOut$ and $\stateIn$ have non-vanishing projections. 
Specifically, if we take $\stateOut$ and $\stateIn$ to be such eigenvectors,
explicitly given
in Propositions~\ref{prop: eigenvector basis for transfer matrix}
and~\ref{prop: eigenvector basis for slit transfer matrix},
\[ \stateOut = \state{\parts} 
\qquad \text{ and } \qquad
\stateIn = \stateLR{\partsR}{\partsL} ,
\]
then the matrix element~\eqref{eq: truncated slit-strip boundary correlation} 
is simply
\begin{align*}
\state{\parts}^\dagger \, (\TmatBasic)^{\heightT} 
			(\TmatLR)^{\heightB} \, \stateLR{\partsR}{\partsL}
= \; & \big( \Teigval{\parts} \big)^{\heightT} \,
		\big( \TeigvalLR{\partsR}{\partsL} \big)^{\heightB} \;
		\state{\parts}^\dagger \, \stateLR{\partsR}{\partsL} .
\end{align*}
The eigenvalues give the rate of exponential growth 
of the quantity, while the height-independent constant 
factor
\begin{align}\label{eq: fusion coefficient}
\fusionIsing{\parts}{\partsR}{\partsL}
:= \; & \state{\parts}^\dagger \, \stateLR{\partsR}{\partsL} \\ \nonumber
= \; & \vac^\dagger \; \big( \prod_{k \in \parts} \dfermode{k} \big) 
    \big( \prod_{k \in \partsL} \dfermodeL{-k} \big)
    \big( \prod_{k \in \partsR} \dfermodeR{-k} \big) \; \vacLR
\end{align}
quantifies how the incoming states at the bottom of the left and right leg of 
the slit-strip (indexed by~$\partsL$, $\partsR$, respectively)
combine to produce an outgoing state (indexed by~$\parts$) in the slit-strip 
geometry. We call these factors~\eqref{eq: fusion coefficient}
the \term{fusion coefficients} of the Ising model in the slit-strip.

Let us concretely exemplify the interpretations of the fusion coefficients
as renormalized boundary correlation functions at the three extremities
of the slit-strip. Fix $\parts \subset \dposhalfint$, $\partsR \subset \dposhalfintR$,
and~$\partsL \subset \dposhalfintL$.
If $f_\topsym \colon \set{\pm 1}^{\crosssec} \to \bC$
is the (unique) function such 
that~$\rvvec{f_{\topsym}} = \state{\parts}$, and if
$f_{\botsym; \rgtsym} \colon \set{\pm 1}^{\crosssecR} \to \bC$ and
$f_{\botsym; \lftsym} \colon \set{\pm 1}^{\crosssecL} \to \bC$
are the functions (unique up to cancelling multiplicative constants
in both) such that~$\rvvec{f_{\botsym; \lftsym};f_{\botsym; \rgtsym}} = 
\stateLR{\partsR}{\partsL}$, then by 
Theorem~\ref{thm: transfer matrix properties slit-strip}(7),
the associated boundary correlation function with locally monochromatic
boundary conditions in the slit-strip has the following
infinite slit-strip renormalized limit
\begin{align*}
& \lim_{\heightT , \heightB \to \infty}
\frac{\EX^{(\width;\heightT,\heightB)}_{\slit}
    \Big[ f_\topsym \big( \spinconf |_{
			\bdry_\topsym \dstriptrTB{\heightT}{\heightB}} \big)
	   \; f_{\botsym;\rgtsym} \big( \spinconf |_{
			\bdry_{\botsym;\rgtsym} \dstriptrTB{\heightT}{\heightB}} \big)
	   \; f_{\botsym;\lftsym} \big( \spinconf |_{
			\bdry_{\botsym;\lftsym} \dstriptrTB{\heightT}{\heightB}} \big)
     \, \Big]}
  {\big( \Teigval{\parts} / \vaceigval \big)^{\heightT} \,
  \big( \TeigvalLR{\partsR}{\partsL} / \vaceigvalLR \big)^{\heightB}} \\
= \, \; & \frac{1}{\PartFren_{\locmonosym}^{(\width)}} \,
    \frac{\fusionIsing{\parts}{\partsR}{\partsL}}
        {\fusionIsing{\emptyset}{\emptyset}{\emptyset}} ,
\end{align*}
where $\PartFren_{\locmonosym}^{(\width)}
= 2 \; \vac^\dagger \rvvec{\cstone} \; (\rvvec{\cstone;\cstone})^\dagger \vacLR$.
Note that the constant prefactors are positive numbers:
$\PartFren_{\locmonosym}^{(\width)} >0$ and~$\fusionIsing{\emptyset}{\emptyset}{\emptyset}>0$,
since they involve inner products of Perron-Frobenius eigenvectors with
other vectors with nonnegative components, and there are overlapping
non-zero components in each case.
For other boundary conditions we get slightly different formulas, but with
the same structure.
For example, consider the minus-plus-plus
boundary conditions (minus on the left 
boundary~$\bdry_{\lftsym} \dstriptrTB{\heightT}{\heightB}$, 
plus on the slit~$\bdry_{\slit} \dstriptrTB{\heightT}{\heightB}$,
plus on the right boundary~$\bdry_{\rgtsym} \dstriptrTB{\heightT}{\heightB}$).
Assume furthermore that $\parts, \partsR, \partsL$ are chosen so that the 
functions~$f_{\topsym}, f_{\botsym; \rgtsym}, f_{\botsym; \lftsym}$
are supported on the spin configurations allowed
by these boundary conditions (this assumption obviously causes no loss of generality
for nontrivial boundary correlation functions, and it lets us avoid certain 
projections).
Then, modifying Theorem~\ref{thm: transfer matrix properties slit-strip}(7)
as indicated in Remark~\ref{rmk: pure boundary conditions}, one finds
\begin{align*}
& \lim_{\heightT , \heightB \to \infty}
\frac{\EX^{(\width;\heightT,\heightB)}_{\bcssMPP;\slit}
    \Big[ f_\topsym \big( \spinconf |_{
			\bdry_\topsym \dstriptrTB{\heightT}{\heightB}} \big)
	   \; f_{\botsym;\rgtsym} \big( \spinconf |_{
			\bdry_{\botsym;\rgtsym} \dstriptrTB{\heightT}{\heightB}} \big)
	   \; f_{\botsym;\lftsym} \big( \spinconf |_{
			\bdry_{\botsym;\lftsym} \dstriptrTB{\heightT}{\heightB}} \big)
     \, \Big]}
  {\big( \Teigval{\parts} / \Teigval{\{\half\}} \big)^{\heightT} \,
  \big( \TeigvalLR{\partsR}{\partsL} 
          / \TeigvalLR{\emptyset}{\{\half\}} \big)^{\heightB}} \\
= \, \; & \frac{1}{\PartFren_{\bcssMPP}^{(\width)}} \,
    \frac{\fusionIsing{\parts}{\partsR}{\partsL}}
        {\fusionIsing{\{\half\}}{\emptyset}{\{\half\}}} ,
\end{align*}
where $\PartFren_{\bcssMPP}^{(\width)}
= \state{\{\half\}}^\dagger \rvvec{\cstone} \;
(\rvvec{\cstone;\cstone})^\dagger \stateLR{\emptyset}{\{\half\}}$.
Note again that the constant prefactors are positive numbers:
$\PartFren_{\bcssMPP}^{(\width)} >0$ and $\fusionIsing{\{\half\}}{\emptyset}{\{\half\}}>0$.

As we vary $\parts, \partsR, \partsL$, the associated 
functions~$f_{\topsym}, f_{\botsym; \rgtsym}, f_{\botsym; \lftsym}$ form a
basis\footnote{Strictly speaking we thus cover basis of
functions supported on the row configurations corresponding to the irreducible
state space~$\statesp$; to include the full state space~$\statespdbl$, a
further global spin-flip is needed,
which would duplicate the basis we have chosen to concentrate on.}
of functions on the spin configurations on the three boundary components,
and the above formulas illustrate the general property
that (ratios of) the fusion coefficients capture
renormalized limits of boundary correlation functions with both the
locally monochromatic as well as any of the fixed monochromatic boundary 
conditions. A minor difference due to the boundary conditions remains in
the prefactors\footnote{Writing the prefactor as a product of two factors 
is merely a natural convention that indicates factors of two kinds, but 
nothing prevents from combining the prefactors into one constant determined 
by the boundary conditions.}
, $\PartFren_{\locmonosym}$ and ${\fusionIsing{\emptyset}{\emptyset}{\emptyset}}$ 
versus $\PartFren_{\bcssMPP}$ and
${\fusionIsing{\{\half\}}{\emptyset}{\{\half\}}}$
in the two examples above. More drastically,
even the appropriate exponential renormalizations
(the denominators on the left hand sides) that are needed
to form nontrivial limits of the boundary correlation functions
are different.
Despite that, the (ratios of) fusion coefficients capture the suitably
renormalized limits of boundary correlation functions in all cases.

Our main result of this second part 
of the series will be that
the fusion coefficients~\eqref{eq: fusion coefficient} converge
in the limit~$\width \to \infty$ of infinite strip width;
this will be proven 
in Section~\ref{sec: continuum fusion coefs}.
In the final part of the series we will show that this limit of the
fusion coefficients fully recovers the
algebraic structure of a (chiral) conformal field theory.
We now proceed to give
a characterization of the fusion coefficients, which lends
itself to those purposes.

\subsubsection*{Recursion for the fusion coefficients}

Recall from
Propositions~\ref{prop: eigenvector basis for transfer matrix}
and~\ref{prop: eigenvector basis for slit transfer matrix}
that the eigenvectors
$\state{\parts}$~\eqref{eq: transfer matrix eigenvectors} 
of the Ising transfer matrix~$\TmatBasic$ are indexed by subsets
\begin{align*}
\parts \subset \dposhalfint 
    = \set{\frac{1}{2}, \;\frac{3}{2}, \; \ldots , \; \width-\frac{1}{2}} ,
\end{align*}
and the eigenvectors
$\stateLR{\partsR}{\partsL}$~\eqref{eq: slit transfer matrix eigenvectors} 
of the Ising
transfer matrix~$\TmatLR$ for the slit part are indexed by pairs
of subsets
\begin{align*}
\partsL \subset \dposhalfintL = \; & 
  \set{\frac{1}{2}, \; \frac{3}{2}, \; \ldots , \; \widthL-\frac{1}{2}} , &
\partsR \subset \dposhalfintR = \; & 
  \set{\frac{1}{2}, \; \frac{3}{2}, \; \ldots , \; \widthR-\frac{1}{2}} .
\end{align*}
We write the elements of the subsets in increasing order
as in~\eqref{eq: canonical ordering of a subset},
\begin{align*}
\parts = \; & \set{k_1 , \ldots, k_m}
& \text{ with } \qquad &
0 < k_1 < \cdots < k_m ,
\end{align*}
and similarly for
$\partsL = \set{k^\lftsym_1 , \ldots, k^\lftsym_{m_\lftsym}}$
and $\partsR = \set{k^\rgtsym_1 , \ldots, k^\rgtsym_{m_\rgtsym}}$.
The number of elements in such a subset
$\parts = \set{k_1 , \ldots, k_m}$ of positive half-integers 
will be denoted by
\begin{align*}
\pLen{\parts} = \; & m .
\end{align*}
We also use the \term{signed indicator} notation
\begin{align}\label{eq: signed indicator}
\pSign{\parts}{k} = \begin{cases}
  (-1)^{m-j} & \text{ if } k = k_j 
				  \\
  0 & \text{ if } k \notin \alpha ,
\end{cases}
\end{align}
whenever $\parts = \set{k_1 , \ldots, k_m}$
is a subset of positive half-integers as above
and $k$ is a given positive half-integer,
and we write $\parts \setminus \set{k}$ for the subset 
where~$k$ has been removed from~$\parts$.
Such notations simplify, for example, the following calculations with 
eigenvectors of the transfer matrices~$\TmatLR$ and~$\TmatBasic$.
\begin{lem}\label{lem: annihilate one in a leg}
Let $\partsL \subset \dposhalfintL$ and $\partsR \subset \dposhalfintR$.

For~$k \in \dposhalfintL$ we have
\begin{align}\label{eq: discrete annihilation formula L}
\dfermodeL{k} \, \stateLR{\partsR}{\partsL}
= \; & (-1)^{\pLen{\partsR}} \, 
        \pSign{\partsL}{k} \; \stateLR{\partsL \setminus \set{k}}{\partsR} ,
\end{align}
and for $k \in \dposhalfintR$ we have
\begin{align}\label{eq: discrete annihilation formula R}
\dfermodeR{k} \, \stateLR{\partsR}{\partsL}
= \; & \pSign{\partsR}{k}
		\; \stateLR{\partsL}{\partsR \setminus \set{k}} .
\end{align}
\end{lem}
\begin{lem}\label{lem: annihilate one at the top}
For any $\parts \subset \dposhalfint$ and~$k \in \dposhalfint$ we have
\begin{align}\label{eq: discrete annihilation formula T}
\state{\parts}^\dagger \, \dfermode{-k} 
= \; & \pSign{\parts}{k} \; \state{\parts \setminus \set{k}}^\dagger .
\end{align}
\end{lem}
\begin{proof}[Proofs of Lemmas~\ref{lem: annihilate one in a leg}
and~\ref{lem: annihilate one at the top}]
The proofs of all three 
formulas~\eqref{eq: discrete annihilation formula L},
\eqref{eq: discrete annihilation formula R}, 
and~\eqref{eq: discrete annihilation formula T} are similar and 
completely standard.
We do the first of these below.

From the defining formula~\eqref{eq: slit transfer matrix eigenvectors}
of~$\stateLR{\partsR}{\partsL}$, we get
\begin{align*}
\dfermodeL{k} \, \stateLR{\partsR}{\partsL}
= \; & \dfermodeL{k} \;\; \, 
        \dfermodeR{-k^\rgtsym_{m_\rgtsym}} \cdots \dfermodeR{-k^\rgtsym_1} \,
        \dfermodeL{-k^\lftsym_{m_\lftsym}} \cdots \dfermodeL{-k^\lftsym_1} \; 
        \vacLR .
\end{align*}
Now by 
Proposition~\ref{prop: creation and annihilation algebras in extremities},
$\dfermodeL{k}$ anticommutes with each~$\dfermodeR{-k^\rgtsym_{i}}$, so we
can first of all rewrite
\begin{align*}
\dfermodeL{k} \, \stateLR{\partsR}{\partsL}
% & \dfermodeL{k} \;\; \, 
%         \dfermodeR{-k^\rgtsym_{m_\rgtsym}} \cdots \dfermodeR{-k^\rgtsym_1} \,
%         \dfermodeL{-k^\lftsym_{m_\lftsym}} \cdots \dfermodeL{-k^\lftsym_1} \; 
%         \vacLR \\
= \; & (-1)^{m_\rgtsym} \;
        \dfermodeR{-k^\rgtsym_{m_\rgtsym}} \cdots \dfermodeR{-k^\rgtsym_1} \,
		\;\; \dfermodeL{k} \;\;
        \dfermodeL{-k^\lftsym_{m_\lftsym}} \cdots \dfermodeL{-k^\lftsym_1} \; 
        \vacLR .
\end{align*}
Moreover, $\dfermodeL{k}$ anticommutes with~$\dfermodeL{-k^\lftsym_{j}}$ except if~$k^\lftsym_{j} = k$, so further anticommuting it to the right, we get
\begin{align*}
\dfermodeL{k} \, \stateLR{\partsR}{\partsL}
% = \; & \dfermodeL{k} \;\; \, 
%         \dfermodeR{-k^\rgtsym_{m_\rgtsym}} \cdots \dfermodeR{-k^\rgtsym_1} \,
%         \dfermodeL{-k^\lftsym_{m_\lftsym}} \cdots \dfermodeL{-k^\lftsym_1} \; 
%         \vacLR \\
% = \; & (-1)^{m_\rgtsym} \;
%         \dfermodeR{-k^\rgtsym_{m_\rgtsym}} \cdots \dfermodeR{-k^\rgtsym_1} \,
% 		\;\; \dfermodeL{k} \;\;
%         \dfermodeL{-k^\lftsym_{m_\lftsym}} \cdots \dfermodeL{-k^\lftsym_1} \; 
%         \vacLR \\
= \; & \sum_{j=1}^{m_{\lftsym}}
    (-1)^{m_\rgtsym+m_\lftsym-j} \, \delta_{k,k^\lftsym_{j}} \;
    \dfermodeR{-k^\rgtsym_{m_\rgtsym}} \cdots \dfermodeR{-k^\rgtsym_1} \,
    \dfermodeL{-k^\lftsym_{m_\lftsym}} \cdots \dfermodeL{-k^\lftsym_{j+1}} \,
    \dfermodeL{-k^\lftsym_{j-1}} \cdots \dfermodeL{-k^\lftsym_1} \,
    \vacLR \\
 & \qquad + (-1)^{m_\rgtsym + m_\lftsym} \;
        \dfermodeR{-k^\rgtsym_{m_\rgtsym}} \cdots \dfermodeR{-k^\rgtsym_1} \,
        \dfermodeL{-k^\lftsym_{m_\lftsym}} \cdots \dfermodeL{-k^\lftsym_1} \,
		\;\; \dfermodeL{k} \;\;
        \vacLR .
% = \; & (-1)^{m_\rgtsym} \, \sum_{j=1}^{m_{\lftsym}}
%     (-1)^{m_\lftsym-j} \, \delta_{k,k^\lftsym_{j}} \;
%     \dfermodeR{-k^\rgtsym_{m_\rgtsym}} \cdots \dfermodeR{-k^\rgtsym_1} \,
%     \dfermodeL{-k^\lftsym_{m_\lftsym}} \cdots \dfermodeL{-k^\lftsym_{j+1}} \,
%     \dfermodeL{-k^\lftsym_{j-1}} \cdots \dfermodeL{-k^\lftsym_1} \,
%     \vacLR \\
%  & \qquad + (-1)^{m_\rgtsym + m_\lftsym} \;
%         \dfermodeR{-k^\rgtsym_{m_\rgtsym}} \cdots \dfermodeR{-k^\rgtsym_1} \,
%         \dfermodeL{-k^\lftsym_{m_\lftsym}} \cdots \dfermodeL{-k^\lftsym_1} \,
% 		\;\; \dfermodeL{k} \;\;
%         \vacLR .
\end{align*}
The last term vanishes in view of the annihilation property
$\dfermodeL{k} \, \vacLR = 0$ of 
Proposition~\ref{prop: positive modes annihilate PF eigenvector of Tslit}.
Since $k^\lftsym_{1} , \ldots , k^\lftsym_{m_\lftsym}$ are distinct,
at most one term in the sum over~$j$ can be non-vanishing, and this happens
if $k \in \partsL$. In that case the sign of the term is
$(-1)^{m_\rgtsym} \, (-1)^{m_\lftsym-j} 
= (-1)^{\pLen{\partsR}} \, \pSign{\partsL}{k}$, and the remaining
creation operators applied to the vacuum yield the vector
$\stateLR{\partsL \setminus \set{k}}{\partsR}$.
Formula~\eqref{eq: discrete annihilation formula L} is thus established.
\end{proof}

We now show how the decompositions of 
Lemma~\ref{lem: decompositions of extremity creations}
yield recursions for the fusion coefficients, which in fact uniquely
determine them apart from an overall multiplicative constant that
cancels in the ratios that we are interested in.

\begin{thm}\label{thm: recursion for fusion coefficients}
The collection $( \fusionIsing{\parts}{\partsR}{\partsL})_{
\parts \subset \dposhalfint ,
\partsL \subset \dposhalfintL , 
\partsR \subset \dposhalfintR}$
of all fusion coefficients satisfies the following properties, 
which furthermore characterize the collection
up to an overall positive multiplicative constant:
\begin{itemize}
\item[\recIsingIC] We have 
$\fusionIsing{\emptyset}{\emptyset}{\emptyset} > 0$.
\item[\recIsingT] If $\parts \subset \dposhalfint,
\partsL \subset \dposhalfintL , 
\partsR \subset \dposhalfintR$ 
and $\parts' = \parts \cup \set{k}$ with $\max(\parts) < k < \width$,
then we have
\begin{align*}
\fusionIsing{\parts'}{\partsR}{\partsL}
= \; & + \sum_{k' \in \partsL} \innprod{\eigfL{k'}}{\poleT{k}}
	\; (-1)^{\pLen{\partsR}} \; \pSign{\partsL}{k'} 
	\; \fusionIsing{\parts}{\partsL\setminus\set{k'}}{\partsR} \\
& + \sum_{k' \in \partsR} \innprod{\eigfR{k'}}{\poleT{k}}
	\; \pSign{\partsR}{k'}
	\; \fusionIsing{\parts}{\partsL}{\partsR\setminus\set{k'}} \\
& - \sum_{k' \in \parts} \innprod{\eigf{-k'}}{\poleT{k}}
	\; \pSign{\parts}{k'}
	\; \fusionIsing{\parts\setminus\set{k'}}{\partsR}{\partsL} .
\end{align*}
\item[\recIsingL] If $\parts \subset \dposhalfint,
\partsL \subset \dposhalfintL , 
\partsR \subset \dposhalfintR$ 
and $\partsL' = \partsL \cup \set{k}$ with $\max(\partsL) < k < \widthL$,
then we have
\begin{align*}
\fusionIsing{\parts}{\partsL'}{\partsR}
= \; & - \sum_{k' \in \partsL} \innprod{\eigfL{k'}}{\poleL{k}}
	\; (-1)^{\pLen{\partsR}} \; \pSign{\partsL}{k'} 
	\; \fusionIsing{\parts}{\partsL\setminus\set{k'}}{\partsR} \\
& - \sum_{k' \in \partsR} \innprod{\eigfR{k'}}{\poleL{k}}
	\; \pSign{\partsR}{k'}
	\; \fusionIsing{\parts}{\partsL}{\partsR\setminus\set{k'}} \\
& + \sum_{k' \in \parts} \innprod{\eigf{-k'}}{\poleL{k}}
	\; \pSign{\parts}{k'}
	\; \fusionIsing{\parts\setminus\set{k'}}{\partsR}{\partsL} .
\end{align*}
\item[\recIsingR] If $\parts \subset \dposhalfint,
\partsL \subset \dposhalfintL , 
\partsR \subset \dposhalfintR$ 
and $\partsR' = \partsR \cup \set{k}$ with $\max(\partsR) < k < \widthR$,
then we have
\begin{align*}
\fusionIsing{\parts}{\partsL}{\partsR'}
= \; & - \sum_{k' \in \partsL} \innprod{\eigfL{k'}}{\poleR{k}}
	\; (-1)^{\pLen{\partsR}} \; \pSign{\partsL}{k'} 
	\; \fusionIsing{\parts}{\partsL\setminus\set{k'}}{\partsR} \\
& - \sum_{k' \in \partsR} \innprod{\eigfR{k'}}{\poleR{k}}
	\; \pSign{\partsR}{k'}
	\; \fusionIsing{\parts}{\partsL}{\partsR\setminus\set{k'}} \\
& + \sum_{k' \in \parts} \innprod{\eigf{-k'}}{\poleR{k}}
	\; \pSign{\parts}{k'}
	\; \fusionIsing{\parts\setminus\set{k'}}{\partsR}{\partsL} .
\end{align*}
\end{itemize}
\end{thm}
\begin{proof}
Let us first address the uniqueness statement.
The three recursions, \recIsingT{}, \recIsingL{}, and \recIsingR{}, allow one 
to express any coefficient
$\fusionIsing{\parts}{\partsR}{\partsL}$
as a linear combination of the coefficients with total size strictly less than
$\pLen{\parts}+\pLen{\partsL}+\pLen{\partsR}$.
The total size is a non-negative 
integer, so inductively these properties allow to write any
$\fusionIsing{\parts}{\partsR}{\partsL}$ as a multiple of
the only coefficient~$\fusionIsing{\emptyset}{\emptyset}{\emptyset}$ with 
total size zero.
The initial condition~\recIsingIC{} specifies the sign of this coefficient,
and therefore the whole collection
$(\fusionIsing{\parts}{\partsR}{\partsL})_{\parts,\partsL,\partsR}$ gets determined
up to a positive overall multiplicative
factor~$\fusionIsing{\emptyset}{\emptyset}{\emptyset}$.

The positivity property~\recIsingIC{} of the initial 
coefficient~$\fusionIsing{\emptyset}{\emptyset}{\emptyset}$
follows directly from the defining formula~\eqref{eq: fusion coefficient},
once one notices that the vectors involved in the inner products are
Perron-Frobenius eigenvectors with non-negative entries, and 
there is non-empty overlap of components where the entries are
non-vanishing.

Consider then~\recIsingT{}. We must calculate
\begin{align*}
\fusionIsing{\parts'}{\partsR}{\partsL}
= \; & \state{\parts'}^\dagger \, \stateLR{\partsR}{\partsL} 
= (\dfermode{-k} \state{\parts})^\dagger \, \stateLR{\partsR}{\partsL}
= \state{\parts}^\dagger \, \dfermode{k} \stateLR{\partsR}{\partsL} .
\end{align*}
To this end, we use
formula~\eqref{eq: decomposition of creation T} of 
Lemma~\ref{lem: decompositions of extremity creations}
for~$\dfermode{k}$,
\begin{align*}
\dfermode{k}
= \; & - \sum_{k' \in \dposhalfint} 
            \innprod{\eigf{-k'}}{\poleT{k}} \; \dfermode{-k'} \,
    + \sum_{k' \in \dposhalfintL} 
            \innprod{\eigfL{k'}}{\poleT{k}} \; \dfermodeL{k'} \,
    + \sum_{k' \in \dposhalfintR} 
            \innprod{\eigfR{k'}}{\poleT{k}} \; \dfermodeR{k'} \, .
\end{align*}
To find the contribution from the second sum above to the fusion 
coefficient~$\fusionIsing{\parts'}{\partsR}{\partsL}$, we 
use~\eqref{eq: discrete annihilation formula L}
from Lemma~\ref{lem: annihilate one in a leg} in
\begin{align*}
\sum_{k' \in \dposhalfintL} \innprod{\eigfL{k'}}{\poleT{k}} \; 
	\state{\parts}^\dagger \;\dfermodeL{k'} \; \stateLR{\partsR}{\partsL}
= \; & \sum_{k' \in \partsL} \innprod{\eigfL{k'}}{\poleT{k}} 
	\; (-1)^{\pLen{\partsR}} \; \pSign{\partsL}{k'} 
	\; \state{\parts}^\dagger \, \stateLR{\partsL\setminus\set{k'}}{\partsR} \\
= \; & \sum_{k' \in \partsL} \innprod{\eigfL{k'}}{\poleT{k}}
	\; (-1)^{\pLen{\partsR}} \; \pSign{\partsL}{k'} 
	\; \fusionIsing{\parts}{\partsL\setminus\set{k'}}{\partsR} .
\end{align*}
Similarly using~\eqref{eq: discrete annihilation formula L}, the
contribution of the third becomes
\begin{align*}
\sum_{k' \in \dposhalfintR} \innprod{\eigfR{k'}}{\poleT{k}} \; 
	\state{\parts}^\dagger \; \dfermodeR{k'} \; \stateLR{\partsR}{\partsL}
= \; & \sum_{k' \in \partsR} \innprod{\eigfR{k'}}{\poleT{k}}
	\; \pSign{\partsR}{k'}
	\; \fusionIsing{\parts}{\partsL}{\partsR\setminus\set{k'}}
\end{align*}
For the contribution of the first sum we 
use~\eqref{eq: discrete annihilation formula T} from
Lemma~\ref{lem: annihilate one at the top}, and get
\begin{align*}
- \sum_{k' \in \dposhalfint} \innprod{\eigf{-k'}}{\poleT{k}} \;
    \state{\parts}^\dagger \, \dfermode{-k'} \; \stateLR{\partsR}{\partsL}
= \; & - \sum_{k' \in \parts} \innprod{\eigf{-k'}}{\poleT{k}}
	\; \pSign{\parts}{k'}
	\; \fusionIsing{\parts\setminus\set{k'}}{\partsR}{\partsL}
\end{align*}
Combining the terms, we obtain the formula asserted in~\recIsingT{}.

The proofs of~\recIsingL{} and~\recIsingR{} are similar.
\end{proof}

\bigskip

\section{Scaling limits of distinguished functions}
\label{sec: limit distinguished functions}

% *******************************************************************
% **   \section{S-hol., RBVs, and distinguished functions}         **
% *******************************************************************

In this section we discuss the continuum analogues of the distinguished
lattice functions of Section~\ref{sec: distinguished functions}, and recall
the relevant scaling limit results from~\cite{part-1}.

\begin{figure}[tb]
\centering
\subfigure[The strip domain~$\cstrip$.] 
{
  \includegraphics[width=.35\textwidth]{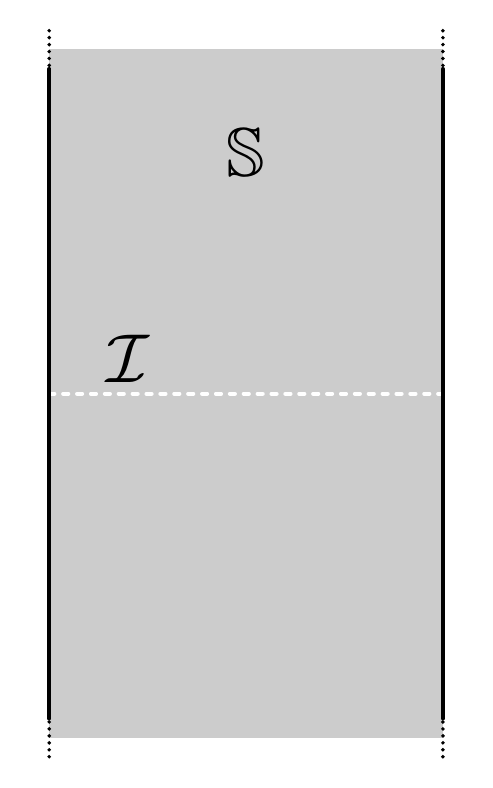}
  \label{sfig: continuum strip with crosssection}
}
\hspace{2.5cm}
\subfigure[The square grid slit-strip~$\dslitstrip$.] 
{
  \includegraphics[width=.35\textwidth]{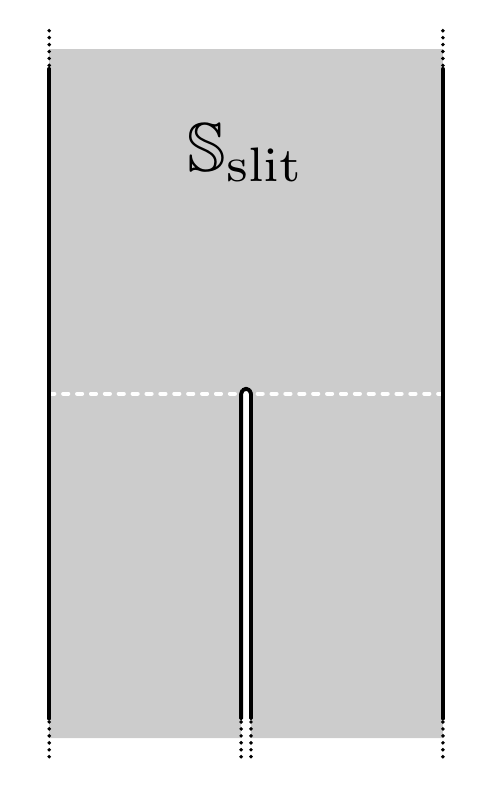}
  \label{sfig: continuum slit strip with crosssection}
}
\caption{The vertically infinite lattice strip and slit-strip.
}
\label{fig: continuum strip and slit-strip with crosssections}
\end{figure}

The appropriate continuum domains are the following two simply connected open
sets of the complex plane: the vertical strip
\begin{align}
\cstrip = \set{ z \in \bC \; \bigg| \; \mhalf < \re(z) < \half}
\end{align}
and the vertical slit-strip
\begin{align}
\cslitstrip = \cstrip \setminus \set{\ii y \; \big| \; y \le 0} .
\end{align}
These are illustrated in 
Figures~\ref{sec: limit distinguished functions}.\ref{sfig: continuum strip with crosssection}
and~\ref{sec: limit distinguished functions}.\ref{sfig: continuum slit strip with crosssection}, respectively. We consider holomorphic functions~$F$ on these domains, occasionally also
just locally defined functions on one of the following three subsets
\begin{align*}
\cslitstripT = \set{z \in \cslitstrip \; \big| \; \im(z) \ge 0} ,
\end{align*}
\begin{align*}
\cslitstripL = \set{z \in \cslitstrip \; \big| \; \im(z) \le 0 , \; \re(z)<0} , \quad
\cslitstripR = \set{z \in \cslitstrip \; \big| \; \im(z) \le 0 , \; \re(z)>0} .
\end{align*}

We again focus on the restrictions of such functions~$F$
to the horizontal cross-section
\begin{align}\label{eq: continuum cross section}
\ccrosssec = \Big[ \, \mhalf , \; \half \, \Big] .
\end{align}
The continuum analogue of the discrete function space~\eqref{eq: discrete function space} is
the real Hilbert space
\begin{align}\label{eq: continuum function space}
\cfunctionsp = L^2_\bR (\ccrosssec, \bC)
\end{align}
of complex-valued square-integrable functions on~$\ccrosssec$, with the
inner product and norm given by
\begin{align}\label{eq: inner product on L2}
\innprod{f}{g} 
= \; & \int_{\aaa}^{\bbb} \re \Big(
    f(x) \,  \overline{g(x)} \Big) \; \ud x , &
\| f \|^2 = \; & \int_{\aaa}^{\bbb} |f(x)|^2 \; \ud x .
\end{align}

The boundary of the strip~$\cstrip$ has two components: the left boundary
$
\big\{\mhalf + \ii y \, \big| \, y \in \bR \big\}$
and the right boundary
$
\big\{\half + \ii y \, \big| \, y \in \bR \big\}$.
The boundary of the slit-strip~$\cslitstrip$ additionally
has the slit component~$\set{ \ii y \, \big| \, y \le 0}$,
and in fact each of the points~$\ii y$ with $y<0$
corresponds to two prime ends, which we denote~$0^- + \ii y$ and $0^+ + \ii y$, 
and interpret as the boundaries as seen from the left and right legs of the 
slit-strip. By the right boundary of the slit strip, we then mean
~$\big\{\half + \ii y \, \big| \, y \in \bR \big\} \cup
\big\{0^- + \ii y \, \big| \, y <0 \big\}$ and by the left boundary
correspondingly
~$\big\{\mhalf + \ii y \, \big| \, y \in \bR \big\} \cup
\big\{0^+ + \ii y \, \big| \, y <0 \big\}$.
A holomorphic function $F$ in any of these domains or their subdomains is
said to have \term{Riemann boundary values} if $F$ has a continuous extension to
the left and right boundaries, and the values on the left boundary are
in~$\eighthrootbar \, \bR$ and the values on the right boundary are
in~$\eighthroot \, \bR$, and additionally the (possible) singularity of~$F$
at the ``tip'' $0 \in \bdry \cslitstrip$ of the slit is such that
the restriction to the cross-section remains square-integrable,
$f = F|_\ccrosssec \in \cfunctionsp$.

\subsubsection*{Indexing sets}
The appropriate indexing sets for the continuum functions
are the set~$\poshalfint$ of positive half-integers~\eqref{eq: positive half integers}
and the set~$\pm\poshalfint = \bZ + \half$ of all half-integers.
In the next section, the continuum fusion coefficients will be indexed
by the set~$\set{\parts \finsubset \poshalfint}$
of all finite subsets of positive half-integers.

\subsection{Distinguished functions in the continuum strip}

The following quarter-integer Fourier modes and their analytic continuations
with Riemann boundary values in the continuum strip~$\cstrip$
are the appropriate continuum counterparts of the discrete
functions~$\eigf{k}, \eigF{k}$ from Section~\ref{sec: distinguished functions},
Proposition~\ref{prop: discrete vertical translation eigenfunctions}.

For $k \in \pm \poshalfint$ a half-integer,
let $C_k = e^{\ii \pi (-k/2-1/4)}$ and define
\begin{align}\label{eq: half integer Fourier mode}
\ccfFun_k \colon & \; \cstrip \to \bC &
\ccfFun_k (x+ \ii y) 
:= \; & C_k \; \exp \big( -\ii \pi k x + \pi k y \big) ,
\\
\nonumber
\ccffun_k \colon & \; \ccrosssec \to \bC &
\ccffun_k (x) 
:= \; & C_k \; e^{ -\ii \pi k x} ,
\end{align}
so that $\ccffun_k = \ccfFun_k |_{\ccrosssec}$.
Note that we have the vertical translation eigenfunction
property, boundary conditions, and normalization
\begin{align}\label{eq: continuum vertical translation eigenfunction}
\ccfFun_k ( z + \ii h ) = e^{\pi k h} \, \ccfFun_k ( z ) , \qquad
\ccfFun_k \big( \frac{-1}{2} + \ii y \big) \in e^{-\ii \pi /4} \, \bR_+ ,
\qquad
\|\ccffun_k\| = 1 .
\end{align}
This collection~$(\ccffun_k)_{k \in \pm\poshalfint}$ of functions forms
an orthonormal basis of~$\cfunctionsp$~\cite[Proposition~2.1]{part-1}.
In analogy with
the reflection formulas~\eqref{eq: opposite index discrete functions},
we also have
\begin{align}\label{eq: opposite index continuum functions}
\ccffun_{-k} (x) 
    = \; & -\ii \; \overline{\ccffun_{k} (x)} , &
\ccfFun_{-k} (x + \ii y) 
    = \; & -\ii \; \overline{\ccfFun_{k} (x - \ii y)} .
\end{align}
It was shown in~\cite[Theorem~4.3]{part-1} that the 
functions~$\eigf{k}, \eigF{k}$ of 
Proposition~\ref{prop: discrete vertical translation eigenfunctions}
converge in the scaling limit to~$\ccffun_k, \ccfFun_k$,
respectively (uniformly on compact subsets, for example).

\subsubsection*{Decomposition to poles and zeroes at the top}

The functions~$\ccffun_{+k}$,
for $k \in \poshalfint$, are exponentially growing in the upwards direction of 
the continuum strip~$\cstrip$, while the functions~$\ccffun_{-k}$
are exponentially decaying. We interpret the growing ones and having a 
pole at the top extremity, and the decaying ones as having a zero.

We decompose the function space~$\cfunctionsp$ into the corresponding 
subspaces. In analogy with~\eqref{eq: discrete top poles and zeros},
this decomposition amounts to defining the closed subspaces
\begin{align}\label{eq: continuum top poles and zeros}
\cfspTpole = \; & 
    \overline{\spn \set{ \ccffun_{+k} \; \big| \; k \in \poshalfint}} , &
\cfspTzero = \; & 
    \overline{\spn \set{ \ccffun_{-k} \; \big| \; k \in \poshalfint}} ,
\end{align}
and the corresponding orthogonal projections
\begin{align*}
\cprTpole \colon \; & \cfunctionsp \to \cfspTpole , &
\cprTzero \colon \; & \cfunctionsp \to \cfspTzero .
\end{align*}

\subsection{Distinguished functions in the slit-strip}
\label{sub: distinguished functions in slit strip}

In the slit-strip there are three infinite extremities: 
the top, the right leg, and the left leg.
In the left and right substrips there are natural functions
defined analogously to the whole strip, which allow us to define
poles and zeroes, and obtain corresponding decompositions. 
The most important distinguished functions in the slit-strip will then
be globally defined
functions which have zeroes (i.e., regular behavior) in two
of the three infinite extremities, and have a pole of a given 
order in the third (i.e., a prescribed singular part).

\subsubsection*{Poles and zeroes in the left and right legs}

In the continuum setting we define the two
halves~$\ccrosssecL := \big[ -\half , 0 \big]$
and~$\ccrosssecR := \big[ 0, \half \big]$ of the cross-section,
and the closed subspaces
\begin{align*}
\cfunctionspL := L^2_\bR(\ccrosssecL , \bC) 
\subset \; & \cfunctionsp , &
\cfunctionspR := L^2_\bR(\ccrosssecR , \bC) 
\subset \; & \cfunctionsp , 
\end{align*}
of functions with support on one of the two halves.
For~$k \in \pm \poshalfint$, let
$C^{\lftsym}_k = \sqrt{2} \, e^{\ii \pi (-k-1/4)}$
and $C^{\rgtsym}_k = \sqrt{2} \, e^{-\ii \pi / 4}$,
and define the substrip functions by
\begin{align}
\label{eq: half integer Fourier mode left}
\ccfFun^{\lftsym}_k (x+ \ii y) 
= \; & C^{\lftsym}_k \; \exp \big(- \ii 2 \pi k x + 2 \pi k y \big) ,
& 
\ccffun^{\lftsym}_k (x) 
= \; & C^{\lftsym}_k \, e^{-\ii 2 \pi k x}
\\
\label{eq: half integer Fourier mode right}
\ccfFun^{\rgtsym}_k (x+ \ii y) 
= \; & C^{\rgtsym}_k \; \exp \big( -\ii 2 \pi k x + 2 \pi k y \big) ,
&
\ccffun^{\rgtsym}_k (x) 
= \; & C^{\rgtsym}_k \, e^{-\ii 2 \pi k x} ,
\end{align}
where $x \in \ccrosssecL$ and $x \in \ccrosssecR$
in~\eqref{eq: half integer Fourier mode left}
and~\eqref{eq: half integer Fourier mode right}, respectively,
and~$y \leq 0$ in both.
The normalization constants~$C^{\lftsym}_k, C^{\rgtsym}_k$
are chosen so as to ensure unit norm and Riemann boundary values
(in the respective substrips).
The collections~$(\ccffunL_{k})_{k \in \pm \poshalfint}$
and~$(\ccffunR_{k})_{k \in \pm \poshalfint}$
form orthonormal bases of the
subspaces~$\cfunctionspL$ and~$\cfunctionspR$, respectively.
In terms of these, we define the closed subspaces\footnote{Just 
like in~\eqref{eq: discrete top poles and zeros},
the signs of the indices in~\eqref{eq: continuum leg poles and zeros}
are again the opposite compared to~\eqref{eq: continuum top poles and zeros}.}
\begin{align}\label{eq: continuum leg poles and zeros}
\cfspLpole = \; & 
    \overline{\spn \set{ \ccffunL_{-k} \; \big| \; k \in \dposhalfint}} , &
\cfspRpole = \; & 
    \overline{\spn \set{ \ccffunR_{-k} \; \big| \; k \in \dposhalfint}} , \\
\nonumber
\cfspLzero = \; & 
    \overline{\spn \set{ \ccffunL_{+k} \; \big| \; k \in \dposhalfint}} , &
\cfspRzero = \; & 
    \overline{\spn \set{ \ccffunR_{+k} \; \big| \; k \in \dposhalfint}} ,
\end{align}
and we denote by
\begin{align*}
\dprLpole \colon \; & \dfunctionsp \to \dfspLpole , &
\dprRpole \colon \; & \dfunctionsp \to \dfspRpole , \\
\dprLzero \colon \; & \dfunctionsp \to \dfspLzero , &
\dprRzero \colon \; & \dfunctionsp \to \dfspRzero ,
\end{align*}
the orthogonal projections onto these subspaces.

\subsubsection*{Distinguished functions in the continuum slit-strip}
The continuum case is exactly parallel to the above discrete case.
For a function~$f \in \cfunctionsp$, we call the projections
$\cprTpole(f)$, $\cprTpole(f)$, $\cprTpole(f)$
its singular parts in the top, right, and left, respectively.
When a singular part vanishes, we say that the function admits
a regular extension in the corresponding extremity.
The continuum distinguished functions are characterized as follows.
\begin{prop}[{\cite[Proposition 2.6]{part-1}}]
\label{prop: continuum pole functions}
For all positive half-integers~$k \in \poshalfint$,
there exist unique functions
$ \cpoleT{k}, \cpoleL{k}, \cpoleR{k} \in \cfunctionsp $
such that
\begin{align}
\nonumber
\cprTpole(\cpoleT{k}) = \; & \ccffun_k , &
\cprLpole(\cpoleT{k}) = \; & 0 , &
\cprRpole(\cpoleT{k}) = \; & 0 , \\ 
\label{eq:cpole-asymptotic}
\cprTpole(\cpoleL{k}) = \; & 0 , &
\cprLpole(\cpoleL{k}) = \; & \ccffun^{\lftsym}_{-k} , &
\cprRpole(\cpoleL{k}) = \; & 0 , \\ 
\nonumber
\cprTpole(\cpoleR{k}) = \; & 0 , &
\cprLpole(\cpoleR{k}) = \; & 0 , &
\cprRpole(\cpoleR{k}) = \; & \ccffun^{\rgtsym}_{-k} .
\end{align}
These functions are the restrictions to the cross-section~$\ccrosssec$
of unique holomorphic functions 
\begin{align*}
\cPoleT{k} , \cPoleL{k} , \cPoleR{k} \colon \; & \cslitstrip \to \bC 
\end{align*}
with Riemann boundary values. 
\end{prop}
It is important that these continuum pole functions
$\cPoleT{k} , \cPoleL{k} , \cPoleR{k}$
are defined globally on the whole slit-strip domain~$\cslitstrip$.

\subsection{Convergence of the inner products of distinguished functions}

In~\cite{part-1} it was shown that the discrete distinguished
functions converge in the scaling limit to their continuum
counterparts uniformly on compact subsets. We do not need 
this form of convergence, but we crucially use its corollary
that the inner products among the discrete distinguished 
functions converge as~$\width \to \infty$ to the inner 
products of the continuum counterparts.

\begin{cor}[{\cite[Corollary~4.8]{part-1}}]
\label{cor:ip-convergence}
Choose sequences $(\lft_{n})_{n \in \bN}$,
$(\rgt_{n})_{n \in \bN}$ of integers
$\lft_{n}, \rgt_{n} \in \bZ$ such that
\begin{itemize}
\item $\lft_{n} < 0 < \rgt_{n}$ for all $n$;
\item $\width_{n} := \rgt_{n} - \lft_{n} \to +\infty$ as $n \to \infty$;
\item $\lft_{n} / \width_{n} \to -\half$
and $\rgt_{n} / \width_{n} \to +\half$ as $n \to \infty$.
\end{itemize}
For~$k \in \poshalfint$, 
let
\begin{align*}
\poleT{k}^{(\width_{n})} , \; 
\poleL{k}^{(\width_{n})} , \;
\poleR{k}^{(\width_{n})} , \;
\eigf{\pm k}^{(\width_{n})} , \; 
\eigfL{\pm k}^{(\width_{n})} , \;
\eigfR{\pm k}^{(\width_{n})} 
    \; \in \; \dfunctionspW{\width_{n}}
\end{align*}
denote the functions defined in Section~\ref{sec: distinguished functions}
in the lattice strips with~$\lft = \lft_n$ and $\rgt = \rgt_n$.
Correspondingly, let
\begin{align*}
\cpoleT{k} , \; 
\cpoleL{k} , \;
\cpoleR{k} , \;
\ccffun_{\pm k} , \; 
\ccffunL_{\pm k} , \;
\ccffunR_{\pm k}
    \; \in \; \cfunctionsp
\end{align*}
be the continuum functions defined above.

Then as $n \to \infty$, we have the convergence of all inner products
in~$\dfunctionspW{\width_{n}}$ to the corresponding ones in~$\cfunctionsp$:
\begin{align*}
\innprod{\eigfX{k}^{(\width_{n})}}{\eigfXbis{k'}^{(\width_{n})}} 
\to \; & \innprod{\ccffunX_{k}}{\ccffunXbis_{k'}}
    & & \text{ for 
        $\wildsym , \wildsymbis \in \set{\topsym, \lftsym, \rgtsym}$
        and $k , k' \in \pm\poshalfint$,} \\
\innprod{\poleX{k}^{(\width_{n})}}{\eigfXbis{k'}^{(\width_{n})}} 
\to \; & \innprod{\cpoleX{k}}{\ccffunXbis_{k'}}
    & & \text{ for 
        $\wildsym , \wildsymbis \in \set{\topsym, \lftsym, \rgtsym}$
        and $k \in \poshalfint$, $ k' \in \pm\poshalfint$,} 
\\
\innprod{\poleX{k}^{(\width_{n})}}{\poleXbis{k'}^{(\width_{n})}} 
\to \; & \innprod{\cpoleX{k}}{\cpoleXbis{k'}}
    & & \text{ for 
        $\wildsym , \wildsymbis \in \set{\topsym, \lftsym, \rgtsym}$
        and $k , k' \in \poshalfint$}
\end{align*}
(where the notation is to be interpreted so
that~$\eigfT{k}^{(\width_{n})} = \eigf{k}^{(\width_{n})}$ 
and~$\ccffunT_{k}=\ccffun_{k}$).
\end{cor}

\bigskip

\section{Continuum fusion coefficients}
\label{sec: continuum fusion coefs}

% *******************************************************************
% **   \section{Continuum fusion coefficients}                     **
% *******************************************************************

We next introduce and study a continuum analogue of the fusion 
coefficients. The fusion coefficients of the Ising model considered
in Section~\ref{sec: transfer matrices} will be shown to converge to these in the scaling 
limit~$\width \to \infty$.
In the final part~\cite{part-3} of this series, 
these continuum analogues of the fusion coefficients will
moreover be related to the structure constants of the vertex 
operator algebra which underlies
the conformal field theory conjectured to describe
the Ising model.

The convergence of the fusion coefficients of the Ising model to their 
continuum analogues (introduced below) could in principle
be derived starting from the defining
Equations~\eqref{eq: discrete fermion mode},
\eqref{eq: transfer matrix eigenvectors},
and~\eqref{eq: fusion coefficient},
using
\cite[Theorem~4.3]{part-1}
about the convergence of the s-holomorphic vertical translation eigenfunctions 
to the quarter-integer Fourier modes
together with the known convergence results
for the discrete fermion multipoint correlation
functions~\cite{Hongler-thesis, HS-energy_density}
(although handling the contributions 
from near the boundaries would require slightly strengthened formulations). 
We choose a slightly different route, however: we prove 
that the continuum fusion coefficients satisfy a recursion with 
similar structure as the fusion coefficients, and we prove convergence of the
coefficients in these recursions. Indeed, the 
coefficients in these two recursions are inner products of distinguished 
functions in the function spaces of 
Sections~\ref{sec: limit distinguished functions} 
and~\ref{sec: discrete complex analysis}, respectively. Therefore the scaling 
limit result for the fusion coefficients will be a consequence of the 
convergence results for the (inner products of the) distinguished functions.
The reasons for our choice of strategy are twofold. First of all, while the 
convergence of the s-holomorphic functions uses 
largely the same discrete complex analysis technology 
as~\cite{Hongler-thesis, HS-energy_density}, our analysis for instance
avoids the notion of s-holomorphic singularities in the bulk (singularities 
only appear in asymptotics) and is in this sense simpler.
Secondly, we view the emphasis 
on the recursion itself natural for our main goal of this series,
since a recursion with a basically similar structure 
arises from the Jacobi identity in the vertex operator algebra
that describes the conformal field theory.

\begin{rmk}
In our normalizations, a few constants in the convergence results end up being
not particularly elegant,
so let us pause to mention the conventions eventually dictating them.
The definitions of the discrete holomorphic 
and antiholomorphic fermions in the discrete setting 
of Section~\ref{sec: operator valued forms} yielded
first of all the straightforward complexified 
s-holomorphicity~\eqref{eq: CSH of fermions}
and Riemann boundary values~\eqref{eq: CRBV of fermions} of the
fermions, and moreover relatively simple
algebraic and functional analytic properties (in 
particular in view of the adjoints and anticommutators
in Lemmas~\ref{lem: adjoints of fermions} 
and~\ref{lem: anticommutators of height zero fermions}).
In the conformal field theory, on the other hand, 
the fields of interest have a conventional normalization which fixes the 
residue of the holomorphic fermion two-point function to unit value.
In addition, we have chosen to use the unit width continuum strip and 
corresponding lattice spacing~$\width^{-1}$.\footnote{
Perhaps the most straightforward 
simplification of the unpleasant constants could be achieved by using
the strips of width~$2\pi$ and lattice spacing~$\frac{2\pi}{\ell}$ instead. 
However, by our judgement the 
simplicity of the lattice spacing is more important.}
With these  
conventional choices,
the unpleasant constants are inevitable: after a renormalization
by the square root of the lattice spacing, the 
discrete holomorphic fermion field basically converges
(whatever the precise notion of convergence) to 
$e^{-\ii\pi/4}/\sqrt{\pi}$ times the holomorphic fermion field of the 
conformal field theory.
\end{rmk}

Before starting, let us fix two conventions for the whole section.

\begin{figure}[tb]
\centering
\subfigure[The slit-strip~$\cslitstrip$.] 
{
\includegraphics[width=.42\textwidth]{pics-lattice_Ising_fusion-pt2-51.pdf}
\label{sfig: continuum slit-strip}
}
\hspace{2.5cm}
\subfigure[Conformal mapping of~$\cslitstrip$.] 
{
\includegraphics[width=.35\textwidth]{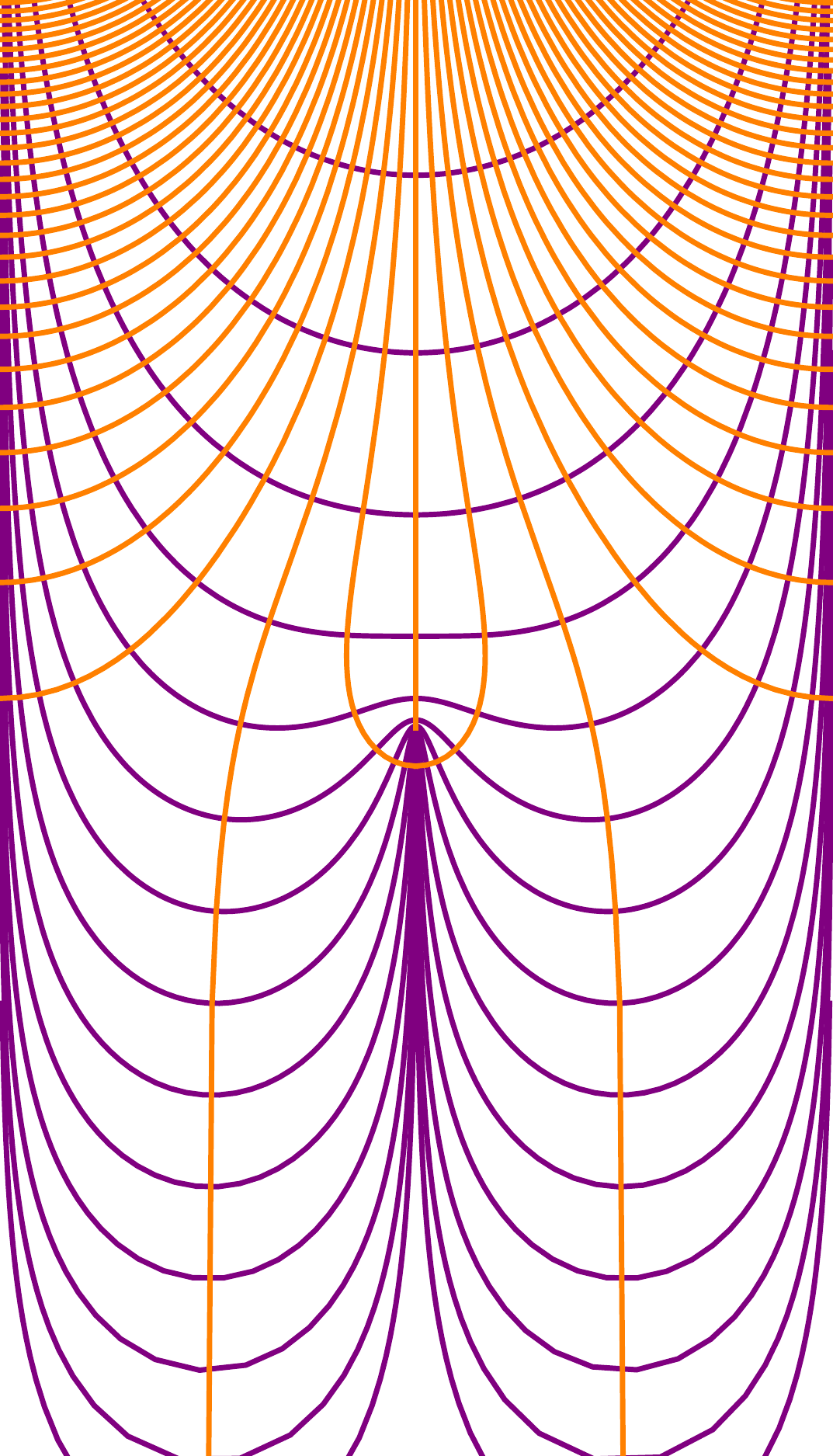}
\label{sfig: slit-strip conformal map}
}
\caption{The continuum slit-strip.}
\label{fig: slit strip}
\end{figure}

Throughout this section, we denote by
\begin{align*}
\mapSS \colon \cslitstrip \to \bH
\end{align*}
the conformal map from the slit-strip~$\cslitstrip$ to the upper 
half-plane~$\bH$ under which the images of the top extremity and the two bottom 
extremities are $\infty$, $-\half$, and $+\half$, respectively. 
This mapping is illustrated in 
Figure~\ref{sec: continuum fusion coefs}.\ref{sfig: slit-strip conformal map},
and an explicit formula for it 
is~$\mapSS(z) = \half \, \sqrt{1-e^{-2\ii\pi z}}$,
where the choice of the branch of square roots is as detailed
in~\cite[Sec.~2]{part-1}.

Formulas in this section involve many complex 
conjugations of sometimes unwieldy expressions. To reduce the 
resulting notational mess, we use $\ccshort{(\cdots)}$ to denote complex 
conjugation. In particular, $\ccshort{\bH}$ denotes the lower half-plane.
As the main exception, in integrations we retain the 
notation~$\ud \overline{z}$ for the antiholomorphic one-form corresponding to 
the coordinate~$z$.

\subsection{Two-point kernels in slit-strip}
\label{ssec: two point kernels}

Now for any given functions $\gencfFun_1 , \gencfFun_2$, each defined 
on the slit-strip or a subset thereof, introduce the 
following four two-point kernels\footnote{
The expressions here  
involve an even number of
square roots of the derivative, so  
the branch choice of~$\sqrt{\mapSS'}$
is inconsequential. 
For definiteness, we could follow the
convention of~\cite[Sec.~2]{part-1}.
}
\begin{align}
\label{eq: two point kernel HH}
\intkerSS{\gencfFun_1 , \gencfFun_2}^{\circ\circ}(z_1 , z_2)
= \; & \frac{\gencfFun_1(z_1) \sqrt{\mapSS'(z_1)} 
		\;\; \gencfFun_2(z_2) \sqrt{\mapSS'(z_2)}}
			{\mapSS(z_1) - \mapSS(z_2)} \\
\label{eq: two point kernel HA}
\intkerSS{\gencfFun_1 , \gencfFun_2}^{\circ\bullet}(z_1 , z_2)
= \; & \frac{\gencfFun_1(z_1) \sqrt{\mapSS'(z_1)}
		\;\; \cclong{\gencfFun_2(z_2) \sqrt{\mapSS'(z_2)}}}
			{\mapSS(z_1) - \ccshort{\mapSS(z_2)}} \\
\label{eq: two point kernel AH}
\intkerSS{\gencfFun_1 , \gencfFun_2}^{\bullet\circ}(z_1 , z_2)
= \; & \frac{\cclong{\gencfFun_1(z_1) \sqrt{\mapSS'(z_1)}}
		\;\; \gencfFun_2(z_2) \sqrt{\mapSS'(z_2)}}
			{\ccshort{\mapSS(z_1)} - \mapSS(z_2)} \\
\label{eq: two point kernel AA}
\intkerSS{\gencfFun_1 , \gencfFun_2}^{\bullet\bullet}(z_1 , z_2)
= \; & \frac{\cclong{\gencfFun_1(z_1) \sqrt{\mapSS'(z_1)}}
		\;\; \cclong{\gencfFun_2(z_2) \sqrt{\mapSS'(z_2)}}}
			{\ccshort{\mapSS(z_1)} - \ccshort{\mapSS(z_2)}} 
\end{align}
One can immediately note some antisymmetry properties of the kernels above:
for example\linebreak[4]
$\intkerSS{\gencfFun_1 , \gencfFun_2}^{\bullet\circ}(z_1 , z_2)
= - \intkerSS{\gencfFun_2 , \gencfFun_1}^{\circ\bullet}(z_2 , z_1)$, etc.

The following lemma is a vertical slidability property for the two-point 
kernels. Despite a few substantial differences, an analogue with
Proposition~\ref{prop: ICRBV implies vertically slidable}
should be apparent.

\begin{lem}\label{lem: continuum vertical slidability for two pt functions}
Suitable Riemann boundary values on the coefficient function~$\gencfFun_1$ 
imply the vanishing of integrals of certain combinations of the above kernels 
along left and right 
vertical boundaries as follows.
\begin{itemize}
\item[(a)]
Let $B^\lftsym \subset \bdry \cslitstrip$ be a boundary segment of 
the left boundary of the slit-strip.
Suppose that the function~$\gencfFun_1$ extends continuously to~$B^\lftsym$ 
and satisfies
\begin{align*}
\gencfFun_1(z_1) \in \eighthrootbar \, \bR
\qquad \text{ for all $z_1 \in B^\lftsym$ .}
\end{align*}
Then for any~$z_2 \in \cslitstrip$ we have 
\begin{align*}
\int_{B^\lftsym} \Big( 
    \intkerSS{\gencfFun_1 , \gencfFun_2}^{\circ\circ}(z_1 , z_2) \, \ud z_1
  + \intkerSS{\gencfFun_1 , \gencfFun_2}^{\bullet\circ}(z_1 , z_2) \, \ud 
    \overline{z}_1 \Big) = \; & 0 \\
\int_{B^\lftsym} \Big( 
    \intkerSS{\gencfFun_1 , \gencfFun_2}^{\circ\bullet}(z_1 , z_2) \, \ud z_1
  + \intkerSS{\gencfFun_1 , \gencfFun_2}^{\bullet\bullet}(z_1 , z_2) \, \ud 
    \overline{z}_1 \Big) = \; & 0 .
\end{align*}
\item[(b)]
Let $B^\rgtsym \subset \bdry \cslitstrip$ be a boundary segment of 
the right boundary of the slit-strip.
Suppose that the function~$\gencfFun_1$ extends continuously to~$B^\rgtsym$ 
and satisfies
\begin{align*}
\gencfFun_1(z_1) \in \eighthroot \, \bR
\qquad \text{ for all $z_1 \in B^\rgtsym$ .}
\end{align*}
Then for any~$z_2 \in \cslitstrip$ we have 
\begin{align*}
\int_{B^\rgtsym} \Big( 
    \intkerSS{\gencfFun_1 , \gencfFun_2}^{\circ\circ}(z_1 , z_2) \, \ud z_1
  + \intkerSS{\gencfFun_1 , \gencfFun_2}^{\bullet\circ}(z_1 , z_2) \, \ud 
    \overline{z}_1 \Big) = \; & 0 \\
\int_{B^\rgtsym} \Big( 
    \intkerSS{\gencfFun_1 , \gencfFun_2}^{\circ\bullet}(z_1 , z_2) \, \ud z_1
  + \intkerSS{\gencfFun_1 , \gencfFun_2}^{\bullet\bullet}(z_1 , z_2) \, \ud 
    \overline{z}_1 \Big) = \; & 0 .
\end{align*}
\end{itemize}
\end{lem}
\begin{proof}
The proofs of the two cases are virtually identical, so let us only 
prove part~(a).

Observe that since~$\mapSS$ maps the slit-strip~$\cslitstrip$ to the upper half 
plane~$\bH$, then a segment of the boundary~$\bdry \cslitstrip$ gets mapped to 
an interval on the real axis~$\bR$.\footnote{The boundary of the slit-strip 
consists of analytic arcs, so continuous and analytic extension of the 
conformal map to 
the boundary is possible.}
Considering orientations, we see that for $z_1 \in B^{\lftsym}$ on a segment 
of the left boundary, the direction of the derivative is
$\mapSS'(z_1) \in \ii \, \bR_+$. Therefore the square root of the derivative
satisfies $\sqrt{\mapSS'(z_1)} \in \eighthroot \, \bR$.
By assumption on the coefficient function
$\gencfFun_1(z_1) \in \eighthrootbar \, \bR$, so we get that
\[ \sqrt{\mapSS'(z_1)} \, \gencfFun_1(z_1) \in \bR . \]
In view of the defining 
formulas~\eqref{eq: two point kernel HH}~--~\eqref{eq: two point kernel AA}, 
this implies that
\begin{align*}
\intkerSS{\gencfFun_1 , \gencfFun_2}^{\circ\circ}(z_1 , z_2)
= \; & \intkerSS{\gencfFun_1 , \gencfFun_2}^{\bullet\circ}(z_1 , z_2) &
\intkerSS{\gencfFun_1 , \gencfFun_2}^{\circ\bullet}(z_1 , z_2)
= \; & \intkerSS{\gencfFun_1 , \gencfFun_2}^{\bullet\bullet}(z_1 , z_2) 
\end{align*}
for $z_1 \in B^{\lftsym}$ and any $z_2 \in \cslitstrip$.
Since 
$\ud \overline{z}_1 = -\ud z_1$ 
along the 
vertical boundary segment~$B^{\lftsym}$, we get cancellations
which ensure the vanishing of the integrals as asserted.
\end{proof}

The following lemma summarizes the holomorphicity and residues at poles for the 
two-point kernels, as well as antiholomorphic variants. 
An analogue to Proposition~\ref{prop: ICSH implies closed} can be noted.

\begin{lem}\label{lem: holomorphicity and poles for two pt functions}
Suppose that the coefficient function~$\gencfFun_1$ is holomorphic.
Then the two-point kernels have the following holomorphicity/antiholomorphicity 
properties and poles.
\begin{itemize}
\item[(a)]
The function~$z_1 \mapsto 
\intkerSS{\gencfFun_1 , \gencfFun_2}^{\circ\circ}(z_1 , z_2)$ is holomorphic 
except at~$z_1=z_2$, where it has a simple pole with residue
\begin{align*}
\frac{1}{2 \pi \ii} \, \oint_{\bdry B_\eps(z_2)}
    \intkerSS{\gencfFun_1 , \gencfFun_2}^{\circ\circ}(z_1 , z_2) \; \ud z_1
= \; & \gencfFun_1(z_2) \, \gencfFun_2(z_2) .
\end{align*}
\item[(b)] The function~$z_1 \mapsto 
\intkerSS{\gencfFun_1 , \gencfFun_2}^{\circ\bullet}(z_1 , z_2)$ is holomorphic.
\item[(c)] The function~$z_1 \mapsto 
\intkerSS{\gencfFun_1 , \gencfFun_2}^{\bullet\circ}(z_1 , z_2)$ is 
antiholomorphic.
\item[(d)]
The function~$z_1 \mapsto 
\intkerSS{\gencfFun_1 , \gencfFun_2}^{\bullet\bullet}(z_1 , z_2)$ is 
antiholomorphic except at~$z_1=z_2$, where it has an antiholomorphic simple 
pole with residue
\begin{align*}
\frac{-1}{2 \pi \ii} \, \oint_{\bdry B_\eps(z_2)}
    \intkerSS{\gencfFun_1 , \gencfFun_2}^{\bullet\bullet}(z_1 , z_2) 
		  \; \ud \overline{z}_1
= \; & \overline{\gencfFun_1(z_2)} \; \overline{\gencfFun_2(z_2)} .
\end{align*}
\end{itemize}
\end{lem}
\begin{proof}
Properties~(c) and~(d) are obtained from~(b) and~(a) by complex 
conjugation, so it suffices to prove the first two statements.

For~(b), observe that the numerator 
of~\eqref{eq: two point kernel HA} is holomorphic as a function of~$z_1$, since 
$z_1 \mapsto \sqrt{\mapSS'(z_1)}$ and $z_1 \mapsto \gencfFun_1(z_1)$ are.
Also the denominator of~\eqref{eq: two point kernel HA} is holomorphic as a 
function of~$z_1$, since $z_1 \mapsto \mapSS(z_1)$ is. Moreover, the 
denominator is non-vanishing, since $\mapSS(z_1) \in \bH$ but
$\ccshort{\mapSS(z_2)} \in \ccshort{\bH}$. Therefore 
the function~$z_1 \mapsto 
\intkerSS{\gencfFun_1 , \gencfFun_2}^{\circ\bullet}(z_1 , z_2)$ defined 
by~\eqref{eq: two point kernel HA} is indeed holomorphic.

For~(a), observe again that the numerator and denominator
of~\eqref{eq: two point kernel HH} are both holomorphic, as functions of~$z_1$.
Since we have~$\mapSS(z_2) \in \bH$,
and since $\mapSS \colon \cslitstrip \to \bH$ is conformal, the 
denominator only vanishes at $z_1 = z_2$, and it has a 
first order zero at that point. The function~$z_1 \mapsto 
\intkerSS{\gencfFun_1 , \gencfFun_2}^{\circ\circ}(z_1 , z_2)$ correspondingly 
is holomorphic except for a simple pole at~$z_1=z_2$. At 
this point the denominator has a Taylor expansion
\begin{align*}
\mapSS(z_1) - \mapSS(z_2) = 0 + (z_1-z_2) \, \mapSS'(z_2)
        + \OO\big( (z_1-z_2)^2 \big) ,
\end{align*}
while the numerator takes the value
\begin{align*}
\sqrt{\mapSS'(z_2)} \, \gencfFun_1(z_2) \;
    \sqrt{\mapSS'(z_2)} \, \gencfFun_2(z_2) 
= \mapSS'(z_2) \; \gencfFun_1(z_2) \, \gencfFun_2(z_2) .
\end{align*}
After cancelling the common factor~$\mapSS'(z_2)$, the assertion about the 
residue follows.
\end{proof}

\subsection{Multi-point kernels in slit-strip}
\label{ssec: multipoint kernels}

In free fermionic theories, multi-point correlations of fermions are obtained 
from two-point correlation functions by Pfaffians. We start by recalling the 
needed properties of Pfaffians, then construct the multi-point kernels from the 
two-point kernels of the previous section, and finally define and study 
integrated multi-point kernels, which (in special cases) give the continuum 
analogues of the fusion coefficients.

\subsubsection*{Pfaffians}

Suppose that~$A = (A_{ij})_{i,j=1}^m \in \bC^{m \times m}$
is a square matrix, which is skew-symmetric, $A_{ij} = - A_{ji}$
for all $i,j=1,\ldots m$. If the dimension is even, $m=2n$ for $n \in \bN$, we 
define the \term{Pfaffian} of~$A$ as
\begin{align}
\label{eq: Pfaffian}
\Pfaff(A) = \frac{1}{2^n \, n!}
    \sum_{\perm \in \SymmGrp_m} \sgn(\perm) \;
            \prod_{p=1}^n A_{\perm(2p-1) \, \perm(2p)} .
\end{align}
If $m$ is odd, we set $\Pfaff(A)=0$.

There is an alternative (less self-explanatory but often more practical) 
expression\footnote{This  
is obtained by combining repeated 
terms in the sum~\eqref{eq: Pfaffian}. The
permutations~$\perm \in \SymmGrp_m = \SymmGrp_{2n}$ can be 
partitioned into equivalence classes of size~$2^n \, n!$ each, with
equivalent permutation contributing equal terms. Representatives of 
the equivalence classes become naturally labeled by pair partitions~$P$ of the 
set~$\set{1,\ldots,m} = \set{1, \ldots, 2n}$. }
for the Pfaffian,
\begin{align*}
\Pfaff(A) = \sum_{P} \sgn(P) \;
            \prod_{\{i,j\} \in P, \; i<j} A_{ij} ,
\end{align*}
where the sum is over pair partitions~$P$ of $\set{1,\ldots,m}$,
and $\sgn(P)$ denotes the signature of any 
permutation~$\perm \in \SymmGrp_n$ such 
that $\set{\perm(2p-1) , \perm(2p)} \in P$ and $\perm(2p-1) < \perm(2p)$ for 
all~$p=1,\ldots,n$.

From the latter expression, for instance the 
following recursion of Pfaffians becomes evident.

\begin{lem}\label{lem: recursion for Pfaffians}
Let $m=2n$, and let
$A \in \bC^{m \times m}$ be a skew-symmetric matrix.
For any $1 \leq i < j \leq m$, 
let $\hat{A}^{(ij)} \in \bC^{(m-2) \times (m-2)}$ denote the matrix obtained 
by erasing rows and columns with indices $i$ and~$j$ from~$A$. Then for any 
fixed~$1 \leq i \leq m$, we have
\begin{align}\label{eq: Pfaffian kernel recursion}
\Pfaff(A) = \sum_{j \neq i} \sgn(i-j)(-1)^{i-j} \, A_{ij}\; \Pfaff \big( \hat{A}^{(ij)} 
\big).
\end{align}
\end{lem}
With the convention\footnote{There exists a unique permutation of zero indices 
and a unique pair partition of the empty set. The empty product is one.}
that the Pfaffian of a $0 \times 0$-matrix is~$1$, this 
recursion is sufficient (and efficient) for computing Pfaffians.
This recursion also features crucially below.

\subsubsection*{Multi-point kernels}
Let $\gencfFun_1 , \ldots , \gencfFun_m$ be functions, each defined 
on the slit-strip or a subset thereof.
In terms of the two-point kernels of Section~\ref{ssec: two point kernels}, we 
then define the $m \times m$ 
matrix~$\matkerSS{\gencfFun_1 , \ldots , \gencfFun_m} (z_1 , \ldots, z_m)$ 
with entries
\begin{align}\label{eq: matrix of 2pt kernels}
& \Big( \matkerSS{\gencfFun_1 , \ldots , \gencfFun_m}
		(z_1 , \ldots, z_m) \Big)_{i,j} \\ \nonumber
= \; & \begin{cases} 
    \intkerSS{\gencfFun_i , \gencfFun_j}^{\circ\circ}(z_i , z_j)
    + \intkerSS{\gencfFun_i , \gencfFun_j}^{\bullet\circ}(z_i , z_j)
    + \intkerSS{\gencfFun_i , \gencfFun_j}^{\circ\bullet}(z_i , z_j)
    + \intkerSS{\gencfFun_i , \gencfFun_j}^{\bullet\bullet}(z_i , z_j)
		& \text{ if $i \neq j$} \\
	0 & \text{ if $i=j$.}
       \end{cases}
\end{align}
This matrix is is skew-symmetric by virtue of the antisymmetry 
properties of the two-point 
kernels~\eqref{eq: two point kernel HH}~--~\eqref{eq: two point kernel AA}.
We then define the $m$-point kernel as its Pfaffian
\begin{align}\label{eq: kernel}
\intkerSS{\gencfFun_1 , \ldots , \gencfFun_m}(z_1 , \ldots, z_m)
= \Pfaff \Big( \matkerSS{\gencfFun_1 , \ldots , \gencfFun_m}(z_1 , \ldots, z_m) 
        \Big) .
\end{align}
This multipoint kernel is totally antisymmetric in the sense that for any 
permutation~$\perm \in \SymmGrp_m$ we have
$\intkerSS{\gencfFun_{\perm(1)} , \ldots , \gencfFun_{\perm(m)}}
        (z_{\perm(1)} , \ldots, z_{\perm(m)})
= \sgn(\perm) \, 
    \intkerSS{\gencfFun_1 , \ldots , \gencfFun_m}(z_1 , \ldots, z_m)$.

As the functions~$\gencfFun_i$, we will use the various
distinguished continuum functions
from Section~\ref{sec: distinguished functions}.
Primarily, we use
the functions $\ccfFun_k \colon \cstrip \to \bC$ in the strip (in fact 
restricted to the top half of it), and the 
functions $\ccfFun^{\lftsym}_k \colon \cstrip^{\lftsym} \to \bC$ and
$\ccfFun^{\rgtsym}_k \colon \cstrip^{\rgtsym} \to \bC$ in the left and the 
right half-strips. However, also the continuous pole functions 
$\cPoleT{k}, \cPoleR{k}, \cPoleL{k} \colon \cslitstrip \to \bC$ will
be used in the derivation of the main recursion.

Now let $\tple{k}, \tple{k}', \tple{k}'' \in 
\bigcup_{m=0}^\infty (\pm \poshalfint)^m$ be three tuples of half-integers,
\begin{align*}
\tple{k} = (k_1 , \ldots, k_m) , \qquad
\tple{k}' = (k'_1 , \ldots, k'_{m'}) ,  \qquad
\tple{k}'' = (k''_1 , \ldots, k''_{m''}) .
\end{align*}
Let
\begin{align}\label{eq: ordering of y coordinates}
0 < y_m < \cdots < y_2 < y_1 , \qquad
& y'_1 < y'_2 < \cdots < y'_{m'}  < 0 , \\ \nonumber
& y''_1 < y''_2 < \cdots < y''_{m''} < 0.
\end{align}
For any $x_1, \ldots, x_m\in (\frac{-1}{2},\frac{+1}{2})$,
$x'_1, \ldots, x'_{m'} \in (\frac{-1}{2},0)$, and
$x''_1, \ldots, x''_{m''} \in (0,\frac{+1}{2})$, use the abbreviated 
notation
\begin{align*}
& (\tple{x} + \ii \,\tple{y} ; \;
			\tple{x}' + \ii \,\tple{y}' , \; \tple{x}'' + \ii \, \tple{y}'') \\
:= \; & (x_1 + \ii y_1  , \ldots, x_m + \ii y_m, 
		 x'_1 + \ii y'_1 , \ldots, x'_{m'} + \ii y'_{m'}, 
		 x''_1 + \ii y''_1 , \ldots, x''_{m''} + \ii y''_{m''} ) 
\end{align*}
for the $(m+m'+m'')$-tuple of points in~$\cslitstrip$. Also use the abbreviated 
notation
\begin{align*}
\intkerSS{\ccfFun_{\tple{k}};
			\ccfFun^\lftsym_{\tple{k}'}, \ccfFun^\rgtsym_{\tple{k}''}}
:= \; & \intkerSS{\ccfFun_{k_1} , \ldots , \ccfFun_{k_m},
		  \ccfFun^\lftsym_{k'_1} , \ldots , \ccfFun^\lftsym_{k'_{m'}},
		  \ccfFun^\rgtsym_{k''_1} , \ldots , \ccfFun^\rgtsym_{k''_{m''}}}
\end{align*}
for the function of type~\eqref{eq: kernel}, with the specific 
choice of coefficient functions 
$\gencfFun_1 = \ccfFun_{k_1}$, \ldots, 
$\gencfFun_{m+m'+m''} = \ccfFun^{\rgtsym}_{k''_{m''}}$.

\subsubsection*{Integrated multipoint kernels}

\begin{figure}
\includegraphics[width=.35\textwidth]{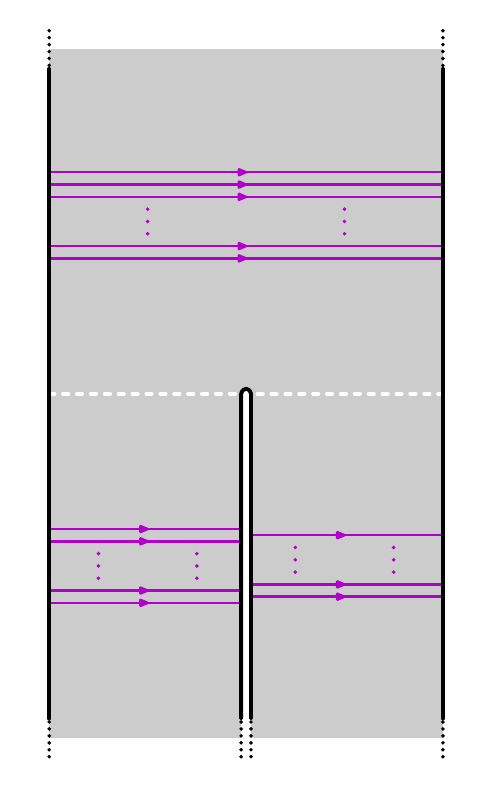}
\label{fig: slit-strip contours}
\caption{Integration contours for the multipoint kernels.}
\end{figure}

We now consider integrated versions of the above multipoint
kernels, with integration contours illustrated in
Figure~\ref{fig: slit-strip contours}.
Define the following integrated $(m+m'+m'')$-point kernel
\begin{align}\label{eq: strip tuple coefficients}
\fusionLimitTuple{\tple{k}}{\tple{k}'}{\tple{k}''} 
:= \; & 
  \Big( \frac{-\ii}{2 \sqrt{\pi}} \Big)^{m+m'+m''}
  \int_{[-\frac{1}{2},\frac{1}{2}]^{m}} \ud^m \tple{x} \;
  \int_{[-\frac{1}{2},0]^{m'}} \ud^{m'} \tple{x}' \;
  \int_{[0,\frac{1}{2}]^{m''}} \ud^{m''} \tple{x}'' \\ \nonumber
& \qquad\qquad\qquad\qquad\qquad\qquad\qquad
  \; \intkerSS{\ccfFun_{\tple{k}};
			\ccfFun^\lftsym_{\tple{k}'}, \ccfFun^\rgtsym_{\tple{k}''}}
    (\tple{x} + \ii \,\tple{y} ;
		\tple{x}' + \ii \,\tple{y}' , \tple{x}'' + \ii \, \tple{y}'') ,
\end{align}
where we also abbreviated $\ud^m \tple{x} = \prod_{i=1}^{m} \ud x_i$, etc.
A priori, the levels $y_1 , \ldots, y''_{m''}$ are free variables on the right 
hand side, but the following lemma shows that the precise choice of them
plays no role.

\begin{lem}\label{lem: independence on level choices}
The quantity~$\fusionLimitTuple{\tple{k}}{\tple{k}'}{\tple{k}''}$
does not depend on $(y_i)_{i=1}^{m}$,
$(y'_i)_{i=1}^{m'}$, $(y''_i)_{i=1}^{m''}$, as long as these 
satisfy the ordering 
constraints~\eqref{eq: ordering of y coordinates}.
\end{lem}
\begin{proof}
Consider, e.g., changing one $y_i$ to another value~$\tilde{y}_i$,
but still so that
$\tilde{y}_i < y_{i-1}$~(if~$i>1$)
and $\tilde{y}_i > y_{i+1}$~(if~$i<m$).
The new integral over the~$x_i$ variable 
which we must consider in~\eqref{eq: strip tuple coefficients} differs from the 
original integral by
\begin{align}\label{eq: difference of integrals at different levels}
\int_{-\frac{1}{2}}^\frac{1}{2} \ud x_i \;
    \intkerSS{\cdots,\ccfFun_{k_i},\cdots}
	  (\cdots, x_i + \ii \tilde{y}_i , \cdots)
- \int_{-\frac{1}{2}}^\frac{1}{2} \ud x_i \;
    \intkerSS{\cdots,\ccfFun_{k_i},\cdots}(\cdots, x_i + \ii y_i , \cdots) .
\end{align}
Applying the recursion for Pfaffians, 
Lemma~\ref{lem: recursion for Pfaffians},
we can isolate factors 
in the
integrand $\intkerSS{\cdots,\ccfFun_{k_i},\cdots}(\cdots, z_i , \cdots)$ 
which depend on the variable~$z_i$, 
and write this integrand as a sum of terms of the form
\begin{align}\label{eq: decomposition to holom and antiholom terms}
\Bigg( \intkerSS{\ccfFun_{k_i} , \gencfFun}^{\circ\circ}(z_i , z)
	+ \intkerSS{\ccfFun_{k_i} , \gencfFun}^{\circ\bullet}(z_i , z) 
	+ \intkerSS{\ccfFun_{k_i} , \gencfFun}^{\bullet\circ}(z_i , z)
	+ \intkerSS{\ccfFun_{k_i} , \gencfFun}^{\bullet\bullet}(z_i , z)
	\Bigg) \times K 
\end{align}
where~$K$ is a Pfaffian of a smaller matrix with no dependence on~$z_i$, 
and~$z$ is one of the other variables.
In particular by the assumed orderings of the imaginary parts, $z$~is always at 
a positive distance from the rectangle
$(-\frac{1}{2},\frac{1}{2}) \times (y_{i},\tilde{y}_{i}) \subset \cslitstrip$.
If we denote by $\chorcontour{y}$ the horizontal path 
from~$-\frac{1}{2}+\ii y$ to~$+\frac{1}{2}+\ii y$, then
the contribution to~\eqref{eq: difference of integrals at different levels} 
from the first two terms 
of~\eqref{eq: decomposition to holom and antiholom terms} is
\begin{align}\label{eq: contribution of holomorphic integrals terms}
K \times \bigg( 
  & \; \phantom{-} \int_{\chorcontour{y_i}} 
  \big( \intkerSS{\ccfFun_{k_i} , \gencfFun}^{\circ\circ}(z_i , z)
	+ \intkerSS{\ccfFun_{k_i} , \gencfFun}^{\circ\bullet}(z_i , z) \big)
	    \; \ud z_i 
\\ \nonumber & \; 
  - \int_{\chorcontour{\tilde{y}_i}} 
	\big( \intkerSS{\ccfFun_{k_i} , \gencfFun}^{\circ\circ}(z_i , z)
	+ \intkerSS{\ccfFun_{k_i} , \gencfFun}^{\circ\bullet}(z_i , z) \big)
	    \; \ud z_i \bigg) ,
\end{align}
since for the integrals along a horizontal segments
we can write~$\ud x_i = \ud z_i$.
The integrand here is holomorphic in the rectangle
$(-\frac{1}{2},\frac{1}{2}) \times (y_i,\tilde{y_i}) \subset \cslitstrip$, and 
the paths $\chorcontour{y_i}$ 
and~$\chorcontour{\tilde{y}_i}$ are the horizontal segments of the boundary of 
this rectangle.
Due to the vanishing by Cauchy's theorem
of the integral along the whole boundary of the rectangle, the
contribution~\eqref{eq: contribution of holomorphic integrals terms} can be 
written alternatively
in terms of the vertical boundary parts $B^\lftsym$ and $B^\rgtsym$
of the rectangle
(from~$\pm\frac{1}{2} + \ii y_i$ to~$\pm\frac{1}{2} + \ii y_i$), as
\begin{align}\label{eq: contribution of holomorphic integrals terms vert}
K \times \bigg( 
  & \; + \int_{B^\lftsym} 
  \big( \intkerSS{\ccfFun_{k_i} , \gencfFun}^{\circ\circ}(z_i , z)
	+ \intkerSS{\ccfFun_{k_i} , \gencfFun}^{\circ\bullet}(z_i , z) \big)
	    \; \ud z_i 
\\ \nonumber & \; 
  - \int_{B^\rgtsym} 
	\big( \intkerSS{\ccfFun_{k_i} , \gencfFun}^{\circ\circ}(z_i , z)
	+ \intkerSS{\ccfFun_{k_i} , \gencfFun}^{\circ\bullet}(z_i , z) \big)
	    \; \ud z_i \bigg) .
\end{align}
Similarly
the contribution to~\eqref{eq: difference of integrals at different levels} 
from the last two terms 
of~\eqref{eq: decomposition to holom and antiholom terms} is
\begin{align}\label{eq: contribution of antiholom integrals terms}
K \times \bigg( 
  & \; \phantom{-} \int_{\chorcontour{y_i}} 
  \big( \intkerSS{\ccfFun_{k_i} , \gencfFun}^{\bullet\circ}(z_i , z)
	+ \intkerSS{\ccfFun_{k_i} , \gencfFun}^{\bullet\bullet}(z_i , z) \big)
	    \; \ud \overline{z}_i 
\\ \nonumber & \; 
  - \int_{\chorcontour{\tilde{y}_i}} 
	\big( \intkerSS{\ccfFun_{k_i} , \gencfFun}^{\bullet\circ}(z_i , z)
	+ \intkerSS{\ccfFun_{k_i} , \gencfFun}^{\bullet\bullet}(z_i , z) \big)
	    \; \ud \overline{z}_i \bigg) ,
\end{align}
since for the integrals along horizontal segments
we can write~$\ud x_i = \ud\overline{z}_i$. By antiholomorphicity of the 
integrand here, we can rewrite this contribution in terms of the vertical 
parts of the boundary of the rectangle as
\begin{align}\label{eq: contribution of antiholom integrals terms vert}
K \times \bigg( 
  & \; + \int_{B^\lftsym} 
  \big( \intkerSS{\ccfFun_{k_i} , \gencfFun}^{\bullet\circ}(z_i , z)
	+ \intkerSS{\ccfFun_{k_i} , \gencfFun}^{\bullet\bullet}(z_i , z) \big)
	    \; \ud \overline{z}_i 
\\ \nonumber & \; 
  - \int_{B^\rgtsym} 
  \big( \intkerSS{\ccfFun_{k_i} , \gencfFun}^{\bullet\circ}(z_i , z)
	+ \intkerSS{\ccfFun_{k_i} , \gencfFun}^{\bullet\bullet}(z_i , z) \big)
	    \; \ud \overline{z}_i \bigg) .
\end{align}

By virtue of
Lemma~\ref{lem: continuum vertical slidability for two pt functions}
and the Riemann boundary values
for~$\ccfFun_{k_i}$, we 
get that the vertical 
integrals
in~\eqref{eq: contribution of holomorphic integrals terms vert}
and~\eqref{eq: contribution of antiholom integrals terms vert}
cancel each other. This proves that
\eqref{eq: difference of integrals at different levels} is zero, and 
consequently that the integrated multi-point 
kernel~$\fusionLimitTuple{\tple{k}}{\tple{k}'}{\tple{k}''}$ does not depend 
on~$y_i$.

The proof
that~$\fusionLimitTuple{\tple{k}}{\tple{k}'}{\tple{k}''}$ does not depend 
on~$y'_i$ and~$y''_i$ is similar.
\end{proof}

The following anticommutation and annihilation properties of the integrated 
multi-point kernels will be a key tool in deriving the 
main recursion.

\begin{lem}\label{lem: anticommutations in slit-strip integrals}
The integrated multi-point 
kernels~$(\fusionLimitTuple{\tple{k}}{\tple{k}'}{\tple{k}''})$ satisfy the 
following. 
\begin{itemize}
\item[(a)] If $k_1 < 0$ or $k'_1>0$ or $k''_1>0$, then we have
$\fusionLimitTuple{\,\tple{k}}{\tple{k}'}{\tple{k}''} = 0$.
\item[(b)] Suppose that~$\widetilde{\tple{k}}$ is obtained from~$\tple{k}$ by 
interchanging the indices at positions~$i$ and~$i+1$, and
$\widehat{\tple{k}}$ is obtained from~$\tple{k}$ by 
removing the indices at positions~$i$ and~$i+1$, i.e.,
\begin{align*}
\tple{k} = \; & 
  (k_1 , \ldots , k_{i-1} , k_i , k_{i+1} , k_{i+2} , \ldots, k_m) \\
\widetilde{\tple{k}} = \; & 
  (k_1 , \ldots , k_{i-1} , k_{i+1} , k_i , k_{i+2} , \ldots, k_m) \\
\widehat{\tple{k}} = \; & \phantom{k_{i+1}}
  (k_1 , \ldots , k_{i-1} , k_{i+2} , \ldots, k_m) .
\end{align*}
Then we have
\begin{align*}
\fusionLimitTuple{\,\tple{k}}{\tple{k}'}{\tple{k}''}
+ \fusionLimitTuple{\,\widetilde{\tple{k}}}{\tple{k}'}{\tple{k}''}
= \; & 	\begin{cases}
	\fusionLimitTuple{\,\widehat{\tple{k}}}{\tple{k}'}{\tple{k}''}
			& \text{ if $k_i + k_{i+1} = 0$} \\
	0 & \text{ if $k_i + k_{i+1} \neq 0$} .
		\end{cases}
\end{align*}
\item[(c)] Similarly, if $\widetilde{\tple{k}'}$ is obtained from~$\tple{k}'$ 
by interchanging the indices at positions~$i$ and~$i+1$, and
$\widehat{\tple{k}'}$  
by removing the indices at positions~$i$ and~$i+1$, then we have
\begin{align*}
\fusionLimitTuple{\,\tple{k}}{\tple{k}'}{\tple{k}''}
+ \fusionLimitTuple{\,\tple{k}}{\widetilde{\tple{k}'}}{\tple{k}''}
= \; & 	\begin{cases}
	\fusionLimitTuple{\,\tple{k}}{\widehat{\tple{k}'}}{\tple{k}''}
			& \text{ if $k'_i + k'_{i+1} = 0$} \\
	0 & \text{ if $k'_i + k'_{i+1} \neq 0$} .
		\end{cases}
\end{align*}
\item[(d)] Similarly, if $\widetilde{\tple{k}''}$ is obtained from~$\tple{k}''$ 
by interchanging the indices at positions~$i$ and~$i+1$, and
$\widehat{\tple{k}''}$  
by removing the indices at positions~$i$ and~$i+1$, then we have
\begin{align*}
\fusionLimitTuple{\,\tple{k}}{\tple{k}'}{\tple{k}''}
+ \fusionLimitTuple{\,\tple{k}}{\tple{k}'}{\widetilde{\tple{k}''}}
= \; & 	\begin{cases}
	\fusionLimitTuple{\,\tple{k}}{\tple{k}''}{\widehat{\tple{k}''}}
			& \text{ if $k''_i + k''_{i+1} = 0$} \\
	0 & \text{ if $k''_i + k''_{i+1} \neq 0$} .
		\end{cases}
\end{align*}
\end{itemize}
\end{lem}
\begin{proof}
Consider, e.g., the case $k_1 < 0$
in the ``annihilation property''~(a).
By the independence on the level choices, we can take~$y_1 \to +\infty$~--- 
ordering~\eqref{eq: ordering of y coordinates} is preserved, as $y_1$ was 
anyway the highest level.
Observe that because of the decaying vertical translation eigenfunction 
${\ccfFun_{k_1}(x_1 + \ii y_1) = e^{\pi k_1 y_1} \, \ccffun_{k_1}(x_1)}$,
the two point kernels
$\intkerSS{\ccfFun_{k_1} , \gencfFun}^{\circ\circ}(x_1 + \ii y_1 , z)$
etc. tend to zero exponentially as~$y_1 \to +\infty$
(from the asymptotics given in
\cite[Sec.~2]{part-1}, one sees
that the factor~$\mapSS(x_1 + \ii y_1) - \mapSS(z)$ in the denominator
grows faster than the factor~$\sqrt{\mapSS'(x_1 + \ii y_1)}$ in the 
numerator). Separate the 
dependence on~$x_1+\ii y_1$ of the integrand
in~$\fusionLimitTuple{\,\tple{k}}{\tple{k}'}{\tple{k}''}$ by writing it as 
a sum of terms of the 
form~\eqref{eq: decomposition to holom and antiholom terms} (with~$i=1$).
This expression shows that the integrand is tending to zero 
exponentially. Thus also the whole
integral~$\fusionLimitTuple{\,\tple{k}}{\tple{k}'}{\tple{k}''}$ tends to 
zero as~$y_1 \to +\infty$, and by its independence on~$y_1$ (Lemma~\ref{lem: 
independence on level choices}),
it in fact must be zero for any~$y_1$. This concludes one of the three cases 
in~(a). The cases $k'_1>0$ and $k''_1>0$ are similarly handled by taking $y'_1 
\to -\infty$ and $y''_1 \to -\infty$, respectively.

The proofs of the ``anticommutation properties''~(b), (c), and (d) are similar, 
so we only prove~(b).
Let us start by considering the integral of the kernel
\begin{align*} 
\intkerSS{\ccfFun_{\tple{k}};
			\ccfFun^\lftsym_{\tple{k}'}, \ccfFun^\rgtsym_{\tple{k}''}}
		  (\tple{x} + \ii \,\tple{y} ;
			\tple{x}' + \ii \,\tple{y}' , \tple{x}'' + \ii \, \tple{y}'')
= \; & \intkerSS{\cdots,\ccfFun_{k_i},\ccfFun_{k_{i+1}},\cdots}
		  (\ldots, x_{i} + \ii y_i ,x_{i+1} + \ii y_{i+1} , \ldots) 
\end{align*}
over the $x_i$ variable, 
and how it changes when the level~$y_i$ is changed 
to another level~$\tilde{y}_i$ which 
satisfies~$\tilde{y}_i < y_{i+1}$ and~$\tilde{y}_i > y_{i+2}$~(if $i+2 \leq m$).
Such a change does not preserve the
ordering of the levels, so the argument of
Lemma~\ref{lem: independence on level choices} has to be modified.
Consider the difference
\begin{align}\label{eq: difference of integrals at very different levels}
 & \int_{-\frac{1}{2}}^\frac{1}{2} \ud x_i
	\intkerSS{\cdots,\ccfFun_{k_i},\ccfFun_{k_{i+1}},\cdots}
		  (\ldots, x_{i} + \ii y_i , x_{i+1} + \ii y_{i+1} , \ldots) 
\\ \nonumber
& - \int_{-\frac{1}{2}}^\frac{1}{2} \ud x_i
	\intkerSS{\cdots,\ccfFun_{k_i},\ccfFun_{k_{i+1}},\cdots}
		  (\ldots, x_{i} + \ii \tilde{y}_i , x_{i+1} + \ii y_{i+1} , \ldots) .
\end{align}
Just as in Lemma~\ref{lem: independence on level choices},
a recursion of the type~\eqref{eq: Pfaffian kernel recursion} can be used to 
isolate the dependence on~$z_i$ by
expressing the integrand
as a sum of terms of the
form~\eqref{eq: decomposition to holom and antiholom terms}. 
Except only for the term which has~$z=z_{i+1} := x_{i+1} + \ii y_{i+1}$,
all these terms can be treated exactly as in
Lemma~\ref{lem: independence on level choices},
and their total contribution to the 
difference~\eqref{eq: difference of integrals at very different levels}
is zero. Thus we are left to consider the term
\begin{align}
\Bigg( \intkerSS{\ccfFun_{k_i} , \ccfFun_{k_{i+1}}}^{\circ\circ}(z_i , z_{i+1})
  + \intkerSS{\ccfFun_{k_i} , \ccfFun_{k_{i+1}}}^{\circ\bullet}(z_i , z_{i+1}) 
  + \intkerSS{\ccfFun_{k_i} , \ccfFun_{k_{i+1}}}^{\bullet\circ}(z_i , z_{i+1})
  + \intkerSS{\ccfFun_{k_i} , \ccfFun_{k_{i+1}}}^{\bullet\bullet}(z_i , z_{i+1})
	\Bigg) \times K
\end{align}
in the integrand,
where $K$ is a Pfaffian of type~\eqref{eq: kernel} with
the indices~$i$ and~$i+1$ omitted.
The only difference to the argument of
Lemma~\ref{lem: independence on level choices} here is that the first term has 
a pole at the point~$z_{i+1}$ inside the rectangle, and the last term has a 
similar antiholomorphic pole. Compared
to~\eqref{eq: contribution of holomorphic integrals terms vert},
the residue of the first term at the pole~$z_i = z_{i+1}$ introduces an extra 
term
\begin{align*}
- K \times \oint_{z_{i+1}} \intkerSS{\ccfFun_{k_i} , 
		\ccfFun_{k_{i+1}}}^{\circ\circ}(z_i , z_{i+1}) \; \ud z_i
= - 2 \pi \ii \, K \; \ccfFun_{k_i}(z_{i+1}) \, \ccfFun_{k_{i+1}}(z_{i+1}) ,
\end{align*}
where we used Lemma~\ref{lem: holomorphicity and poles for two pt functions} to 
explicitly calculate the residue. Similarly compared
to~\eqref{eq: contribution of antiholom integrals terms vert},
the antiholomorphic residue of the last term at the pole~$z_i = z_{i+1}$ 
introduces an extra term
\begin{align*}
- K \times \oint_{z_{i+1}} \intkerSS{\ccfFun_{k_i} , 
  \ccfFun_{k_{i+1}}}^{\bullet\bullet}(z_i , z_{i+1}) \; \ud \overline{z}_i
= + 2 \pi \ii \, K \; \overline{\ccfFun_{k_i}(z_{i+1})} \; 
		\overline{\ccfFun_{k_{i+1}}(z_{i+1})} .
\end{align*}
In the end, the 
difference~\eqref{eq: difference of integrals at very different levels} 
therefore simplifies to
\begin{align*}
\text{\eqref{eq: difference of integrals at very different levels}}
= \; - 2 \pi \ii \, K \; \Big(
    \ccfFun_{k_i}(z_{i+1}) \, \ccfFun_{k_{i+1}}(z_{i+1})
    - \overline{\ccfFun_{k_i}(z_{i+1})} \; 
		\overline{\ccfFun_{k_{i+1}}(z_{i+1})} \Big) .
\end{align*}
Denoting $z_{i+1} = x_{i+1} + \ii y_{i+1}$, and
using the vertical translation eigenfunction properties of 
both~$\ccfFun_{k_{i}}$ and~$\ccfFun_{k_{i+1}}$ as well as the property
$\ccfFun_{k_{i+1}} (z_{i+1}) 
= -\ii \, e^{2 \pi k_{i+1} \im (z_{i+1})} \,
    \overline{\ccfFun_{-k_{i+1}} (z_{i+1})} $
from~\eqref{eq: opposite index continuum functions},
we rewrite this as
\begin{align*}
\text{\eqref{eq: difference of integrals at very different levels}}
= \; & - 2 \pi \ii \, K \, e^{2 \pi k_{i+1} \, \im(z_{i+1})} \; \bigg(
    - \ii \, \ccfFun_{k_i}(z_{i+1}) \, 
				\overline{\ccfFun_{-k_{i+1}}(z_{i+1})}
    - \ii \, \overline{\ccfFun_{k_i}(z_{i+1})} \; 
				\ccfFun_{-k_{i+1}}(z_{i+1}) \bigg)  \\
= \; & - 4 \pi \, K \, e^{\pi (k_i + k_{i+1}) y_{i+1}} \; \re \Big(
    \ccffun_{k_i}(x_{i+1}) \, 
				\overline{\ccffun_{-k_{i+1}}(x_{i+1})} \Big) .
\end{align*}
The difference~\eqref{eq: difference of integrals at very different levels} is 
still to be integrated over the other variables, including in 
particular~$x_{i+1}$. In view of the above, the integral over~$x_{i+1}$ becomes 
simply
\begin{align*}
\int_{-1/2}^{+1/2} \big[
    \text{\eqref{eq: difference of integrals at very different levels}}
    \big]\; \ud x_{i+1} 
= \; & - 4 \pi \, K \, e^{\pi (k_i + k_{i+1}) y_{i+1}} \;
	\int_{-1/2}^{+1/2} \re \Big(
		  \ccffun_{k_i}(x_{i+1}) \, 
				\overline{\ccffun_{-k_{i+1}}(x_{i+1})} \Big) \; \ud x_{i+1} \\
= \; & - 4 \pi \, K \; \delta_{k_i, - k_{i+1}} ,
\end{align*}
using the orthonormality of the 
functions~$(\ccffun_{k})_{k \in \bZ + \frac{1}{2}}$~\cite[Proposition~2.1]{part-1}.
Now note that the factor~$K$ which depends on the other variables is simply the 
Pfaffian
\begin{align*}
K = \intkerSS{\ccfFun_{\,\widehat{\tple{k}}};
			\ccfFun^\lftsym_{\tple{k}'}, \ccfFun^\rgtsym_{\tple{k}''}}
		  (\widehat{\tple{x}} + \ii \, \widehat{\tple{y}} ;
			\tple{x}' + \ii \,\tple{y}' , \tple{x}'' + \ii \, \tple{y}'')
\end{align*}
with indices at positions $i$ and~$i+1$ omitted. Therefore performing the 
remaining integrations over the variables $x_j$, $j \neq i, i+1$, yields 
\begin{align*}
- 4 \pi \; \delta_{k_i+ k_{i+1},0} \;
   \big( 2 \sqrt{\pi} \, \ii \big)^{m+m'+m''-2} \;
    \fusionLimitTuple{\,\widehat{\tple{k}}}{\tple{k}'}{\tple{k}''} .
\end{align*}
On the other hand, we can instead integrate the 
difference~\eqref{eq: difference of integrals at very different levels} 
directly over all the remaining variables $x_j$, $j \neq i$. Calculating this 
way, we obtain the sum
\begin{align*}
\big( 2 \sqrt{\pi} \, \ii \big)^{m+m'+m''}
    \Big( \fusionLimitTuple{\,\tple{k}}{\tple{k}'}{\tple{k}''}
    + \fusionLimitTuple{\,\widetilde{\tple{k}}}{\tple{k}'}{\tple{k}''} \Big) ,
\end{align*}
since the relabeling of the integration variables needed to recover the desired
ordering of levels in the second term is an odd permutation (transposition 
of~$i$ and~$i+1$), which introduces a sign change of the completely 
antisymmetric integral kernel.
By comparing these two equal expressions we conclude the asserted 
anticommutation property.
\end{proof}

\subsection{Scaling limit of the fusion coefficients}

Using the properties of 
Lemma~\ref{lem: anticommutations in slit-strip integrals}, it is possible
to rewrite any integrated multi-point 
kernel~$\fusionLimitTuple{\tple{k}}{\tple{k}'}{\tple{k}''}$
in terms of only those, where the indices in the tuple~$\tple{k}$ are positive 
and in increasing order, and the indices in the tuples~$\tple{k}'$ 
and~$\tple{k}''$ are negative and in decreasing order. In this form they are 
more directly analogous to the fusion coefficients of the Ising model, and we 
therefore give the following definition.

Let $\parts, \partsR, \partsL \finsubset \poshalfint$. Write these as
\begin{align*}
\parts =  \; & \set{k_1 , \ldots , k_m} , &
\partsR = \; & \set{k'_1 , \ldots , k'_{m'}} , &
\partsL = \; & \set{k''_1 , \ldots , k''_{m''}} \quad \text{ with} \\
0 < \; & k_1 < \cdots < k_m , &
0 < \; & k'_1 < \cdots < k'_{m'} , &
0 < \; & k''_1 < \cdots < k''_{m''} .
\end{align*}
Introduce the corresponding tuples $\tple{k} = (k_1 , \ldots, k_m)$,
$-\tple{k}' = (-k'_1 , \ldots, -k'_{m'})$, and \linebreak[4]
${-\tple{k}'' = (-k''_1 , \ldots, -k''_{m''})}$. The 
corresponding \term{continuum fusion coefficient} is defined as
\begin{align}\label{eq: continuum fusion coefficient}
\fusionLimit{\parts}{\partsR}{\partsL}
:= \fusionLimitTuple{\tple{k}}{-\tple{k}'}{-\tple{k}''} .
\end{align}

These continuum fusion coefficients satisfy a recursion analogous to
Theorem~\ref{thm: recursion for fusion coefficients}. Here
$\pSign{\parts}{k}$ again denotes the signed indicator given
by~\eqref{eq: signed indicator}.

\begin{thm}\label{thm: recursion for continuum fusion coefficients}
The collection $( \fusionLimit{\parts}{\partsR}{\partsL})_{
\parts , \partsR , \partsL \finsubset \poshalfint}$
of all continuum fusion coefficients satisfies the following properties, 
which furthermore uniquely characterize the collection:
\begin{itemize}
\item[\recLimitIC] We have $\fusionLimit{\emptyset}{\emptyset}{\emptyset} = 1$.
\item[\recLimitT] If $\parts, \partsR, \partsL \finsubset \poshalfint$ 
and $\parts' = \parts \cup \set{k}$ with $\max(\parts) < k$,
then we have
\begin{align*}
\fusionLimit{\parts'}{\partsL}{\partsR}
= \; & \phantom{+}\; \sum_{k' \in \partsL} 
	\innprod{\ccffunL_{k'}}{\cpoleT{k}}
	\; (-1)^{\pLen{\partsR}} \; \pSign{\partsL}{k'} 
	\; \fusionLimit{\parts}{\partsL\setminus\set{k'}}{\partsR} \\
& + \sum_{k' \in \partsR} \innprod{\ccffunR_{k'}}{\cpoleT{k}}
	\; \pSign{\partsR}{k'}
	\; \fusionLimit{\parts}{\partsL}{\partsR\setminus\set{k'}} \\
& - \sum_{k' \in \parts} \innprod{\ccffun_{-k'}}{\cpoleT{k}}
	\; \pSign{\parts}{k'}
	\; \fusionLimit{\parts\setminus\set{k'}}{\partsL}{\partsR} .
\end{align*}
\item[\recLimitL] If $\parts, \partsR, \partsL \finsubset \poshalfint$ 
and $\partsL' = \partsL \cup \set{k}$ with $\max(\partsL) < k$,
then we have
\begin{align*}
\fusionLimit{\parts}{\partsL'}{\partsR}
= \; & - \sum_{k' \in \partsL} \innprod{\ccffunL_{k'}}{\cpoleL{k}}
	\; (-1)^{\pLen{\partsR}} \; \pSign{\partsL}{k'} 
	\; \fusionLimit{\parts}{\partsL\setminus\set{k'}}{\partsR} \\
& - \sum_{k' \in \partsR} \innprod{\ccffunR_{k'}}{\cpoleL{k}}
	\; \pSign{\partsR}{k'}
	\; \fusionLimit{\parts}{\partsL}{\partsR\setminus\set{k'}} \\
& + \sum_{k' \in \parts} \innprod{\ccffun_{-k'}}{\cpoleL{k}}
	\; \pSign{\parts}{k'}
	\; \fusionLimit{\parts\setminus\set{k'}}{\partsL}{\partsR} .
\end{align*}
\item[\recLimitR] If $\parts, \partsR, \partsL \finsubset \poshalfint$ 
and $\partsR' = \partsR \cup \set{k}$ with $\max(\partsR) < k$,
then we have
\begin{align*}
\fusionLimit{\parts}{\partsL}{\partsR'}
= \; & - \sum_{k' \in \partsL} 
    \innprod{\ccffunL_{k'}}{\cpoleR{k}}
	\; (-1)^{\pLen{\partsR}} \; \pSign{\partsL}{k'} 
	\; \fusionLimit{\parts}{\partsL\setminus\set{k'}}{\partsR} \\
& - \sum_{k' \in \partsR} \innprod{\ccffunR_{k'}}{\cpoleR{k}}
	\; \pSign{\partsR}{k'}
	\; \fusionLimit{\parts}{\partsL}{\partsR\setminus\set{k'}} \\
& + \sum_{k' \in \parts} \innprod{\ccffun_{-k'}}{\cpoleR{k}}
	\; \pSign{\parts}{k'}
	\; \fusionLimit{\parts\setminus\set{k'}}{\partsL}{\partsR} .
\end{align*}
\end{itemize}
\end{thm}
The idea of the proof is exactly parallel to that of
Theorem~\ref{thm: recursion for fusion coefficients}, so we content 
ourselves to sketching the strategy.
The uniqueness of the solution to the recursion goes through verbatim, and the
initial condition~\recLimitIC{} is direct by definition/convention, so the main 
task is to prove the three recursive properties~\recLimitT{}, \recLimitR{}, and 
\recLimitL{}.
Each of these is proved by replacing
a singular Fourier mode of index~$\pm k$ in the appropriate extremity by the 
corresponding continuous pole function, up to regular Fourier modes. The 
integrations of the regular Fourier modes can be anticommuted and eventually 
annihilated with the help of
Lemma~\ref{lem: anticommutations in slit-strip integrals}.
These anticommutations result in 
one of the sums on the right hand side.
Then the integration of the pole function can first of all be considered at
height zero by virtue of Lemma~\ref{lem: independence on level choices}.
It can be split to two contributions, over the crosscuts of the two other 
extremities. The pole function can moreover be expanded in regular Fourier 
modes in each of these other extremities. The integrations of these 
regular Fourier modes can then again be anticommuted and eventually 
annihilated, 
and the anticommutations from each extremity produces one of the two 
remaining terms on the 
right hand side. To finish the calculation one needs to notice that the 
expansion coefficients of the pole functions are suitable inner products with 
regular Fourier modes, and that there are signs resulting from the
anticommutations, the integration directions, and sign changes 
needed in the kernel when permuting the integration variables to the 
desired order.

We can now derive the following scaling limit result for the Ising model
fusion coefficients.

\begin{thm}
\label{thm: scaling limit of fusion coefficients}
Choose sequences $(\lft_{n})_{n \in \bN}$,
$(\rgt_{n})_{n \in \bN}$ of integers
$\lft_{n}, \rgt_{n} \in \bZ$ such that
\begin{itemize}
\item $\lft_{n} < 0 < \rgt_{n}$ for all $n$;
\item $\width_{n} := \rgt_{n} - \lft_{n} \to +\infty$ as $n \to \infty$;
\item $\lft_{n} / \width_{n} \to -\half$
and $\rgt_{n} / \width_{n} \to +\half$ as $n \to \infty$.
\end{itemize}
Denote by $(\fusionIsingW{\width_{n}}{\parts}{\partsR}{\partsL})$
the 
Ising model fusion 
coefficients~\eqref{eq: fusion coefficient} on the discrete 
slit-strip with left and right
boundaries at horizontal positions~$\lft_{n}$ and~$\rgt_{n}$,
respectively, and by
$(\fusionLimit{\parts}{\partsR}{\partsL})$ the continuum fusion
coefficients~\eqref{eq: continuum fusion coefficient}. Then for 
any given $\parts, \partsR, \partsL \finsubset \poshalfint$,
the fusion coefficient~$\fusionIsingW{\width_{n}}{\parts}{\partsR}{\partsL}$
is defined for all sufficiently large~$n$, and we have
\begin{align*}
\lim_{n \to \infty} 
    \frac{\fusionIsingW{\width_{n}}{\parts}{\partsR}{\partsL}}
        {\fusionIsingW{\width_{n}}{\emptyset}{\emptyset}{\emptyset}}
= \; & \fusionLimit{\parts}{\partsR}{\partsL} .
\end{align*}
\end{thm}
\begin{proof}
Fix $\parts, \partsR, \partsL \finsubset \poshalfint$.
The assumptions imply $-\lft_{n} , \rgt_n , \width_n \to +\infty$
as $n \to \infty$. The fusion coefficient
$\fusionIsingW{\width_{n}}{\parts}{\partsR}{\partsL}$
is defined as soon as $-\lft_{n} > \max \partsL$,
$\rgt_{n} > \max \partsR$, and
$\width_{n} > \max \parts$; in particular for all sufficiently large~$n$.

By Theorem~\ref{thm: recursion for fusion coefficients},
a finite recursion allows to express
$\fusionIsingW{\width_{n}}{\parts}{\partsR}{\partsL}$ in terms of
the overall multiplicative factor
$\fusionIsingW{\width_{n}}{\emptyset}{\emptyset}{\emptyset}$
(coming from the initial condition) times a polynomial in
the inner products
$\innprod{\eigf{-k'}}{\poleT{k}}$,
$\innprod{\eigfR{-k'}}{\poleT{k}}$,
$\innprod{\eigfL{-k'}}{\poleT{k}}$,
$\innprod{\eigf{-k'}}{\poleR{k}}$,
$\innprod{\eigfR{-k'}}{\poleR{k}}$,
$\innprod{\eigfL{-k'}}{\poleR{k}}$,
$\innprod{\eigf{-k'}}{\poleL{k}}$,
$\innprod{\eigfR{-k'}}{\poleL{k}}$,
$\innprod{\eigfL{-k'}}{\poleL{k}}$,
in the discrete function 
space ~$\bC^{\dintervaldual{\lft_{n}}{\rgt_{n}}}$
of type~\eqref{eq: discrete function space}.

Similarly, by
Theorem~\ref{thm: recursion for continuum fusion coefficients},
a finite recursion allows to express
$\fusionLimit{\parts}{\partsR}{\partsL}$ as a polynomial in
the inner products of
$\innprod{\ccffun_{k'}}{\cpoleT{k}}$,
$\innprod{\ccffunR_{k'}}{\cpoleT{k}}$,
$\innprod{\ccffunL_{k'}}{\cpoleT{k}}$,
$\innprod{\ccffun_{k'}}{\cpoleR{k}}$,
$\innprod{\ccffunR_{k'}}{\cpoleR{k}}$,
$\innprod{\ccffunL_{k'}}{\cpoleR{k}}$,
$\innprod{\ccffun_{k'}}{\cpoleL{k}}$,
$\innprod{\ccffunR_{k'}}{\cpoleL{k}}$,
$\innprod{\ccffunL_{k'}}{\cpoleL{k}}$,
in the function space~$\cfunctionsp$ of~\eqref{eq: continuum function space}
(the multiplicative factor from the initial condition is 
just~$\fusionLimit{\emptyset}{\emptyset}{\emptyset} = 1$ here).

The two recursions have exactly the same structure, so the 
polynomials in both cases are the same. The asserted convergence 
therefore follows from Corollary~\ref{cor:ip-convergence},
which gives the convergence of the former inner products to the latter 
ones.
\end{proof}

\bigskip

\section{Conclusions and outlook}

In this article we developed tools based on Clifford algebra valued 
discrete $1$-forms for the analysis of the critical
Ising model with locally monochromatic boundary conditions.
These tools are transparent counterparts of objects in boundary 
conformal field theory. We showed how to use them for calculations in
the Ising model; in particular in order to obtain a characterization of
the fusion coefficients that arise as renormalized boundary correlation
functions for the Ising model in the lattice slit strip. Discrete
complex analysis techniques and judiciously chosen s-holomorphic
solutions to the Riemann boundary value problem in the lattice strip and
slit strip \cite{part-1} played a key role,
and enabled in particular the proof of our main result, 
Theorem~\ref{thm: scaling limit of fusion coefficients},
stating the convergence of the fusion coefficients in the scaling limit.

In the last part~\cite{part-3} of this series, we will relate the continuum fusion 
coefficients~$\fusionLimit{\parts}{\partsR}{\partsL}$ to the structure constants
\begin{align*}
\ldual \state{\parts}' , Y(\state{\partsR},\voavar) \, \state{\partsL} \rdual \Big|_{\voavar = 1} 
\end{align*}
of the appropriate fermionic vertex operator algebra~$(V,Y,\cstone,\omega)$
(this still involves a transformation
from the slit strip geometry to the half-plane geometry, since the
VOA axioms in their standard formulation are only directly appropriate
for the latter).
With that connection to the correct VOA,
Theorem~\ref{thm: scaling limit of fusion coefficients} rigorously
relates the scaling limit of boundary correlation functions of the Ising model
with locally monochromatic boundary conditions to the algebraic
axiomatization of the boundary conformal field theory.


\begin{thebibliography}{AbMa73}

\bibitem[AbMa73]{AM-transfer_matrix_for_a_pure_phase}
D.~B.~Abraham and A.~Martin-L\"of. 
The Transfer Matrix for a Pure Phase in the Two-dimensional Ising Model.
\emph{Comm. Math. Phys.}, 1973, 32, 245-268

\bibitem[AKPR20]{part-1}
T.~Ameen, K.~Kyt\"ol\"a, S.~C.~Park, and D.~Radnell.
Slit-strip Ising boundary conformal field theory 1:
Discrete and continuous function spaces.
Preprint \url{https://arxiv.org/abs/2009.13624}, 2020.

\bibitem[Bax82]{Baxter-exactly_solved_models}
%{baxter}
R.~J.~Baxter.
\emph{Exactly solved models in statistical mechanics.}
Academic Press, 1982.

\bibitem[BPZ84a]{BPZ-infinite_conformal_symmetry_in_QFT}
A.~A.~Belavin, A.~M.~Polyakov, and A.~B.~Zamolodchikov.
Infinite conformal symmetry
in two-dimensional quantum field theory.
\emph{Nucl. Phys. B}, 241(2):333\textendash 380, 1984.

\bibitem[BPZ84b]{BPZ-infinite_conformal_symmetry_of_critical_fluctuations}
A.~A.~Belavin, A.~M.~Polyakov, and A.~B.~Zamolodchikov.
Infinite conformal symmetry
of critical fluctuations in two dimensions.
\emph{J. Stat. Phys.}, 34(5-6):763\textendash 774, 1984.

\bibitem[BeHo19]{BH-scaling_limit_of_Ising_interfaces}
S.~Benoist and C.~Hongler.
The scaling limit of critical Ising interfaces is 
CLE(3).
\emph{Ann. Probab.} 47:2049--2086, 2019.

\bibitem[BDH16]{BDH-crossing_probabilities_with_free_bc}
S.~Benoist H.~Duminil-Copin, and C.~Hongler.
Conformal invariance of crossing 
probabilities for the Ising model with free boundary conditions.
\emph{Ann. Inst. H. Poincar\'e Probab. Statist.} 52:1784--1798, 2016.

\bibitem[Bil99]{Billingsley-convergence_of_probability_measures}
%{Billingsley}
P.~Billingsley.
\emph{Convergence of Probability Measures}.
Wiley Series in Probability and Statistics, 1999.

\bibitem[Car84]{Cardy-surface_critical_behavior}
J. Cardy
Conformal invariance and Surface Critical Behavior.
\emph{Nucl. Phys. B} 240:514--532, 1984.

\bibitem[Car86]{Cardy-effect_of_boundary_conditions}
J. Cardy.
Effect of boundary conditions on the operator content of two-dimensional 
conformally invariant theories.
\emph{Nucl. Phys. B} 275(2):200--218, 1986.

\bibitem[Car89]{Cardy-Verlinde_formula}
J. Cardy.
Boundary conditions, fusion rules and the Verlinde formula.
\emph{Nucl. Phys. B} 324(3):581--596, 1989.

\bibitem[Car06]{Cardy-BCFT_encyclopedia}
J. Cardy.
Boundary Conformal Field Theory.
In \emph{Encyclopedia of Mathematical Physics},
Elsevier, 2006.

% \bibitem[CaLe91]{CL-bulk_and_boundary_operators}
% J.~L.~Cardy and D.~C.~Lewellen.
% Bulk and boundary operators in conformal field theory.
% \emph{Phys. Lett.},
% B259:274--278, 1991.

\bibitem[CGN15]{CGN-planar_Ising_magnetization_field}
F.~Camia, C.~Garban, and C.~M.~Newman. 
Planar Ising magnetization field 
I. Uniqueness of the critical scaling limit.
\emph{Ann. Probab.} 43:528--571, 2015.

\bibitem[CGN16]{CGN-planar_Ising_magnetization_field_2}
F.~Camia, C.~Garban, and C.~M.~Newman. 
Planar Ising magnetization field 
II. Properties of the critical and near-critical scaling limits.
Ann. Inst. Henri Poincaré, Probab. Stat., 2016, 52, 146-161

\bibitem[Che18]{Chelkak-state_of_the_art_and_perspectives}
D.~Chelkak.
Planar Ising model at criticality: state-of-the-art and perspectives.
\emph{Proceedings of the International Congress of 
Mathematicians 2018}~3:2789--2816.
World Scientific Publishing Company Inc., 2018.

\bibitem[CCK17]{CCK-revisiting}
D.~Chelkak, D.~Cimasoni, and A.~Kassel.
Revisiting the combinatorics of the 2D Ising model.
\emph{Ann. Inst. Henri Poincar\'e (D)} 4(3):309--385, 2017.

\bibitem[CDHKS14]{CDHKS-convergence_of_ising_interfaces}
D.~Chelkak, H.~Duminil-Copin, C.~Hongler, A.~Kemppainen, and S.~Smirnov.
Convergence of Ising interfaces to Schramm's SLE curves.
C.~R. Acad. Sci. Paris Ser. I 352:157--161, 2014.

\bibitem[CHI15]{CHI-conformal_invariance_of_spin_correlations}
D.~Chelkak, C.~Hongler, and K.~Izyurov.
Conformal invariance of spin correlations in the planar Ising model.
\emph{Annals of Math.}, 181(3):1087\textendash 1138,
2015.

\bibitem[CHI21]{CHI-primary_field_correlations}
D.~Chelkak, C.~Hongler, and K.~Izyurov.
Correlations of primary fields in the critical Ising model.
Preprint \url{https://arxiv.org/abs/2103.10263}, 2021.

\bibitem[ChIz13]{CI-holomorphic_spinor_observables}
D.~Chelkak and K.~Izyurov.
Holomorphic Spinor Observables in the 
Critical Ising Model.
\emph{Comm. Math. Phys.} 322:303--332, 2013.

\bibitem[ChSm11]{CS-discrete_complex_analysis_on_isoradial_graphs}
D.~Chelkak and S.~Smirnov.
Discrete complex analysis on isoradial graphs.
\emph{Advances in Mathematics},
228:1590\textendash 1630, 2011.

\bibitem[ChSm12]{CS-universality_in_Ising}
%{chelkak-smirnov}
D.~Chelkak, S.~Smirnov.
Universality in the 2D Ising model and conformal invariance of fermionic 
observables.
\emph{Invent math.}, 189(3):515\textendash 580, 2012.

\bibitem[DMS97]{DMS-CFT}
P.~Di~Francesco, P.~Mathieu, and D.~S\'en\'echal.
\emph{Conformal Field Theory}. 
Graduate texts in contemporary physics.
Springer-Verlag New York, 1997.

\bibitem[Dub11]{Dubedat-exact_bosonization}
J.~Dub\'edat.
Exact bosonization of the Ising model.
\url{https://arxiv.org/abs/1112.4399}, 2011.

\bibitem[FFS92]{FFS-triviality_in_QFT}
R.~Fern{\'a}ndez, J.~Fr{\"o}hlich, and A.~D.~Sokal.
\emph{Random walks, critical phenomena, and triviality in quantum field theory}.
Springer Verlag, 1992.

\bibitem[FLM88]{FLM-VOAs_and_Monster}
I.~B.~Frenkel, J.~Lepowsky, and A.~Meurman.
\emph{Vertex Operator Algebras and the Monster}, 
Academic Press, 1988.

\bibitem[GJRSV13]{GJRSV-inspiration}
A.~M.~Gainutdinov, J.~L.~Jacobsen, N.~Read, H.~Saleur, and R.~Vasseur.
Logarithmic Conformal Field Theory: a Lattice Approach.
\emph{J. Phys. A} 46(49), 2013.

\bibitem[GHP19]{GHP-Ising_local_spin_correlations}
R.~Gheissari, C.~Hongler, S.~C.~Park.
Ising Model: Local Spin Correlations and Conformal Invariance.
\emph{Comm. Math. Phys.} 367(3):771\textendash 833, 2019.

\bibitem[Geo11]{Georgii-Gibbs_measures}
H.-O.~Georgii.
\emph{Gibbs measures and phase transitions}.
De Gruyter, 2011.

\bibitem[Hon10]{Hongler-thesis}
C.~Hongler.
\emph{Conformal Invariance of
Ising Model Correlations}, Ph.D. thesis, Universit\'e de Gen\`eve, 
https://archive-ouverte.unige.ch/unige:18163,
2010.

\bibitem[HGS21]{Hongler_et_al-CFT_book}
C.~Hongler, F.~Gabriel, and F.~Spadaro.
\emph{Lattice models and conformal field theory.}
In preparation, 2021.

\bibitem[HoKy13]{HK-Ising_interfaces_and_free_bc}
C.~Hongler and K.~Kyt\"ol\"a.
Ising Interfaces and Free Boundary Conditions.
\emph{J. Amer. Math. Soc.} 26:1107--1189, 2013.

\bibitem[HKV13]{HKV-CFT_at_the_lattice_level}
C.~Hongler, K.~Kyt\"ol\"a, and F.~Viklund.
Conformal Field Theory at the Lattice Level:
Discrete Complex Analysis and Virasoro Structure.
\url{https://arxiv.org/abs/1307.4104}, 2019.

\bibitem[HKZ14]{HKZ-discrete_holomorphicity_and_operator_formalism}
C.~Hongler, K.~Kyt\"ol\"a, and A.~Zahabi.
Discrete Holomorphicity and Ising model Operator Formalism.
\emph{Analysis, Complex Geometry, and Mathematical Physics:
A Conference in Honor of Duong H. Phong, Columbia}, 
NYC, USA, 2014.

\bibitem[HoSm13]{HS-energy_density}
C.~Hongler and S.~Smirnov
The energy density in the planar Ising model.
\emph{Acta Math.} 211:191\textendash 225, 2013.

\bibitem[Hua12]{Huang-CFT_and_VOA}
Y.-Z.~Huang.
Two-dimensional conformal geometry and 
vertex operator algebras.
Springer, 2012.

\bibitem[Isi25]{Ising-beitrag}
E.~Ising.
Beitrag zur Theorie des Ferromagnetismus.
\emph{Zeitschrift f\"ur Physik}~31, 1925.

\bibitem[ItDr91]{ID-statistical_field_theory}
C.~Itzykson and J.-M.~Drouffe
Statistical field theory. Vol 1 \& 2.
Cambridge University Press, 1991.

\bibitem[Izy11]{Izyurov-thesis}
K.~Izyurov.
\emph{Holomorphic spinor observables and interfaces 
in the critical Ising model.}
PhD thesis, Universit\'e de Gen\`eve, 2011.

\bibitem[Izy15]{Izyurov-Smirnovs_observable_for_free_boundary_conditions}
K.~Izyurov.
Smirnov's observable for free boundary conditions, 
interfaces and crossing probabilities.
\emph{Comm. Math. Phys.} 337:225-252, 2015.

\bibitem[Izy17]{Izyurov-Ising_interfaces_in_multiply_connected_domains}
K.~Izyurov.
Critical Ising interfaces in multiply-connected domains.
\emph{Probab. Th. Rel. Fields} 167:379--415, 2017.

\bibitem[JiMi80]{JM-studies_on_holonomic_quantum_fields_17}
M.~Jimbo and T.~Miwa.
Studies on Holonomic Quantum Fields. XVII 
\emph{Proc. Japan Acad.} Ser A 56:405--410, 1980,
57:347 (1981).

\bibitem[Kac97]{Kac-vertex_algebras_for_beginners}
V.~Kac Vertex algebras for beginners American Mathematical Society, 1997, 10, 
viii+141

\bibitem[KaCe71]{KC-determination_of_an_operator_algebra}
L.~Kadanoff and H.~Ceva.
Determination of an operator algebra for the 
two-dimensional Ising model.
\emph{Phys. Rev.} B3:3918--3939, 1971.

\bibitem[Kau49]{Kaufman-crystal_statistics_2}%{kaufman}
B.~Kaufman.
Crystal statistics II. Partition function evaluated by spinor analysis.
\emph{Phys. Rev., II. Ser.} 76:1232--1243, 1949.

\bibitem[KeSm19]{KS-boundary_touching_loops}
A.~Kemppainen and S.~Smirnov.
Conformal invariance of boundary touching loops of 
FK Ising model.
\emph{Comm. Math. Phys.} 369:49--98, 2019.

\bibitem[KeSm16]{KS-conformal_invariance_of_RCM_2}
A.~Kemppainen and S.~Smirnov.
Conformal invariance in random cluster models. 
II. Full scaling limit as a branching SLE.
\url{https://arxiv.org/pdf/1609.08527}, 2016.

\bibitem[KeSm17b]{KS-configurations_of_FK_Ising_interfaces}
A.~Kemppainen and S.~Smirnov.
Configurations of FK Ising interfaces and 
hypergeometric SLE.
\url{https://arxiv.org/pdf/1704.02823}, 2017.

\bibitem[KrWa41]{KW-statistics_of_the_2D_ferromagnet}
H.~A.~Kramers, and G.~H.~Wannier.
Statistics of the two-dimensional ferromagnet .
\emph{Phys. Rev.} 60:252--262, 1941.

\bibitem[KPR21]{part-3}
K.~Kyt\"ol\"a, S.~C.~Park, and D.~Radnell.
Slit-strip Ising boundary conformal field theory 3:
The vertex operator algebra in the scaling limit.
In preparation, 2021.

\bibitem[Lan71]{Lanford}
O.~E.~Lanford.
Entropy and equilibrium states in classical statistical mechanics.
In \emph{Statistical mechanics and mathematical problems}, 
A.~Lenard~(ed.),
Battelle Seattle Rencontres, LNPh 20:1--113, 1971.

\bibitem[Len20]{Lenz-beitrag}
W.~Lenz.
Beitrag zum Verst\"andnis der magnetischen Erscheinungen in 
festen K\"orpern Physikalische.
\emph{Physikalische Zeitschrift}~21:613--615, 1920.

\bibitem[LeLi04]{LL-introduction_to_VOAs}
J.~Lepowsky and H.~Li. 
\emph{Introduction to Vertex Operator Algebras and their Representations.}
Birkh\"auser, Boston, 2004.

\bibitem[McC95]{McCoy-connection_between_statistical_mechanics_and_QFT}
B.~M.~McCoy.
\emph{The connection between statistical mechanics and quantum field theory}.
In  \emph{Statistical Mechanics and Field Theory}: 26--128
(Eds. V.~V.~Bazhanov and C.J.~Burden).
World Scientific (Singapore), 1995.

\bibitem[McWu73]{MW-two_dimensional_Ising_model}
B.~M.~McCoy and T.~T.~Wu. 
\emph{The two-dimensional Ising model.} 
Harvard Univ. Press, 1973.

\bibitem[Mer01]{Mercat-discrete_Riemann_surfaces}
C.~Mercat.
Discrete Riemann Surfaces and the Ising Model.
\emph{Comm. Math. Phys.} 218:177--216, 2001.

\bibitem[Mus09]{Mussardo-statistical_field_theory}
G.~Mussardo.
\emph{Statistical Field Theory:
An Introduction to Exactly Solved Models in Statistical Physics}.
Oxford University Press, 2009.

\bibitem[Ons44]{Onsager-crystal_statistics}
L.~Onsager. 
Crystal statistics. I. 
A two-dimensional model with order-disorder transition.
\emph{Phys. Rev.} 65:117--149, 1944.

\bibitem[OsSc73]{OS-axioms_for_Euclidean_Greens_functions_1}
K.~Osterwalder and R.~Schrader.
Axioms for Euclidean Green's Functions.
\emph{Comm. Math. Phys.} 31:83--112, 1973.

\bibitem[OsSc75]{OS-axioms_for_Euclidean_Greens_functions_2}
K.~Osterwalder and R.~Schrader.
Axioms for Euclidean Green's Functions.
\emph{Comm. Math. Phys.} 42:281--305, 1975.

\bibitem[Pal07]{Palmer-planar_Ising_correlations}%{palmer}
J.~Palmer.
\emph{Planar Ising correlations}.
Birkh\"auser, 2007.

\bibitem[Pei36]{Peierls-on_Ising_model_of_ferromagnetism}
R.~Peierls.
On Ising's model of ferromagnetism.
\emph{Proc. Camb. Philos. Soc.}~32, 1936.

\bibitem[PeWu18]{PW-crossing_probabilities_of_multiple_Ising_interfaces}
E.~Peltola and H.~Wu.
Crossing probabilities of multiple Ising interfaces.
\url{https://arxiv.org/pdf/1808.09438}, 2018.

\bibitem[Per80]{Perk-quadratic_identities_for_Ising_model_correlations}
J.~H.~H.~Perk. 
Quadratic identities for Ising model correlations.
\emph{Phys. Lett.} 79A:3--5, 1980.

\bibitem[SML64]{SML-Ising_model_as_a_problem_of_fermions}%{SML}
T.~D.~Schultz, D.~C.~Mattis, and E.~H.~Lieb.
Two-dimensional Ising model as a soluble problem of many fermions.
\emph{Rev. Mod. Phys.} 36:856--871, 1964.

\bibitem[Seg88]{Segal-definition_of_CFT}
G.~Segal.
The definition of conformal field theory.
\emph{Differential geometrical methods in theoretical physics}
250:165\textendash 171, 1988.

\bibitem[Seg04]{Segal-definition_of_CFT-new}
G.~Segal.
The definition of conformal field theory.
\emph{Topology, geometry and quantum field theory},
Cambridge University Press, 308:421\textendash 577, 2004.

\bibitem[Smi06]{Smirnov-towards_conformal_invariance}
S. Smirnov.
Towards conformal invariance of 2D lattice models.
\emph{%Sanz-Sol\'e, Marta (ed.) et al.,
Proceedings of the international congress of mathematicians (ICM),
Madrid, Spain, August 22\textendash 30, 2006}.
Vol. II: Invited lectures,
1421--1451. Zürich: European Mathematical Society (EMS), 2006.

\bibitem[Smi10a]{Smirnov-conformal_invariance_in_RCM_1}
S.~Smirnov.
Conformal invariance in random cluster models. I. 
Holomorphic fermions in the Ising model.
\emph{Ann. Math.} 172(2):1435-1467, 2010.

\bibitem[Smi10b]{Smirnov-discrete_complex_analysis_and_probability}
S. Smirnov.
Discrete complex analysis and probability.
\emph{Proceedings of the international congress of mathematicians (ICM), 
Hyderabad, India}, 2010.

\bibitem[Wil71]{Wilson-renormalization_group_and_critical_phenomena_1_and_2}
K.~G.~Wilson.
Renormalization group and critical phenomena, I and II.
\emph{Phys. Rev. B} 4:3174 and 4:3184, 1971.

\bibitem[WMTB76]{WMTB-spin_spin}
T.~T.~Wu, B.~M.~McCoy, C.~A.~Tracy, E.~Barouch. 
Spin-spin correlation functions for the 
two-dimensional Ising model: exact theory in the
scaling region.
\emph{Phys. Rev. B} 13:316--374, 1976.

\bibitem[Yan52]{Yang-spontaneous_magnetization}
C.~N.~Yang. 
The spontaneous magnetization of a two-dimensional Ising model.
\emph{Phys. Rev.} 85:808--816, 1952.

\end{thebibliography}
\end{document}